\documentclass[aps,prd,12pt,superscriptaddress,tightenlines]{revtex4-2}
\pdfoutput=1
\usepackage{hyperref}
\usepackage{amsmath}
\usepackage{graphicx}
\usepackage[export]{adjustbox} 
\usepackage{grffile}
\usepackage{xcolor}
\usepackage{multirow}
\usepackage{dcolumn}
\usepackage{bm}
\newcolumntype{d}[1]{D{.}{.}{#1}}

\begin{document}
\preprint{KEK-CP-382}
\title{Form factors of $B\to\pi\ell\nu$ and a determination of $|V_{ub}|$ with M\"{o}bius domain-wall fermions}
\author{Brian Colquhoun}
\altaffiliation[Current address: ]{SUPA, School of Physics and Astronomy, University of Glasgow, Glasgow, G12 8QQ, UK}
\email{brian.colquhoun@glasgow.ac.uk}
\affiliation{Department of Physics and Astronomy, York University, Toronto, Ontario, M1J 1P3, Canada}
\affiliation{High Energy Accelerator Research Organization (KEK), Tsukuba 305-0801, Japan}
\author{Shoji Hashimoto}
\email{shoji.hashimoto@kek.jp}
\affiliation{High Energy Accelerator Research Organization (KEK), Tsukuba 305-0801, Japan}
\affiliation{School of High Energy Accelerator Science, SOKENDAI (The Graduate University for Advanced Studies), Tsukuba 305-0801, Japan}
\author{Takashi Kaneko}
\email{takashi.kaneko@kek.jp}
\affiliation{High Energy Accelerator Research Organization (KEK), Tsukuba 305-0801, Japan}
\affiliation{School of High Energy Accelerator Science, SOKENDAI (The Graduate University for Advanced Studies), Tsukuba 305-0801, Japan}
\author{Jonna Koponen}
\email{jkoponen@uni-mainz.de}
\affiliation{PRISMA+ Cluster of Excellence \& Institut f\"{u}r Kernphysik, Johannes Gutenberg-Universit\"{a}t Mainz, D-55128 Mainz, Germany}
\affiliation{High Energy Accelerator Research Organization (KEK), Tsukuba 305-0801, Japan}
\collaboration{JLQCD Collaboration}
\noaffiliation

\begin{abstract}
  Using a fully relativistic lattice fermion action, we compute the form factors of the semileptonic decay $B\to\pi\ell\nu$, which is required for the determination of the Cabibbo-Kobayashi-Maskawa matrix element $|V_{ub}|$. We employ the M\"{o}bius domain-wall fermion formalism for the generation of lattice ensembles with 2+1 sea quark flavours as well as for the valence heavy and light quarks. We compute the form factors at various values of the lattice spacing and multiple light and heavy quark masses, and extrapolate the results to the physical point. We combine our lattice results with the available experimental data to obtain $|V_{ub}| = (3.93\pm 0.41)\times 10^{-3}$.
\end{abstract}

\maketitle
\vfill
\pagebreak

\section{Introduction}
The determination of the Cabibbo-Kobayashi-Maskawa (CKM) matrix element $|V_{ub}|$ from the measurement of the exclusive semileptonic decay $B\to\pi\ell\nu$ requires precise knowledge of the corresponding decay form factors, which can be obtained using lattice simulations of Quantum Chromodynamics (QCD). For the first time in lattice QCD we calculate these quantities using a fully relativistic approach.  $|V_{ub}|$ is an important Standard Model parameter, and the ratio $|V_{ub}|/|V_{cb}|$ is a particularly sought-after result that requires continued refining of both these elements of the CKM matrix. 

Heavy quarks require special consideration in lattice QCD since, on coarse lattices, discretization errors from large masses in lattice units, $am_Q$, become uncontrollable. Therefore, $B\to\pi\ell\nu$ calculations typically use effective actions for $b$ quarks, such as Nonrelativistic QCD (NRQCD)~\cite{Colquhoun:2015mfa,Hughes:2017spc,Dalgic:2006dt}, the Columbia interpretation of Relativistic Heavy Quarks (RHQ)~\cite{Flynn:2015mha} and the Fermilab interpretation of the Sheikholeslami-Wohlert clover action~\cite{FermilabLattice:2015mwy}.
Alternatively, it is possible to use multiple values of the heavy quark mass $am_Q<am_b$ in a fully relativistic action and extrapolate to the physical mass.
This requires that sufficiently fine lattices are available to keep $am_Q$ small enough that discretization effects can be controlled when combining the data at various lattice spacings.
We take the latter approach in this work using the M\"{o}bius domain-wall fermion action \cite{Ginsparg:1981bj, Kaplan:1992bt, Shamir:1993zy, Furman:1994ky, Brower:2012vk}.
We use the same action for the heavy and light quarks, and for both valence and sea light quarks.

With the M\"{o}bius domain-wall fermion formalism, the leading discretization effects are of $\mathcal{O}(a^2)$.
In our analysis we extrapolate the results at finite lattice spacing to the continuum limit assuming that there are effects of $\mathcal{O}(a^2)$ as well as a term proportional to $(am_Q)^2$, which is specific to the heavy quark.
The maximum value of $am_Q$ used in this work is $0.688$ so that discretization effects are kept under control.
The continuum extrapolation is combined with the extrapolation to the physical heavy and light quark masses in a global fit function.
The associated systematic errors are estimated by introducing higher order dependences on the lattice spacing and quark masses.

The momentum transfer range in $B\to\pi\ell\nu$ decays is large, owing to the large energy release from the $b$ quark.
The most precise experimentally available data points are in the small momentum transfer $q^2\ll m_b^2$ corresponding to the kinematics where the pion recoil momentum is large.
The small recoil data near maximum momentum transfer $q^2\approx 26.46~\mathrm{GeV^2}$ is less copious and the relative statistical error is larger.
On the lattice QCD side, the most accurate form factor results are obtained at large momentum transfer when the recoil momentum is small.
In order to make most use of the available information from both experiment and lattice calculations, one can combine the data to constrain the $q^2$ dependence using the so-called $z$-parameter expansion~\cite{Okubo:1971my, Okubo:1971jf, Boyd:1995sq, Boyd:1997kz, Caprini:1997mu, Bourrely:1980gp, Bourrely:2005hp, Bourrely:2008za}, as first applied to the $B\to\pi\ell\nu$ process in Ref.~\cite{Bailey:2008wp}.
This approach only makes assumptions about the analytic structure of the form factors and, because it only involves an expansion about a small parameter $z$, the results are robust.
We follow this strategy in this work and estimate the associated errors.

In calculations of both $|V_{cb}|$ and $|V_{ub}|$ there exist persistent tensions between their exclusive and inclusive determinations~\cite{HFLAV:2019otj,Aoki:2021kgd,*FlavourLatticeAveragingGroup:2019iem}.
The cause(s) of this tension is still unclear, although new theoretical and experimental analyses for $|V_{cb}|$ are revealing potential problems in previous analyses, such as the assumed functional form of the form factors.
A recent review of the $|V_{cb}|$ puzzle can be found in Ref.~\cite{Gambino:2019sif}, while general overviews of the CKM matrix elements from a lattice perspective can be found in Refs.~\cite{Wingate:2021ycr,Gottlieb:2020zsa}.

A more elaborate analysis of $|V_{ub}|$ is premature due to the small branching fraction, but care is needed to ensure that the choice of the parametrization of the form factors allows systematic improvement when more data becomes available.
On the exclusive side, the model-independent lattice calculation is a key element in the combined analysis with experimental data.
In this work we provide a fully nonperturbative computation of the $B\to\pi\ell\nu$ form factors with controlled extrapolation to the physical mass parameters for both heavy and light quarks as well as to vanishing lattice spacing.
A discussion of the inclusive determination of $|V_{ub}|$ is beyond the scope of this paper as it involves very different theoretical methods, such as perturbative QCD and the heavy quark expansion, but we note that a promising new direction for tackling the problem using lattice QCD is also being developed~\cite{Hashimoto:2017wqo,Gambino:2020crt}.

The rest of this paper is organised as follows.
In Sec.~\ref{sec:form_factors} we discuss the relevant background, including details on the form factors obtained from the calculation and how they are extracted using the appropriate matrix elements.
The lattice setup and procedure for our calculation is described in Sec.~\ref{sec:lattice_calculation}, while further details of the ensemble generation and the properties of the generated ensembles are described in the supplemental material~\footnote{See Supplemental Material at \url{http://link.aps.org/supplemental/10.1103/PhysRevD.106.054502} or appended to the end of this paper for details on the generation of the gauge fields and their properties, and measurements of basic physical quantities.}.
We discuss the results of the lattice form factors and the estimation of various sources of systematic uncertainties in Sec.~\ref{sec:latt_results}.
In Sec.~\ref{sec:continuum_results_vub} we discuss the continuum results for the form factors, the use of the $z$-parameter expansion to obtain results across the entire $q^2$ range, and our main result: the determination of $|V_{ub}|$ when our lattice form factors are combined with differential branching fractions from experiment.
Finally, we conclude in Sec.~\ref{sec:conclusions}.

\section{Form factors}
\label{sec:form_factors}
Form factors to describe the semileptonic decay of a $B$ meson to a pion can be defined for the transition matrix element
$\langle{\pi(p_\pi)} |V^{\mu}|{B(p_B)}\rangle$ of the flavor-changing vector current $V^\mu=\bar{q}\gamma^\mu Q$ as
\begin{equation}
\langle{\pi(p_\pi)} |V^{\mu}|{B(p_B)}\rangle =
f_{+}(q^2) \left[p_B^{\mu} + p^{\mu}_\pi - \frac{M_B^2 - M_\pi^2}{q^2} q^{\mu} \right] +
f_{0} (q^2) \, \frac{M^2_B - M^2_\pi}{q^2} \, q^{\mu} \, ,
\end{equation}
where $f_{+}(q^2)$ and $f_{0}(q^2)$ are the vector and scalar form factors of this process, $p_B$ and $p_\pi$ are the four-momenta of the $B$ and $\pi$ respectively, and $M_B$ and $M_\pi$ are their masses.
The momentum transfer is $q^\mu=p^\mu_B-p^\mu_\pi$.
At $q^2=0$ there exists a kinematic constraint, $f_+(0)=f_0(0)$.

A common alternative parametrization that is useful for lattice calculations relates the matrix elements to parallel and perpendicular form factors, $f_\parallel(E_\pi)$ and $f_\perp(E_\pi)$, through
\begin{equation}
  \langle \pi(p_\pi) | V^{\mu} | B(p_B) \rangle = \sqrt{2M_B}\left[v^\mu f_\parallel + p^\mu_{\pi,\perp} f_\perp\right],
\end{equation}
where $v^\mu=p_B^\mu/M_B$ is the velocity of the $B$ meson, and
$p^\mu_{\pi,\perp} \equiv p^\mu_\pi-(v \cdot p_\pi)v^\mu$.
The pion energy $E_\pi$ is related to the momentum transfer of the leptons by
\begin{equation}
  E_\pi \equiv v\cdot p_\pi = \frac{M^2_B+M^2_\pi-q^2}{2M_B}.
\end{equation}
Throughout this paper we keep the $B$ meson on the lattice at rest and so can use the relations
\begin{align}
  f_\parallel(E_\pi) =& \frac{\langle \pi(p_\pi) | V^{0} | B(p_B) \rangle}{\sqrt{2M_B}},
  \label{eq:f_para}\\
  f_\perp(E_\pi) =& \frac{\langle \pi(p_\pi) | V^{i} | B(p_B) \rangle}{\sqrt{2M_B}}\frac{1}{p_\pi^{i}},
  \label{eq:f_perp}
\end{align}
where the temporal, $\mu=0$, and spatial, $\mu=i$, components of the vector current $V^\mu$ are considered, respectively.

Another possible parametrization---motivated by heavy quark effective theory---is~\cite{Burdman:1993es}
\begin{equation}
  \langle \pi(p_\pi) | V^{\mu} | B(v) \rangle =2\left[f_1\left(v\cdot p_\pi\right)v^\mu+f_2\left(v \cdot p_\pi\right)\frac{p_\pi^\mu}{v \cdot p_\pi}\right],
\end{equation}
where the $B$ meson state is defined as $|B(v)\rangle=(1/\sqrt{M_B})|B(p_B)\rangle$ such that it is properly defined in the heavy quark limit.
The form factors $f_1(v\cdot p_\pi)$ and $f_2(v\cdot p_\pi)$ are also consistently defined in the heavy quark limit and the heavy quark mass dependence would start from $1/m_b$.
Comparing with Eqs.~\eqref{eq:f_para} and~\eqref{eq:f_perp}, we get
\begin{align}
f_1(v\cdot p_\pi)+f_2(v\cdot p_\pi) &= \dfrac{f_\parallel(E_\pi)}{\sqrt{2}},\\
f_2(v\cdot p_\pi)&=f_\perp(E_\pi) \left(\dfrac{v\cdot p_\pi}{\sqrt{2}}\right).
\end{align}
The relation to the conventionally defined form factors $f_+(q^2)$ and $f_0(q^2)$ is given by
\begin{align}
  f_+(q^2) =& \sqrt{M_B} \left\{
             \frac{f_2(v\cdot p_\pi)}{v\cdot p_\pi} +
             \frac{f_1(v\cdot p_\pi)}{M_B} \right\},
  \\
  f_0(q^2) =& \frac{2}{\sqrt{M_B}}\frac{M_B^2}{M_B^2-M_\pi^2} \biggl\{
              \left[f_1\left(v\cdot p_\pi\right)+f_2\left(v\cdot p_\pi\right)\right]
              \nonumber\\
            & -\frac{v\cdot p_\pi}{M_B} \left[
              f_1(v\cdot p_\pi)+\frac{M_\pi^2}{(v\cdot p_\pi)^2}f_2(v\cdot p_\pi) \right] \biggr\},
\end{align}
or, equivalently, by
\begin{align}
f_+(q^2) =& \frac{1}{\sqrt{2M_B}}\big[f_{\parallel}(E_{\pi})+(M_B-E_{\pi})f_{\perp}(E_{\pi})\big],\\
f_0(q^2) =& \frac{\sqrt{2M_B}}{M_B^2-M_{\pi}^2}\big[(M_B-E_{\pi})f_{\parallel}(E_{\pi})+(E_{\pi}^2-M_{\pi}^2)f_{\perp}(E_{\pi})\big].
\end{align}

In the limit $v \cdot p_\pi \rightarrow 0$, the soft pion theorem and the pole dominance ansatz is justified using the heavy meson chiral Lagrangian approach and one obtains~\cite{Burdman:1993es}
\begin{equation}\label{eq:pole_dominance}
  \lim_{v \cdot p_\pi \to 0} f_2\left(v \cdot p_\pi\right) =
  g_{B^{*} B\pi} \frac{f_{B^*}\sqrt{M_{B^*}}}{2f_\pi}
  \frac{v \cdot p_\pi}{v \cdot p_\pi + \Delta_B},
\end{equation}
with $M_{B^*}$ the mass of the vector meson $B^*$,  $\Delta_B=M_{B^*}-M_B$ the hyperfine splitting, $f_{B^*}$ and $f_\pi$ the $B^*$ and $\pi$ decay constants respectively, and $g_{B^* B\pi}$ the $B^*B\pi$ coupling.

\section{Lattice Calculation}
\label{sec:lattice_calculation}

\subsection{Ensembles and correlators}
We use the M\"{o}bius domain-wall fermion action~\cite{Brower:2012vk} in this work for both heavy and light quarks. The gauge ensembles were generated with $2+1$ flavours of dynamical quarks by the JLQCD Collaboration.
The tree-level Symanzik-improved gauge action is employed, and stout smearing~\cite{Morningstar:2003gk} is applied to the gauge fields when coupled to fermions.
The lattice ensembles used in this work are summarized in Table~\ref{tab:gauge_params}. They form a subset of those generated by the JLQCD Collaboration. (The full list is found in the supplemental material~\cite{Note1}.) Each ensemble is given an ID of the form ``X-$ud$\#-$s$a'', where X ($=$ C, M, or F) denotes the lattice spacing, the number after $ud$ represents the pion mass in units of $100\;\mathrm{MeV}$, and the letter after $s$ distinguishes whether the strange quark mass is above (a) or below (b) its physical value.

\begin{table}[tbp]
  \begin{tabular}{c c c c c c c c c}
    \hline
    ID & $a~(\mathrm{fm})$ & $\beta$ & $L^3\times N_T\times L_s$ & $N_{\mathrm{cfg}}$ & $am_{l}$ & $am_s$ & $am_Q$ & $N_{\mathrm{tsrc}}$ \\
    \hline
    C-$ud$5-$s$a & $0.080$ & $4.17$ & $32^3\times 64 \times 12 $ & $100$ & $0.019$ & $0.04$ & $0.44037$ & $2$\\
    & & & & & & & $0.68808$ & $2$ \\
    C-$ud$5-$s$b & $0.080$ & $4.17$ & $32^3\times 64 \times 12 $ & $100$ & $0.019$ & $0.03$ & $0.44037$ & $2$ \\
    & & & & & & & $0.68808$ & $1$ \\
    C-$ud$4-$s$a & $0.080$ & $4.17$ & $32^3\times 64 \times 12 $ & $100$ & $0.012$ & $0.04$ & $0.44037$ & $2$ \\
    & & & & & & & $0.68808$ & $2$ \\
    C-$ud$4-$s$b & $0.080$ & $4.17$ & $32^3\times 64 \times 12 $ & $100$ & $0.012$ & $0.03$ & $0.44037$ & $2$ \\
    & & & & & & & $0.68808$ & $1$ \\
    C-$ud$3-$s$a & $0.080$ & $4.17$ & $32^3\times 64 \times 12 $ & $100$ & $0.007$ & $0.04$ & $0.44037$ & $4$ \\
    & & & & & & & $0.68808$ & $4$ \\
    C-$ud$3-$s$b & $0.080$ & $4.17$ & $32^3\times 64 \times 12 $ & $100$ & $0.007$ & $0.03$ & $0.44037$ & $4$ \\
    & & & & & & & $0.68808$ & $1$ \\
    C-$ud$2-$s$a-L & $0.080$ & $4.17$ & $48^3\times 96 \times 12 $ & $100$ & $0.0035$ & $0.04$ & $0.44037$ & $4$ \\
    & & & & & & & $0.68808$ & $2$ \\
    \hline
    M-$ud$5-$s$a & $0.055$ & $4.35$ & $48^3\times 96 \times 8 $ & $50$ & $0.012$  & $0.025$ & $0.27287$ & $2$ \\
    & & & & & & & $0.42636$ & $2$ \\
    & & & & & & & $0.66619$ & $2$ \\
    M-$ud$4-$s$a & $0.055$ & $4.35$ & $48^3\times 96 \times 8 $ & $50$ & $0.008$  & $0.025$ & $0.27287$ & $2$ \\
    & & & & & & & $0.42636$ & $2$ \\
    & & & & & & & $0.66619$ & $2$ \\
    M-$ud$3-$s$a & $0.055$ & $4.35$ & $48^3\times 96 \times 8 $ & $42$ & $0.0042$ & $0.025$ & $0.27287$ & $4$ \\
    & & & & & & & $0.42636$ & $2$ \\
    & & & & & & & $0.66619$ & $2$ \\
    \hline                                               
    F-$ud$3-$s$a & $0.044$ & $4.47$ & $64^3\times 128 \times 8 $ & $50$ & $0.003$ & $0.015$ & $0.210476$ & $4$ \\
    & & & & & & & $0.328869$ & $2$ \\
    & & & & & & & $0.5138574$ & $1$ \\
    \hline
  \end{tabular}
  \caption{Parameters of the gauge configurations used in this analysis. We give the ID, the lattice spacing, coupling and dimensions in the first four columns. The number of configurations, $N_{\mathrm{cfg}}$, are given in column five. We then provide the light, strange and heavy quark masses in lattice units in the next three columns respectively. Finally, we note the number of times sources $N_{\mathrm{tsrc}}$ used for each set of parameters, where a time source at $t=0$ is always employed, and additional time sources are evenly spaced in the time direction.}
  \label{tab:gauge_params}
\end{table}

The simulation parameters are chosen as follows.
The lattice spacings for coarse ``C'', middle ``M'' and fine ``F'' lattices are $0.0804(1)$, $0.0547(1)$ and $0.0439(1)\;\mathrm{fm}$, corresponding to lattice cutoffs $a^{-1} = 2.453(4)$, $3.610(9)$ and $4.496(9)\;\mathrm{GeV}$, respectively.
We use a range of light quark masses that correspond to pion masses from $500\;\mathrm{MeV}$ down to $230\;\mathrm{MeV}$.
They are roughly tuned to 500 ($ud$5), 400 ($ud$4), 300 ($ud$3) and 230 ($ud$2)$\;\mathrm{MeV}$.
Two values of strange quark mass are taken to sandwich its physical
value on the coarsest lattice, i.e., above ($s$a) or below ($s$b) the physical strange quark mass.
Lattice volumes are $32^3\times 64$, $48^3\times 96$ and $64^3\times 128$
for the three lattice spacings, respectively.
They are chosen such that the spatial extent $L$ of the lattice is
kept constant, $\sim 2.6\;\mathrm{fm}$, in physical units.
The only exception is for the ``C''
ensemble with the lightest pion mass, ``C-$ud$2-$s$a-L'', which has a larger volume of $48^3\times 96$.
The temporal extent $N_T$ is chosen as $N_T=2L$.
All ensembles satisfy the condition $M_\pi L>4$,
which is often required to suppress the finite volume effects to a
sufficient level, i.e., below a few per cent level for meson masses, decay constants, and form factors.
We summarize the parameters of the gauge configurations including the light and strange sea quark masses, $m_l$ and $m_s$, in Table~\ref{tab:gauge_params}. The ID for each ensemble is the same as those in the supplemental material where further details about the lattice ensembles, including the measurement of the lattice spacing through the gradient flow, the observation of the topology tunnelings, and the light pseudoscalar meson masses and decay constants, are discussed~\cite{Note1}.

The chiral symmetry of M\"{o}bius domain-wall fermions is not exact due to the finite fifth dimension $L_s$.
The resulting residual mass depends on the lattice spacing and the details of the implementation of the domain-wall fermion.
In our case the residual mass on the coarsest lattice ($\beta = 4.17$, the ``C'' lattices) is at the level of $1\;\mathrm{MeV}$ and an order of magnitude smaller on finer lattices (``M'' and ``F''). Detailed measurements are described in the supplemental material~\cite{Note1}.
The residual mass, however, does not directly affect the analysis of the $B\to\pi\ell\nu$ form factors because we use the pion and kaon masses as parameters to control the chiral extrapolation.

In addition to this work, the ensembles have so far been used for 
a determination of the renormalization constants~\cite{Tomii:2016xiv}, 
a calculation of the charmonium correlator and the extraction of the
charm quark mass and the strong coupling constant~\cite{Nakayama:2016atf}, 
and a calculation of the $D$ semileptonic decay form factors~\cite{Kaneko:2017xgg}.
The lattice data have also been applied to
a calculation of the topological susceptibility in QCD~\cite{Aoki:2017paw},
a study of the Dirac eigenvalue spectrum and a precise calculation
of the chiral condensate~\cite{Cossu:2016eqs},
another study of the Dirac eigenvalue spectrum but in the high energy
region~\cite{Nakayama:2018ubk},
the short-distance current correlator and its comparison with
experimental data~\cite{Tomii:2017cbt},
and a proposal for lattice calculations of inclusive $B$ meson decays~\cite{Hashimoto:2017wqo,Gambino:2020crt}.

The valence sector also uses the M\"{o}bius domain-wall fermion action and the light quark masses are the same as used in the sea.
Heavy quark masses were chosen as $am_Q=1.25^{2n}\times am_c$, for $n\geq0$ and limited to values $am_Q\leq0.7$ to keep discretization errors under proper control.
This results in mass values $am_c\leq am_Q \leq 2.44\times am_c$.
The charm quark mass in lattice units, $am_c$, is tuned such that the spin-averaged Charmonium $1S$ state reproduces its physical mass. (Details are discussed in Ref.~\cite{Nakayama:2016atf}.)
Since the lowest heavy quark mass used is that of the charm, we have the process $D\to\pi\ell\nu$ included as part of our dataset.
We can then use form factors from that decay plus the additional heavier quark masses to extrapolate to the physical $b$ quark mass.

In order to extract the form factors, we compute the three-point functions of the form
\begin{equation}
  C_{\mathrm{3pt}}^{\pi V^\mu B}(t,T) = \sum_{\bm{x},\bm{y},\bm{z}} e^{i(\bm{p}_\pi\cdot\bm{x}+\bm{q}\cdot\bm{y})}
  \langle P_\pi^S(\bm{x},0) V^\mu(\bm{y},t) P_Q^S(\bm{z},T)\rangle,
\end{equation}
where $P_\pi^S$ and $P_Q^S$ are interpolating operators to create or annihilate the pseudoscalar pion and heavy mesons. These operators are smeared to enhance the overlap with the corresponding ground state. The smearing is applied in a gauge invariant manner using an operator $(1-(\alpha/N)\Delta)^N$ with a discretized Laplacian $\Delta$ and parameters $\alpha$ = 20 and $N$ = 200.
The source of the quark propagator is generated on the entire source time slice with random $Z_2$ noise, and then the smearing is applied.
The $B$ meson is always set at rest so that $\bm{q}=-\bm{p}_\pi$.
The source-sink separation in the temporal direction $T$ is kept approximately fixed in physical units across all lattice spacings. We use $T$ = $28$, $42$ and $56$ on ensembles with $\beta=4.17$, $4.35$, and $4.47$, respectively.
The ground state can then be well isolated by the fits as described in Sec.~\ref{subsec:corr_fits}.

The heavy-to-light vector current is defined as $V^\mu=\bar{q}\gamma^\mu Q$.
Both light ($q$) and heavy ($Q$) quark fields are described by the M\"{o}bius domain-wall fermion action, and the current is local on the lattice.
The renormalization constant $Z_V$ is multiplied with the current afterwards, as discussed in Sec.~\ref{sec:renormalization}.

We also compute the pion and heavy meson two-point functions. These are used to constrain the energies of the initial and final states in the combined fit, as discussed in Sec.~\ref{subsec:corr_fits}.

These measurements are performed on $N_{\mathrm{cfg}}$ gauge configurations for each ensemble and repeated $N_{\mathrm{tsrc}}$ times by always using time source $t=0$ and then shifting the source and sink time slices by $N_T/N_{\mathrm{tsrc}}$. The number of measurements is thus $N_{\mathrm{cfg}}\times N_{\mathrm{tsrc}}$ per ensemble. Details are listed in Table~\ref{tab:gauge_params} for each choice of ensemble and valence heavy quark mass.

\subsection{Two-point \& three-point correlator fits}
\label{subsec:corr_fits}
To extract the required form factors, we perform simultaneous fits of all two-point and three-point correlators for each set of ensembles and quark mass parameters using a constrained multi-exponential fit~\cite{Lepage:2001ym}. Doing so allows us to fit the majority of the time extent of the correlators while isolating the ground states---needed to determine the form factors---from the excited states, which can be discarded. We include data starting from time slice $t_{\mathrm{min}}=2$, $3$ or $4$ in the fit, depending on the ensemble. The two-point correlators are fit to the cosh form
\begin{equation}
  C_P(t) = \sum_{n=0}^{n_{\mathrm{exp}}-1} a_{P,n}b^*_{P,n}
  \left(e^{-E_{P,n}t}+e^{-E_{P,n}(N_T-t)}\right),
\end{equation}
where the subscripts $P,n$ correspond to state $n$ of pseudoscalar $P$, such that $n=0$ is the ground state. The interpolating operators are always smeared at the source, and are either local or have the same smearing parameters at the sink.
The amplitudes $a_{P,n}$ and $b_{P,n}$ are then equal for the smeared sink or different if the sink is local. We always fit both cases simultaneously to improve the determination of the ground-state energy.
We use $n_{\mathrm{exp}}=3$ and, since we only require ground-state energies and amplitudes for our calculation, we simply check that fits with two or four exponentials give consistent ground-state results. This multiexponential approach to our fits ensures the uncertainty due to contamination of excited states is taken into account.

For the three-point correlators, we fit to the form 
\begin{equation}
  C_{3\mathrm{pt}}(t,T) = \sum_{n,m=0}^{n_{\mathrm{exp}}-1} a_{\pi,n} V_{n,m} a^*_{B,m}e^{-E_{\pi,n}t}e^{-E_{B,m}(T-t)}.
  \label{eq:C3pt}
\end{equation}
The energies, $E_{\pi,n}$ and $E_{B,m}$, and smeared amplitudes, $a_{\pi,n}$ and $a_{B,m}$, are the same as those from the pion and heavy meson two-point correlator fit form. The amplitude $V_{0,0}$, which connects the ground-state heavy meson to the ground-state pion, is needed to determine the form factors. It relates to the corresponding matrix element by
\begin{equation}
  V^\mu_{0,0} = \frac{\langle\pi|V^\mu |B\rangle}{2\sqrt{E_\pi M_B}}.
\end{equation}
As with the two-point correlators, we use $n_{\mathrm{exp}}=3$.

We use the Python packages Gvar~\cite{peter_lepage_2021_5202756}, Lsqfit~\cite{peter_lepage_2021_5202760} and Corrfitter~\cite{peter_lepage_2020_4281296} to fit our correlators. The fit parameters are given Bayesian priors as follows. The magnitudes of meson two-point amplitudes can be estimated by fitting that correlator alone with a single exponential at large time $t$, which leaves only the ground-state contribution. The heavy meson and pion two-point amplitudes are found to be of order $0.1$--$1.0$ (smeared) and $10$--$30$ (local), depending on the lattice spacing of the ensemble. These are taken as the central values, and the priors are given very conservative widths that are $5$ times these values. Similarly, one can extract estimates of the magnitudes of the three-point amplitudes, which are found to be of order $1.0$ ($\mu=0$) and $0.4$ ($\mu=1,2,3$), and we again assign widths $5$ times these values. Priors for the energies of ground states with zero momentum are given 10\% widths. The energies of ground states with nonzero momentum get their priors according to the dispersion relation for energy using the prior of the zero momentum ground state. The gaps between energies of two consecutive states are given priors of $\approx 0.7$~GeV with 70\% widths.

We simultaneously fit a substantial amount of two-point and three-point correlator data, including multiple $am_Q$ and $q^2$ values. This can be difficult as we have to invert large covariance matrices in our fits. If the available statistics is limited, as it is in our case, the eigenvalues of the matrices tend to be underestimated and driven to zero. A standard way to deal with this is to impose singular value decomposition (svd) cuts $c_{\mathrm{svd}}$. In this procedure any eigenvalue smaller than $c_{\mathrm{svd}}$ times the largest eigenvalue $e_{\mathrm{max}}$ is replaced by $c_{\mathrm{svd}}e_{\mathrm{max}}$. The use of the svd cuts makes the matrices less singular. This is a conservative approach since it can only serve to increase the final error. We have chosen the value of $c_{\mathrm{svd}}$ for each ensemble such that the fit quality is good while keeping as many eigenvalues as possible. 

As we have to use fairly large svd cuts in these fits, using $\chi^2$ per degree of freedom ($\chi^2/N_{\mathrm{dof}}$) as a measure of goodness-of-fit becomes less reliable. An svd cut increases the uncertainties in the data without increasing the random fluctuations in the data means. This tends to make the contributions from the parts of the $\chi^2$ function affected by the svd cut much smaller than naively expected, which pulls $\chi^2/N_{\mathrm{dof}}$ down artificially. We therefore check the fits and the final $\chi^2/N_{\mathrm{dof}}$ by adding extra noise to the priors and svd cut, which does not change the fits significantly. 

\begin{figure}[tbp]
\includegraphics[width=0.49\textwidth]{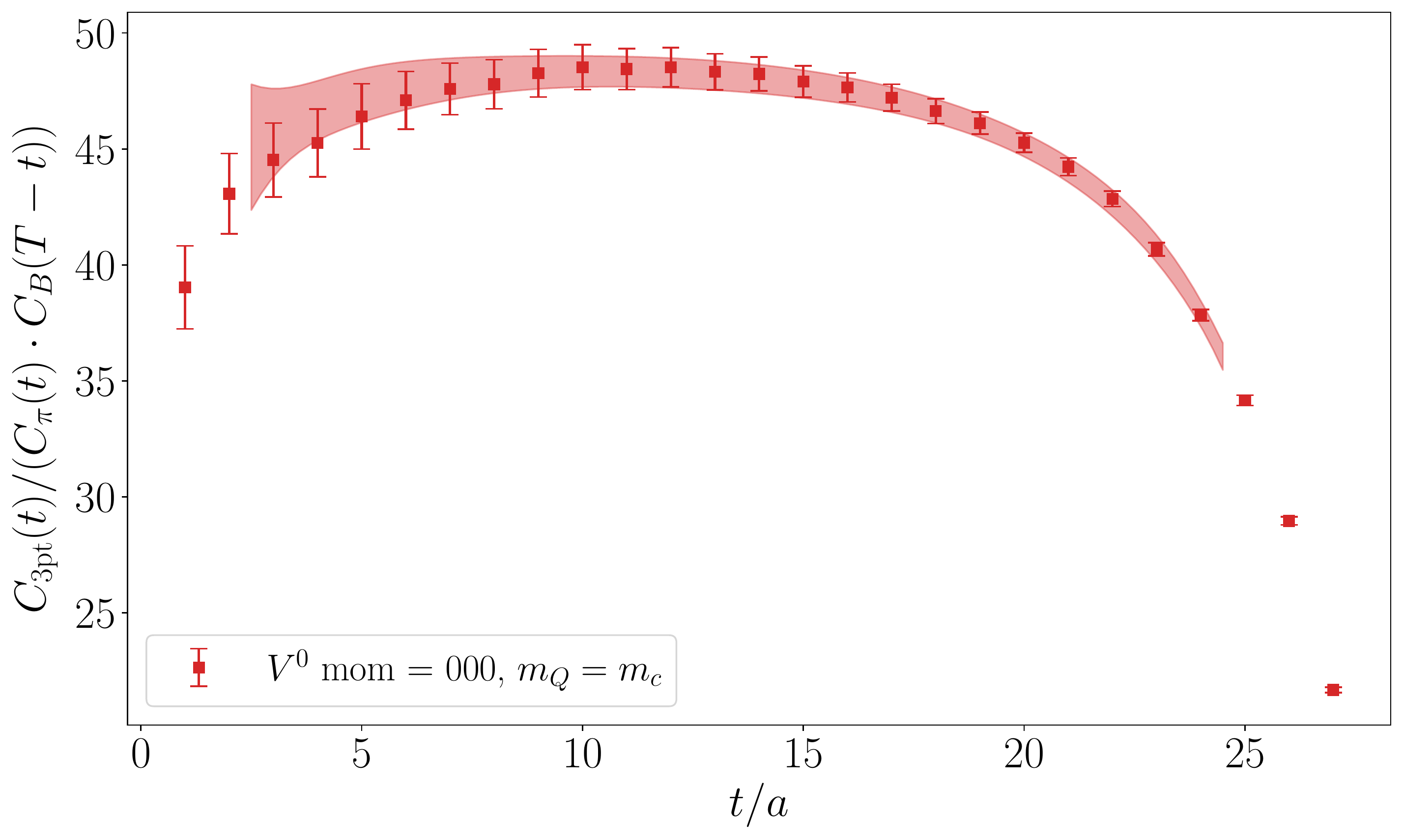}
\includegraphics[width=0.49\textwidth]{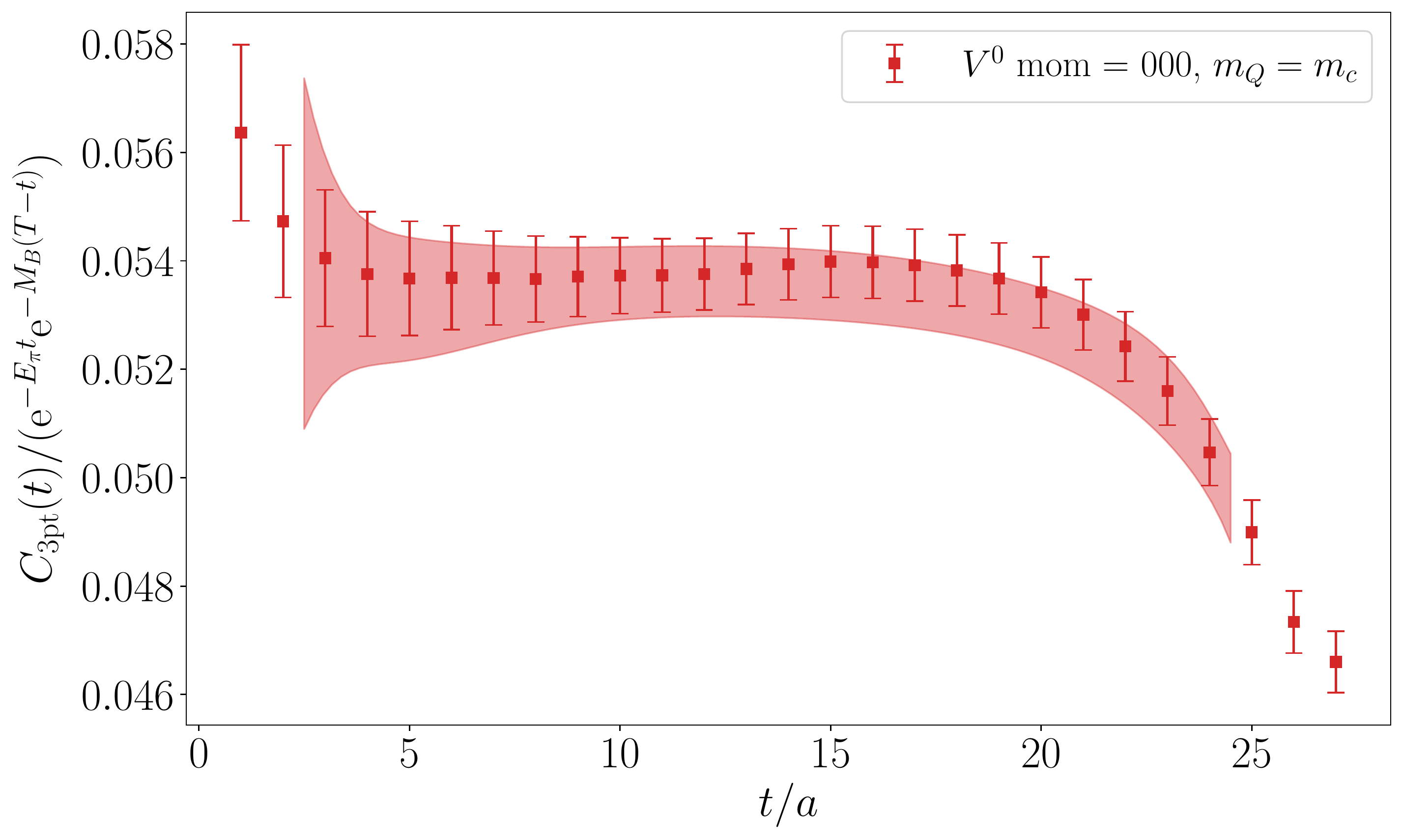}
\caption{Three-point correlator data with a $V^0$ insertion on the ensemble with $\beta=4.17$, ${am_{u,d}=0.007}$ and $am_Q=0.44037$. The bands represent the fit results and their fit range. The pion is created at $t=0$ while the $B$ meson is annihilated at $t=T$.  Left panel: the three-point correlator is divided by the pion and $B$ meson two-point correlators. Right panel: the three-point correlator data is divided by the exponential function corresponding to the meson ground-state energies extracted from our fits.}
\label{fig:corr_ratios}
\end{figure}

\begin{figure}[tbp]
\includegraphics[width=0.49\textwidth,valign=t]{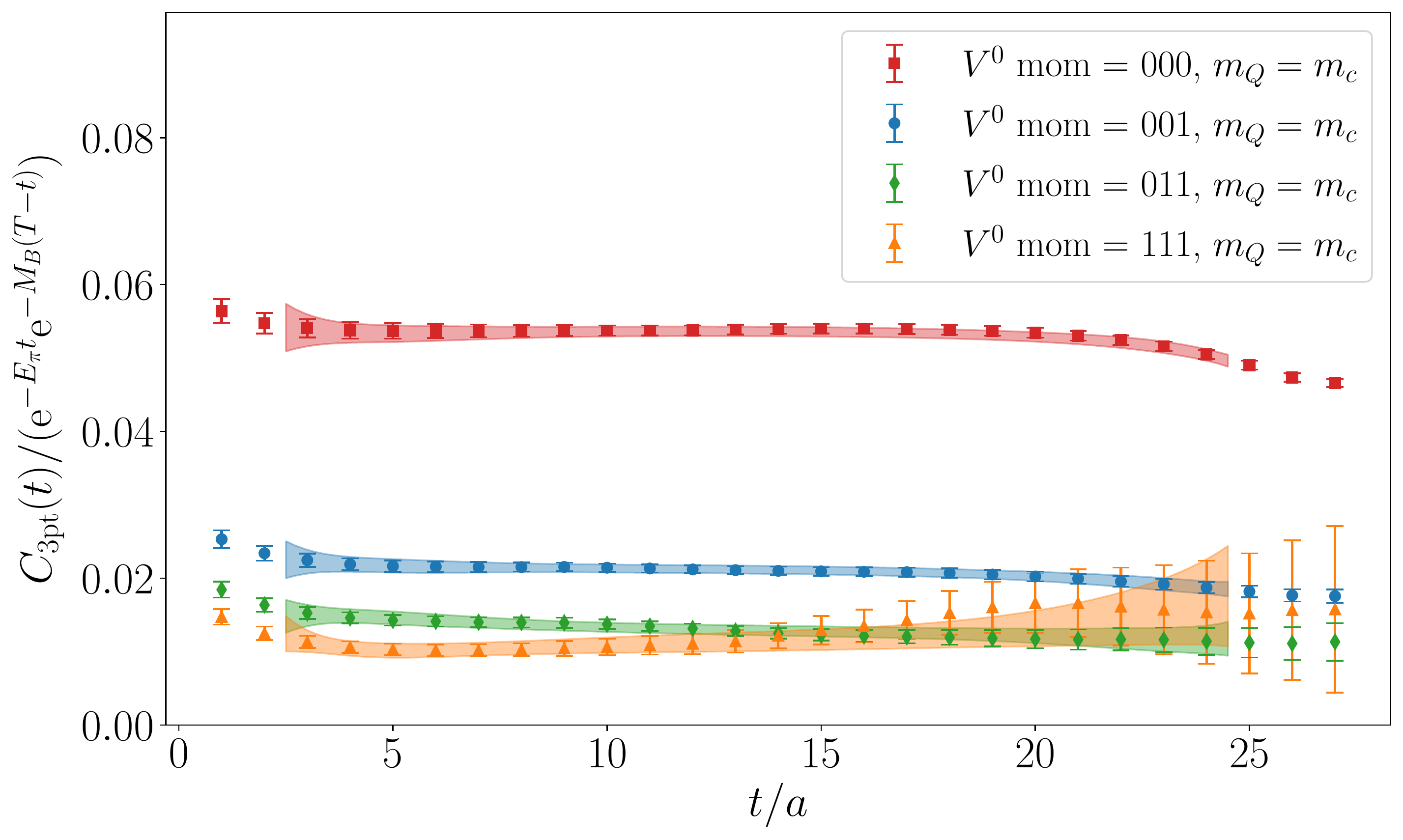}
\includegraphics[width=0.49\textwidth,valign=t]{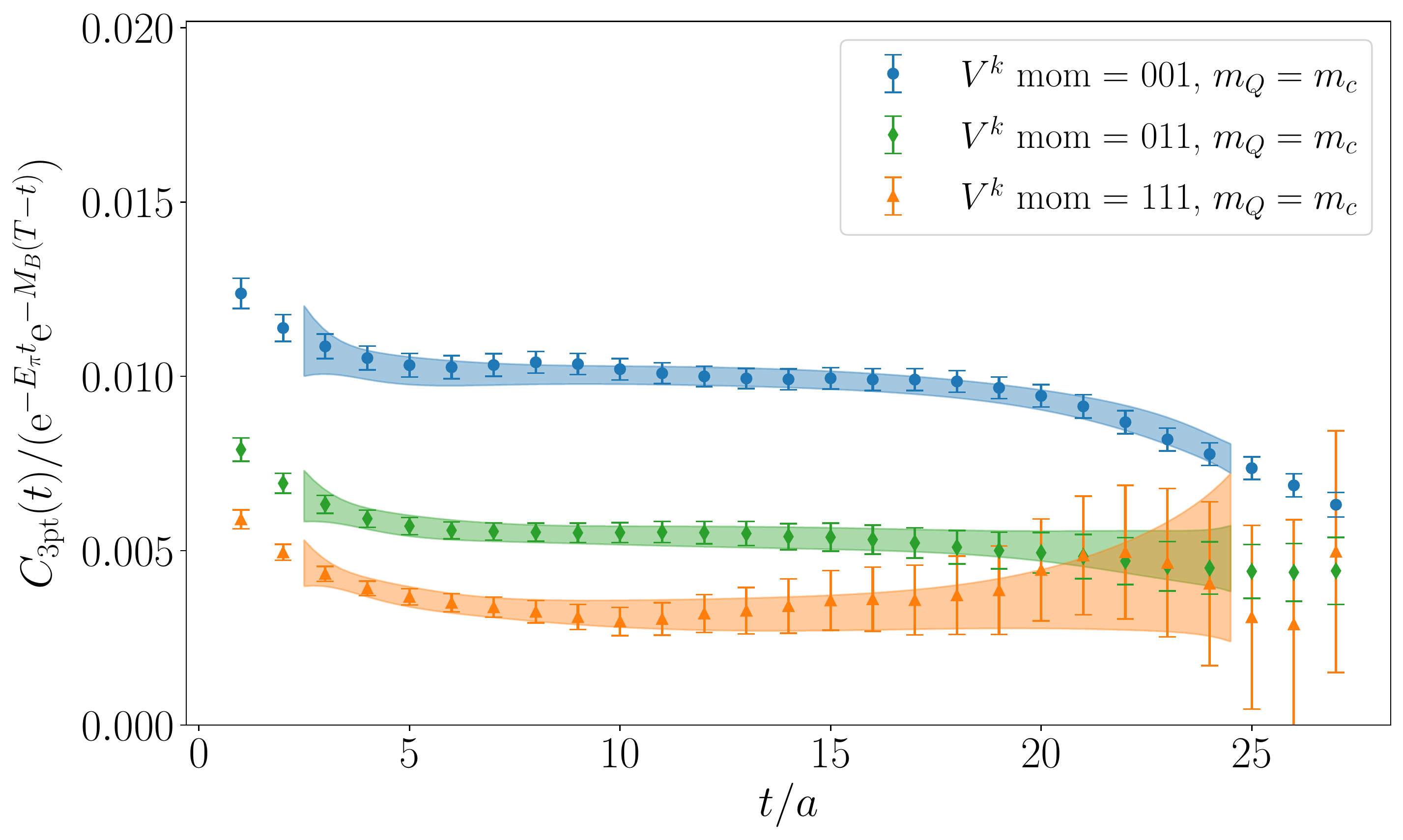}
\caption{Three-point correlators $V^{\mu}$ divided by corresponding ground-state exponentials. Data is from the ensemble with $\beta=4.17$, $am_{u,d}=0.007$ and $am_Q=0.44037$. The pion is created at $t=0$ while the $B$ meson annihilated at $t=T$. Results for the temporal (left) and spatial (right) vector currents are shown for all available momenta: $(0,0,0)$, $(0,0,1)$, $(0,1,1)$, $(1,1,1)$ in units of $2\pi/L$.}
\label{fig:corr_ratios_beta4.17}
\end{figure}

\begin{figure}[tbp]
\includegraphics[width=0.49\textwidth,valign=t]{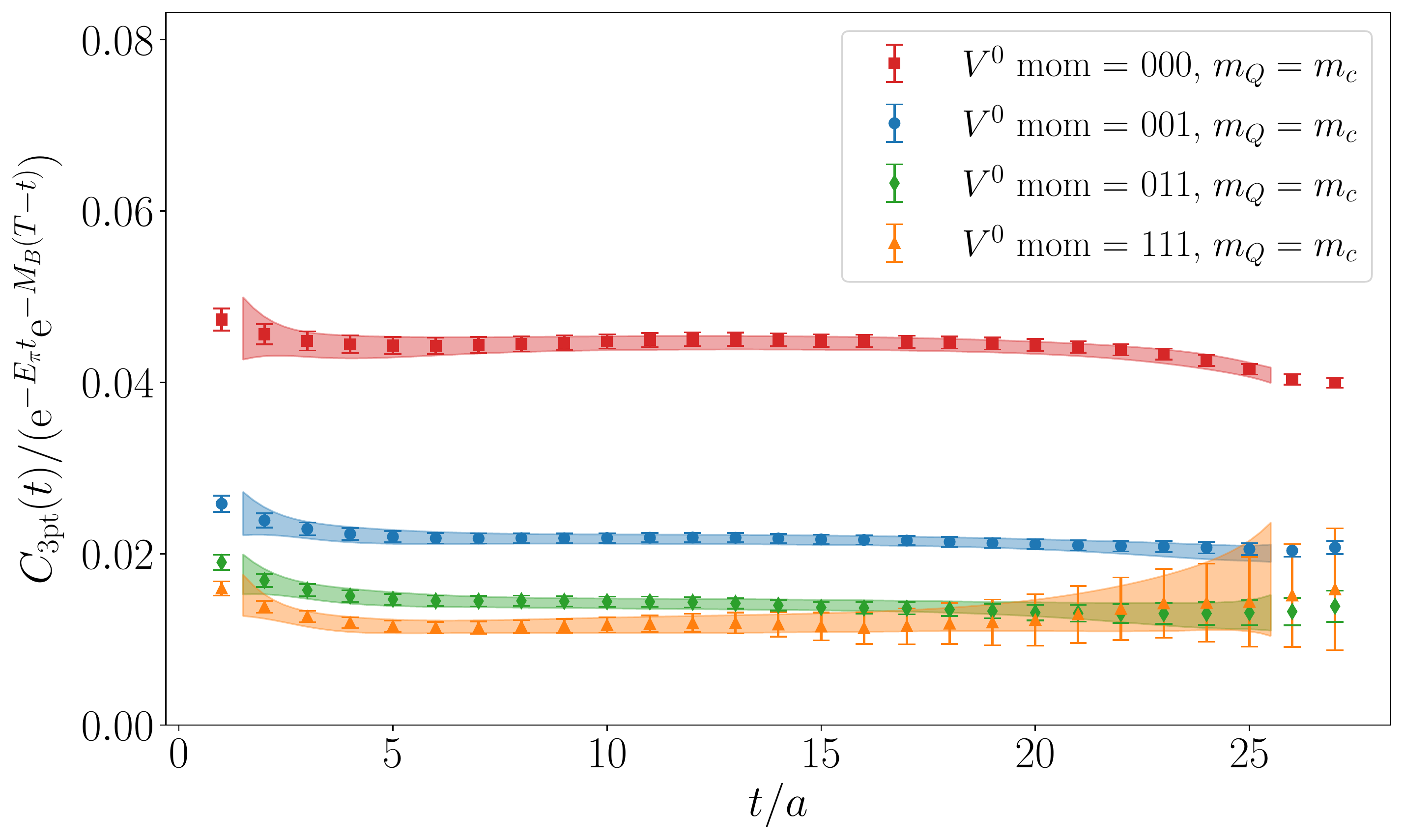}
\includegraphics[width=0.49\textwidth,valign=t]{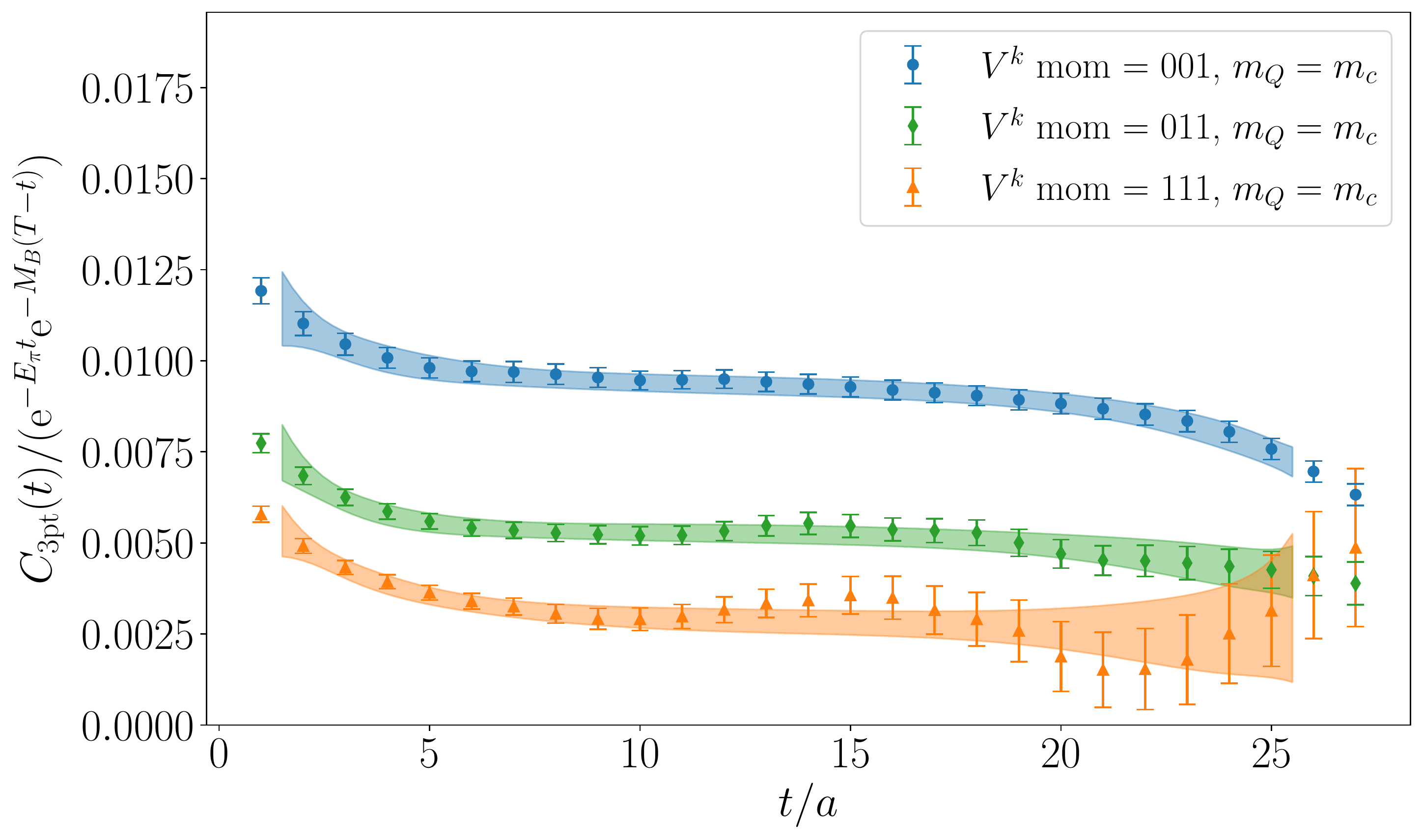}
\caption{Same as Fig.~\ref{fig:corr_ratios_beta4.17}, but at heavier light quark mass $am_{u,d}=0.012$ while keeping the heavy quark mass as $am_Q=0.44037$.}
\label{fig:corr_ratios_mpi}
\end{figure}

\begin{figure}[tbp]
\includegraphics[width=0.49\textwidth,valign=t]{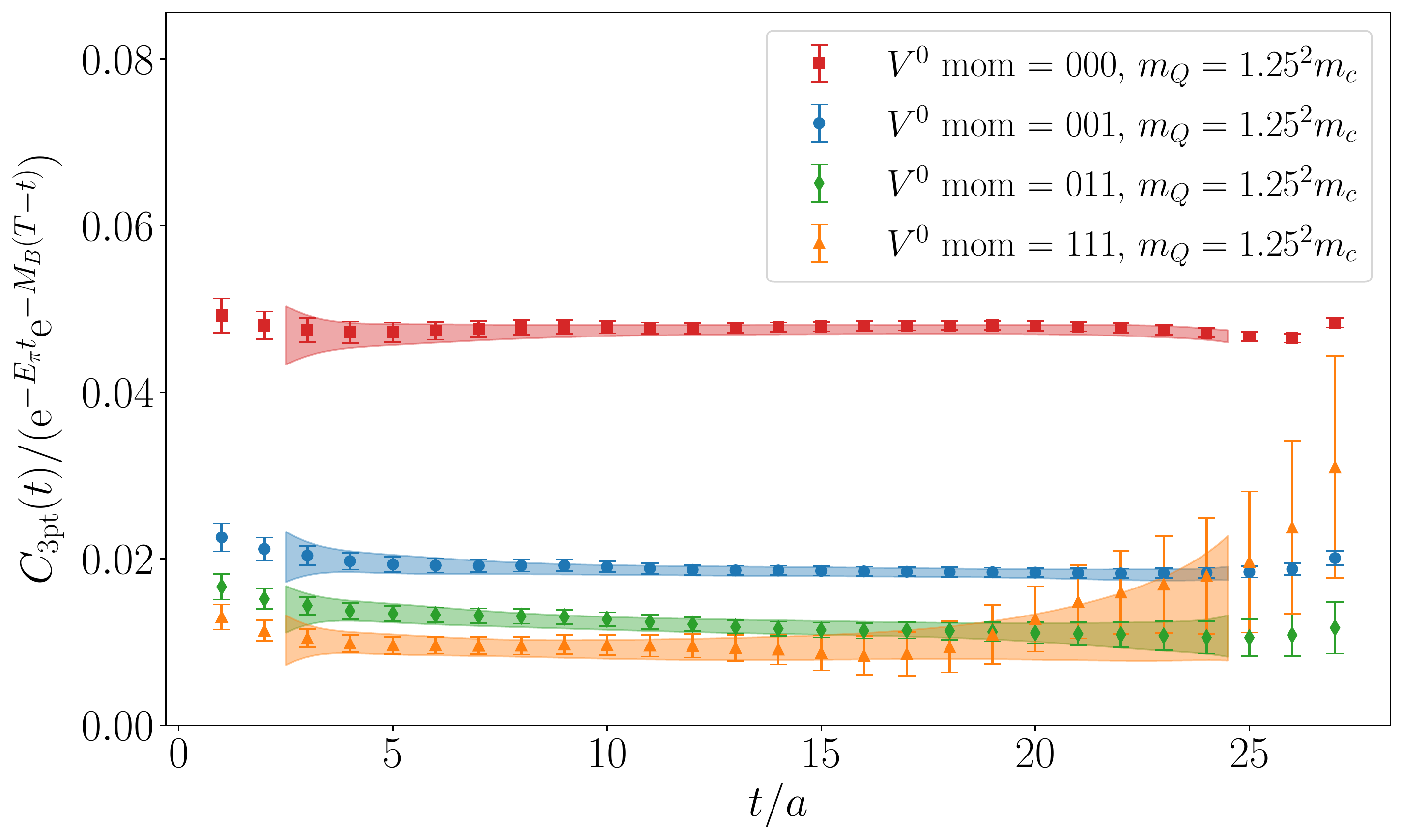}
\includegraphics[width=0.49\textwidth,valign=t]{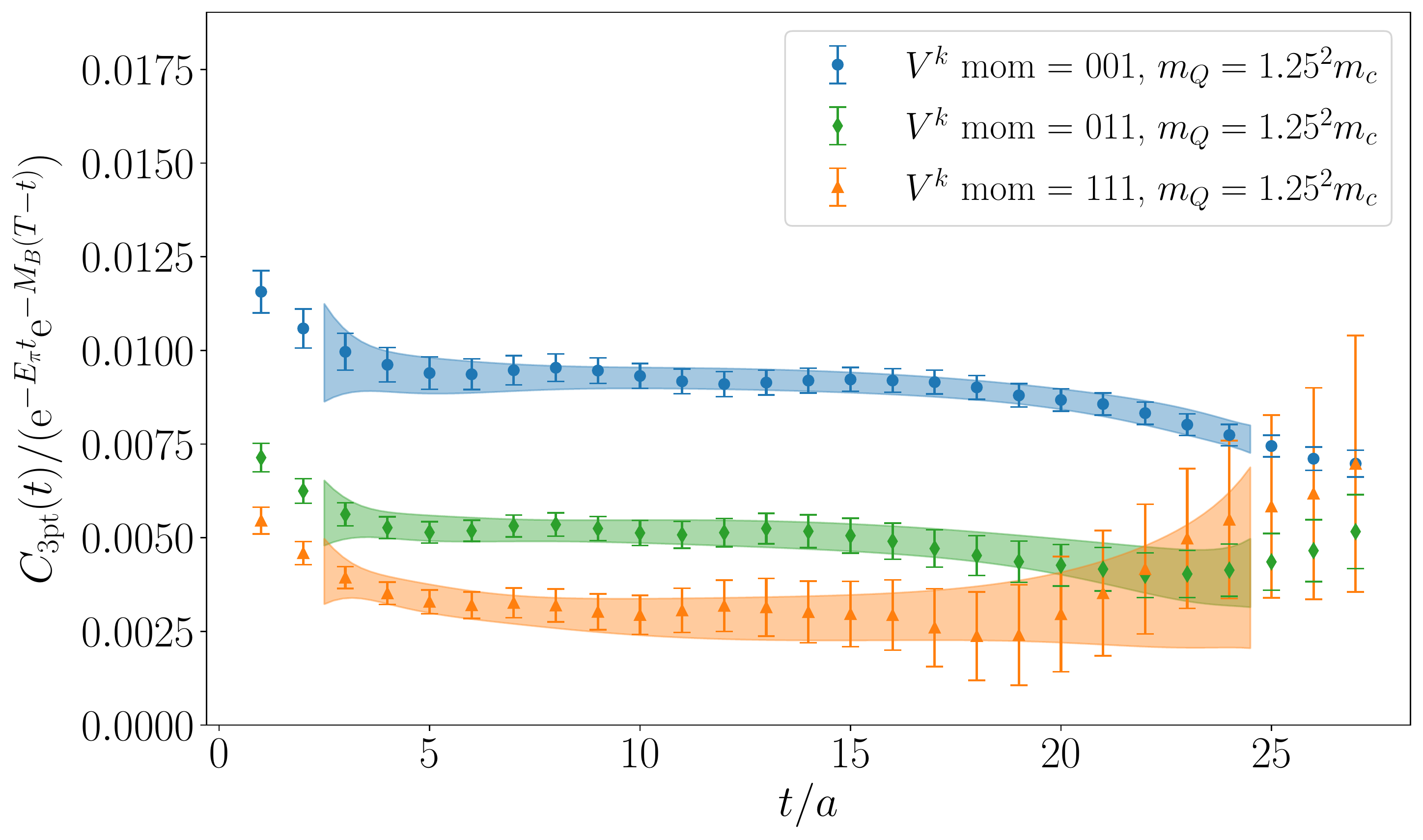}
\caption{Same as Fig.~\ref{fig:corr_ratios_beta4.17}, but at heavier heavy quark mass $am_Q=0.68808$ while keeping the light quark mass as $am_{u,d}=0.007$.}
\label{fig:corr_ratios_mh}
\end{figure}

\begin{figure}[tbp]
\includegraphics[width=0.49\textwidth,valign=t]{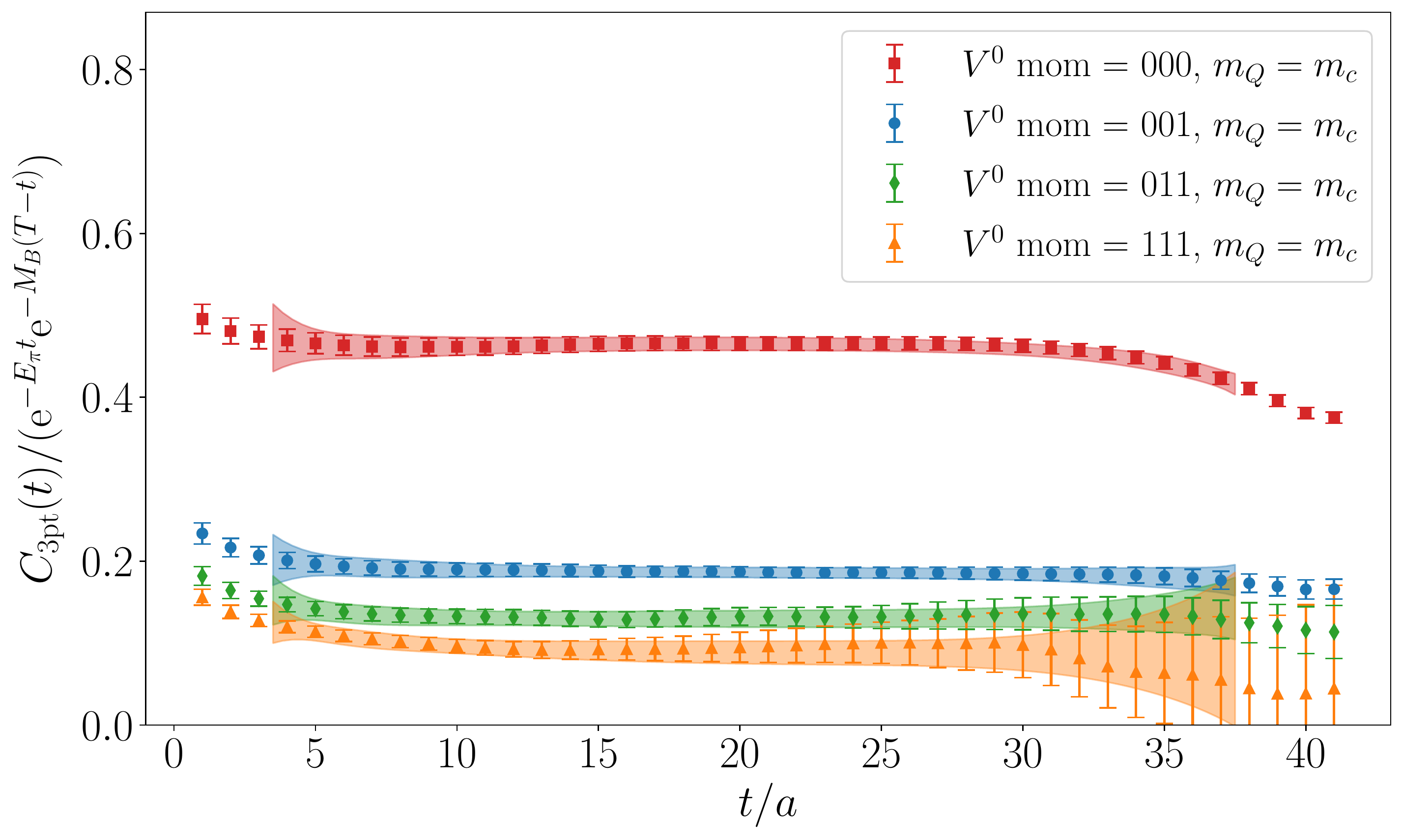}
\includegraphics[width=0.49\textwidth,valign=t]{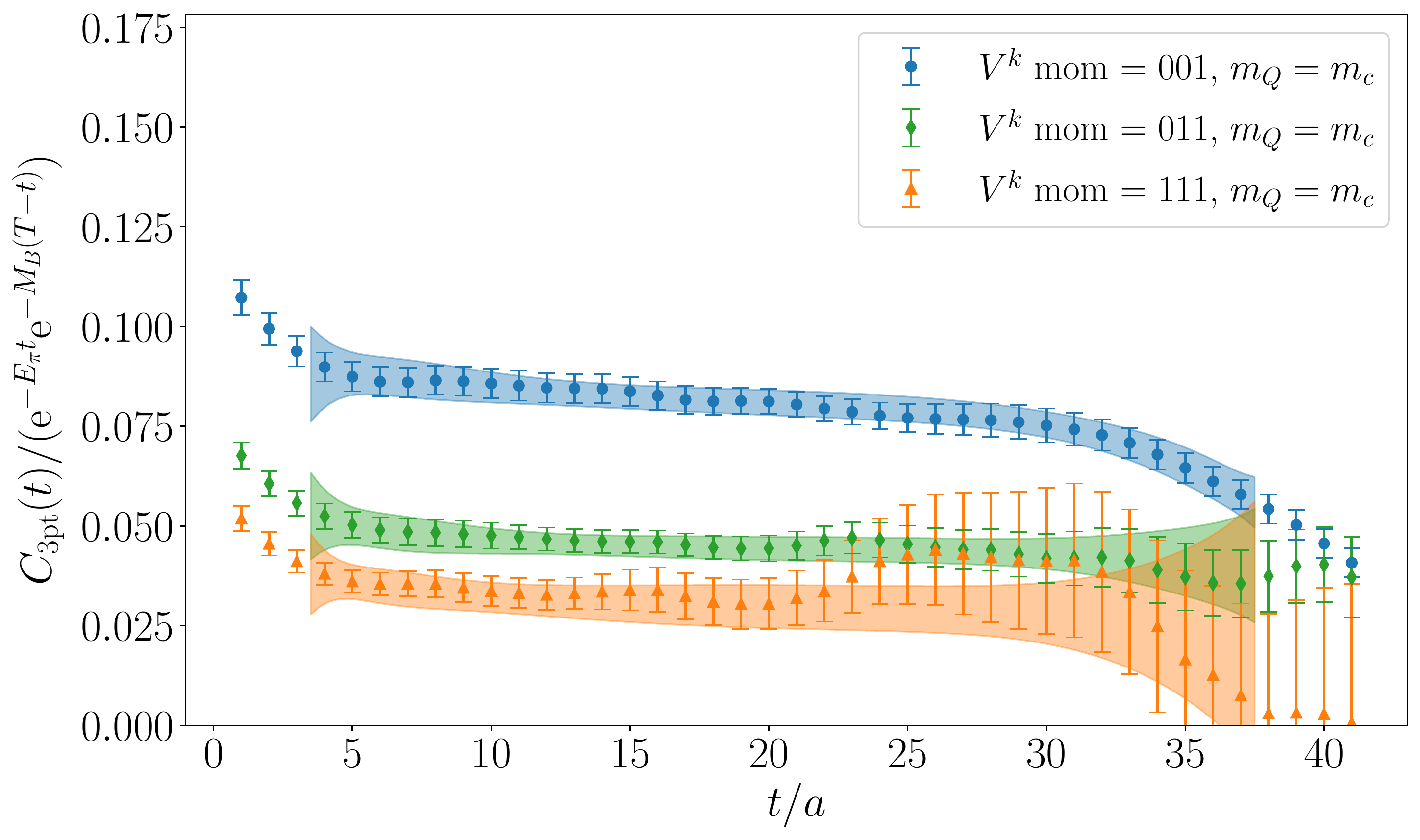}
\caption{Same as Fig.~\ref{fig:corr_ratios_beta4.17}, but on a finer lattice, $\beta=4.35$, and at $am_{u,d}=0.0042$ and $am_Q=0.27287$.}
\label{fig:corr_ratios_beta4.35}
\end{figure}

\begin{figure}[tbp]
\includegraphics[width=0.49\textwidth,valign=t]{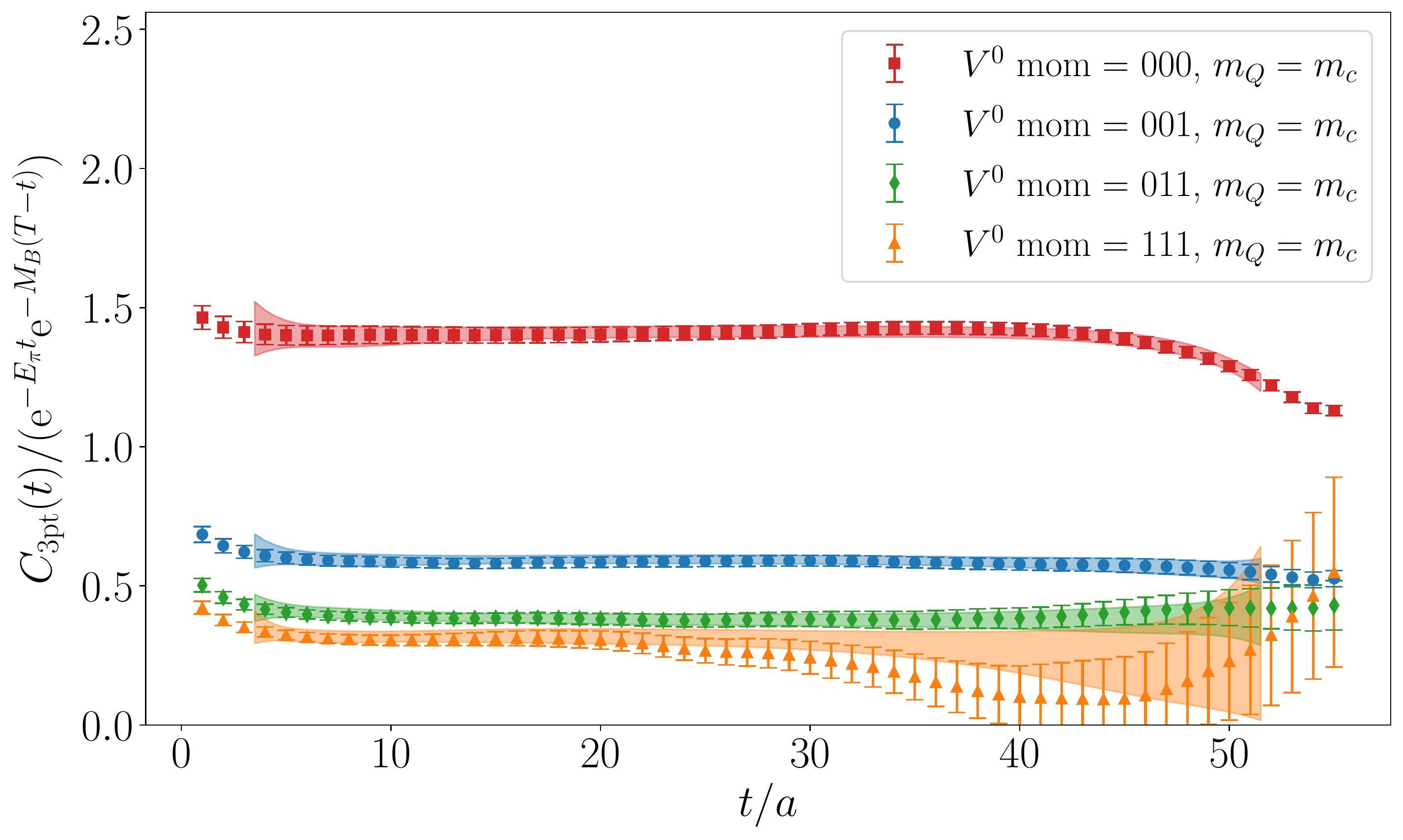}
\includegraphics[width=0.49\textwidth,valign=t]{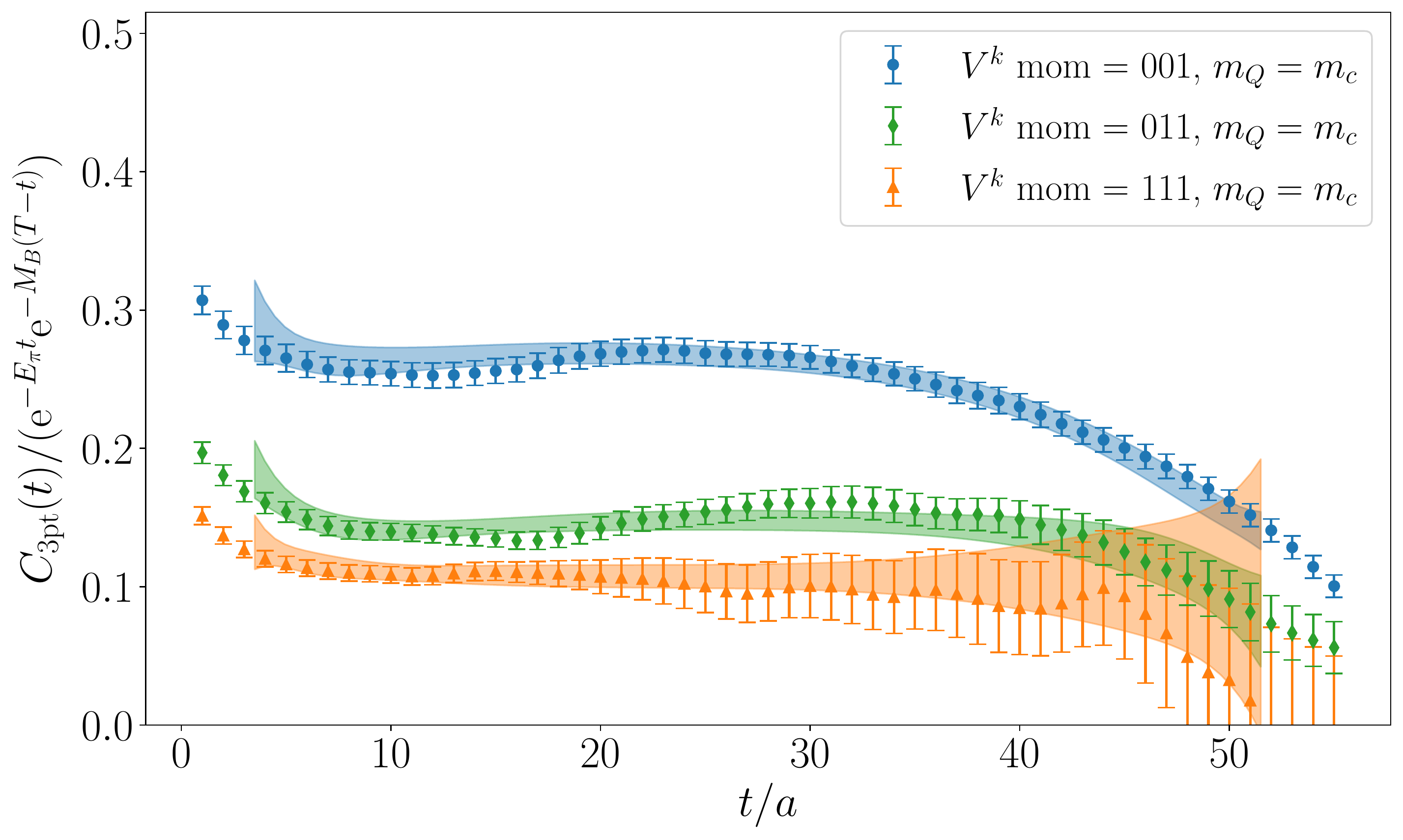}
\caption{Same as Fig.~\ref{fig:corr_ratios_beta4.17}, but on the finest lattice,
  $\beta=4.47$, and at $am_{u,d}=0.003$ and $am_Q=0.210476$.}
\label{fig:corr_ratios_beta4.47}
\end{figure}

In Figs.~\ref{fig:corr_ratios}--\ref{fig:corr_ratios_beta4.47} we show how well our fit results agree with the correlator data for various values of lattice spacing, pion mass and heavy quark mass. In Fig.~\ref{fig:corr_ratios} (left panel) we plot a representative example of the ratio of three-point and two-point correlators, $C_{V^0}(t)/(C_{\pi}(t) C_B(T-t))$, alongside the fit result. The data at $\beta=4.17$, $am_{u,d}=0.007$ and $am_Q=0.44037$ with zero momentum insertion are shown. In the time range where the ground states dominate, this ratio will be a constant: the three-point ground-state amplitude divided by the two-point ground-state amplitudes. Towards $T=28$ we observe a significant curvature of the correlator ratio downward. If, on the other hand, we plot the ratio of the three-point correlator to the leading exponential functions $\mathrm{e}^{-E_{\pi}t}$ and $\mathrm{e}^{-M_B(T-t)}$, a much longer plateau is evident as shown in Fig.~\ref{fig:corr_ratios} (right panel). This implies that the significant excited state contribution comes from the $B$ meson two-point function. The plateau represents $a_{\pi,0} V^{\mu}_{0,0} a^*_{B,0}$ in Eq.~\eqref{eq:C3pt}. In either case, the fit results capture the excited-state effects in the data very well. We emphasize that we do \rm{not} fit these correlator ratios. Rather, we use the simultaneous, multi-exponential fits to two-point and three-point correlators described earlier in this section for each ensemble.

Similar plots of the three-point function divided by the ground-state exponentials are shown in Figs.~\ref{fig:corr_ratios_beta4.17}, \ref{fig:corr_ratios_mpi} and~\ref{fig:corr_ratios_mh} for the lattice data obtained at the coarsest lattice, $\beta$ = 4.17. Here the data are shown for both temporal (left) and spatial (right) vector-current components for all available momentum insertions: $(0,0,0)$, $(0,0,1)$, $(0,1,1)$, $(1,1,1)$ in units of $2\pi/L$. Figs.~\ref{fig:corr_ratios_beta4.17} and~\ref{fig:corr_ratios_mpi} should be compared for the effect of different light quark masses, while Figs.~\ref{fig:corr_ratios_beta4.17} and~\ref{fig:corr_ratios_mh} should be compared for the effect of different heavy quark masses. In all cases, the fit results closely follow the lattice data.

The correlators computed on finer lattices are shown in Figs.~\ref{fig:corr_ratios_beta4.35} and~\ref{fig:corr_ratios_beta4.47}. General observations are the same as those on the coarse lattice, but we observe larger noise due to limited statistics, especially on the finest lattice at $\beta=4.47$ (Fig.~\ref{fig:corr_ratios_beta4.47}).

\subsection{Current renormalization}
\label{sec:renormalization}
For the lightest heavy quark mass, i.e., when $am_Q=am_c$, we find that it is sufficient to renormalize our currents using results from the massless coordinate space current correlators as described in Ref.~\cite{Tomii:2017cbt}. However, as discussed in Refs.~\cite{Hashimoto:2017wqo,Hashimoto:2018gld}, discretization effects arising from larger quark masses can lead to the renormalization constant $Z_V$ from vector currents $\bar{Q}\gamma_\mu Q$ deviating substantially from $1$. We therefore consider it prudent to use the matrix element $\langle B_s | \bar{Q}\gamma_\mu Q | B_s \rangle$ to partially renormalize our vector current alongside the massless renormalization results. (Here $B_s$ stands for the pseudoscalar state comprising the heavy quark $Q$ and the strange quark.)

By calculating three-point $B_s\to B_s$ correlators and demanding that the inserted temporal vector current matrix element is $1$---since it is conserved in the continuum---we can obtain the renormalization constant
\begin{equation}
  Z_{V_{QQ}}^{-1}=\langle B_s | \bar{Q}\gamma_0 Q | B_s \rangle.
\end{equation}
We then take the overall renormalization constant for the heavy-light current $Z_V=\sqrt{Z_{V_{QQ}}Z_{V_{qq}}}$ where the renormalization constant for light-light current are determined as
$Z_{V_{qq}}^{-1} = 1.047(10)$, $1.038(6)$, $1.031(5)$ at $\beta$ = 4.17, 4.35, 4.47, respectively~\cite{Tomii:2016xiv}.

We generated three-point correlators on each of the ensembles with the heavier of the available strange quark masses. On each ensemble and for each value of $am_Q>am_c$ we used smeared sources and sinks with time separation $T$. We averaged over two time sources that were separated by half the temporal extent of the lattice. The exception was on the finest ensemble for which we used only a single time source. We also generated two-point correlators with the same sources so that we could extract the required matrix element by
\begin{equation}
  \label{eq:renorm_ratio}
  \langle B_s | \bar{Q}\gamma_0 Q | B_s \rangle = \dfrac{C_{\mathrm{3pt}}(t)}{C_{\mathrm{2pt}}(T)}.
\end{equation}

\begin{figure}[tbp]
\includegraphics[width=0.62\textwidth]{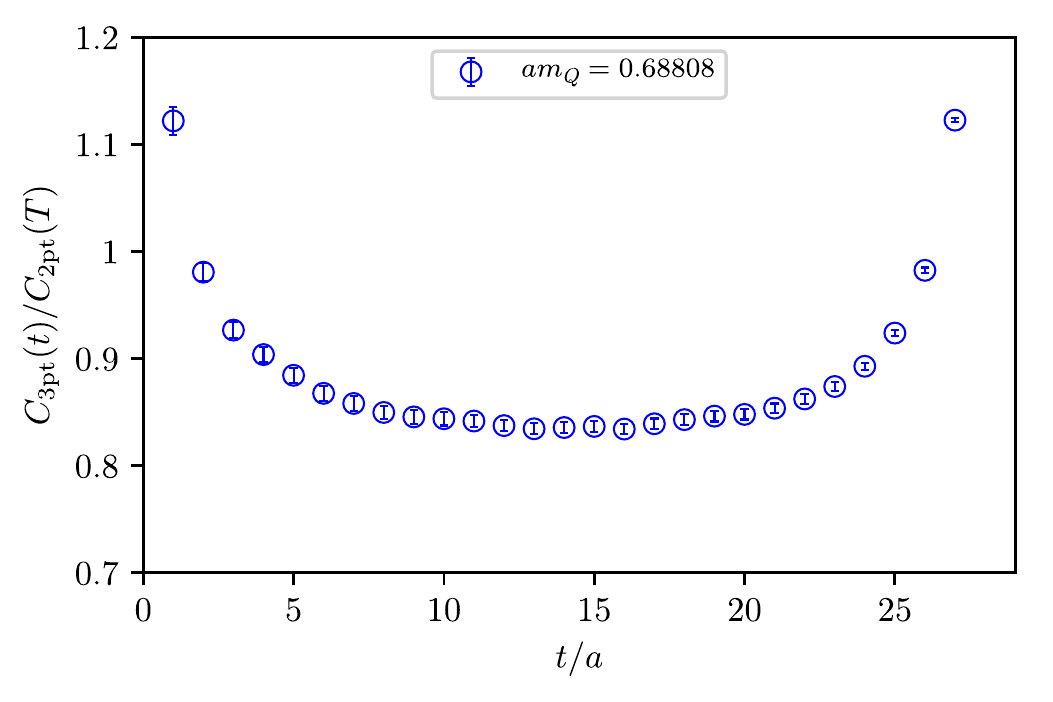}
\includegraphics[width=0.62\textwidth]{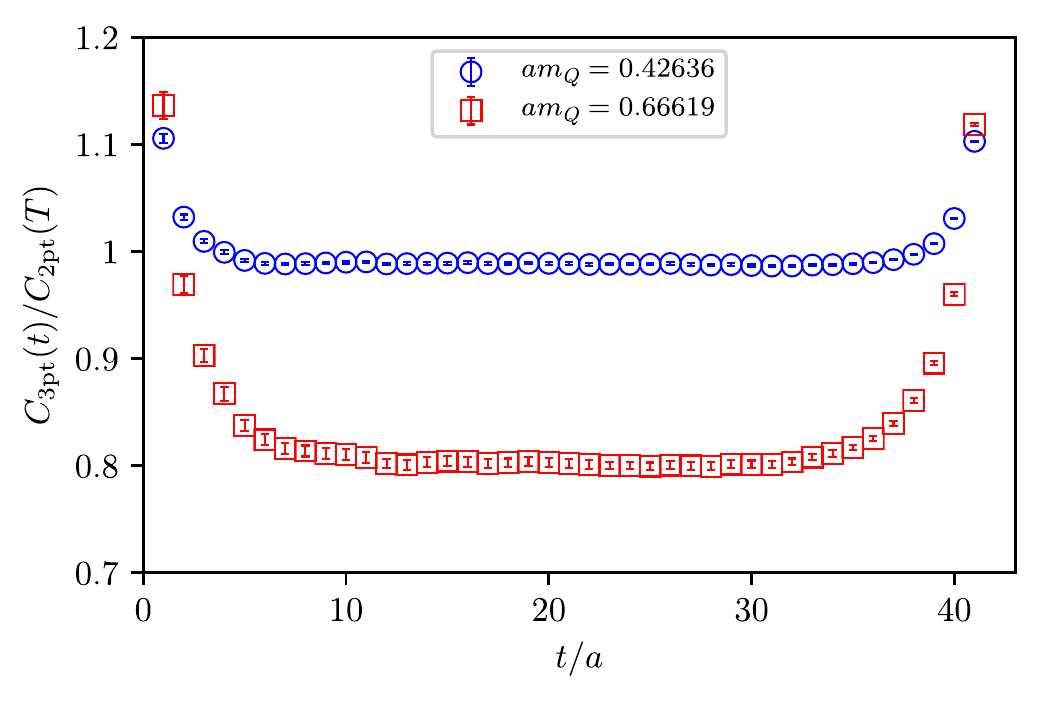}
\includegraphics[width=0.62\textwidth]{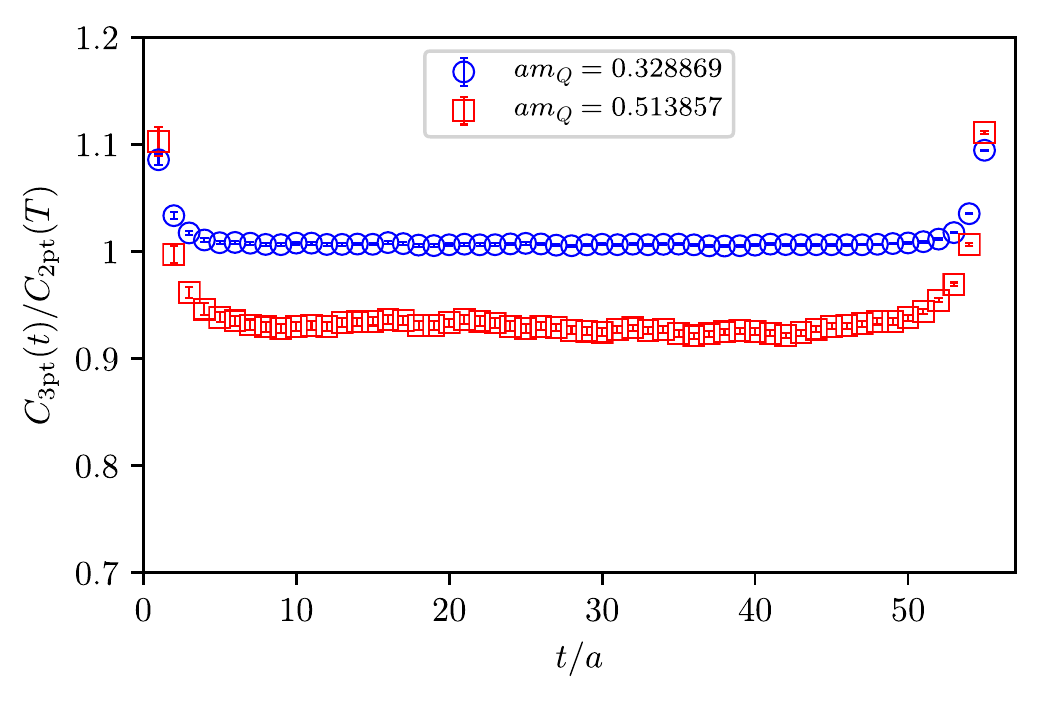}
\caption{Ratio of the $B_s\to B_s$ three-point correlators at time slices $t$ to the $B_s$ two-point correlators. The data is from the ensembles with $\beta=4.17$ and $am_l=0.007$ (top),  $\beta=4.35$ and $am_l=0.0042$ (middle), and $\beta=4.47$ (bottom). The heavy quark masses are shown in the plots. They correspond to $m_Q=1.25^2m_c$ and $1.25^4m_c$.}
\label{fig:renorm_ratio}
\end{figure}

We show plots of the ratio from Eq.~\eqref{eq:renorm_ratio} in Fig.~\ref{fig:renorm_ratio} for ensembles with $\beta=4.17$, 4.35 and 4.47. We are able to find plateaus in all cases and thus simply fit to a constant in these regions. Table~\ref{tab:zVhh_results} gives the results for $Z_{V_{QQ}}^{-1}$.

\setlength\tabcolsep{6pt}
\begin{table}[tbp]
  \begin{tabular}{c|cc|c|c|c}
    \hline
    \hline
    $\beta$ & $am_l$   & $am_Q$       & $[t_{\mathrm{min}},t_{\mathrm{max}}]$  & $Z^{-1}_{V_{QQ}}$ & $\chi^2/N_{\mathrm{dof}}$ \\
    \hline            
    \multirow{3}{*}{$4.17$}  & $0.019$  & $0.68808$   & $[11,14]$ & 0.8342(29) & 0.45 \\
            & $0.012$  &    $0.68808$         & $[11,14]$ & 0.8382(35) & 0.36  \\
    & $0.007$  &     $0.68808$        & $[11,14]$ & 0.8396(28) & 0.46 \\
    \hline
    \multirow{6}{*}{$4.35$}  & \multirow{2}{*}{$0.012$}  & $0.42636$   & $[8,21]$  & 0.9878(6)  & 0.38  \\
            &          & $0.66619$   & $[14,21]$ & 0.8013(24) & 0.73  \\
            & \multirow{2}{*}{$0.008$}  & $0.42636$   & $[8,21]$  & 0.9886(6)  & 0.83  \\
            &          & $0.66619$   & $[14,21]$ & 0.8031(27) & 0.89  \\
            & \multirow{2}{*}{$0.0042$} & $0.42636$   & $[8,21]$  & 0.9877(4)  & 0.73  \\
            &          & $0.66619$   & $[14,21]$ & 0.8020(26) & 0.37  \\
    \hline
    \multirow{2}{*}{$4.47$}  & \multirow{2}{*}{$0.003$}  & $0.328869$  & $[12,28]$ & 1.0062(7)  & 0.96 \\
            &          & $0.513857$  & $[10,28]$ & 0.9267(14) & 0.94 \\
    \hline
  \end{tabular}
  \caption{Results for the inverse of the heavy-heavy renormalization constant $Z^{-1}_{V_{QQ}}$ with statistical errors for each $\beta$ value. In columns two and three we give the light and heavy quark masses respectively. We provide the fit ranges we used in column four.}\label{tab:zVhh_results}
\end{table}

\section{Results}
\label{sec:latt_results}

\subsection{Global fit to form factors}
\label{subsec:global_fit}
In order to obtain the $B$ to $\pi$ form factors ${f_1(v\cdot p_\pi)+f_2(v\cdot p_\pi)}$ and ${f_2(v\cdot p_\pi)}$ at the physical quark masses and in the continuum limit, we perform a global fit.
The form factors are functions of $v\cdot p_\pi=E_\pi$, which should also be parametrized. We assume the energy dependence
of the form factor $f_1(v\cdot p_\pi)+f_2(v\cdot p_\pi)$ is described by a simple polynomial, and use a fit function
\begin{multline}\label{eq:f1_f2_fit}
  f_1(v\cdot p_\pi)+f_2(v\cdot p_\pi)=
    C_0
    \left(1+\sum^3_{n=1}C_{E^n}N^n_{E}E^n_\pi\right)
    \left(1+C_{\chi \mathrm{log}}\delta f^{B\to\pi}+C_{M^2_{\pi}}N_{M^2_{\pi}}M^2_\pi\right)\\
    \times \left(1+\frac{C_{m_Q}N_{m_Q}}{m_Q}\right)
    \left(1+C_{m_{s\bar{s}}^2}\delta m_{s\bar{s}}^2\right)\left(1+C_{a^2}(\Lambda_{\mathrm{QCD}}a)^2+C_{(am_Q)^2}(am_Q)^2\right).
\end{multline}
For $f_2(v\cdot p_\pi)$, since we expect a contribution from the vector meson ($B^*$) pole as described in Eq.~\eqref{eq:pole_dominance}, we use
\begin{multline}\label{eq:f2_fit}
  f_2(v\cdot p_\pi)=
    D_0
    \left[\frac{E_\pi}{E_\pi+\Delta_{B}}
    \left(1+D_{E_\pi}N_{E}E_\pi\right)\right]
    \left(1+D_{\chi \mathrm{log}}\delta f^{B\to\pi}+D_{M^2_\pi}N_{M^2_{\pi}} M^2_\pi\right)\\
    \times \left(1+\frac{D_{m_Q}N_{m_Q}}{m_Q}\right)
    \left(1+D_{m_{s\bar{s}}^2}\delta m_{s\bar{s}}^2\right)\left(1+D_{a^2}(\Lambda_{\mathrm{QCD}} a)^2+D_{(am_Q)^2}(am_Q)^2\right).
  \end{multline}
Here $C_x$ and $D_x$ are fit parameters, and $N_x$ are normalization constants that fix the units for energies and masses.
These have been chosen so that $C_x$ and $D_x$ are $\sim \mathcal{O}(1)$. We choose $N_{E}=1/(0.3\;\mathrm{GeV})$ and $N_{M^2_{\pi}}=1/(0.3\;\mathrm{GeV})^2$,
where $0.3\;\mathrm{GeV}$ is a typical pion mass/energy, and $N_{m_Q}=1\;\mathrm{GeV}^{-1}$.
We take $\Lambda_{\mathrm{QCD}}=0.5\;\mathrm{GeV}$.

The heavy quark mass dependence as an expansion in terms of $1/m_Q$ is justified because the form factors $f_1(v\cdot p_\pi)$
and $f_2(v\cdot p_\pi)$ can be defined even in the heavy quark limit. The $1/m_Q$ term represents the first correction to
that limit.

The strange quark masses have been set such that they are close to the physical strange quark mass. They are not, however,
exactly tuned so we include the term
\begin{align}
  \delta m_{s\bar{s}}^2 =&
  \left((m_{s\bar{s}}^{\mathrm{lat}})^2-(m_{s\bar{s}}^{\mathrm{phys}})^2\right)/(m_{s\bar{s}}^{\mathrm{phys}})^2 \nonumber\\
  \equiv &
  \left[\left(2\left(M^{\mathrm{lat}}_K\right)^2-\left(M^{\mathrm{lat}}_{\pi}\right)^2\right)
  -
           \left(2\left(M^{\mathrm{phys}}_K\right)^2-\left(M^{\mathrm{phys}}_{\pi}\right)^2\right)\right]\bigg/
           \left[2\left(M^{\mathrm{phys}}_K\right)^2-\left(M^{\mathrm{phys}}_{\pi}\right)^2\right] 
\end{align}
in our fit to take this into account. Having two strange quark masses on either side of the physical mass on the
coarsest lattice allows the fit to determine the coefficient of this correction term.

For the light quark mass dependence, we take the expectation from $SU(2)$ ``hard-pion'' chiral perturbation theory
for heavy-light mesons~\cite{Bijnens:2010ws} (see also Ref.~\cite{Becirevic:2002sc}):
\begin{align}
  \delta f^{B\to\pi}=-\frac{3}{4}(3g_{B^\ast B\pi}^2+1)\left(\frac{M_{\pi}}{4\pi f_{\pi}}\right)^2\ln\frac{M_\pi^2}{\Lambda^2},
\end{align}
plus a term linear in $M_\pi^2$. We take $1.0\;\mathrm{GeV}$ as the value for the scale $\Lambda$ appearing in the chiral logarithm terms. For the pion decay constant $f_\pi$ appearing in the denominator, we take $f_\pi = 130.4\;\mathrm{MeV}$. The logarithmic dependence expected from chiral effective theory is not very significant with the precision of the current lattice data, and in our main fit we use the result from $SU(2)$ chiral perturbation theory by fixing $C_{\chi \mathrm{log}}=D_{\chi \mathrm{log}}=1$. However, this depends on the value we choose for the $B^*B\pi$ coupling $g_{B^\ast B\pi}$. In the literature, the extracted values cover a wide range~\cite{Detmold:2012ge,Ohki:2008py,Becirevic:2009yb,Becirevic:2012pf,Bernardoni:2014kla,Flynn:2015xna}, and it is not straightforward to assess the overall uncertainty. On the other hand, it is not clear whether we can see the chiral log in our data. We therefore estimate the systematic uncertainty related to this term by setting $g_{B^\ast B\pi}=0.45$~\cite{Detmold:2012ge} as a representative value in our main fit with fixed $C_{\chi \mathrm{log}}$ and $D_{\chi \mathrm{log}}=1$, followed by another fit where $C_{\chi \mathrm{log}}$ and $D_{\chi \mathrm{log}}$ are free fit parameters. In this way the uncertainty due to $g_{B^\ast B\pi}$ is taken into account in the estimated systematic error. This is discussed in Sec.~\ref{sec:systematics}.

We assume that the leading discretization effects appear as an overall factor of the form $(1+C_{a^2}(\Lambda_{\mathrm{QCD}}a)^2+C_{(am_Q)^2}(am_Q)^2)$, and do not consider cross terms, e.g., a term of the form $E_\pi a^2$ with independent parameters. This is justified because the dependence on the lattice spacing is small. We confirmed that adding such cross terms with free fit parameters has a negligible effect on the fit. 

We find a good fit when simply fitting up to the quadratic term in pion energy for ${f_1(v\cdot p_\pi)+f_2(v\cdot p_\pi)}$, but larger uncertainties in data points with large pion momentum make it unclear what behaviour is exhibited at higher pion energies. For this reason we include the cubic term in Eq.~\eqref{eq:f1_f2_fit}. The impact of the choice to include this higher order term is minimal since, as we will discuss in Sec.~\ref{subsec:form_factor_shape}, when extrapolating towards $q^2=0$ we restrict our choice of synthetic data for the $z$-expansion to the region of pion energies covered by our simulation data. For $f_2(v \cdot p_\pi)$ we only include a term linear in the pion energy.

Fitting both form factors $f_1(v\cdot p_\pi)+f_2(v\cdot p_\pi)$ and $f_2(v\cdot p_\pi)$ simultaneously, we obtain a fit with $\chi^2/N_{\mathrm{dof}}=0.59$ ($N_{\mathrm{dof}}=182$). We use Bayesian priors for the fit parameters: we
choose $1.0\pm 2.0$ for $C_0$ and $D_0$, and $0.0\pm 2.0$ for all other fit parameters.
Results for the parameters from the global fit are given in Table~\ref{tab:global_fit_param}.

\begin{table}[tbp]
\begin{tabular}{c@{\hspace{10pt}}c@{\hspace{10pt}}c@{\hspace{10pt}}c@{\hspace{10pt}}c@{\hspace{10pt}}c@{\hspace{10pt}}c@{\hspace{10pt}}c@{\hspace{10pt}}c}
\hline
\hline
 $C_0$ & $C_E$ & $C_{E^2}$ & $C_{E^3}$ & $C_{M_\pi^2}$ & $C_{m_Q}$ & $C_{m^2_{s\bar{s}}}$ & $C_{a^2}$ & $C_{(am_Q)^2}$ \\
\hline
 $1.33(8)$ & $-0.37(5)$ & $0.09(3)$ & $-0.009(6)$ & $0.096(10)$ & $-0.34(6)$ & $0.06(4)$ & $-0.6(6)$ & $0.04(7)$ \\
\hline
 $D_0$ & $D_E$ & $D_{E^2}$ & $D_{E^3}$ & $D_{M_\pi^2}$ & $D_{m_Q}$ & $D_{m^2_{s\bar{s}}}$ & $D_{a^2}$ & $D_{(am_Q)^2}$ \\
\hline
 $0.52(5)$ & $-0.086(14)$ & -- & -- & $0.026(15)$ & $-0.09(14)$ & $0.10(7)$ & $0.03(1.09)$ & $0.14(12)$ \\
\hline
\hline
\end{tabular}
\caption{Our best fit parameters from the global fit functions [Eqs.~\eqref{eq:f1_f2_fit} and~\eqref{eq:f2_fit}].}
\label{tab:global_fit_param}
\end{table}

\begin{figure}[tbp]
\includegraphics[width=0.65\textwidth]{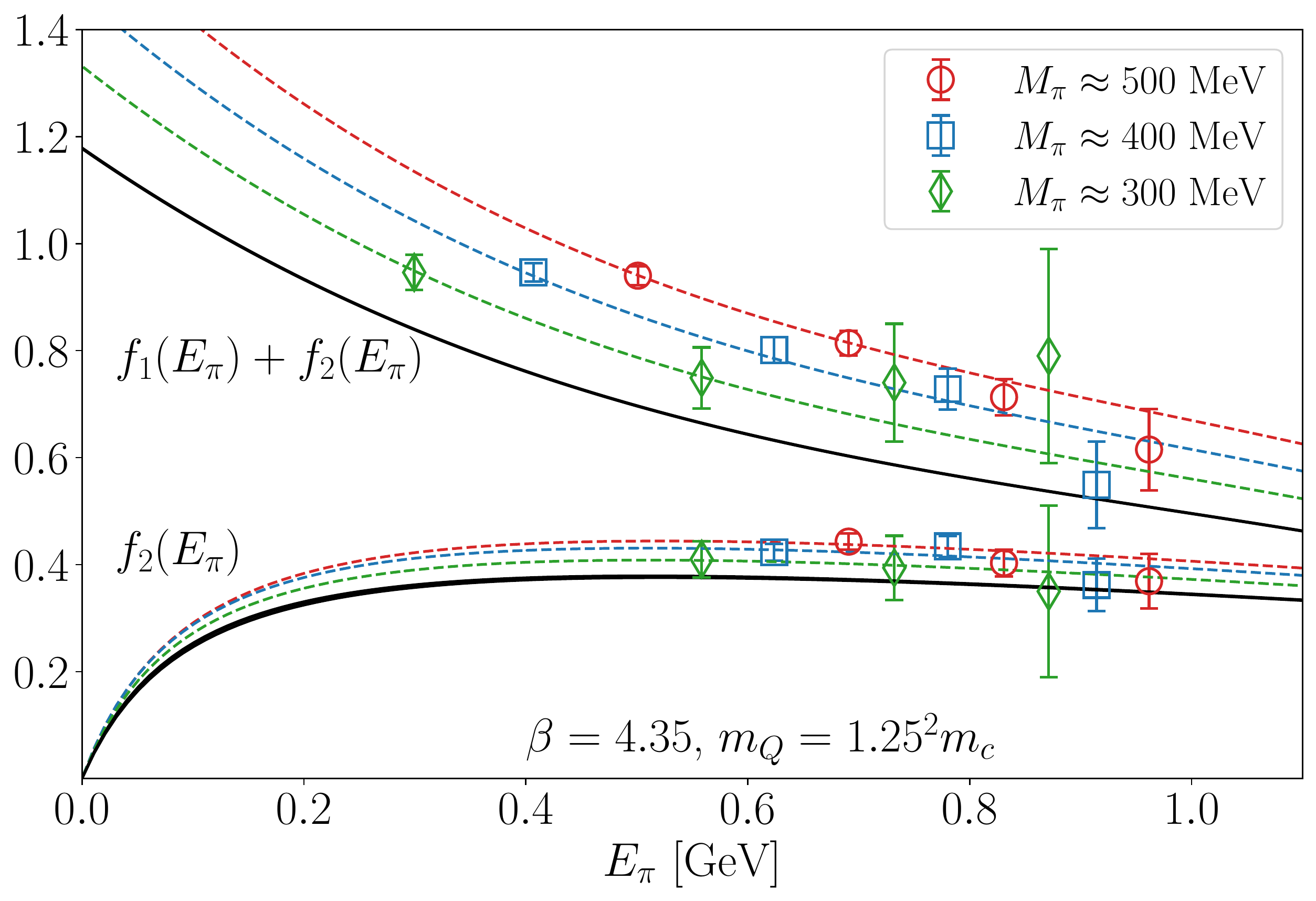}
\caption{
  Heavy-to-light form factors $f_1(v\cdot p_\pi)+f_2(v\cdot p_\pi)$ and $f_2(v\cdot p_\pi)$ at light quark masses corresponding to $M_\pi^2\simeq 300\;\mathrm{MeV}$ (diamonds), $400\;\mathrm{MeV}$ (squares) and $500\;\mathrm{MeV}$ (circles). Data at $\beta=4.35$ ($1/a\simeq 3.6\;\mathrm{GeV}$) and at $m_Q=1.56 m_c$.
  Dashed curves are the results of the global fit at corresponding pion masses, and the solid curves show the fit results extrapolated to the physical pion mass.
}
\label{fig:global_fit_f1_f2_extrapolation_mpi}
\end{figure}

\begin{figure}[tbp]
\includegraphics[width=0.65\textwidth]{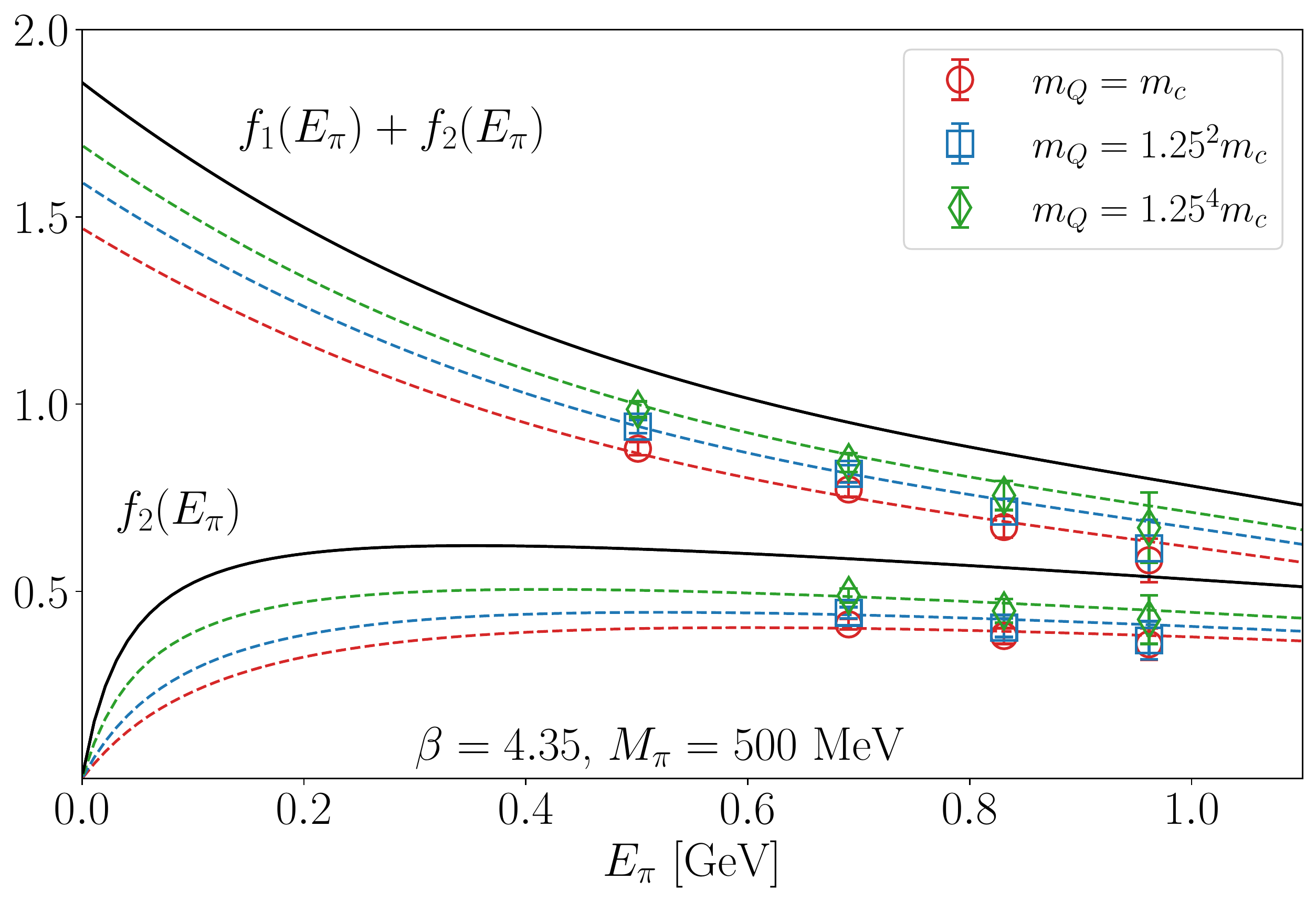}
\caption{
  Heavy-to-light form factors $f_1(v\cdot p_\pi)+f_2(v\cdot p_\pi)$ and $f_2(v\cdot p_\pi)$ at three different heavy quark masses: $m_c$ (diamonds); $1.56m_c$ (squares); and $2.44m_c$ (circles). Data at $\beta=4.35$ ($1/a\simeq 3.6\;\mathrm{GeV}$) and at a fixed light quark mass corresponding to $M_\pi\simeq 500\;\mathrm{MeV}$. Dashed curves are the results of the global fit at corresponding heavy quark masses, and the solid curves show the fit results extrapolated to the physical $b$ quark mass.
  }
\label{fig:global_fit_f1_f2_extrapolation_mh}
\end{figure}

\begin{figure}[tbp]
\includegraphics[width=0.65\textwidth]{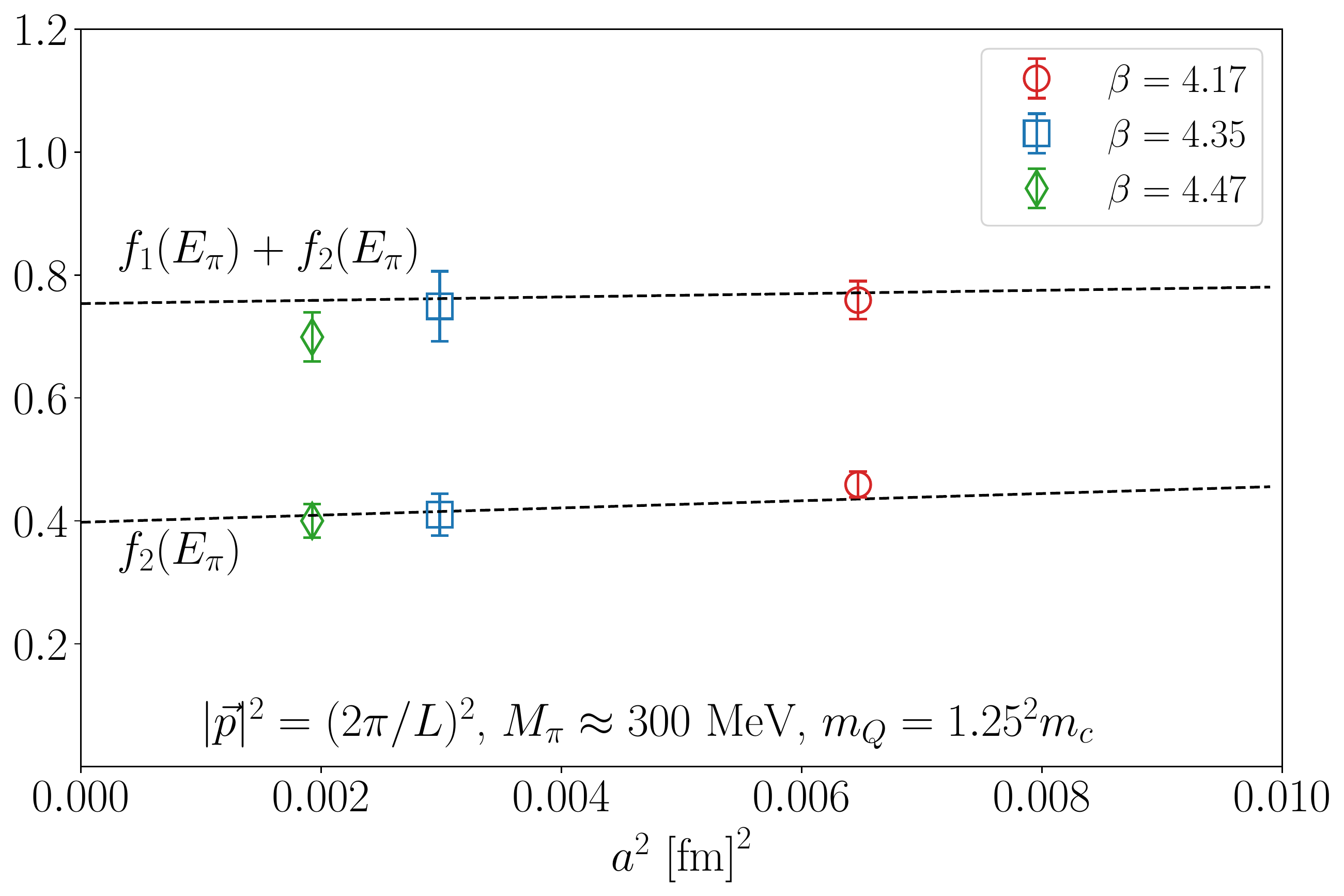}
\caption{
  Continuum extrapolation of the form factors $f_1(v\cdot p_\pi)+f_2(v\cdot p_\pi)$ and $f_2(v\cdot p_\pi)$ evaluated with a typical parameter choice:
  $\bm{p}_\pi^2=(2\pi/La)^2$ (note that the physical volumes of the three lattices are similar); $M_\pi\simeq 300\;\mathrm{MeV}$; and $m_Q=1.56m_c$.
  }
\label{fig:global_fit_f1_f2_extrapolation_a2}
\end{figure}

We illustrate the extrapolations in pion mass, heavy quark mass and lattice spacing in Figs.~\ref{fig:global_fit_f1_f2_extrapolation_mpi}, \ref{fig:global_fit_f1_f2_extrapolation_mh} and~\ref{fig:global_fit_f1_f2_extrapolation_a2}, respectively. Figure~\ref{fig:global_fit_f1_f2_extrapolation_mpi} shows the form factors $f_1(v\cdot p_\pi)+f_2(v\cdot p_\pi)$ and $f_2(v\cdot p_\pi)$ as functions of $v\cdot p_\pi=E_\pi$ computed at different light quark masses corresponding to $M_\pi = 300$, $400$ and $500\;\mathrm{MeV}$. The extrapolations to the chiral limit (or to the physical pion mass) are performed using the fit to Eqs.~\eqref{eq:f1_f2_fit} and~\eqref{eq:f2_fit}. One can see that the values of the form factors are rather stable as a function of the quark mass.
The data points are well described by the global fit shown by dashed curves. The thick curves represent the results corresponding to the physical pion mass.

The heavy quark mass extrapolation is demonstrated in Fig.~\ref{fig:global_fit_f1_f2_extrapolation_mh}, which shows the form factors computed for three different heavy quark masses: $m_Q$ = $m_c$; $1.56\times m_c$; and $2.44\times m_c$. We find that both form factors increase towards the physical $b$ quark mass. As represented in Eqs.~\eqref{eq:f1_f2_fit} and~\eqref{eq:f2_fit}, we extrapolate assuming dependence of the form $1/m_Q$, and the results at the physical point are represented by the solid curves. The systematic error due to the effect of neglecting a $1/m_Q^2$ term is estimated in the next subsection.

The continuum extrapolation is shown in Fig.~\ref{fig:global_fit_f1_f2_extrapolation_a2} for a typical parameter choice ($\bm{p}_\pi^2=(2\pi/La)^2$, $M_\pi\simeq 300\;\mathrm{MeV}$ and $m_Q=1.56\times m_c$). Since the physical volumes of the three lattices are similar, so too are the values of the physical momenta of the three points shown. We find that the continuum extrapolation in $a^2$ is also mild, even though a potentially significant discretization effect due to the heavy quark mass of the form $(am_Q)^2$ is expected. This is partly because the renormalization factor discussed in the previous section absorbs the bulk of the discretization effects. The global fit forms of Eqs.~\eqref{eq:f1_f2_fit} and~\eqref{eq:f2_fit} assume that the discretization effect applies as an overall factor $(1+C_{a^2}(\Lambda_{\mathrm{QCD}}a)^2+C_{(am_Q)^2}(am_Q)^2)$, independent of light quark masses and energies $v\cdot p_\pi=E_\pi$. This choice is justified because the dependence on each such parameter is small as we saw above. In principle this allows the global fit to discriminate between the $(am_Q)^2$ and $(\Lambda_{\mathrm{QCD}}a)^2$ effects; in practice, both terms in our fits return coefficients consistent with zero.

\begin{figure}[tbp]
\includegraphics[width=0.65\textwidth]{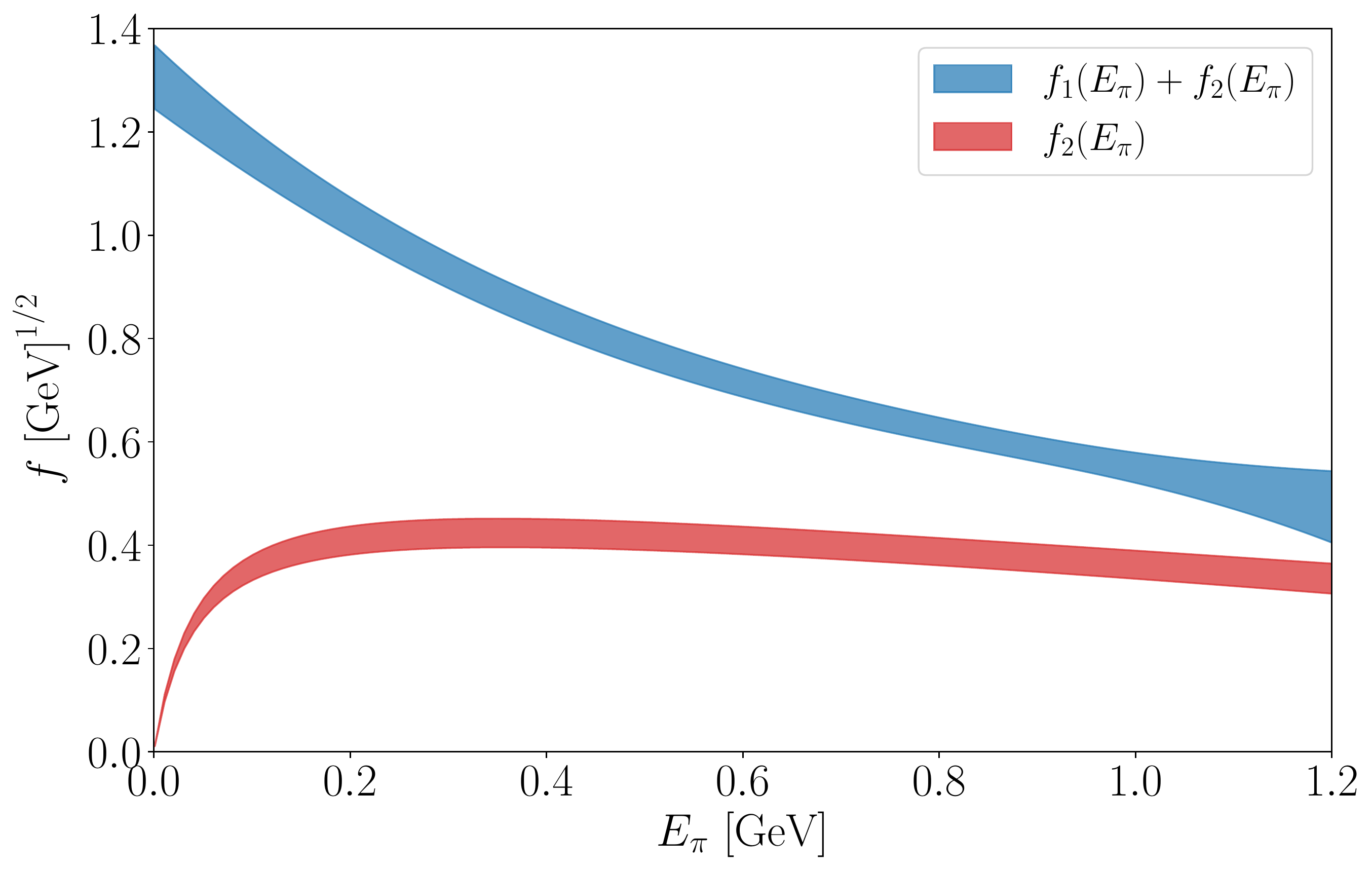}
\caption{Results of the global fit of the data for ${f_1(v\cdot p_\pi)+f_2(v\cdot p_\pi)}$ (upper curve) to Eq.~\eqref{eq:f1_f2_fit} and ${f_2(v\cdot p_\pi)}$ (lower curve) to Eq.~\eqref{eq:f2_fit}. The data from which these are obtained exist in the region $0.225\;\mathrm{GeV}<E_\pi<0.975\;\mathrm{GeV}$.}
\label{fig:global_fit_f1_f2}
\end{figure}

The final results for $f_1(v\cdot p_\pi)+f_2(v\cdot p_\pi)$ and $f_2(v\cdot p_\pi)$ at the physical quark masses and in the continuum limit are shown in Fig.~\ref{fig:global_fit_f1_f2} as a function of $v\cdot p_\pi=E_\pi$. The bands represent the one standard deviation regions with only the statistical uncertainties included.
The region that our lattice data cover is from $0.225~\mathrm{GeV}$ to $0.975~\mathrm{GeV}$.  The results outside of this region are obtained from the fit functions in Eqs.~\eqref{eq:f1_f2_fit} and~\eqref{eq:f2_fit}. In the soft pion limit, the form factor $f_2(v \cdot p_\pi)$ rapidly goes to zero as a result of the pole term included in Eq.~\eqref{eq:f2_fit}, and is not directly confirmed by the lattice data.

\subsection{Estimation of systematic errors}
\label{sec:systematics}
\begin{figure}
  \includegraphics[width=0.48\textwidth,valign=t]{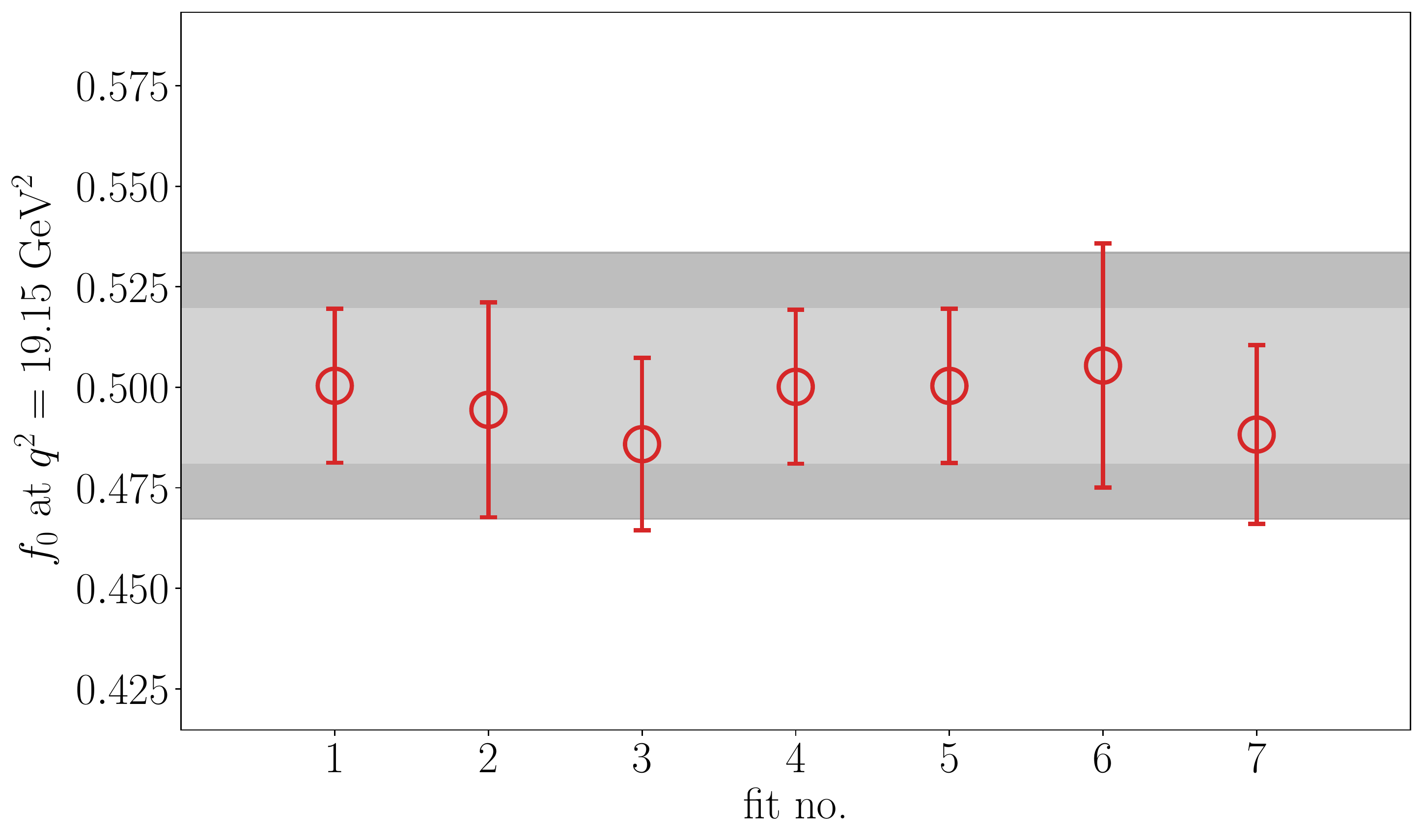}
  \includegraphics[width=0.48\textwidth,valign=t]{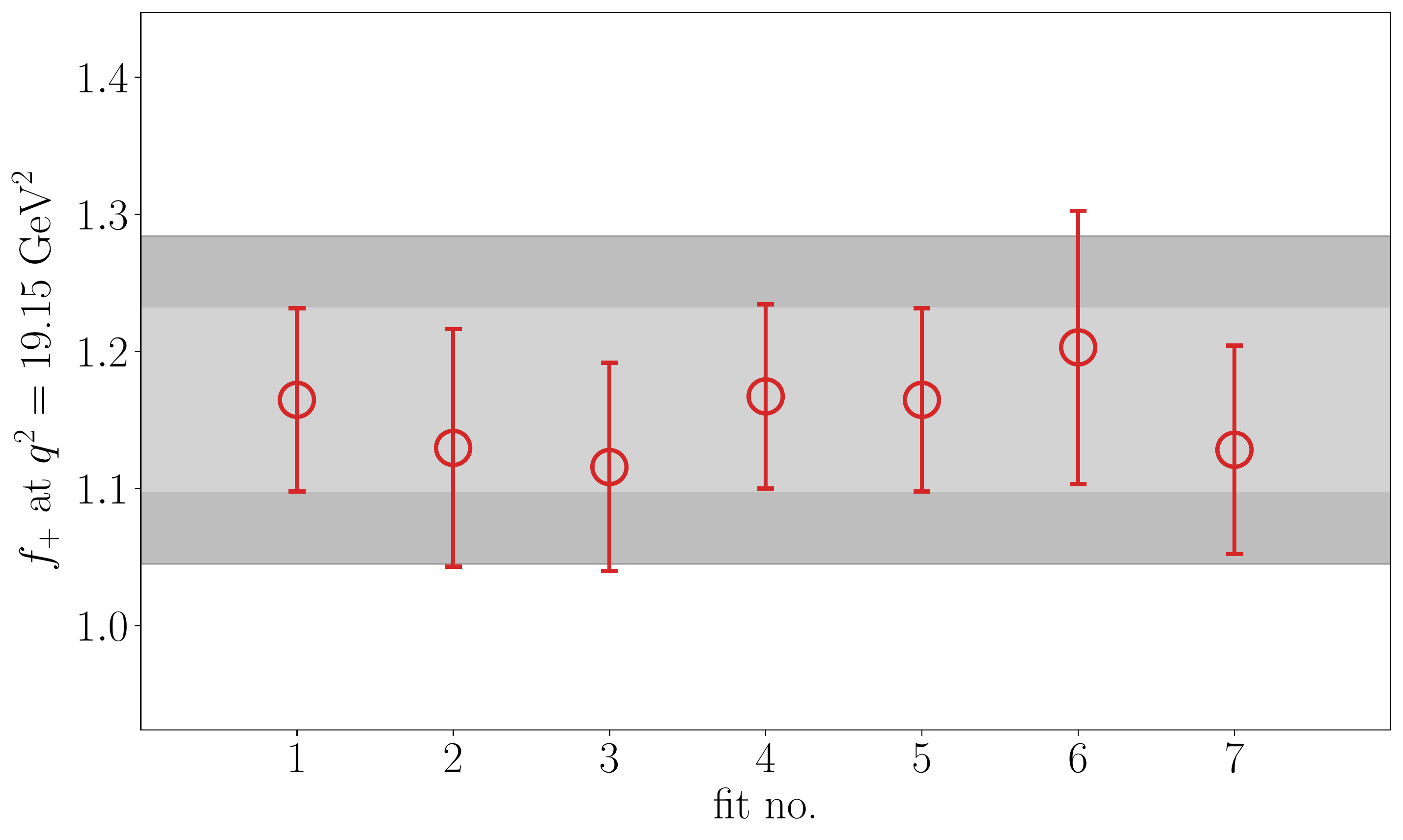}\\
  \includegraphics[width=0.48\textwidth,valign=t]{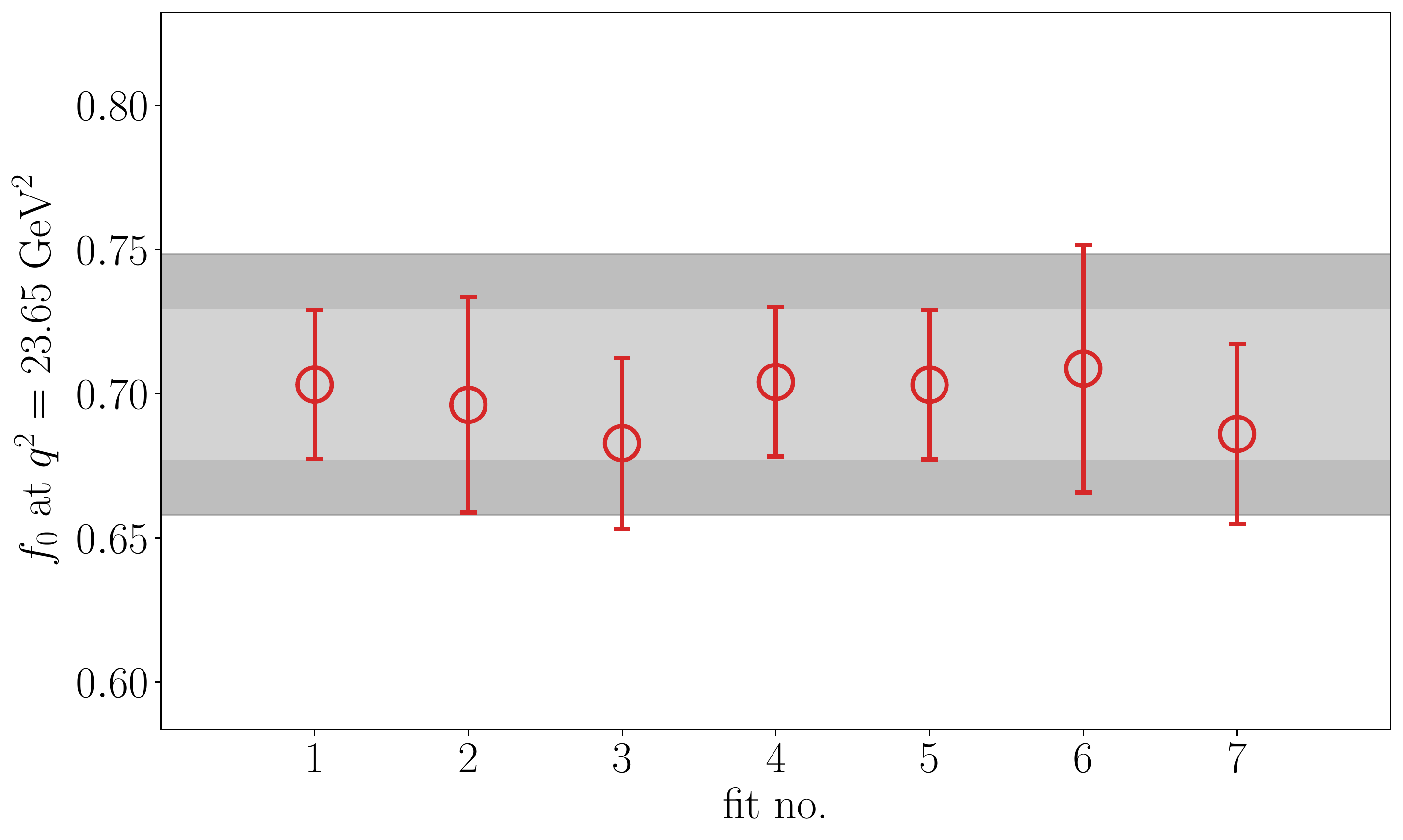}
  \includegraphics[width=0.48\textwidth,valign=t]{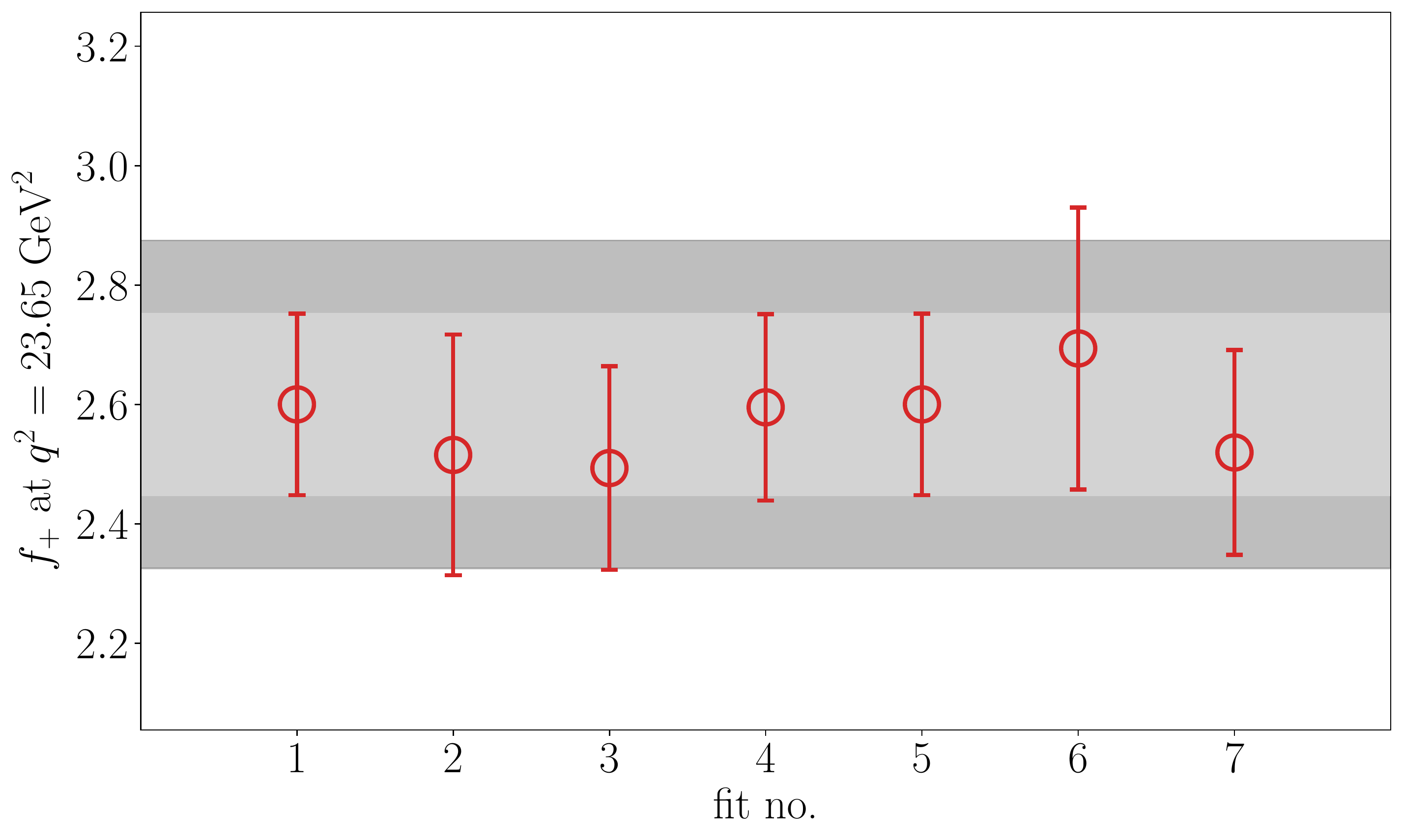}\\
  \includegraphics[width=0.48\textwidth,valign=t]{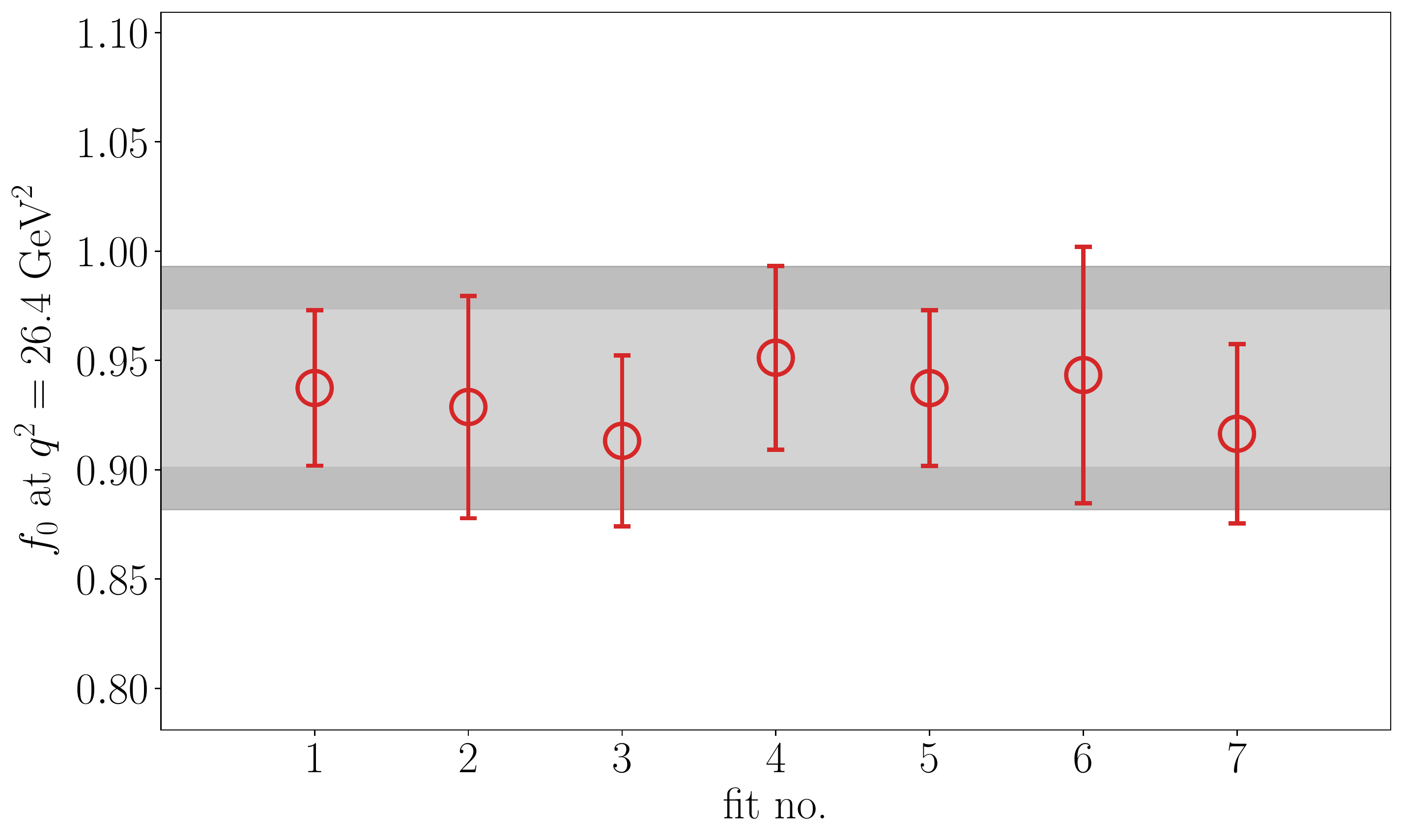}
  \includegraphics[width=0.48\textwidth,valign=t]{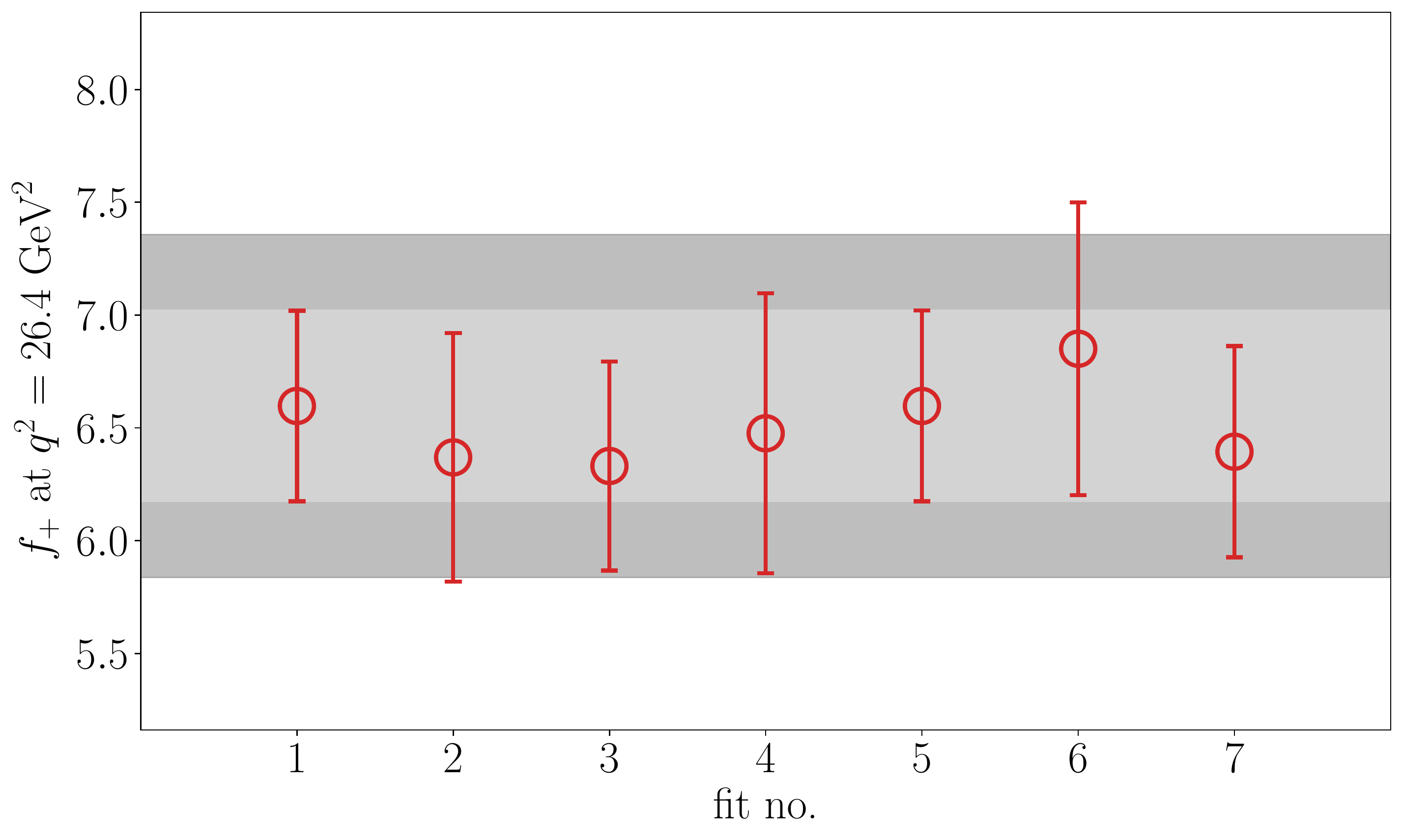}
  \caption{Results for $f_0(q^2)$ (left panels) and $f_+(q^2)$ (right panels) for each of our fits (numbered according to the list in the text) in tests of systematic uncertainties. The results at representative values of $q^2$ are shown: $19.15\;\mathrm{GeV}^2$, $23.65\;\mathrm{GeV}^2$ and $26.40\;\mathrm{GeV}^2$. The inner grey bands are the statistical errors only, while the outer bands show the total statistical plus systematic uncertainties.}
  \label{fig:systematic_fits}
\end{figure}

We now turn to the analysis of systematic uncertainties. To make an assessment of their impact we perform additional fits with particular terms added or amended. We attempt the following variations of the fits:

\begin{enumerate}
\item
  The original fit using the form of Eqs.~\eqref{eq:f1_f2_fit} and~\eqref{eq:f2_fit}.
\item
  Adding a $1/m^2_Q$ term such that the heavy quark dependence of $f_1(v\cdot p_\pi)+f_2(v\cdot p_\pi)$ is parametrized by a factor $(1+C_{m_Q}N_{m_Q}/m_Q+C_{m_Q^2}N^2_{m_Q}/m_Q^2)$ instead of $(1+C_{m_Q}N_{m_Q}/m_Q)$. Similarly for $f_2(v\cdot p_\pi)$.
\item
  Adding $M^4_\pi$ terms such that the pion mass dependence of $f_1(v\cdot p_\pi)+f_2(v\cdot p_\pi)$ is parametrized by a factor $(1+C_{\chi\mathrm{log}}\delta f^{B\to\pi}/(4\pi f_\pi)^2+C_{M_\pi^2}N_{M^2_{\pi}}M_\pi^2+C_{M_\pi^4}N^2_{M^2_{\pi}}M_\pi^4)$ instead of $(1+C_{\chi\mathrm{log}}\delta f^{B\to\pi}/(4\pi f_\pi)^2+C_{M_\pi^2}N_{M^2_{\pi}}M_\pi^2)$. Similarly for $f_2(v\cdot p_\pi)$.
\item
  Adding the next order term in $E_\pi$, so that $f_1(v\cdot p_\pi)+f_2(v\cdot p_\pi)$ is parametrized by $(1+\sum_{n=1}^4 C_{E^n}N^n_{E}E_\pi^n)$ and $f_2(v\cdot p_\pi)$ by $(1+\sum_{n=1}^2 D_{E^n}N^n_{E}E_\pi^n)$.
\item
  Adding $a^4$ terms such that the discretization effects of $f_1(v\cdot p_\pi)+f_2(v\cdot p_\pi)$ are parametrized by a factor $(1+C_{a^2}(\Lambda_{\mathrm{QCD}}a)^2+C_{a^4}(\Lambda_{\mathrm{QCD}}a)^4+C_{(am_Q)^2}(am_Q)^2)$ instead of $(1+C_{a^2}(\Lambda_{\mathrm{QCD}}a)^2+C_{(am_Q)^2}(am_Q)^2)$. Similarly for $f_2(v\cdot p_\pi)$.
\item
  Adding $(am_Q)^4$ terms such that the discretization effects of $f_1(v\cdot p_\pi)+f_2(v\cdot p_\pi)$ are parametrized by a factor $(1+C_{a^2}(\Lambda_{\mathrm{QCD}}a)^2+C_{(am_Q)^2}(am_Q)^2+C_{(am_Q)^4}(am_Q)^4)$ instead of $(1+C_{a^2}(\Lambda_{\mathrm{QCD}}a)^2+C_{(am_Q)^2}(am_Q)^2)$. Similarly for $f_2(v\cdot p_\pi)$.
\item
  Allowing the fit to determine the coefficient in front of the chiral log, i.e., letting $C_{\chi \mathrm{log}}$ and $D_{\chi \mathrm{log}}$ be free fit parameters instead of fixing them to $1$.
\end{enumerate}
We plot the result of these alternative fits in Fig.~\ref{fig:systematic_fits} at three representative $q^2$ values ($19.15\;\mathrm{GeV}^2$, $23.65\;\mathrm{GeV}^2$ and $26.40\;\mathrm{GeV}^2$) after converting to $f_0(q^2)$ and $f_+(q^2)$. The results are very stable across the alternative fits. The inner, lighter grey band shows our statistical uncertainty only, which is exactly the result from fit 1. The outer, darker grey band displays our total error, which includes systematic effects that come from the deviation from fit 1 of each of fits 2--7 added in quadrature.

\begin{figure}
  \includegraphics[width=0.65\textwidth,valign=t]{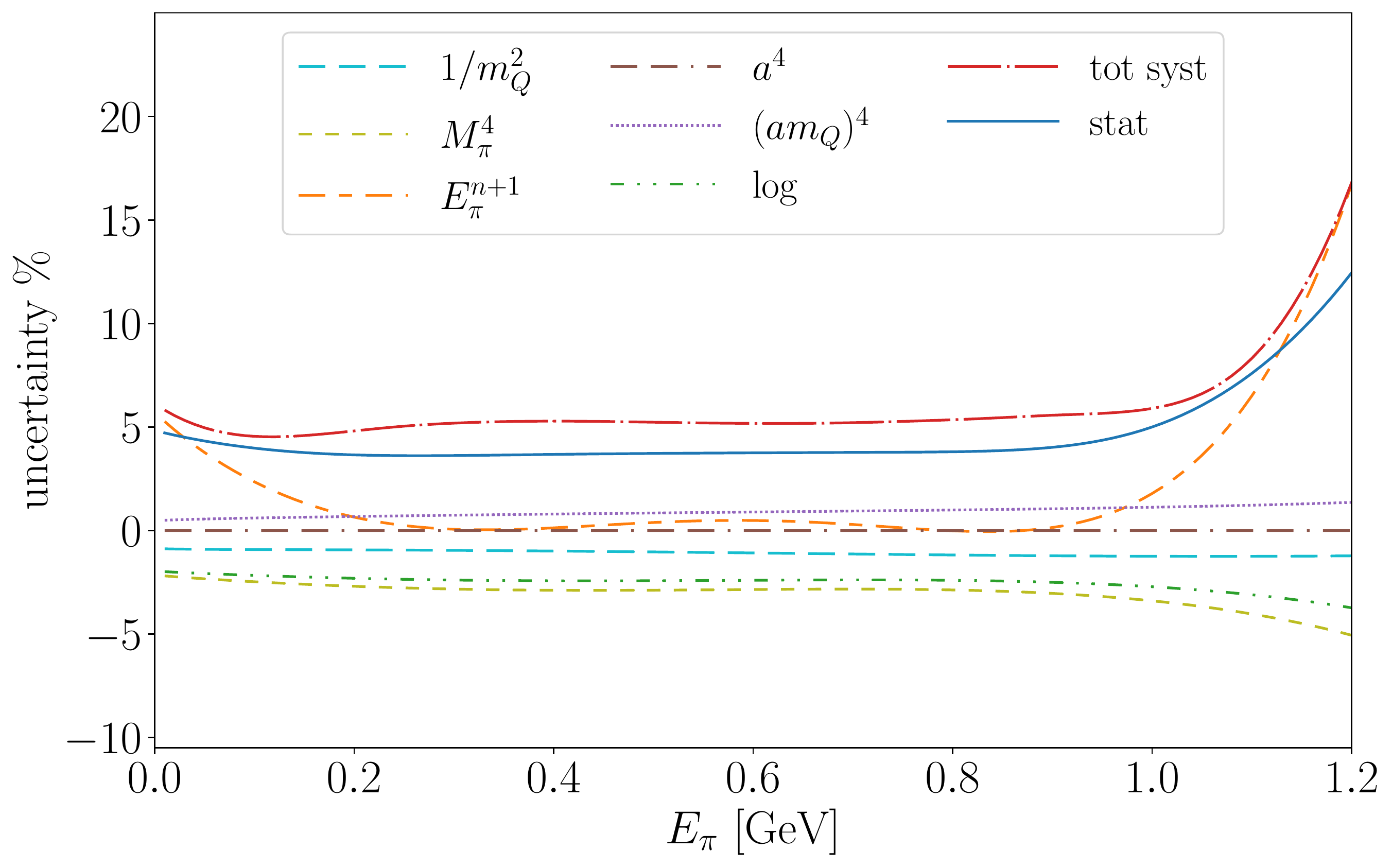}\\
  \includegraphics[width=0.65\textwidth,valign=t]{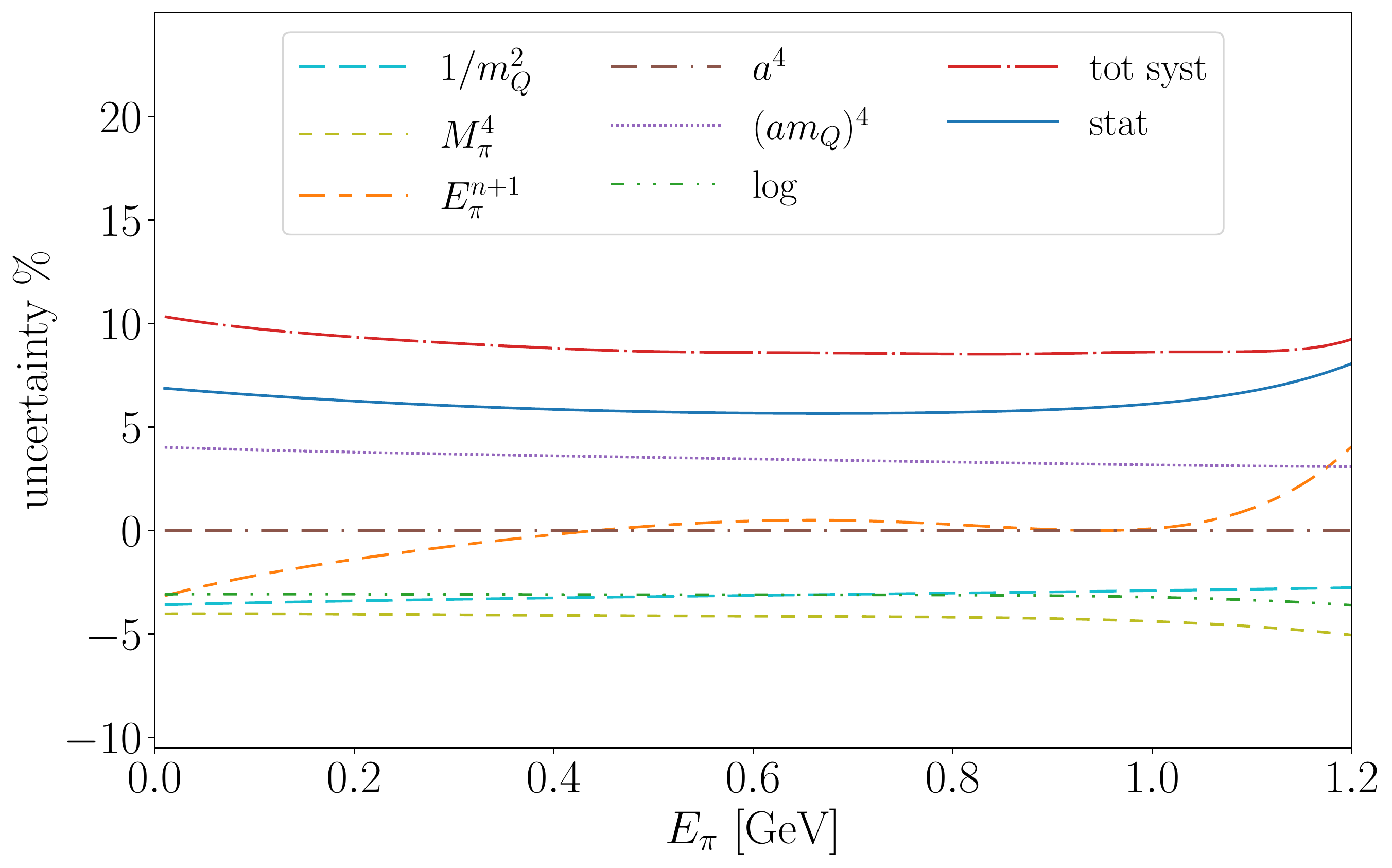}
  \caption{Systematic errors as a function of pion energy for form factors, $f_0$ (top panel) and $f_+$ (bottom panel). Individual contributions are estimated using the fits as described in the text. The total systematic errors (red dash-dot lines) are obtained by adding the other systematic uncertainties in quadrature. The statistical errors are depicted by the blue solid lines.}
  \label{fig:systematic_errorbudget}
\end{figure}

We also plot the systematic uncertainty coming from each of the listed sources as a function of pion energy in Fig.~\ref{fig:systematic_errorbudget} for both form factors $f_0$ and $f_+$, covering the $q^2$ range where we have data. They are estimated using the fits as described above, i.e., the deviation from the main fit ``1'' is plotted. They can therefore be either positive or negative. The estimated total systematic errors (red dash-dot lines), calculated from all sources of systematic uncertainty added in quadrature, are comparable in size to the statistical errors (blue solid lines).

\section{Form factors in the continuum and $|V_{ub}|$}
\label{sec:continuum_results_vub}
The differential decay width relates to the form factors $f_+(q^2)$ and $f_0(q^2)$, and $|V_{ub}|$ through
\begin{align}
  \dfrac{d\Gamma(B\to\pi\ell\nu)}{d q^2} = &\dfrac{G^2_F|V_{ub}|^2}{24\pi^3}
  \dfrac{(q^2-m^2_{\ell})^2\sqrt{E^2_{\pi}-M^2_{\pi}}}{q^4M^2_B} \nonumber\\
  \times&\left[\left(1+\dfrac{m^2_{\ell}}{2q^2} \right)M^2_B\left(E^2_{\pi}-M^2_{\pi}\right)|f_+(q^2)|^2 
  +\dfrac{3m^2_{\ell}}{8q^2}\left(E^2_{\pi}-M^2_{\pi}\right)^2|f_0(q^2)|^2\right],
\end{align}
where $G_F$ is Fermi's constant and $m_{\ell}$ is the lepton mass. For electrons and muons the terms
suppressed by $m^2_{\ell}$ can be discarded (at least at the current theoretical and experimental precision),
which means that the contribution from the scalar form factor $f_0$ can be neglected.
Thus the relation between the differential decay width and the form factors is reduced to a much simpler form:
\begin{equation}
  \dfrac{d\Gamma(B\to\pi\ell\nu)}{d q^2} = \dfrac{G^2_F|V_{ub}|^2}{24\pi^3}|\bm{p}_\pi(q^2)|^3|f_+(q^2)|^2,
\end{equation}
where the pion momentum in the rest frame of the $B$ meson is
\begin{equation}
  |\bm{p}_\pi|=\dfrac{1}{2M_B}\sqrt{\left(M_B^2+M^2_\pi-q^2\right)^2-4M^2_BM^2_\pi}.
\end{equation}
To determine $|V_{ub}|$, we need the branching fractions obtained from experiment as well as form factors from our lattice calculation. In this section, we first discuss the parametrization of the $q^2$ dependence of the form factors. The treatment of the experimental data is then described so that we can combine this with our lattice data to make a determination of $|V_{ub}|$.

\subsection{Form factor shape}
\label{subsec:form_factor_shape}
We use the $z$-parameter expansion to parametrize the shape of the form factors. Here, $q^2$ is transformed to a small parameter $z$ as
\begin{equation}
z(q^2,t_0) = \dfrac{\sqrt{t_+-q^2}-\sqrt{t_++t_0}}{\sqrt{t_+-q^2}+\sqrt{t_++t_0}},
\end{equation}
where $t_+=(M_{B^0}+M_{\pi^+})^2$ is the $B\pi$ threshold. We are free to choose the value of $t_0\leq t_+$. We choose $t_0=(M_B+M_\pi)(\sqrt{M_B}-\sqrt{M_\pi})^2$ since this symmetrizes the values of $z$ around 0, with $|z|<0.28$. 

For our final results of the $f_+(q^2)$ form factor we fit our data to the Bourrely-Caprini-Lellouch (BCL) expansion~\cite{Bourrely:2008za},
\begin{equation}
\label{eq:bcl}
f_+(q^2)=\dfrac{1}{1-q^2/M^2_{B^*}}\sum^{N_z-1}_{k= 0}b^+_k\left[z^k-(-1)^{k-N_z}\dfrac{k}{N_z}z^{N_z}\right],
\end{equation}
where the denominator on the right hand side addresses a pole at $q^2=M^2_{B^*}$. The second term in parentheses is introduced to ensure that the form factor satisfies the appropriate asymptotic form near the threshold. For the scalar form factor, $f_0(q^2)$, we fit to a simple series expansion in $z$:
\begin{equation}
\label{eq:bcl_f0}  
  f_0(q^2)=\sum^{N_z-1}_{k=0}b^0_kz^k.
\end{equation}

Another widely used parametrization is the Boyd-Grinstein-Lebed (BGL) expansion~\cite{Boyd:1995sq,Boyd:1997kz}:
\begin{equation}
  f_0(q^2)=\dfrac{1}{\mathcal{P}_0(q^2)\phi_0(q^2,t_0)}\sum_{n=0}^{N_z}a_n^0z^n,\quad f_+(q^2)=\dfrac{1}{\mathcal{P}_+(q^2)\phi_+(q^2,t_0)}\sum_{n=0}^{N_z}a_n^+z^n,
\end{equation}
where $\mathcal{P}_0(q^2)$ is usually taken as $1$, and the pole in the vector form factor is taken care of by the Blaschke factor $\mathcal{P}_+=z(q^2,M^2_{B^\ast})$. The outer functions $\phi_0(q^2,t_0)$ and $\phi_+(q^2,t_0)$ are analytic. Often, the outer function for the scalar form factor is chosen as $\phi_0(q^2,t_0)= 1$. For the vector form factor we follow~\cite{BaBar:2010efp} and choose
\begin{align}
  \phi_+(q^2,t_0)=&\sqrt{\dfrac{1}{32\pi \chi^{(0)}_J}}\left(\sqrt{t_+-q^2}+\sqrt{t_+-t_0}\right)
  \left(\sqrt{t_+-q^2}+\sqrt{t_+-t_-}\right)^{3/2} \nonumber \\
  &\times \left(\sqrt{t_+-q^2}+\sqrt{t_+}\right)^{-5}\dfrac{(t_+-q^2)}{(t_+-t_0)^{1/4}},
\end{align}
where $t_\pm=(M_{B^0}\pm M_{\pi^+})^2$, $t_0=0.65t_-$ and $\chi^{(0)}_J=6.9\times 10^{-4}\mathrm{ GeV}^{-2}$. 
Note that the choice of $t_0$ differs between the BCL and BGL $z$-expansion parametrizations in our analysis. Although our final results use the BCL parametrization, we confirmed that the BGL parametrization produces entirely consistent results.

The coefficients of the BCL ansatze in Eqs.~\eqref{eq:bcl} and~\eqref{eq:bcl_f0} obey the unitarity constraint~\cite{Boyd:1997qw, Bourrely:2008za}
\begin{equation}
  \label{eq:unitarity_bounds}
\sum_{m,n=0}^{N_z}B_{mn}b_mb_n \lesssim 1.
\end{equation}
This holds for both $b^+_k$ and $b^0_k$. The coefficients $B_{mn}$ are symmetric in the indices, $B_{mn}=B_{nm}$, and satisfy
the relation $B_{mn}=B_{0|m-n|}$. They depend on the choice of $t_0$, and we list them for our choice
$t_0=(M_B+M_{\pi})(\sqrt{M_B}-\sqrt{M_{\pi}})^2$ for both form factors $f_+$ and $f_0$ in Table~\ref{tab:unitarity_bound_constants}.
We do not implement these constraints explicitly in our fits, but we do check that they are satisfied by our results.

\begin{table}[tbp]
\begin{tabular}{c|cccccc}
\hline
\hline
& $B_{00}$ & $B_{01}$ & $B_{02}$ & $B_{03}$ & $B_{04}$ & $B_{05}$ \\
\hline
$f_0$ & $0.1032$ & $0.0408$ & $-0.0357$ & $-0.0394$ & $-0.0195$ & $-0.0055$\\
$f_+$ & $0.0198$ & $0.0042$ & $-0.0109$ & $-0.0059$ & $-0.0002$ & $0.0012$ \\
\hline
\hline
\end{tabular}
\caption{Constants used to estimate the unitarity bound for the BCL ansatz, taken from Refs.~\cite{Bourrely:2008za} and~\cite{FermilabLattice:2015mwy}.}
\label{tab:unitarity_bound_constants}
\end{table}

From the results of the global fit, we generate synthetic data for a range of $q^2$ values. Note that we have six degrees of freedom left after the extrapolations, so we can pick six data points (choosing more would result in a singular correlation matrix). We choose to generate three data points for each $f_+(q^2)$ and $f_0(q^2)$ at $q^2$ values $q^2_1=19.15$~GeV$^2$, $q^2_2=23.65$~GeV$^2$, and $q^2_3=26.40$~GeV$^2$. We pick these so that they are approximately evenly spaced in $z$. The values of the form factors are given in Table~\ref{tab:synthetic_lattice_points} together with the statistical and systematic errors at each point. The correlation matrices of the statistical and systematic errors are provided in Table~\ref{tab:lattdata_corr}. The systematic covariance matrix is calculated as follows. For each reference $q^2$ value, we first add all systematic effects listed in Sec.~\ref{sec:systematics} in quadrature, including correlations between different effects. We can then calculate the (statistical) correlations between the total systematic effects (for both form factors $f_+$ and $f_0$) at different reference $q^2$ values.

\newcommand\mc[1]{\multicolumn{1}{c}{#1}} 

\begin{table}[tbp]
\begin{tabular}{c|rrrrrr}
\hline
\hline
& \mc{$f_+(q^2_1)$} & \mc{$f_+(q^2_2)$} & \mc{$f_+(q^2_3)$} & \mc{$f_0(q^2_1)$} & \mc{$f_0(q^2_2)$} & \mc{$f_0(q^2_3)$} \\
\hline
 mean      & $1.165$ & $2.600$ & $6.597$ & $0.500$ & $0.703$ & $0.937$ \\
 stat. err & $0.067$ & $0.152$ & $0.423$ & $0.019$ & $0.026$ & $0.036$ \\
 syst. err & $0.099$ & $0.229$ & $0.631$ & $0.027$ & $0.037$ & $0.043$ \\
\hline
 tot. err  & $0.120$ & $0.275$ & $0.760$ & $0.033$ & $0.045$ & $0.056$ \\
\hline
\hline
\end{tabular}
\caption{Synthetic data points for $f_+(q^2)$ and $f_0(q^2)$ at $q^2_1=19.15$~GeV$^2$, $q^2_2=23.65$~GeV$^2$, and $q^2_3=26.40$~GeV$^2$. Their statistical and systematic errors are listed together with the total errors estimated by adding them in quadrature.}
\label{tab:synthetic_lattice_points}
\end{table}

\begin{table}[tbp]
\begin{tabular}{c|rrrrrr}
\hline
\hline
& \mc{$f_+(q^2_1)$} & \mc{$f_+(q^2_2)$} & \mc{$f_+(q^2_3)$} & \mc{$f_0(q^2_1)$} & \mc{$f_0(q^2_2)$} & \mc{$f_0(q^2_3)$}  \\
\hline
$f_+(q^2_1)$ & $1.000$ & $0.957$ & $0.901$ & $0.799$ & $0.728$ & $0.663$ \\
$f_+(q^2_2)$ & $0.957$ & $1.000$ & $0.989$ & $0.758$ & $0.720$ & $0.662$ \\
$f_+(q^2_3)$ & $0.901$ & $0.989$ & $1.000$ & $0.708$ & $0.682$ & $0.639$ \\
$f_0(q^2_1)$ & $0.799$ & $0.758$ & $0.708$ & $1.000$ & $0.971$ & $0.921$ \\
$f_0(q^2_2)$ & $0.728$ & $0.720$ & $0.682$ & $0.971$ & $1.000$ & $0.943$ \\
$f_0(q^2_3)$ & $0.663$ & $0.662$ & $0.639$ & $0.921$ & $0.943$ & $1.000$ \\
\hline
\end{tabular}

\begin{tabular}{c|rrrrrrrrrr}
\hline
& \mc{$f_+(q^2_1)$} & \mc{$f_+(q^2_2)$} & \mc{$f_+(q^2_3)$} & \mc{$f_0(q^2_1)$} & \mc{$f_0(q^2_2)$} & \mc{$f_0(q^2_3)$}  \\
\hline
$f_+(q^2_1)$ & $1.000$ & $0.996$ & $0.969$ & $0.761$ & $0.675$ & $0.692$ \\
$f_+(q^2_2)$ & $0.996$ & $1.000$ & $0.981$ & $0.737$ & $0.650$ & $0.663$ \\
$f_+(q^2_3)$ & $0.969$ & $0.981$ & $1.000$ & $0.682$ & $0.590$ & $0.604$ \\
$f_0(q^2_1)$ & $0.761$ & $0.737$ & $0.682$ & $1.000$ & $0.992$ & $0.996$ \\
$f_0(q^2_2)$ & $0.675$ & $0.650$ & $0.590$ & $0.992$ & $1.000$ & $0.996$ \\
$f_0(q^2_3)$ & $0.692$ & $0.663$ & $0.604$ & $0.996$ & $0.996$ & $1.000$ \\
\hline
\hline
\end{tabular}
\caption{Statistical (upper panel) and systematic (lower panel) correlation matrix for the synthetic data points at $q^2_1=19.15\;\mathrm{GeV}^2$, $q^2_2=23.65\;\mathrm{GeV}^2$, and $q^2_3=26.40\;\mathrm{GeV^2}$.}
\label{tab:lattdata_corr}
\end{table}

\begin{figure}[tbp]
\centering
\includegraphics[width=0.6\textwidth]{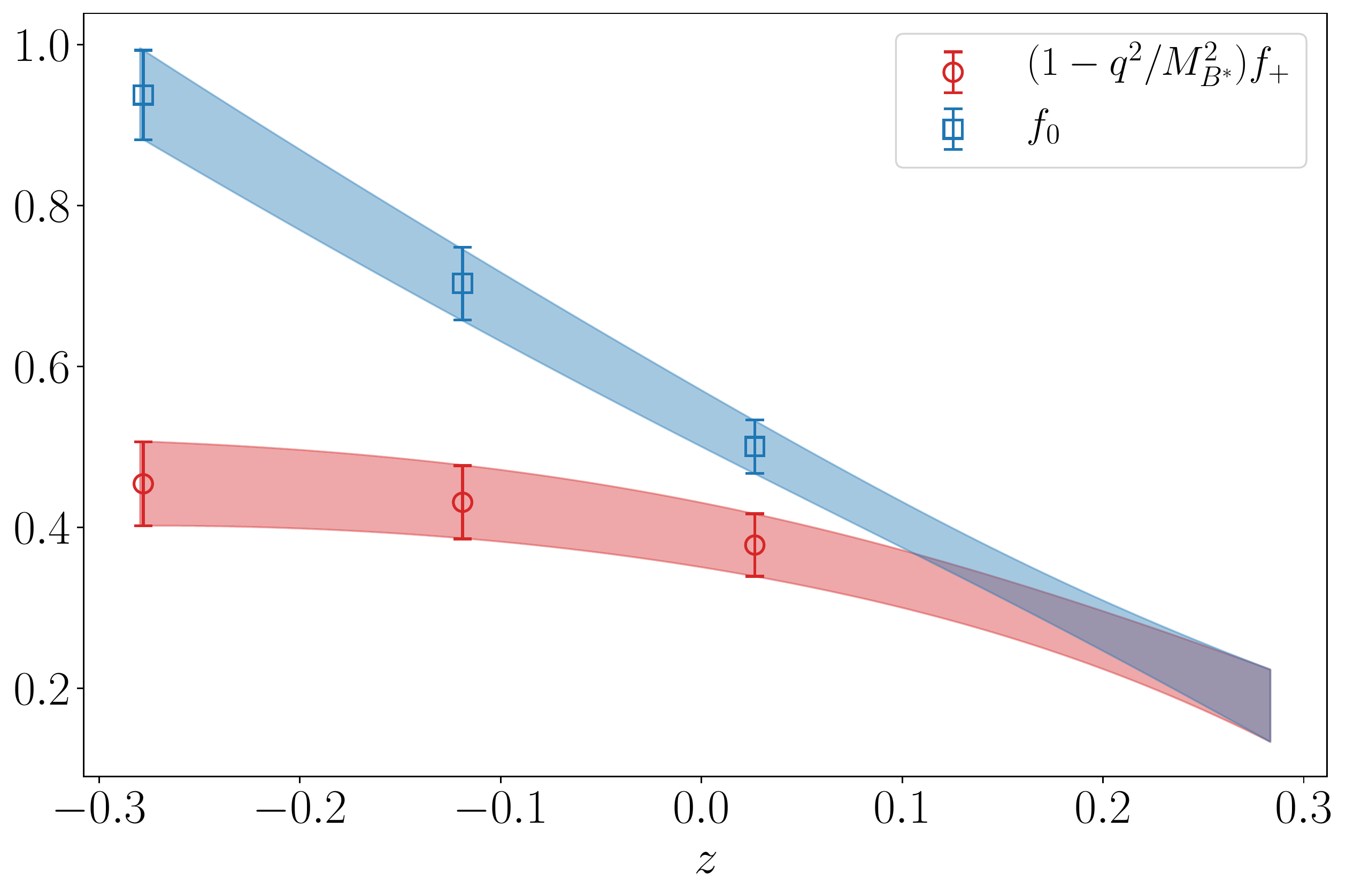}
\caption{Form factors using the BCL form of the $z$-parameter expansion. Lattice data for $f_0$ (blue squares) and $(1-q^2/M^2_{B^*})f_+$ (red circles) are shown with corresponding fit bands covering the entire $z$ region.}
\label{fig:fp_f0_vs_z}
\end{figure}

Our results for a fit to the BCL form of the $z$-expansion are given in Table~\ref{tab:form_factors_lattice_only}. The correlation matrix of the resulting parameters $b_k^+$ and $b_k^0$ are in Table~\ref{tab:app_lattice_only_cov}. We do not use priors in this fit. We obtain a good fit when the order of the polynomial is chosen as $N_z=3$. Here we impose the kinematic constraint $f_+(0)=f_0(0)$, i.e., we have six data points and five fit parameters. If we do not include the constraint then we have six data points and six fit parameters so cannot use $\chi^2/N_{\mathrm{dof}}$ as a measure of goodness of the fit. The fit result, however, remains unchanged. Although we do not impose them explicitly, we find that the unitarity constraints from Eq.~\eqref{eq:unitarity_bounds} are satisfied and we get $0.034(16)$ and $0.122(44)$ for $f_+$ and $f_0$, respectively. We find that $N_z=2$ is insufficient for a good fit. We also test fitting the form factor $f_+(q^2)$ alone using five synthetic data points. This makes very little difference to the $f_+(q^2)$ results. We plot results of the form factors across the entire $z$ range in Fig.~\ref{fig:fp_f0_vs_z}. The blue squares show $f_0$ and the red circles show $(1-q^2/M^2_{B^*})f_+$, while the bands are their corresponding fit results.

\begin{table}[tbp]
  \setlength{\tabcolsep}{2pt}
  \begin{tabular}{ccccc}
    \hline
    \hline
 \multicolumn{1}{c}{$b^+_0$} & \multicolumn{1}{c}{$b^+_1$} & \multicolumn{1}{c}{$b^+_2$} & \multicolumn{1}{c}{$b^0_1$} & \multicolumn{1}{c}{$b^0_2$} \\
\hline
 $0.391(40)$ & $-0.450(92)$ & $-0.92(29)$ & $-1.35(11)$ & $0.33(31)$ \\
\hline
  \end{tabular}
  \caption{Fit results from the BCL $z$-expansion parametrization with $N_z=3$ on our synthetic lattice data.
    Coefficient $b^0_0$ is fixed by the kinematic constraint $f_+(0)=f_0(0)$. The value is $b^0_0=0.535(35)$.}
  \label{tab:form_factors_lattice_only}
\end{table}

\begin{table}[tbp]
\begin{tabular}{c|rrrrr}
\hline
\hline
 & $b^+_0$ & $b^+_1$ & $b^+_2$ & $b^0_1$ & $b^0_2$ \\
\hline
$b^+_0$ & $1.000$  & $-0.515$ & $-0.281$ & $-0.100$ & $0.102$ \\
$b^+_1$ & $-0.515$ & $1.000$  & $0.496$  & $0.447$  & $0.531$ \\
$b^+_2$ & $-0.281$ & $0.496$  & $1.000$  & $0.606$  & $0.790$ \\
$b^0_1$ & $-0.100$ & $0.447$  & $0.606$  & $1.000$  & $0.638$ \\
$b^0_2$ & $0.102 $ & $0.531$  & $0.790$  & $0.638$  & $1.000$ \\
\hline
\hline
\end{tabular}
\caption{Correlation matrix from the $z$-expansion fit to our synthetic lattice data only with $N_z=3$ using the BCL parametrization. The constraint $f_+(0)=f_0(0)$ has been applied (this determines $b^0_0$).}
\label{tab:app_lattice_only_cov}
\end{table}

\begin{figure}
  \centering
  \includegraphics[width=0.49\textwidth]{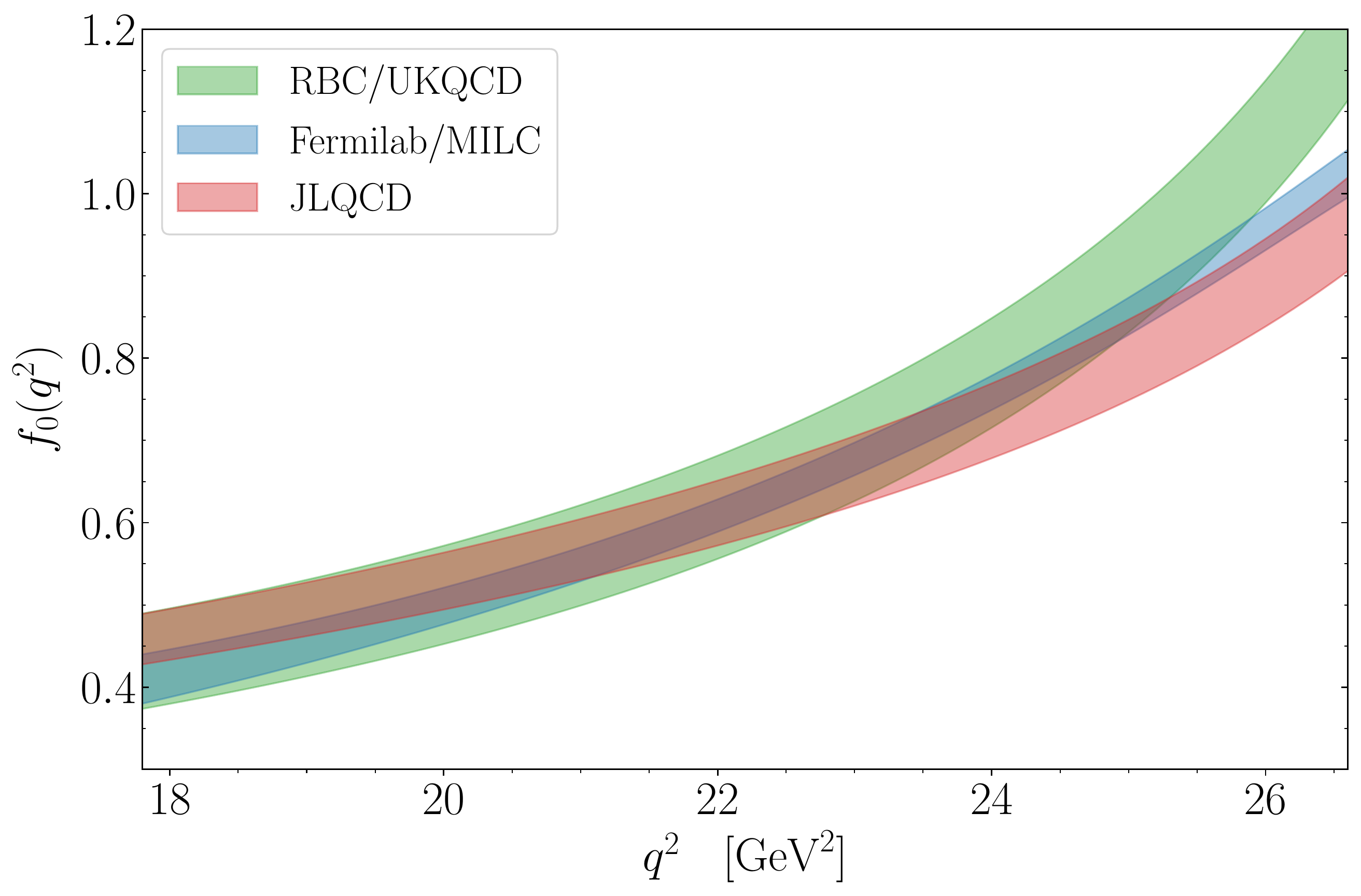}
  \includegraphics[width=0.49\textwidth]{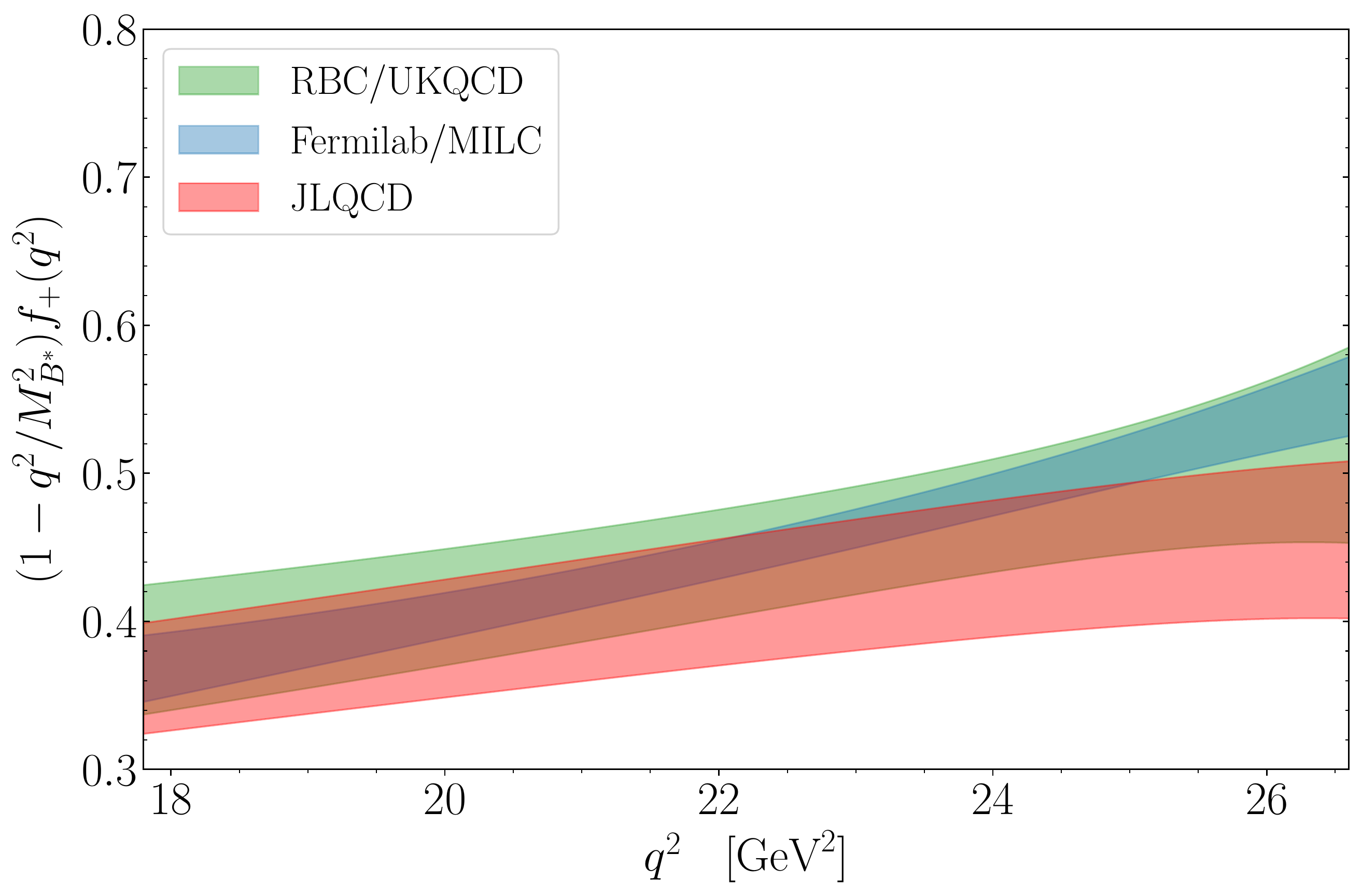}
\caption{Comparison of the physical form factors $f_0(q^2)$ and $f_+(q^2)$ with results from other lattice QCD calculations. Results from the RBC and UKQCD Collaborations are from Ref.~\cite{Flynn:2015mha} and results from the Fermilab Lattice and MILC Collaborations are from Ref.~\cite{FermilabLattice:2015mwy}.}
\label{fig:f0_fp_JLQCD_RBC_MILC}
\end{figure}

We can compare the form factors $f_0(q^2)$ and $f_+(q^2)$ to the results from other lattice QCD calculations when both statistical and systematic uncertainties are included.
Results from the RBC and UKQCD Collaborations~\cite{Flynn:2015mha} and the Fermilab lattice and MILC Collaborations~\cite{FermilabLattice:2015mwy} are plotted alongside our results in Fig.~\ref{fig:f0_fp_JLQCD_RBC_MILC}.
We restrict this comparison to the $q^2$ region that approximately corresponds to the inserted pion momentum in the lattice calculations and find general agreement for both form factors.
Near $q^2_{\mathrm{max}}$ there are slight discrepancies with RBC/UKQCD for $f_0(q^2)$ and Fermilab/MILC for $f_+(q^2)$.
This may hint at some systematic effects, although the statistical significance is limited.

\begin{figure}[tbp]
\centering
\includegraphics[width=0.6\textwidth]{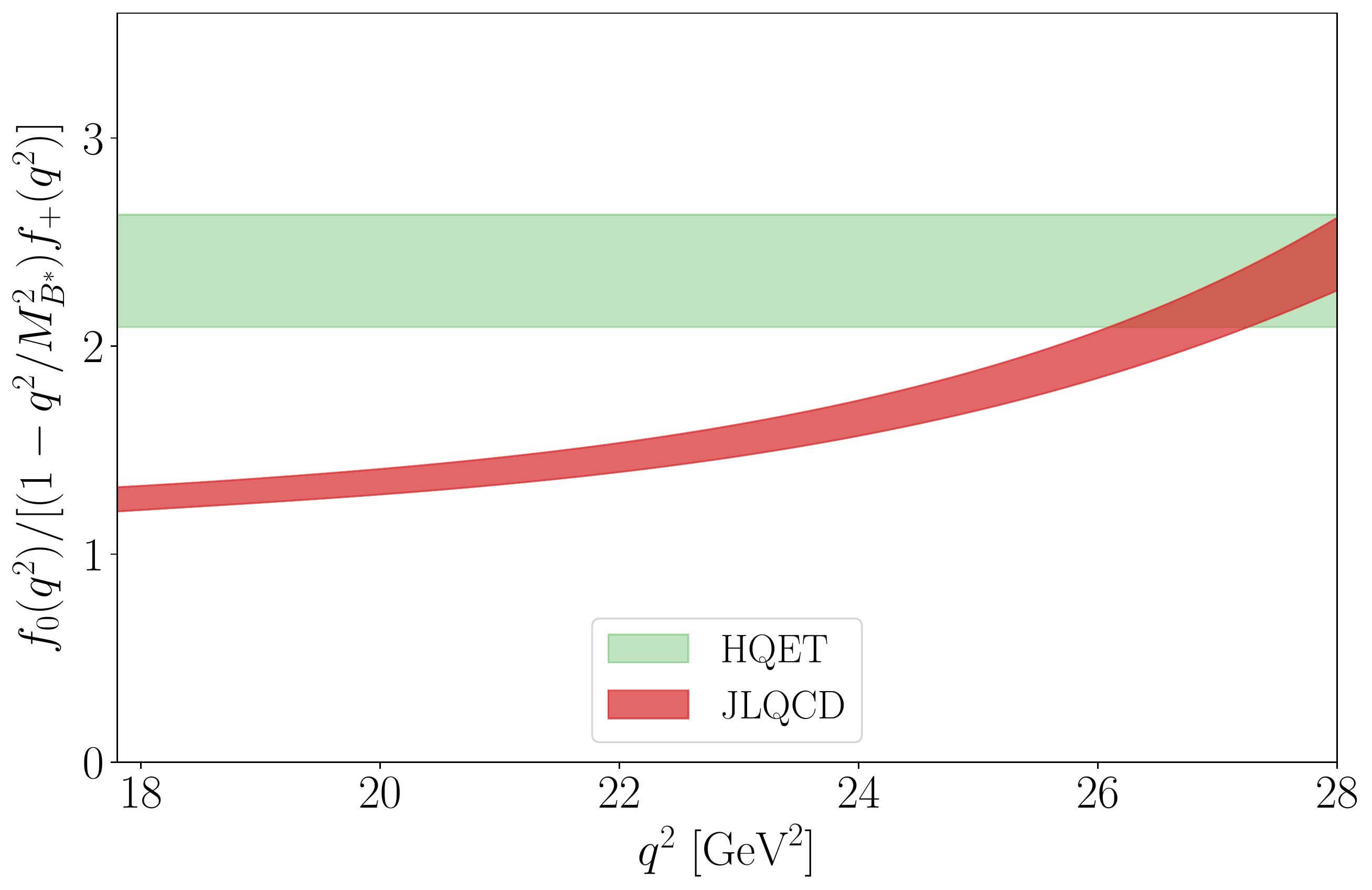}
\caption{Form factor ratio $f_0(q^2)/\left[(1 - q^2/M^2_{B^\ast})f_+(q^2)\right]$ as a function of $q^2$ compared with the prediction in the soft-pion limit from heavy-quark symmetry and $\chi\mathrm{PT}$~\cite{Burdman:1993es}. The width of the green error band reflects only the uncertainties from $g_{B^\ast B\pi} = 0.45(5)$ (from Ref.~\cite{Detmold:2012ge}) and $f_{B^\ast}/f_B=0.941(26)$ (from Ref.~\cite{Colquhoun:2015oha}), and not any other theoretical errors.}
\label{fig:softpion}
\end{figure}

It is also interesting to compare the lattice form factors with theoretical expectations from heavy-quark symmetry. In the soft-pion limit, the vector and scalar form factors, $f_+(q^2)$ and $f_0(q^2)$, are related by~\cite{Burdman:1993es}
\begin{equation}
\lim_{q^2\to M^2_B}\frac{f_0(q^2)}{f_+(q^2)} = \frac{f_B}{f_{B^\ast}}\frac{1-q^2/M^2_{B^\ast}}{g_{B^\ast B\pi}},
\end{equation}
up to corrections of $\mathcal{O}(1/m_b^2)$. This ratio is plotted in Fig.~\ref{fig:softpion} along with the theoretical expectation. We take $g_{B^\ast B\pi} = 0.45(5)$ (from Ref.~\cite{Detmold:2012ge}) and $f_{B^\ast}/f_B=0.941(26)$ (from Ref.~\cite{Colquhoun:2015oha}). The width of the green error band that represents the Heavy Quark Effective Theory (HQET) expectation reflects only the uncertainties from $g_{B^\ast B\pi}$ and $f_{B^\ast}/f_B$, and not any other theoretical errors. For the lattice data, we take our result of fit ``1'' extrapolated to the chiral limit $M_\pi^2=0$, showing only the statistical uncertainty. The agreement with the theoretical expectation in the soft pion limit and $q^2\to M_B^2$, which is at the rightmost end of the plot, is excellent.

\subsection{Branching fractions from experiment}
For the experimental results we use the following sets of data: the BaBar 2010 untagged analysis in 6 bins~\cite{BaBar:2010efp}; the Belle 2010 untagged analysis in 13 bins~\cite{Belle:2010hep}; the BaBar 2012 untagged analysis in 12 bins~\cite{BaBar:2012thb}; and the Belle 2013 tagged analysis in which the $B^0\to\pi^+\ell\nu$ process was measured in 13 bins and the $B^-\to\pi^0\ell\nu$ process was measured in 7 bins~\cite{Belle:2013hlo}. We deal with this last set of data by assuming isospin symmetry, which allows us to convert the $B^-$ decay to the $B^0$ decay through
\begin{equation}
  \Delta \mathcal{B} (B^0\to\pi^+\ell\nu) = 2 \dfrac{\tau_{B^0}}{\tau_{B^-}}\Delta \mathcal{B}(B^-\to\pi^0\ell\nu),
\end{equation}
where the mean life of the neutral and charged $B$ mesons are $\tau_{B^0}=1.519(4)\;\mathrm{ps}$ and $\tau_{B^-}=1.638(4)\;\mathrm{ps}$, respectively~\cite{ParticleDataGroup:2020ssz}. These are the same sets of data as used by the Heavy Flavour Averaging Group (HFLAV)~\cite{HFLAV:2019otj}, the Flavour Lattice Averaging Group (FLAG)~\cite{Aoki:2021kgd,*FlavourLatticeAveragingGroup:2019iem} and in the analysis presented in Ref.~\cite{Biswas:2021qyq}, as well as in the most recent lattice calculations of $|V_{ub}|$~\cite{Flynn:2015mha,FermilabLattice:2015mwy}.

We assume that systematic correlations between each of the individual datasets are negligible. We do, however, include correlations from the systematic uncertainties in the Belle 2013 analysis between the 13-bin and 7-bin data. The Belle collaboration indicated systematic correlations of $49\%$. We construct a total covariance matrix for the $B^0$ and $B^-$ data (after conversion to the isospin symmetric $B^0$ mode) by taking the direct sum of the statistical covariance matrices (where the off-diagonal blocks are ${\bf 0}$) and of the systematic covariance matrices (with $49\%$ correlation between each of the bins in the off-diagonal blocks), and then summing these two $20\times20$ matrices. The inclusion of these systematic correlations was found to have a negligible effect on the parameters and fit quality.

Our first step is fitting the four sets of data individually and then collectively without any lattice input. Using the BCL parametrization, we fit for the branching fraction in the $i\mathrm{th}$ bin through
\begin{equation}
\label{eq:branching_fraction}  
\Delta\mathcal{B}_i =  \dfrac{G_F^2|V_{ub}|^2}{24\pi^3}\int^{q^2_{i+1}}_{q^2_i} |\bm{p}_\pi(q^2)|^3|f_+(q^2)|^2dq^2,
\end{equation}
so that the combination of the form factor and CKM matrix element results in an overall normalization of $b^+_0|V_{ub}|$. 

The slope and the curvature from the $z$-expansion fits are captured in the ratios $b_1^+/b_0^+$ and $b_2^+/b_0^+$, respectively. Table~\ref{tab:z_only_experiment} gives our results of fits to each of the branching fraction results with $N_z=3$. We find that the fit quality is acceptable for each set of data when fitted individually, but that fitting all data simultaneously (``All'') results in a relatively poor fit. This is due to a tension between the BaBar 2010 data and the other results. We confirm this by fitting various combinations of datasets, finding poor fit quality whenever BaBar 2010 is included. Therefore, we also give results for the case where BaBar 2010 is dropped (``Excl. BaBar 2010''), which results in an acceptable fit.

\begin{table}[tbp]
  \begin{tabular}{c|cccc|cc}
    \hline
    \hline
    \multirow{2}{*}{Experiment} & BaBar & BaBar & Belle & Belle & \multirow{2}{*}{All} & Excl.  \\
                               & 2010  & 2012  & 2010  & 2013 &  & BaBar 2010  \\
    \hline
    $b^+_1/b^+_0$                  & $-0.85(47)$  & $-0.24(44)$  & $-1.25(26)$  & $-1.79(51)$  & $-0.96(19)$  & $-1.05(21)$  \\
    $b^+_2/b^+_0$                  &  $0.4(1.5)$  & $-3.8(1.3)$  & $-0.90(88)$  & $1.1(1.6)$   & $-1.37(60)$  & $-1.42(65)$  \\
    $b^+_0|V_{ub}|\times 10^{3}$  & $1.360(74)$  & $1.499(59)$  & $1.602(62)$  & $1.558(85)$  & $1.518(33)$  & $1.557(36)$  \\
    $\chi^2/N_{\mathrm{dof}}$      & $1.99$       & $0.45$       & $1.18$       &  $1.26$      & $1.39$       & $1.07$ \\
    $p$-value                  & $0.11$       & $0.91$       & $0.30$       & $0.21$       & $0.04$       & $0.36$\\
    \hline
    \hline
  \end{tabular}
  \caption{Results of the fits to the branching fractions obtained from experiments.}
  \label{tab:z_only_experiment}
\end{table}

Fitting with $N_z=3$ is sufficient, and higher order fits do not improve the fit quality. Although we agree with the values of the fitted parameters for the BaBar 2012 data reported by the Fermilab Lattice and MILC Collaborations in Ref.~\cite{FermilabLattice:2015mwy}, we find that the fit quality is actually better. Our result is in agreement with that found by the RBC and UKQCD Collaborations~\cite{Flynn:2015mha} and the result presented in Ref.~\cite{Biswas:2021qyq} where they each find a similar discrepancy with the fit quality reported by the Fermilab and MILC Collaborations.

\begin{figure}[tbp]
  \includegraphics[width=0.8\textwidth]{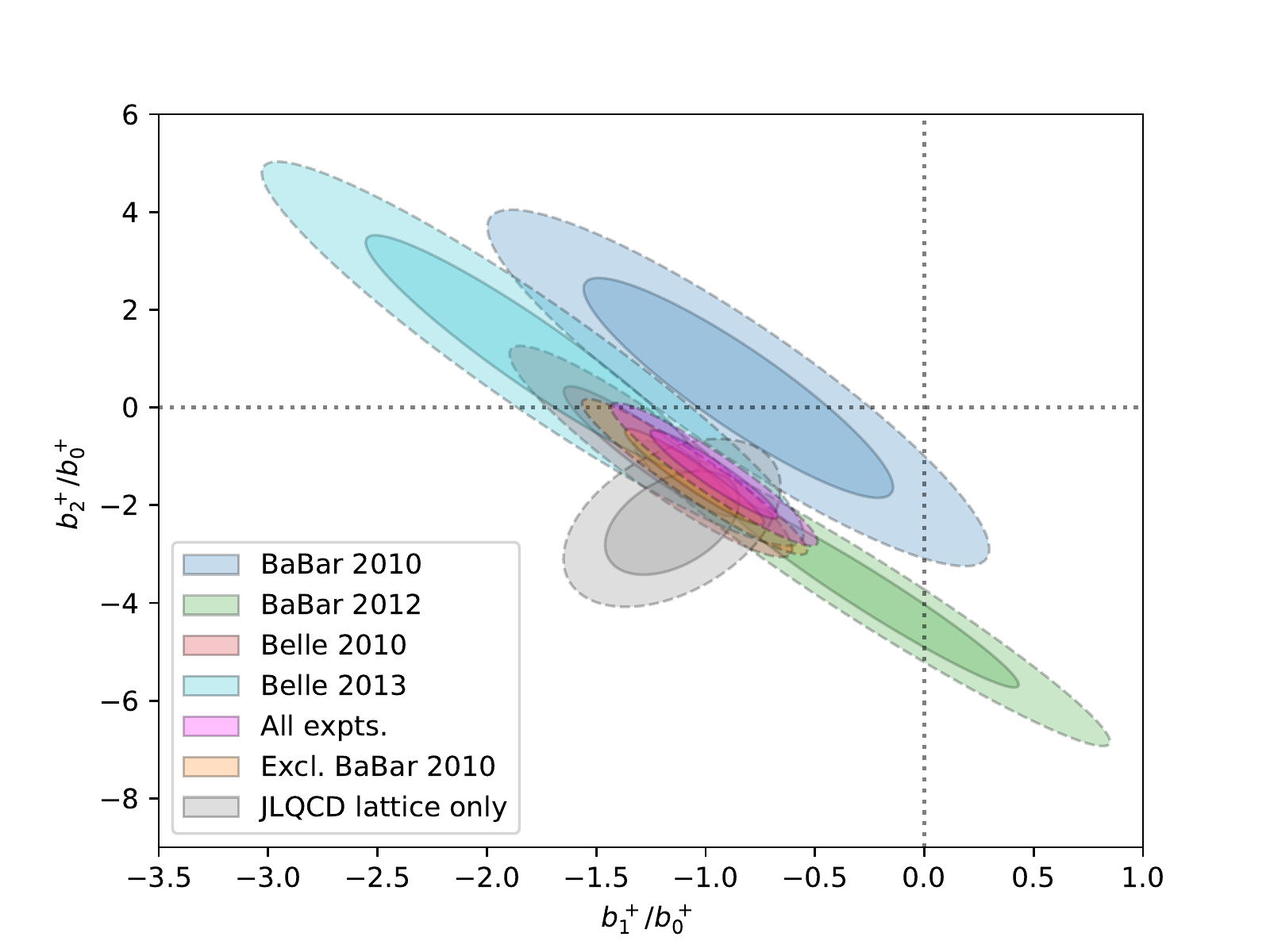}
  \caption{Contour plots for the shape parameters $b^+_1/b^+_0$ and $b^+_2/b^+_0$. We show $68\%$ confidence regions with a solid outline, and $95\%$ regions with a dashed outline.}
  \label{fig:shape_contours}
\end{figure}

In Fig.~\ref{fig:shape_contours} we plot 68\% and 95\% confidence regions for $b^+_1/b^+_0$ and $b^+_2/b^+_0$ for each of the cases listed in Table~\ref{tab:z_only_experiment}. This visually demonstrates the tension between the BaBar 2010 dataset and the other measurements. We also show the consistency between these shapes and with the shapes determined from our lattice only fit to the form factors using the BCL parametrization with $N_z=3$ and with the kinematic constraint $f_+(0)=f_0(0)$ imposed.

\subsection{Determination of $|V_{ub}|$}
\label{subsec:vub}

We now turn to fitting the above branching fraction results alongside our form factor results from the lattice. In this way we can determine the $z$-expansion parameters $b_n^+$ and our main result of $|V_{ub}|$, which appeared in the normalization of the experiment-only fits above. As discussed earlier, the contribution from the scalar form factor $f_0(q^2)$ to the branching fraction is suppressed by the squared lepton mass, and we neglect it. Therefore only $f_+(q^2)$ appears in Eq.~\eqref{eq:branching_fraction}. However, we do include lattice data for both form factors in the fit, and fit $f_+(q^2)$, $f_0(q^2)$ and experimental branching fraction data simultaneously. We impose the constraint $f_+(0)=f_0(0)$ explicitly, although this makes a negligible difference to our final results since the low-$q^2$ region is primarily controlled by the branching fraction data.

As we have only three data points for $f_0$, $N_z=4$ gives the maximum number of fit parameters we can use for $f_0$ if the constraint $f_+(0)=f_0(0)$ is imposed ($N_z=3$ without the constraint). For $f_+$ we have data points from lattice and experiment, and are not limited to $N_z=4$. We therefore choose $(N_z^{f_+},N_z^{f_0}) = (3,3)$, $(4,4)$ and $(5,4)$ for our main fits (imposing the constraint at $q^2=0$), and $(N_z^{f_+},N_z^{f_0}) = (3,3)$, $(4,3)$ and $(5,3)$ for test fits without the constraint. We find that all these choices give a reasonable fit quality and the parameters are stable. We take $(N_z^{f_+},N_z^{f_0}) = (4,4)$ for our accepted final result.

Numerical results for our combined lattice and experiment fits are given in Table~\ref{tab:final_fits}. We first fit the lattice form factors with each of the experimental branching fraction analyses in turn and find acceptable fit quality in each case. Next, we fit the lattice data alongside all experimental datasets simultaneously. As in the experiment-only fit, we do not find that the fit quality is particularly good when all experimental analyses are included. We therefore provide a further set of numerical values for the case where the BaBar 2010 analysis is excluded. This improves the fit quality while all parameters are consistent with the all-experiment fit. It should be noted that when BaBar 2010 is excluded, the value of $|V_{ub}|$ is determined to be marginally higher. The unitarity constraints from Eq.~\eqref{eq:unitarity_bounds} are satisfied in each case, although we stress again that they are not explicitly imposed on the fits. The correlation matrices for the combined fit of all lattice and experimental data are in Table~\ref{tab:app_final_fit_cov}, while those without BaBar 2010 are in Table~\ref{tab:app_final_fit_cov_exl_babar10}.

\begin{table}[tbp]
  \begin{tabular}{c|cccc|cc}
    \hline
    \hline
    \multirow{2}{*}{Experiment} & BaBar & BaBar & Belle & Belle & \multirow{2}{*}{All} & Excl.  \\
                               & 2010  & 2012  & 2010  & 2013 &  & BaBar 2010  \\
    \hline
    $b^+_0$ & $0.388(40)$  & $0.385(40)$  & $0.390(40)$  & $0.388(40)$  & $0.389(40)$  & $0.390(40)$  \\
    $b^+_1$ & $-0.389(80)$ & $-0.350(78)$ & $-0.438(76)$ & $-0.469(84)$ & $-0.391(66)$ & $-0.411(69)$  \\
    $b^+_2$ & $-0.20(18)$  & $-0.72(16)$  & $-0.66(16)$  & $-0.57(18)$  & $-0.62(15)$  & $-0.66(15)$ \\
    $b^+_3$ & $1.79(77)$   & $-0.40(64)$  & $0.23(65)$  & $0.96(76)$   & $0.22(52)$   & $0.09(5)$ \\
    $b^0_0$ & $0.535(35)$  & $0.536(35)$  & $0.535(35)$  & $0.533(35)$  & $0.536(35)$  & $0.536(35)$\\
    $b^0_1$ & $-1.31(12)$  & $-1.33(12)$  & $-1.35(12)$  & $-1.35(12)$  & $-1.33(12)$  & $-1.34(12)$\\
    $b^0_2$ & $1.16(23)$   & $0.56(17)$   & $0.59(17)$   & $0.71(21)$   & $0.68(16)$   & $0.60(16)$\\
    $b^0_3$ & $2.4(1.1)$   & $0.63(97)$  & $0.88(98)$   & $1.3(1.0)$   & $1.03(96)$   & $0.85(96)$\\
    $|V_{ub}|\times 10^{3}$ & $3.58(41)$ & $4.04(43)$ & $4.10(45)$ & $3.91(45)$ & $3.93(41)$ & $4.01(42)$  \\
    $\sum B^+_{mn}b^+_mb^+_n$ & $0.075(59)$ & $0.027(14)$ & $0.023(9)$ & $0.038(31)$ & $0.020(8)$ & $0.022(7)$  \\
    $\sum B^0_{mn}b^0_mb^0_n$ & $1.07(70)$ & $0.21(24)$ & $0.28(29)$ & $0.44(42)$ & $0.32(32)$ & $0.27(28)$  \\
    $\chi^2/N_{\mathrm{dof}}$ & $1.43$ & $0.77$ & $1.13$ &  $1.22$ & $1.37$ & $1.05$ \\
    $p$-value & $0.22$ & $0.66$ & $0.33$ & $0.23$ & $0.04$ & $0.38$ \\
    \hline
    \hline
  \end{tabular}
  \caption{Results of the simultaneous fits to form factors from our lattice calculation and experimental branching fractions, with $(N_z^{f_+},N_z^{f_0}) = (4,4)$.
  We list $b^0_0$ here for completeness, but it is fixed through the constraint $f_+(0)=f_0(0)$.}
  \label{tab:final_fits}
\end{table}

\begin{table}[tbp]
\begin{tabular}{c|rrrrrrrrrr}

\hline
\hline
 & $|V_{ub}|$ & $b^+_0$ & $b^+_1$ & $b^+_2$ & $b^+_3$ & $b^0_1$ & $b^0_2$ & $b^0_3$  \\
\hline
$|V_{ub}|$& $ 1.000$ & $-0.980$ & $ 0.568$ & $ 0.346$ & $ 0.007$ & $ 0.051$ & $-0.409$ & $-0.060$ \\
$b^+_0$  & $-0.980$ & $ 1.000$ & $-0.652$ & $-0.379$ & $ 0.048$ & $-0.067$ & $ 0.392$ & $ 0.064$ \\
$b^+_1$  & $ 0.568$ & $-0.652$ & $ 1.000$ & $-0.024$ & $-0.570$ & $ 0.093$ & $-0.349$ & $-0.159$ \\
$b^+_2$  & $ 0.346$ & $-0.379$ & $-0.024$ & $ 1.000$ & $-0.192$ & $ 0.153$ & $ 0.066$ & $-0.050$ \\
$b^+_3$  & $ 0.007$ & $ 0.048$ & $-0.570$ & $-0.192$ & $ 1.000$ & $-0.158$ & $ 0.126$ & $ 0.251$ \\
$b^0_1$  & $ 0.051$ & $-0.067$ & $ 0.093$ & $ 0.153$ & $-0.158$ & $ 1.000$ & $ 0.388$ & $-0.647$ \\
$b^0_2$  & $-0.409$ & $ 0.392$ & $-0.349$ & $ 0.066$ & $ 0.126$ & $ 0.388$ & $ 1.000$ & $-0.376$ \\
$b^0_3$  & $-0.060$ & $ 0.064$ & $-0.159$ & $-0.050$ & $ 0.251$ & $-0.647$ & $-0.376$ & $ 1.000$ \\
\hline
\hline
  \end{tabular}
  \caption{Correlation matrix from the $z$-expansion fit to all experiments and our synthetic lattice data with $(N_z^{f_+},N_z^{f_0}) = (4,4)$ parameters.
   Note that $b^0_0$ is fixed by the constraint $f_+(0)=f_0(0)$.}
\label{tab:app_final_fit_cov}
\end{table}

\begin{table}[tbp]
\begin{tabular}{c|rrrrrrrrrr}
\hline
\hline
 & $|V_{ub}|$ & $b^+_0$ & $b^+_1$ & $b^+_2$ & $b^+_3$ & $b^0_1$ & $b^0_2$ & $b^0_3$  \\
\hline
$|V_{ub}|$ & $ 1.000$ & $-0.977$ & $ 0.552$ & $ 0.379$ & $ 0.039$ & $ 0.056$ & $-0.367$ & $-0.040$ \\
$b^+_0$ & $-0.977$ & $ 1.000$ & $-0.643$ & $-0.412$ & $ 0.019$ & $-0.073$ & $ 0.348$ & $ 0.045$ \\
$b^+_1$ & $ 0.552$ & $-0.643$ & $ 1.000$ & $-0.008$ & $-0.573$ & $ 0.102$ & $-0.326$ & $-0.157$ \\
$b^+_2$ & $ 0.379$ & $-0.412$ & $-0.008$ & $ 1.000$ & $-0.141$ & $ 0.153$ & $ 0.079$ & $-0.035$ \\
$b^+_3$ & $ 0.039$ & $ 0.019$ & $-0.573$ & $-0.141$ & $ 1.000$ & $-0.159$ & $ 0.132$ & $ 0.260$ \\
$b^0_1$ & $ 0.056$ & $-0.073$ & $ 0.102$ & $ 0.153$ & $-0.159$ & $ 1.000$ & $ 0.393$ & $-0.646$ \\
$b^0_2$ & $-0.367$ & $ 0.348$ & $-0.326$ & $ 0.079$ & $ 0.132$ & $ 0.393$ & $ 1.000$ & $-0.382$ \\
$b^0_3$ & $-0.040$ & $ 0.045$ & $-0.157$ & $-0.035$ & $ 0.260$ & $-0.646$ & $-0.382$ & $ 1.000$ \\
\hline
\hline
  \end{tabular}
\caption{Correlation matrix from the $z$-expansion fit of our synthetic lattice data and experiment excluding BaBar 2010 with $(N_z^{f_+},N_z^{f_0}) = (4,4)$ parameters. Note that $b^0_0$ is fixed by the constraint $f_+(0)=f_0(0)$.}
\label{tab:app_final_fit_cov_exl_babar10}
\end{table}

\begin{figure}[tbp]
  \includegraphics[width=0.8\textwidth]{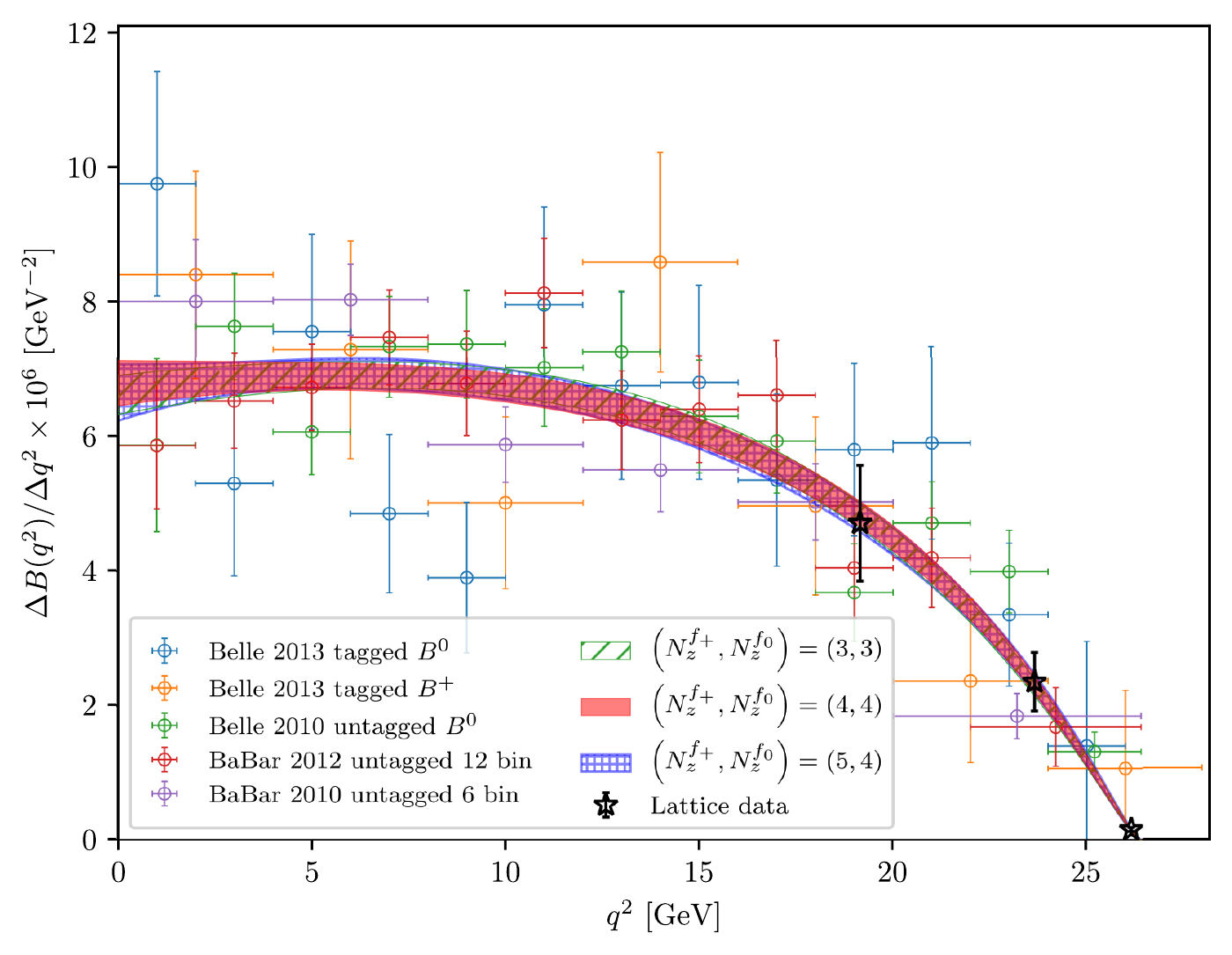}
  \caption{Fitting experimental branching fractions together with form factors from lattice QCD to extract $|V_{ub}|$. The error bands show our fit results when we include $\left(N^{f_+}_z,N^{f_0}_z\right)$ terms in the $z$-expansion. We find that $N^{f_+}_z \geq 3$ gives a reasonable fit quality, and take $\left(N_z^{f_+},N_z^{f_0}\right) = (4,4)$ as our main result.}
  \label{fig:branching_fractions}
\end{figure}

\begin{figure}[tbp]
\centering
\includegraphics[width=0.6\textwidth]{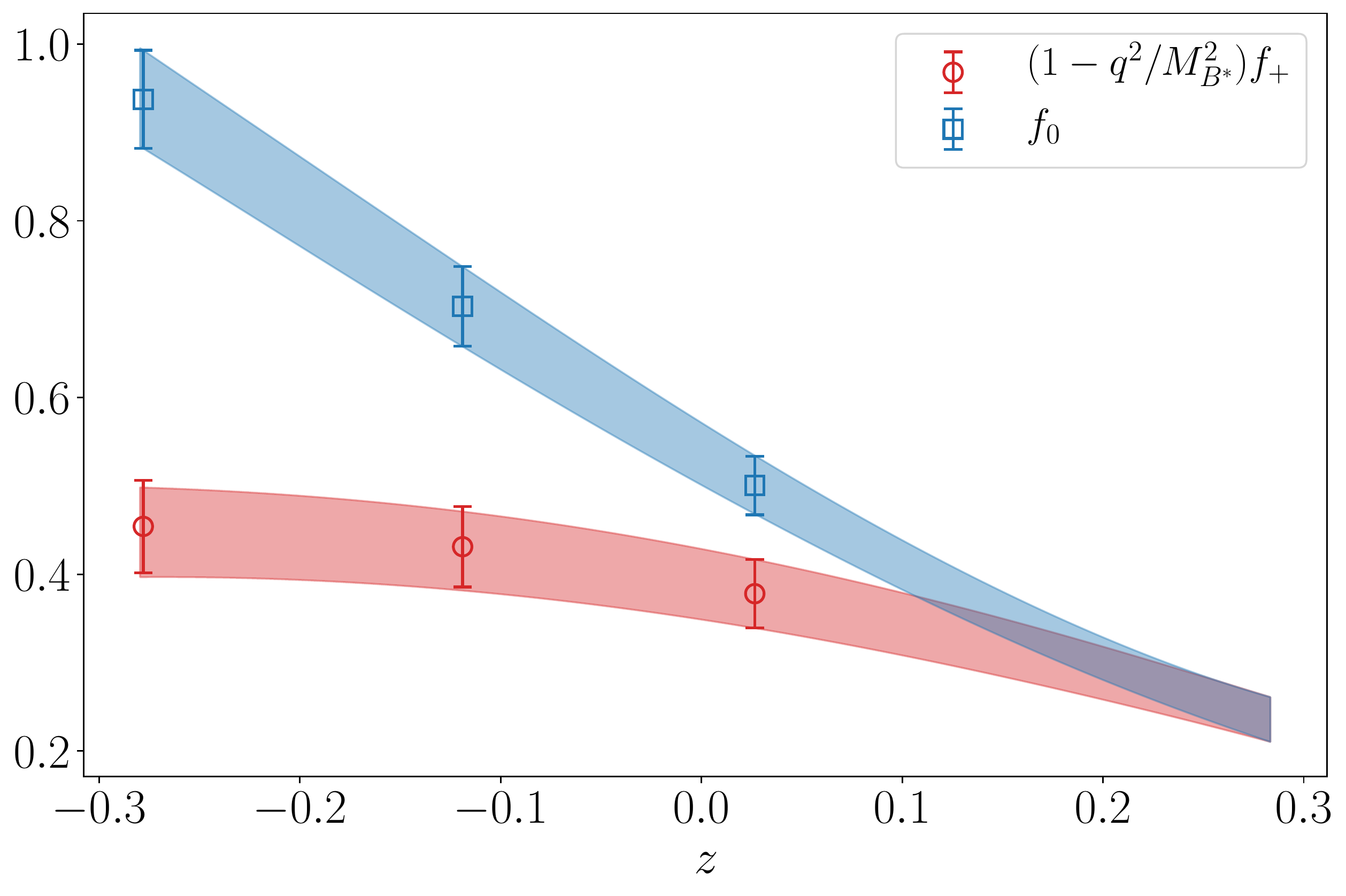}
\caption{Form factors using the BCL form of the $z$-parameter expansion determined from a combined fit of lattice data and branching fractions from experiment. Lattice data for $f_0$ (blue squares) and $(1-q^2/M^2_{B^*})f_+$ (red circles) are shown with fit bands covering the entire $z$ region.}
\label{fig:fp_f0_vs_z_wexperiment}
\end{figure}

The differential branching fraction data from experiments, our lattice data (converted using $|V_{ub}|$ from our accepted fit) and bands representing our $z$-expansion fit results with all errors included are plotted in Fig.~\ref{fig:branching_fractions}. The differences among the results with different $(N^{f_+}_z,N^{f_0}_z)$ are hardly visible, and they give essentially the same result for $|V_{ub}|$. We reiterate that we take $\left(N_z^{f_+},N_z^{f_0}\right) = (4,4)$ as our main result. In Fig.~\ref{fig:fp_f0_vs_z_wexperiment} we again show the form factors across the entire $z$ range, this time using the above BCL fits combining lattice form factor data and branching fractions from experiment. The lattice data for $f_0$ (blue squares) and $(1-q^2/M^2_{B^*})f_+$ (red circles) are shown with corresponding fit bands from the combined fit.

Our final result for $|V_{ub}|$ is thus from the combined fit with all experimental data:
\begin{equation}
  |V_{ub}| = (3.93 \pm 0.41) \times 10^{-3}.
\end{equation}
The uncertainty includes the statistical and systematic errors originating from our lattice calculation as well as the total errors from the experimental data.
If we exclude the BaBar 2010 data set from the analysis, we obtain
$|V_{ub}| = (4.01 \pm 0.42) \times 10^{-3}$ with a much improved $p$-value (see Table~\ref{tab:final_fits}).

\begin{figure}[tbp]
  \includegraphics[width=0.8\textwidth]{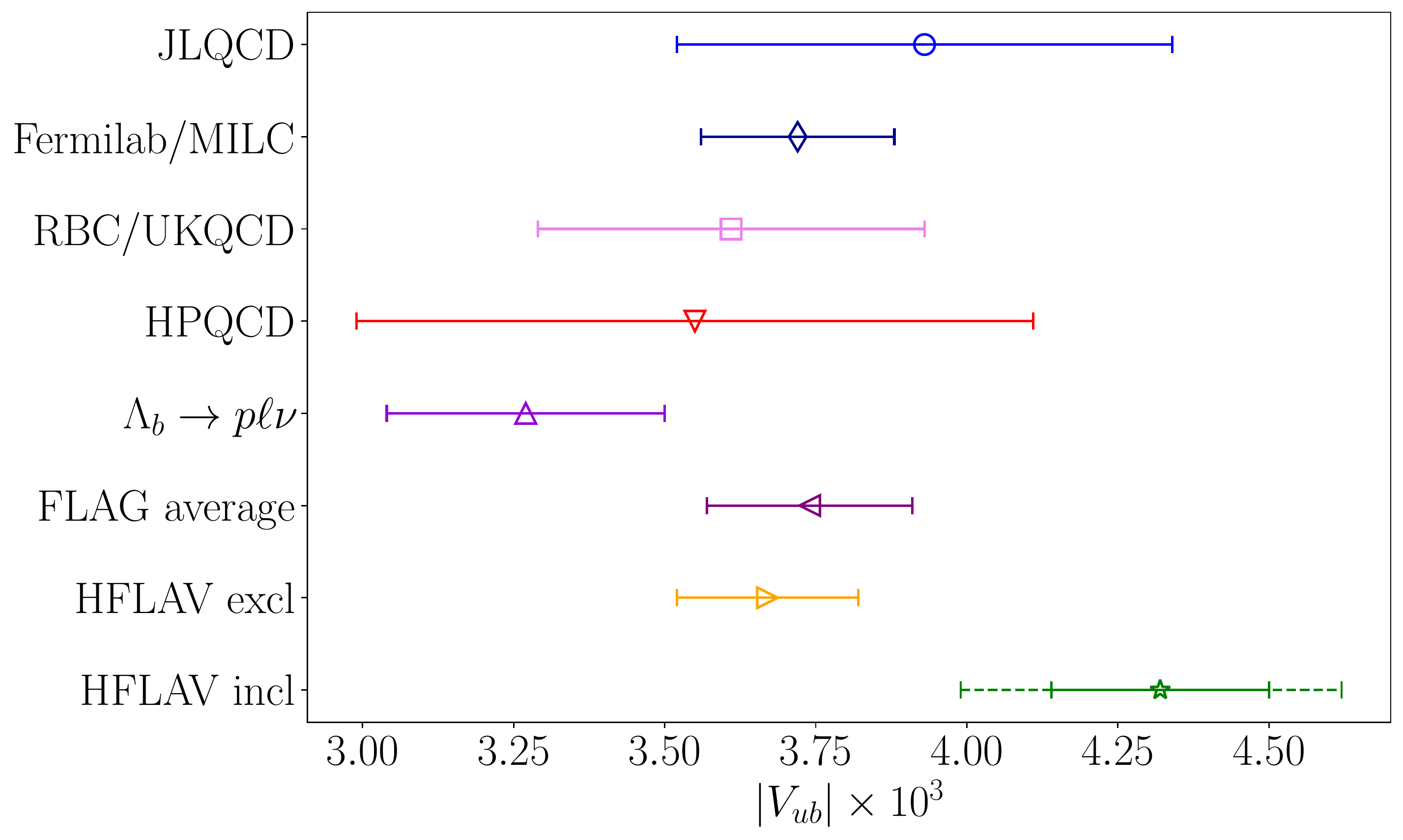}
  \caption{Comparison of our result for $|V_{ub}|$ with other lattice QCD calculations and exclusive and inclusive determinations by HFLAV and FLAG. The data point labelled `JLQCD' is our final result (this work). Other results are from the following publications: the Fermilab Lattice and MILC Collaborations~\cite{FermilabLattice:2015mwy}; the RBC and UKQCD Collaborations~\cite{Flynn:2015mha}; and the HPQCD Collaboration~\cite{Dalgic:2006dt}. The value tagged $\Lambda_b\to p\ell\nu$ is from Refs.~\cite{Detmold:2015aaa,LHCb:2015eia}. This combines a lattice QCD calculation of the form factors of the $\Lambda_b$ to $p$ process with experimental measurement of the ratio $\mathcal{B}(\Lambda_b^0\to p\mu^-\bar{\nu}_{\mu})/\mathcal{B}(\Lambda_b^0\to \Lambda_c^+\mu^-\bar{\nu}_{\mu})$ presented by the LHCb Collaboration, which allows the extraction of the ratio $|V_{ub}|/|V_{cb}|$. Using $|V_{cb}|=(39.5\pm 0.8)\times 10^{-3}$ from exclusive decays~\cite{LHCb:2015eia,ParticleDataGroup:2014cgo}, the authors quoted a value for $|V_{ub}|$. The FLAG average is from the 2021 report~\cite{Aoki:2021kgd,*FlavourLatticeAveragingGroup:2019iem}, and the HFLAV exclusive and inclusive results are from Ref.~\cite{HFLAV:2019otj}. The inclusive data point is from their GGOU analysis, with a second (dashed) error bar to represent the spread of values from other frameworks.}
  \label{fig:Vub_comparison}
\end{figure}

Our result for $|V_{ub}|$ is compared with other lattice QCD calculations and exclusive and inclusive determinations by HFLAV and FLAG in Fig.~\ref{fig:Vub_comparison}. Compared with other lattice QCD computations of the $B\to\pi\ell\nu$ process (Fermilab/MILC~\cite{FermilabLattice:2015mwy}, RBC/UKQCD~\cite{Flynn:2015mha} and HPQCD~\cite{Dalgic:2006dt}) our result is slightly higher but still consistent within the estimated errors. Our result is also compatible with the inclusive determination, which we have taken from HFLAV~\cite{HFLAV:2019otj} using the ``GGOU'' analysis. We include dashed error bars to indicate the spread of results from other methods. We also note that our value is in good agreement with those of Refs.~\cite{Biswas:2021qyq,Gonzalez-Solis:2021pyh,Leljak:2021vte}, while moderately higher than---but still consistent with---that in Ref~\cite{Dingfelder:2016twb}, all of which use lattice form factor results as input.

\section{Conclusions}\label{sec:conclusions}
For the determination of $|V_{ub}|$, the combination of the lattice computation of form factors and the experimental measurements of the differential cross section is crucial. This is not solely because the experiments can only measure the product of the form factor $f_+(q^2)$ and $|V_{ub}|$, but because they provide complementary information about the form factor shape. The lattice calculation provides the form factor in the large $q^2$ region with controlled errors, while the experimental data are more sensitive to the low $q^2$ region. As one can see from the fit results, by combining the data from both experiment and lattice QCD, the form factor shape is much better controlled.

Our combined result for $|V_{ub}|$ is $3.93(41)\times 10^{-3}$ when including data from all experiments, and $4.01(42)\times 10^{-3}$ when excluding the 6-bin untagged BaBar 2010 analysis. In both cases these results are consistent with the inclusive determination of $|V_{ub}|$ and with previous results on the exclusive $B\to\pi\ell\nu$ process.

The advantage of our lattice calculation over previous work is the use of a fully relativistic lattice fermion formulation, with which no extra matching procedure is required. (For the renormalization constant, we employed a strategy to eliminate the bulk of the large discretization effects appearing in the wave-function renormalization by making a non-perturbative determination of $Z_V$ using heavy-to-heavy three-point functions.) Our analysis therefore becomes rather straightforward: we simply assume the discretization effects are of $\mathcal{O}(a^2)$ and $\mathcal{O}((am_Q)^2)$ and let the numerical data determine their size by combining the lattice data at various $a$ and $am_Q$. We also explore the dependence on the heavy quark mass and find that it is consistent with a leading $1/m_Q$ correction to the heavy quark limit. 

A major challenge in this analysis was due to the multiple extrapolations that have to be performed at the same time in three parameter dimensions: the light quark mass; the heavy quark mass; and the lattice spacing. We find that these limits are reached rather smoothly with our global fit function. We estimate systematic errors due to potentially missing higher order terms in the ansatze by attempting the fit including one such term at a time. There is no single dominant source of error, but after adding them in quadrature the total systematic error is comparable to the statistical error in our calculation. The inclusion of heavier masses for $m_Q$ and smaller pion masses would help further control systematic effects, while additional statistics is the key to improving the calculation of these form factors in the future.

We anticipate more lattice calculations of the $B\to\pi\ell\nu$ process using fully relativistic actions in the near future. Crucially, this includes cases where the heavy quark is tuned to the physical $b$ quark mass on the finest lattices, allowing for an improved approach to the physical point, and therefore even better control of systematic effects.

\begin{acknowledgments}
  We would like to thank Zechariah Gelzer and Justus Tobias Tsang for useful discussions during this project. We also thank members of the JLQCD Collaboration for discussions and for providing the computational framework. In particular, we are grateful to Brendan Fahy for his participation in the early stage of this work, and to Guido Cossu and Jun-Ichi Noaki for the development of the IroIro++ code set~\cite{Cossu:2013ola}, which was used to generate the data. Numerical calculations were performed on the Blue Gene/Q supercomputer at KEK under its Large Scale Simulation Program (No. 16/17-14). This work is supported in part by JSPS KAKENHI Grant Numbers 18H03710 and 21H01085, and by MEXT as ``Program for Promoting Researches on the Supercomputer Fugaku'' (Simulation for basic science: from fundamental laws of particles to creation of nuclei, JPMXP1020200105) through the Joint Institute for Computational Fundamental Science (JICFuS). The reference list was automatically populated using \textit{filltex}~\cite{Gerosa:2017xrm}.
\end{acknowledgments}

\bibliographystyle{apsrev4-2}
\bibliography{main}

\begin{thebibliography}{66}%
\makeatletter
\providecommand \@ifxundefined [1]{%
 \@ifx{#1\undefined}
}%
\providecommand \@ifnum [1]{%
 \ifnum #1\expandafter \@firstoftwo
 \else \expandafter \@secondoftwo
 \fi
}%
\providecommand \@ifx [1]{%
 \ifx #1\expandafter \@firstoftwo
 \else \expandafter \@secondoftwo
 \fi
}%
\providecommand \natexlab [1]{#1}%
\providecommand \enquote  [1]{``#1''}%
\providecommand \bibnamefont  [1]{#1}%
\providecommand \bibfnamefont [1]{#1}%
\providecommand \citenamefont [1]{#1}%
\providecommand \href@noop [0]{\@secondoftwo}%
\providecommand \href [0]{\begingroup \@sanitize@url \@href}%
\providecommand \@href[1]{\@@startlink{#1}\@@href}%
\providecommand \@@href[1]{\endgroup#1\@@endlink}%
\providecommand \@sanitize@url [0]{\catcode `\\12\catcode `\$12\catcode
  `\&12\catcode `\#12\catcode `\^12\catcode `\_12\catcode `\%12\relax}%
\providecommand \@@startlink[1]{}%
\providecommand \@@endlink[0]{}%
\providecommand \url  [0]{\begingroup\@sanitize@url \@url }%
\providecommand \@url [1]{\endgroup\@href {#1}{\urlprefix }}%
\providecommand \urlprefix  [0]{URL }%
\providecommand \Eprint [0]{\href }%
\providecommand \doibase [0]{https://doi.org/}%
\providecommand \selectlanguage [0]{\@gobble}%
\providecommand \bibinfo  [0]{\@secondoftwo}%
\providecommand \bibfield  [0]{\@secondoftwo}%
\providecommand \translation [1]{[#1]}%
\providecommand \BibitemOpen [0]{}%
\providecommand \bibitemStop [0]{}%
\providecommand \bibitemNoStop [0]{.\EOS\space}%
\providecommand \EOS [0]{\spacefactor3000\relax}%
\providecommand \BibitemShut  [1]{\csname bibitem#1\endcsname}%
\let\auto@bib@innerbib\@empty
\bibitem [{\citenamefont {Colquhoun}\ \emph {et~al.}(2016)\citenamefont
  {Colquhoun}, \citenamefont {Dowdall}, \citenamefont {Koponen}, \citenamefont
  {Davies},\ and\ \citenamefont {Lepage}}]{Colquhoun:2015mfa}%
  \BibitemOpen
  \bibfield  {author} {\bibinfo {author} {\bibfnamefont {B.}~\bibnamefont
  {Colquhoun}}, \bibinfo {author} {\bibfnamefont {R.~J.}\ \bibnamefont
  {Dowdall}}, \bibinfo {author} {\bibfnamefont {J.}~\bibnamefont {Koponen}},
  \bibinfo {author} {\bibfnamefont {C.~T.~H.}\ \bibnamefont {Davies}},\ and\
  \bibinfo {author} {\bibfnamefont {G.~P.}\ \bibnamefont {Lepage}},\ }\href
  {https://doi.org/10.1103/PhysRevD.93.034502} {\bibfield  {journal} {\bibinfo
  {journal} {Phys. Rev. D}\ }\textbf {\bibinfo {volume} {93}},\ \bibinfo
  {pages} {034502} (\bibinfo {year} {2016})},\ \Eprint
  {https://arxiv.org/abs/1510.07446} {arXiv:1510.07446 [hep-lat]} \BibitemShut
  {NoStop}%
\bibitem [{\citenamefont {Hughes}\ \emph {et~al.}(2018)\citenamefont {Hughes},
  \citenamefont {Davies},\ and\ \citenamefont {Monahan}}]{Hughes:2017spc}%
  \BibitemOpen
  \bibfield  {author} {\bibinfo {author} {\bibfnamefont {C.}~\bibnamefont
  {Hughes}}, \bibinfo {author} {\bibfnamefont {C.~T.~H.}\ \bibnamefont
  {Davies}},\ and\ \bibinfo {author} {\bibfnamefont {C.~J.}\ \bibnamefont
  {Monahan}},\ }\href {https://doi.org/10.1103/PhysRevD.97.054509} {\bibfield
  {journal} {\bibinfo  {journal} {Phys. Rev. D}\ }\textbf {\bibinfo {volume}
  {97}},\ \bibinfo {pages} {054509} (\bibinfo {year} {2018})},\ \Eprint
  {https://arxiv.org/abs/1711.09981} {arXiv:1711.09981 [hep-lat]} \BibitemShut
  {NoStop}%
\bibitem [{\citenamefont {Dalgic}\ \emph {et~al.}(2006)\citenamefont {Dalgic},
  \citenamefont {Gray}, \citenamefont {Wingate}, \citenamefont {Davies},
  \citenamefont {Lepage},\ and\ \citenamefont {Shigemitsu}}]{Dalgic:2006dt}%
  \BibitemOpen
  \bibfield  {author} {\bibinfo {author} {\bibfnamefont {E.}~\bibnamefont
  {Dalgic}}, \bibinfo {author} {\bibfnamefont {A.}~\bibnamefont {Gray}},
  \bibinfo {author} {\bibfnamefont {M.}~\bibnamefont {Wingate}}, \bibinfo
  {author} {\bibfnamefont {C.~T.~H.}\ \bibnamefont {Davies}}, \bibinfo {author}
  {\bibfnamefont {G.~P.}\ \bibnamefont {Lepage}},\ and\ \bibinfo {author}
  {\bibfnamefont {J.}~\bibnamefont {Shigemitsu}},\ }\href
  {https://doi.org/10.1103/PhysRevD.75.119906} {\bibfield  {journal} {\bibinfo
  {journal} {Phys. Rev. D}\ }\textbf {\bibinfo {volume} {73}},\ \bibinfo
  {pages} {074502} (\bibinfo {year} {2006})},\ \bibinfo {note} {[Erratum:
  Phys.Rev.D 75, 119906 (2007)]},\ \Eprint
  {https://arxiv.org/abs/hep-lat/0601021} {arXiv:hep-lat/0601021} \BibitemShut
  {NoStop}%
\bibitem [{\citenamefont {Flynn}\ \emph {et~al.}(2015)\citenamefont {Flynn},
  \citenamefont {Izubuchi}, \citenamefont {Kawanai}, \citenamefont {Lehner},
  \citenamefont {Soni}, \citenamefont {Van~de Water},\ and\ \citenamefont
  {Witzel}}]{Flynn:2015mha}%
  \BibitemOpen
  \bibfield  {author} {\bibinfo {author} {\bibfnamefont {J.~M.}\ \bibnamefont
  {Flynn}}, \bibinfo {author} {\bibfnamefont {T.}~\bibnamefont {Izubuchi}},
  \bibinfo {author} {\bibfnamefont {T.}~\bibnamefont {Kawanai}}, \bibinfo
  {author} {\bibfnamefont {C.}~\bibnamefont {Lehner}}, \bibinfo {author}
  {\bibfnamefont {A.}~\bibnamefont {Soni}}, \bibinfo {author} {\bibfnamefont
  {R.~S.}\ \bibnamefont {Van~de Water}},\ and\ \bibinfo {author} {\bibfnamefont
  {O.}~\bibnamefont {Witzel}},\ }\href
  {https://doi.org/10.1103/PhysRevD.91.074510} {\bibfield  {journal} {\bibinfo
  {journal} {Phys. Rev. D}\ }\textbf {\bibinfo {volume} {91}},\ \bibinfo
  {pages} {074510} (\bibinfo {year} {2015})},\ \Eprint
  {https://arxiv.org/abs/1501.05373} {arXiv:1501.05373 [hep-lat]} \BibitemShut
  {NoStop}%
\bibitem [{\citenamefont {Bailey}\ \emph {et~al.}(2015)\citenamefont {Bailey}
  \emph {et~al.}}]{FermilabLattice:2015mwy}%
  \BibitemOpen
  \bibfield  {author} {\bibinfo {author} {\bibfnamefont {J.~A.}\ \bibnamefont
  {Bailey}} \emph {et~al.} (\bibinfo {collaboration} {Fermilab Lattice,
  MILC}),\ }\href {https://doi.org/10.1103/PhysRevD.92.014024} {\bibfield
  {journal} {\bibinfo  {journal} {Phys. Rev. D}\ }\textbf {\bibinfo {volume}
  {92}},\ \bibinfo {pages} {014024} (\bibinfo {year} {2015})},\ \Eprint
  {https://arxiv.org/abs/1503.07839} {arXiv:1503.07839 [hep-lat]} \BibitemShut
  {NoStop}%
\bibitem [{\citenamefont {Ginsparg}\ and\ \citenamefont
  {Wilson}(1982)}]{Ginsparg:1981bj}%
  \BibitemOpen
  \bibfield  {author} {\bibinfo {author} {\bibfnamefont {P.~H.}\ \bibnamefont
  {Ginsparg}}\ and\ \bibinfo {author} {\bibfnamefont {K.~G.}\ \bibnamefont
  {Wilson}},\ }\href {https://doi.org/10.1103/PhysRevD.25.2649} {\bibfield
  {journal} {\bibinfo  {journal} {Phys. Rev. D}\ }\textbf {\bibinfo {volume}
  {25}},\ \bibinfo {pages} {2649} (\bibinfo {year} {1982})}\BibitemShut
  {NoStop}%
\bibitem [{\citenamefont {Kaplan}(1992)}]{Kaplan:1992bt}%
  \BibitemOpen
  \bibfield  {author} {\bibinfo {author} {\bibfnamefont {D.~B.}\ \bibnamefont
  {Kaplan}},\ }\href {https://doi.org/10.1016/0370-2693(92)91112-M} {\bibfield
  {journal} {\bibinfo  {journal} {Phys. Lett. B}\ }\textbf {\bibinfo {volume}
  {288}},\ \bibinfo {pages} {342} (\bibinfo {year} {1992})},\ \Eprint
  {https://arxiv.org/abs/hep-lat/9206013} {arXiv:hep-lat/9206013} \BibitemShut
  {NoStop}%
\bibitem [{\citenamefont {Shamir}(1993)}]{Shamir:1993zy}%
  \BibitemOpen
  \bibfield  {author} {\bibinfo {author} {\bibfnamefont {Y.}~\bibnamefont
  {Shamir}},\ }\href {https://doi.org/10.1016/0550-3213(93)90162-I} {\bibfield
  {journal} {\bibinfo  {journal} {Nucl. Phys. B}\ }\textbf {\bibinfo {volume}
  {406}},\ \bibinfo {pages} {90} (\bibinfo {year} {1993})},\ \Eprint
  {https://arxiv.org/abs/hep-lat/9303005} {arXiv:hep-lat/9303005} \BibitemShut
  {NoStop}%
\bibitem [{\citenamefont {Furman}\ and\ \citenamefont
  {Shamir}(1995)}]{Furman:1994ky}%
  \BibitemOpen
  \bibfield  {author} {\bibinfo {author} {\bibfnamefont {V.}~\bibnamefont
  {Furman}}\ and\ \bibinfo {author} {\bibfnamefont {Y.}~\bibnamefont
  {Shamir}},\ }\href {https://doi.org/10.1016/0550-3213(95)00031-M} {\bibfield
  {journal} {\bibinfo  {journal} {Nucl. Phys. B}\ }\textbf {\bibinfo {volume}
  {439}},\ \bibinfo {pages} {54} (\bibinfo {year} {1995})},\ \Eprint
  {https://arxiv.org/abs/hep-lat/9405004} {arXiv:hep-lat/9405004} \BibitemShut
  {NoStop}%
\bibitem [{\citenamefont {Brower}\ \emph {et~al.}(2017)\citenamefont {Brower},
  \citenamefont {Neff},\ and\ \citenamefont {Orginos}}]{Brower:2012vk}%
  \BibitemOpen
  \bibfield  {author} {\bibinfo {author} {\bibfnamefont {R.~C.}\ \bibnamefont
  {Brower}}, \bibinfo {author} {\bibfnamefont {H.}~\bibnamefont {Neff}},\ and\
  \bibinfo {author} {\bibfnamefont {K.}~\bibnamefont {Orginos}},\ }\href
  {https://doi.org/10.1016/j.cpc.2017.01.024} {\bibfield  {journal} {\bibinfo
  {journal} {Comput. Phys. Commun.}\ }\textbf {\bibinfo {volume} {220}},\
  \bibinfo {pages} {1} (\bibinfo {year} {2017})},\ \Eprint
  {https://arxiv.org/abs/1206.5214} {arXiv:1206.5214 [hep-lat]} \BibitemShut
  {NoStop}%
\bibitem [{\citenamefont {Okubo}(1971{\natexlab{a}})}]{Okubo:1971my}%
  \BibitemOpen
  \bibfield  {author} {\bibinfo {author} {\bibfnamefont {S.}~\bibnamefont
  {Okubo}},\ }\href {https://doi.org/10.1103/PhysRevD.4.725} {\bibfield
  {journal} {\bibinfo  {journal} {Phys. Rev. D}\ }\textbf {\bibinfo {volume}
  {4}},\ \bibinfo {pages} {725} (\bibinfo {year}
  {1971}{\natexlab{a}})}\BibitemShut {NoStop}%
\bibitem [{\citenamefont {Okubo}(1971{\natexlab{b}})}]{Okubo:1971jf}%
  \BibitemOpen
  \bibfield  {author} {\bibinfo {author} {\bibfnamefont {S.}~\bibnamefont
  {Okubo}},\ }\href {https://doi.org/10.1103/PhysRevD.3.2807} {\bibfield
  {journal} {\bibinfo  {journal} {Phys. Rev. D}\ }\textbf {\bibinfo {volume}
  {3}},\ \bibinfo {pages} {2807} (\bibinfo {year}
  {1971}{\natexlab{b}})}\BibitemShut {NoStop}%
\bibitem [{\citenamefont {Boyd}\ \emph {et~al.}(1996)\citenamefont {Boyd},
  \citenamefont {Grinstein},\ and\ \citenamefont {Lebed}}]{Boyd:1995sq}%
  \BibitemOpen
  \bibfield  {author} {\bibinfo {author} {\bibfnamefont {C.~G.}\ \bibnamefont
  {Boyd}}, \bibinfo {author} {\bibfnamefont {B.}~\bibnamefont {Grinstein}},\
  and\ \bibinfo {author} {\bibfnamefont {R.~F.}\ \bibnamefont {Lebed}},\ }\href
  {https://doi.org/10.1016/0550-3213(95)00653-2} {\bibfield  {journal}
  {\bibinfo  {journal} {Nucl. Phys. B}\ }\textbf {\bibinfo {volume} {461}},\
  \bibinfo {pages} {493} (\bibinfo {year} {1996})},\ \Eprint
  {https://arxiv.org/abs/hep-ph/9508211} {arXiv:hep-ph/9508211} \BibitemShut
  {NoStop}%
\bibitem [{\citenamefont {Boyd}\ \emph {et~al.}(1997)\citenamefont {Boyd},
  \citenamefont {Grinstein},\ and\ \citenamefont {Lebed}}]{Boyd:1997kz}%
  \BibitemOpen
  \bibfield  {author} {\bibinfo {author} {\bibfnamefont {C.~G.}\ \bibnamefont
  {Boyd}}, \bibinfo {author} {\bibfnamefont {B.}~\bibnamefont {Grinstein}},\
  and\ \bibinfo {author} {\bibfnamefont {R.~F.}\ \bibnamefont {Lebed}},\ }\href
  {https://doi.org/10.1103/PhysRevD.56.6895} {\bibfield  {journal} {\bibinfo
  {journal} {Phys. Rev. D}\ }\textbf {\bibinfo {volume} {56}},\ \bibinfo
  {pages} {6895} (\bibinfo {year} {1997})},\ \Eprint
  {https://arxiv.org/abs/hep-ph/9705252} {arXiv:hep-ph/9705252} \BibitemShut
  {NoStop}%
\bibitem [{\citenamefont {Caprini}\ \emph {et~al.}(1998)\citenamefont
  {Caprini}, \citenamefont {Lellouch},\ and\ \citenamefont
  {Neubert}}]{Caprini:1997mu}%
  \BibitemOpen
  \bibfield  {author} {\bibinfo {author} {\bibfnamefont {I.}~\bibnamefont
  {Caprini}}, \bibinfo {author} {\bibfnamefont {L.}~\bibnamefont {Lellouch}},\
  and\ \bibinfo {author} {\bibfnamefont {M.}~\bibnamefont {Neubert}},\ }\href
  {https://doi.org/10.1016/S0550-3213(98)00350-2} {\bibfield  {journal}
  {\bibinfo  {journal} {Nucl. Phys. B}\ }\textbf {\bibinfo {volume} {530}},\
  \bibinfo {pages} {153} (\bibinfo {year} {1998})},\ \Eprint
  {https://arxiv.org/abs/hep-ph/9712417} {arXiv:hep-ph/9712417} \BibitemShut
  {NoStop}%
\bibitem [{\citenamefont {Bourrely}\ \emph {et~al.}(1981)\citenamefont
  {Bourrely}, \citenamefont {Machet},\ and\ \citenamefont
  {de~Rafael}}]{Bourrely:1980gp}%
  \BibitemOpen
  \bibfield  {author} {\bibinfo {author} {\bibfnamefont {C.}~\bibnamefont
  {Bourrely}}, \bibinfo {author} {\bibfnamefont {B.}~\bibnamefont {Machet}},\
  and\ \bibinfo {author} {\bibfnamefont {E.}~\bibnamefont {de~Rafael}},\ }\href
  {https://doi.org/10.1016/0550-3213(81)90086-9} {\bibfield  {journal}
  {\bibinfo  {journal} {Nucl. Phys. B}\ }\textbf {\bibinfo {volume} {189}},\
  \bibinfo {pages} {157} (\bibinfo {year} {1981})}\BibitemShut {NoStop}%
\bibitem [{\citenamefont {Bourrely}\ and\ \citenamefont
  {Caprini}(2005)}]{Bourrely:2005hp}%
  \BibitemOpen
  \bibfield  {author} {\bibinfo {author} {\bibfnamefont {C.}~\bibnamefont
  {Bourrely}}\ and\ \bibinfo {author} {\bibfnamefont {I.}~\bibnamefont
  {Caprini}},\ }\href {https://doi.org/10.1016/j.nuclphysb.2005.06.013}
  {\bibfield  {journal} {\bibinfo  {journal} {Nucl. Phys. B}\ }\textbf
  {\bibinfo {volume} {722}},\ \bibinfo {pages} {149} (\bibinfo {year}
  {2005})},\ \Eprint {https://arxiv.org/abs/hep-ph/0504016}
  {arXiv:hep-ph/0504016} \BibitemShut {NoStop}%
\bibitem [{\citenamefont {Bourrely}\ \emph {et~al.}(2009)\citenamefont
  {Bourrely}, \citenamefont {Caprini},\ and\ \citenamefont
  {Lellouch}}]{Bourrely:2008za}%
  \BibitemOpen
  \bibfield  {author} {\bibinfo {author} {\bibfnamefont {C.}~\bibnamefont
  {Bourrely}}, \bibinfo {author} {\bibfnamefont {I.}~\bibnamefont {Caprini}},\
  and\ \bibinfo {author} {\bibfnamefont {L.}~\bibnamefont {Lellouch}},\ }\href
  {https://doi.org/10.1103/PhysRevD.82.099902} {\bibfield  {journal} {\bibinfo
  {journal} {Phys. Rev. D}\ }\textbf {\bibinfo {volume} {79}},\ \bibinfo
  {pages} {013008} (\bibinfo {year} {2009})},\ \bibinfo {note} {[Erratum:
  Phys.Rev.D 82, 099902 (2010)]},\ \Eprint {https://arxiv.org/abs/0807.2722}
  {arXiv:0807.2722 [hep-ph]} \BibitemShut {NoStop}%
\bibitem [{\citenamefont {Bailey}\ \emph {et~al.}(2009)\citenamefont {Bailey}
  \emph {et~al.}}]{Bailey:2008wp}%
  \BibitemOpen
  \bibfield  {author} {\bibinfo {author} {\bibfnamefont {J.~A.}\ \bibnamefont
  {Bailey}} \emph {et~al.} (\bibinfo {collaboration} {Fermilab Lattice,
  MILC}),\ }\href {https://doi.org/10.1103/PhysRevD.79.054507} {\bibfield
  {journal} {\bibinfo  {journal} {Phys. Rev. D}\ }\textbf {\bibinfo {volume}
  {79}},\ \bibinfo {pages} {054507} (\bibinfo {year} {2009})},\ \Eprint
  {https://arxiv.org/abs/0811.3640} {arXiv:0811.3640 [hep-lat]} \BibitemShut
  {NoStop}%
\bibitem [{\citenamefont {Amhis}\ \emph {et~al.}(2021)\citenamefont {Amhis}
  \emph {et~al.}}]{HFLAV:2019otj}%
  \BibitemOpen
  \bibfield  {author} {\bibinfo {author} {\bibfnamefont {Y.~S.}\ \bibnamefont
  {Amhis}} \emph {et~al.} (\bibinfo {collaboration} {HFLAV}),\ }\href
  {https://doi.org/10.1140/epjc/s10052-020-8156-7} {\bibfield  {journal}
  {\bibinfo  {journal} {Eur. Phys. J. C}\ }\textbf {\bibinfo {volume} {81}},\
  \bibinfo {pages} {226} (\bibinfo {year} {2021})},\ \Eprint
  {https://arxiv.org/abs/1909.12524} {arXiv:1909.12524 [hep-ex]} \BibitemShut
  {NoStop}%
\bibitem [{\citenamefont {Aoki}\ \emph {et~al.}(2021)\citenamefont {Aoki} \emph
  {et~al.}}]{Aoki:2021kgd}%
  \BibitemOpen
  \bibfield  {author} {\bibinfo {author} {\bibfnamefont {Y.}~\bibnamefont
  {Aoki}} \emph {et~al.},\ }\href@noop {} {\  (\bibinfo {year} {2021})},\
  \Eprint {https://arxiv.org/abs/2111.09849} {arXiv:2111.09849 [hep-lat]}
  \BibitemShut {NoStop}%
\bibitem [{\citenamefont {Aoki}\ \emph {et~al.}(2020)\citenamefont {Aoki} \emph
  {et~al.}}]{FlavourLatticeAveragingGroup:2019iem}%
  \BibitemOpen
  \bibfield  {author} {\bibinfo {author} {\bibfnamefont {S.}~\bibnamefont
  {Aoki}} \emph {et~al.} (\bibinfo {collaboration} {Flavour Lattice Averaging
  Group}),\ }\href {https://doi.org/10.1140/epjc/s10052-019-7354-7} {\bibfield
  {journal} {\bibinfo  {journal} {Eur. Phys. J. C}\ }\textbf {\bibinfo {volume}
  {80}},\ \bibinfo {pages} {113} (\bibinfo {year} {2020})},\ \Eprint
  {https://arxiv.org/abs/1902.08191} {arXiv:1902.08191 [hep-lat]} \BibitemShut
  {NoStop}%
\bibitem [{\citenamefont {Gambino}\ \emph {et~al.}(2019)\citenamefont
  {Gambino}, \citenamefont {Jung},\ and\ \citenamefont
  {Schacht}}]{Gambino:2019sif}%
  \BibitemOpen
  \bibfield  {author} {\bibinfo {author} {\bibfnamefont {P.}~\bibnamefont
  {Gambino}}, \bibinfo {author} {\bibfnamefont {M.}~\bibnamefont {Jung}},\ and\
  \bibinfo {author} {\bibfnamefont {S.}~\bibnamefont {Schacht}},\ }\href
  {https://doi.org/10.1016/j.physletb.2019.06.039} {\bibfield  {journal}
  {\bibinfo  {journal} {Phys. Lett. B}\ }\textbf {\bibinfo {volume} {795}},\
  \bibinfo {pages} {386} (\bibinfo {year} {2019})},\ \Eprint
  {https://arxiv.org/abs/1905.08209} {arXiv:1905.08209 [hep-ph]} \BibitemShut
  {NoStop}%
\bibitem [{\citenamefont {Wingate}(2021)}]{Wingate:2021ycr}%
  \BibitemOpen
  \bibfield  {author} {\bibinfo {author} {\bibfnamefont {M.}~\bibnamefont
  {Wingate}},\ }\href {https://doi.org/10.1140/epja/s10050-021-00547-z}
  {\bibfield  {journal} {\bibinfo  {journal} {Eur. Phys. J. A}\ }\textbf
  {\bibinfo {volume} {57}},\ \bibinfo {pages} {239} (\bibinfo {year} {2021})},\
  \Eprint {https://arxiv.org/abs/2103.17224} {arXiv:2103.17224 [hep-lat]}
  \BibitemShut {NoStop}%
\bibitem [{\citenamefont {Gottlieb}(2020)}]{Gottlieb:2020zsa}%
  \BibitemOpen
  \bibfield  {author} {\bibinfo {author} {\bibfnamefont {S.}~\bibnamefont
  {Gottlieb}},\ }\href {https://doi.org/10.22323/1.363.0275} {\bibfield
  {journal} {\bibinfo  {journal} {PoS}\ }\textbf {\bibinfo {volume}
  {LATTICE2019}},\ \bibinfo {pages} {275} (\bibinfo {year} {2020})},\ \Eprint
  {https://arxiv.org/abs/2002.09013} {arXiv:2002.09013 [hep-lat]} \BibitemShut
  {NoStop}%
\bibitem [{\citenamefont {Hashimoto}(2017)}]{Hashimoto:2017wqo}%
  \BibitemOpen
  \bibfield  {author} {\bibinfo {author} {\bibfnamefont {S.}~\bibnamefont
  {Hashimoto}},\ }\href {https://doi.org/10.1093/ptep/ptx052} {\bibfield
  {journal} {\bibinfo  {journal} {PTEP}\ }\textbf {\bibinfo {volume} {2017}},\
  \bibinfo {pages} {053B03} (\bibinfo {year} {2017})},\ \Eprint
  {https://arxiv.org/abs/1703.01881} {arXiv:1703.01881 [hep-lat]} \BibitemShut
  {NoStop}%
\bibitem [{\citenamefont {Gambino}\ and\ \citenamefont
  {Hashimoto}(2020)}]{Gambino:2020crt}%
  \BibitemOpen
  \bibfield  {author} {\bibinfo {author} {\bibfnamefont {P.}~\bibnamefont
  {Gambino}}\ and\ \bibinfo {author} {\bibfnamefont {S.}~\bibnamefont
  {Hashimoto}},\ }\href {https://doi.org/10.1103/PhysRevLett.125.032001}
  {\bibfield  {journal} {\bibinfo  {journal} {Phys. Rev. Lett.}\ }\textbf
  {\bibinfo {volume} {125}},\ \bibinfo {pages} {032001} (\bibinfo {year}
  {2020})},\ \Eprint {https://arxiv.org/abs/2005.13730} {arXiv:2005.13730
  [hep-lat]} \BibitemShut {NoStop}%
\bibitem [{Note1()}]{Note1}%
  \BibitemOpen
  \bibinfo {note} {See Supplemental Material at \protect \url
  {http://link.aps.org/supplemental/10.1103/PhysRevD.106.054502} or the end of
  the arXiv version of this document for details on the generation of the gauge
  fields and their properties, and measurements of basic physical
  quantities.}\BibitemShut {Stop}%
\bibitem [{\citenamefont {Burdman}\ \emph {et~al.}(1994)\citenamefont
  {Burdman}, \citenamefont {Ligeti}, \citenamefont {Neubert},\ and\
  \citenamefont {Nir}}]{Burdman:1993es}%
  \BibitemOpen
  \bibfield  {author} {\bibinfo {author} {\bibfnamefont {G.}~\bibnamefont
  {Burdman}}, \bibinfo {author} {\bibfnamefont {Z.}~\bibnamefont {Ligeti}},
  \bibinfo {author} {\bibfnamefont {M.}~\bibnamefont {Neubert}},\ and\ \bibinfo
  {author} {\bibfnamefont {Y.}~\bibnamefont {Nir}},\ }\href
  {https://doi.org/10.1103/PhysRevD.49.2331} {\bibfield  {journal} {\bibinfo
  {journal} {Phys. Rev. D}\ }\textbf {\bibinfo {volume} {49}},\ \bibinfo
  {pages} {2331} (\bibinfo {year} {1994})},\ \Eprint
  {https://arxiv.org/abs/hep-ph/9309272} {arXiv:hep-ph/9309272} \BibitemShut
  {NoStop}%
\bibitem [{\citenamefont {Morningstar}\ and\ \citenamefont
  {Peardon}(2004)}]{Morningstar:2003gk}%
  \BibitemOpen
  \bibfield  {author} {\bibinfo {author} {\bibfnamefont {C.}~\bibnamefont
  {Morningstar}}\ and\ \bibinfo {author} {\bibfnamefont {M.~J.}\ \bibnamefont
  {Peardon}},\ }\href {https://doi.org/10.1103/PhysRevD.69.054501} {\bibfield
  {journal} {\bibinfo  {journal} {Phys. Rev. D}\ }\textbf {\bibinfo {volume}
  {69}},\ \bibinfo {pages} {054501} (\bibinfo {year} {2004})},\ \Eprint
  {https://arxiv.org/abs/hep-lat/0311018} {arXiv:hep-lat/0311018} \BibitemShut
  {NoStop}%
\bibitem [{\citenamefont {Tomii}\ \emph {et~al.}(2016)\citenamefont {Tomii},
  \citenamefont {Cossu}, \citenamefont {Fahy}, \citenamefont {Fukaya},
  \citenamefont {Hashimoto}, \citenamefont {Kaneko},\ and\ \citenamefont
  {Noaki}}]{Tomii:2016xiv}%
  \BibitemOpen
  \bibfield  {author} {\bibinfo {author} {\bibfnamefont {M.}~\bibnamefont
  {Tomii}}, \bibinfo {author} {\bibfnamefont {G.}~\bibnamefont {Cossu}},
  \bibinfo {author} {\bibfnamefont {B.}~\bibnamefont {Fahy}}, \bibinfo {author}
  {\bibfnamefont {H.}~\bibnamefont {Fukaya}}, \bibinfo {author} {\bibfnamefont
  {S.}~\bibnamefont {Hashimoto}}, \bibinfo {author} {\bibfnamefont
  {T.}~\bibnamefont {Kaneko}},\ and\ \bibinfo {author} {\bibfnamefont
  {J.}~\bibnamefont {Noaki}} (\bibinfo {collaboration} {JLQCD}),\ }\href
  {https://doi.org/10.1103/PhysRevD.94.054504} {\bibfield  {journal} {\bibinfo
  {journal} {Phys. Rev. D}\ }\textbf {\bibinfo {volume} {94}},\ \bibinfo
  {pages} {054504} (\bibinfo {year} {2016})},\ \Eprint
  {https://arxiv.org/abs/1604.08702} {arXiv:1604.08702 [hep-lat]} \BibitemShut
  {NoStop}%
\bibitem [{\citenamefont {Nakayama}\ \emph {et~al.}(2016)\citenamefont
  {Nakayama}, \citenamefont {Fahy},\ and\ \citenamefont
  {Hashimoto}}]{Nakayama:2016atf}%
  \BibitemOpen
  \bibfield  {author} {\bibinfo {author} {\bibfnamefont {K.}~\bibnamefont
  {Nakayama}}, \bibinfo {author} {\bibfnamefont {B.}~\bibnamefont {Fahy}},\
  and\ \bibinfo {author} {\bibfnamefont {S.}~\bibnamefont {Hashimoto}},\ }\href
  {https://doi.org/10.1103/PhysRevD.94.054507} {\bibfield  {journal} {\bibinfo
  {journal} {Phys. Rev. D}\ }\textbf {\bibinfo {volume} {94}},\ \bibinfo
  {pages} {054507} (\bibinfo {year} {2016})},\ \Eprint
  {https://arxiv.org/abs/1606.01002} {arXiv:1606.01002 [hep-lat]} \BibitemShut
  {NoStop}%
\bibitem [{\citenamefont {Kaneko}\ \emph {et~al.}(2018)\citenamefont {Kaneko},
  \citenamefont {Colquhoun}, \citenamefont {Fukaya},\ and\ \citenamefont
  {Hashimoto}}]{Kaneko:2017xgg}%
  \BibitemOpen
  \bibfield  {author} {\bibinfo {author} {\bibfnamefont {T.}~\bibnamefont
  {Kaneko}}, \bibinfo {author} {\bibfnamefont {B.}~\bibnamefont {Colquhoun}},
  \bibinfo {author} {\bibfnamefont {H.}~\bibnamefont {Fukaya}},\ and\ \bibinfo
  {author} {\bibfnamefont {S.}~\bibnamefont {Hashimoto}} (\bibinfo
  {collaboration} {JLQCD}),\ }\href
  {https://doi.org/10.1051/epjconf/201817513007} {\bibfield  {journal}
  {\bibinfo  {journal} {EPJ Web Conf.}\ }\textbf {\bibinfo {volume} {175}},\
  \bibinfo {pages} {13007} (\bibinfo {year} {2018})},\ \Eprint
  {https://arxiv.org/abs/1711.11235} {arXiv:1711.11235 [hep-lat]} \BibitemShut
  {NoStop}%
\bibitem [{\citenamefont {Aoki}\ \emph {et~al.}(2018)\citenamefont {Aoki},
  \citenamefont {Cossu}, \citenamefont {Fukaya}, \citenamefont {Hashimoto},\
  and\ \citenamefont {Kaneko}}]{Aoki:2017paw}%
  \BibitemOpen
  \bibfield  {author} {\bibinfo {author} {\bibfnamefont {S.}~\bibnamefont
  {Aoki}}, \bibinfo {author} {\bibfnamefont {G.}~\bibnamefont {Cossu}},
  \bibinfo {author} {\bibfnamefont {H.}~\bibnamefont {Fukaya}}, \bibinfo
  {author} {\bibfnamefont {S.}~\bibnamefont {Hashimoto}},\ and\ \bibinfo
  {author} {\bibfnamefont {T.}~\bibnamefont {Kaneko}} (\bibinfo {collaboration}
  {JLQCD}),\ }\href {https://doi.org/10.1093/ptep/pty041} {\bibfield  {journal}
  {\bibinfo  {journal} {PTEP}\ }\textbf {\bibinfo {volume} {2018}},\ \bibinfo
  {pages} {043B07} (\bibinfo {year} {2018})},\ \Eprint
  {https://arxiv.org/abs/1705.10906} {arXiv:1705.10906 [hep-lat]} \BibitemShut
  {NoStop}%
\bibitem [{\citenamefont {Cossu}\ \emph {et~al.}(2016)\citenamefont {Cossu},
  \citenamefont {Fukaya}, \citenamefont {Hashimoto}, \citenamefont {Kaneko},\
  and\ \citenamefont {Noaki}}]{Cossu:2016eqs}%
  \BibitemOpen
  \bibfield  {author} {\bibinfo {author} {\bibfnamefont {G.}~\bibnamefont
  {Cossu}}, \bibinfo {author} {\bibfnamefont {H.}~\bibnamefont {Fukaya}},
  \bibinfo {author} {\bibfnamefont {S.}~\bibnamefont {Hashimoto}}, \bibinfo
  {author} {\bibfnamefont {T.}~\bibnamefont {Kaneko}},\ and\ \bibinfo {author}
  {\bibfnamefont {J.-I.}\ \bibnamefont {Noaki}},\ }\href
  {https://doi.org/10.1093/ptep/ptw129} {\bibfield  {journal} {\bibinfo
  {journal} {PTEP}\ }\textbf {\bibinfo {volume} {2016}},\ \bibinfo {pages}
  {093B06} (\bibinfo {year} {2016})},\ \Eprint
  {https://arxiv.org/abs/1607.01099} {arXiv:1607.01099 [hep-lat]} \BibitemShut
  {NoStop}%
\bibitem [{\citenamefont {Nakayama}\ \emph {et~al.}(2018)\citenamefont
  {Nakayama}, \citenamefont {Fukaya},\ and\ \citenamefont
  {Hashimoto}}]{Nakayama:2018ubk}%
  \BibitemOpen
  \bibfield  {author} {\bibinfo {author} {\bibfnamefont {K.}~\bibnamefont
  {Nakayama}}, \bibinfo {author} {\bibfnamefont {H.}~\bibnamefont {Fukaya}},\
  and\ \bibinfo {author} {\bibfnamefont {S.}~\bibnamefont {Hashimoto}},\ }\href
  {https://doi.org/10.1103/PhysRevD.98.014501} {\bibfield  {journal} {\bibinfo
  {journal} {Phys. Rev. D}\ }\textbf {\bibinfo {volume} {98}},\ \bibinfo
  {pages} {014501} (\bibinfo {year} {2018})},\ \Eprint
  {https://arxiv.org/abs/1804.06695} {arXiv:1804.06695 [hep-lat]} \BibitemShut
  {NoStop}%
\bibitem [{\citenamefont {Tomii}\ \emph {et~al.}(2017)\citenamefont {Tomii},
  \citenamefont {Cossu}, \citenamefont {Fahy}, \citenamefont {Fukaya},
  \citenamefont {Hashimoto}, \citenamefont {Kaneko},\ and\ \citenamefont
  {Noaki}}]{Tomii:2017cbt}%
  \BibitemOpen
  \bibfield  {author} {\bibinfo {author} {\bibfnamefont {M.}~\bibnamefont
  {Tomii}}, \bibinfo {author} {\bibfnamefont {G.}~\bibnamefont {Cossu}},
  \bibinfo {author} {\bibfnamefont {B.}~\bibnamefont {Fahy}}, \bibinfo {author}
  {\bibfnamefont {H.}~\bibnamefont {Fukaya}}, \bibinfo {author} {\bibfnamefont
  {S.}~\bibnamefont {Hashimoto}}, \bibinfo {author} {\bibfnamefont
  {T.}~\bibnamefont {Kaneko}},\ and\ \bibinfo {author} {\bibfnamefont
  {J.}~\bibnamefont {Noaki}} (\bibinfo {collaboration} {JLQCD}),\ }\href
  {https://doi.org/10.1103/PhysRevD.96.054511} {\bibfield  {journal} {\bibinfo
  {journal} {Phys. Rev. D}\ }\textbf {\bibinfo {volume} {96}},\ \bibinfo
  {pages} {054511} (\bibinfo {year} {2017})},\ \Eprint
  {https://arxiv.org/abs/1703.06249} {arXiv:1703.06249 [hep-lat]} \BibitemShut
  {NoStop}%
\bibitem [{\citenamefont {Lepage}\ \emph {et~al.}(2002)\citenamefont {Lepage},
  \citenamefont {Clark}, \citenamefont {Davies}, \citenamefont {Hornbostel},
  \citenamefont {Mackenzie}, \citenamefont {Morningstar},\ and\ \citenamefont
  {Trottier}}]{Lepage:2001ym}%
  \BibitemOpen
  \bibfield  {author} {\bibinfo {author} {\bibfnamefont {G.~P.}\ \bibnamefont
  {Lepage}}, \bibinfo {author} {\bibfnamefont {B.}~\bibnamefont {Clark}},
  \bibinfo {author} {\bibfnamefont {C.~T.~H.}\ \bibnamefont {Davies}}, \bibinfo
  {author} {\bibfnamefont {K.}~\bibnamefont {Hornbostel}}, \bibinfo {author}
  {\bibfnamefont {P.~B.}\ \bibnamefont {Mackenzie}}, \bibinfo {author}
  {\bibfnamefont {C.}~\bibnamefont {Morningstar}},\ and\ \bibinfo {author}
  {\bibfnamefont {H.}~\bibnamefont {Trottier}},\ }\href
  {https://doi.org/10.1016/S0920-5632(01)01638-3} {\bibfield  {journal}
  {\bibinfo  {journal} {Nucl. Phys. B Proc. Suppl.}\ }\textbf {\bibinfo
  {volume} {106}},\ \bibinfo {pages} {12} (\bibinfo {year} {2002})},\ \Eprint
  {https://arxiv.org/abs/hep-lat/0110175} {arXiv:hep-lat/0110175} \BibitemShut
  {NoStop}%
\bibitem [{\citenamefont {Lepage}\ \emph {et~al.}(2021)\citenamefont {Lepage},
  \citenamefont {Gohlke},\ and\ \citenamefont
  {Hackett}}]{peter_lepage_2021_5202756}%
  \BibitemOpen
  \bibfield  {author} {\bibinfo {author} {\bibfnamefont {P.}~\bibnamefont
  {Lepage}}, \bibinfo {author} {\bibfnamefont {C.}~\bibnamefont {Gohlke}},\
  and\ \bibinfo {author} {\bibfnamefont {D.}~\bibnamefont {Hackett}},\ }\href
  {https://doi.org/10.5281/zenodo.5202756} {\bibinfo {title} {gplepage/gvar:
  gvar version 11.9.3}} (\bibinfo {year} {2021})\BibitemShut {NoStop}%
\bibitem [{\citenamefont {Lepage}\ and\ \citenamefont
  {Gohlke}(2021)}]{peter_lepage_2021_5202760}%
  \BibitemOpen
  \bibfield  {author} {\bibinfo {author} {\bibfnamefont {P.}~\bibnamefont
  {Lepage}}\ and\ \bibinfo {author} {\bibfnamefont {C.}~\bibnamefont
  {Gohlke}},\ }\href {https://doi.org/10.5281/zenodo.5202760} {\bibinfo {title}
  {gplepage/lsqfit: lsqfit version 12.0}} (\bibinfo {year} {2021})\BibitemShut
  {NoStop}%
\bibitem [{\citenamefont {Lepage}(2020)}]{peter_lepage_2020_4281296}%
  \BibitemOpen
  \bibfield  {author} {\bibinfo {author} {\bibfnamefont {P.}~\bibnamefont
  {Lepage}},\ }\href {https://doi.org/10.5281/zenodo.4281296} {\bibinfo {title}
  {gplepage/corrfitter: corrfitter version 8.1.1}} (\bibinfo {year}
  {2020})\BibitemShut {NoStop}%
\bibitem [{\citenamefont {Hashimoto}\ \emph {et~al.}(2018)\citenamefont
  {Hashimoto}, \citenamefont {Colquhoun}, \citenamefont {Izubuchi},
  \citenamefont {Kaneko},\ and\ \citenamefont {Ohki}}]{Hashimoto:2018gld}%
  \BibitemOpen
  \bibfield  {author} {\bibinfo {author} {\bibfnamefont {S.}~\bibnamefont
  {Hashimoto}}, \bibinfo {author} {\bibfnamefont {B.}~\bibnamefont
  {Colquhoun}}, \bibinfo {author} {\bibfnamefont {T.}~\bibnamefont {Izubuchi}},
  \bibinfo {author} {\bibfnamefont {T.}~\bibnamefont {Kaneko}},\ and\ \bibinfo
  {author} {\bibfnamefont {H.}~\bibnamefont {Ohki}},\ }\href
  {https://doi.org/10.1051/epjconf/201817513006} {\bibfield  {journal}
  {\bibinfo  {journal} {EPJ Web Conf.}\ }\textbf {\bibinfo {volume} {175}},\
  \bibinfo {pages} {13006} (\bibinfo {year} {2018})}\BibitemShut {NoStop}%
\bibitem [{\citenamefont {Bijnens}\ and\ \citenamefont
  {Jemos}(2010)}]{Bijnens:2010ws}%
  \BibitemOpen
  \bibfield  {author} {\bibinfo {author} {\bibfnamefont {J.}~\bibnamefont
  {Bijnens}}\ and\ \bibinfo {author} {\bibfnamefont {I.}~\bibnamefont
  {Jemos}},\ }\href {https://doi.org/10.1016/j.nuclphysb.2010.06.021}
  {\bibfield  {journal} {\bibinfo  {journal} {Nucl. Phys. B}\ }\textbf
  {\bibinfo {volume} {840}},\ \bibinfo {pages} {54} (\bibinfo {year} {2010})},\
  \bibinfo {note} {[Erratum: Nucl.Phys.B 844, 182--183 (2011)]},\ \Eprint
  {https://arxiv.org/abs/1006.1197} {arXiv:1006.1197 [hep-ph]} \BibitemShut
  {NoStop}%
\bibitem [{\citenamefont {Becirevic}\ \emph {et~al.}(2003)\citenamefont
  {Becirevic}, \citenamefont {Prelovsek},\ and\ \citenamefont
  {Zupan}}]{Becirevic:2002sc}%
  \BibitemOpen
  \bibfield  {author} {\bibinfo {author} {\bibfnamefont {D.}~\bibnamefont
  {Becirevic}}, \bibinfo {author} {\bibfnamefont {S.}~\bibnamefont
  {Prelovsek}},\ and\ \bibinfo {author} {\bibfnamefont {J.}~\bibnamefont
  {Zupan}},\ }\href {https://doi.org/10.1103/PhysRevD.67.054010} {\bibfield
  {journal} {\bibinfo  {journal} {Phys. Rev. D}\ }\textbf {\bibinfo {volume}
  {67}},\ \bibinfo {pages} {054010} (\bibinfo {year} {2003})},\ \Eprint
  {https://arxiv.org/abs/hep-lat/0210048} {arXiv:hep-lat/0210048} \BibitemShut
  {NoStop}%
\bibitem [{\citenamefont {Detmold}\ \emph {et~al.}(2012)\citenamefont
  {Detmold}, \citenamefont {Lin},\ and\ \citenamefont
  {Meinel}}]{Detmold:2012ge}%
  \BibitemOpen
  \bibfield  {author} {\bibinfo {author} {\bibfnamefont {W.}~\bibnamefont
  {Detmold}}, \bibinfo {author} {\bibfnamefont {C.~J.~D.}\ \bibnamefont
  {Lin}},\ and\ \bibinfo {author} {\bibfnamefont {S.}~\bibnamefont {Meinel}},\
  }\href {https://doi.org/10.1103/PhysRevD.85.114508} {\bibfield  {journal}
  {\bibinfo  {journal} {Phys. Rev. D}\ }\textbf {\bibinfo {volume} {85}},\
  \bibinfo {pages} {114508} (\bibinfo {year} {2012})},\ \Eprint
  {https://arxiv.org/abs/1203.3378} {arXiv:1203.3378 [hep-lat]} \BibitemShut
  {NoStop}%
\bibitem [{\citenamefont {Ohki}\ \emph {et~al.}(2008)\citenamefont {Ohki},
  \citenamefont {Matsufuru},\ and\ \citenamefont {Onogi}}]{Ohki:2008py}%
  \BibitemOpen
  \bibfield  {author} {\bibinfo {author} {\bibfnamefont {H.}~\bibnamefont
  {Ohki}}, \bibinfo {author} {\bibfnamefont {H.}~\bibnamefont {Matsufuru}},\
  and\ \bibinfo {author} {\bibfnamefont {T.}~\bibnamefont {Onogi}},\ }\href
  {https://doi.org/10.1103/PhysRevD.77.094509} {\bibfield  {journal} {\bibinfo
  {journal} {Phys. Rev. D}\ }\textbf {\bibinfo {volume} {77}},\ \bibinfo
  {pages} {094509} (\bibinfo {year} {2008})},\ \Eprint
  {https://arxiv.org/abs/0802.1563} {arXiv:0802.1563 [hep-lat]} \BibitemShut
  {NoStop}%
\bibitem [{\citenamefont {Becirevic}\ \emph {et~al.}(2009)\citenamefont
  {Becirevic}, \citenamefont {Blossier}, \citenamefont {Chang},\ and\
  \citenamefont {Haas}}]{Becirevic:2009yb}%
  \BibitemOpen
  \bibfield  {author} {\bibinfo {author} {\bibfnamefont {D.}~\bibnamefont
  {Becirevic}}, \bibinfo {author} {\bibfnamefont {B.}~\bibnamefont {Blossier}},
  \bibinfo {author} {\bibfnamefont {E.}~\bibnamefont {Chang}},\ and\ \bibinfo
  {author} {\bibfnamefont {B.}~\bibnamefont {Haas}},\ }\href
  {https://doi.org/10.1016/j.physletb.2009.07.031} {\bibfield  {journal}
  {\bibinfo  {journal} {Phys. Lett. B}\ }\textbf {\bibinfo {volume} {679}},\
  \bibinfo {pages} {231} (\bibinfo {year} {2009})},\ \Eprint
  {https://arxiv.org/abs/0905.3355} {arXiv:0905.3355 [hep-ph]} \BibitemShut
  {NoStop}%
\bibitem [{\citenamefont {Becirevic}\ and\ \citenamefont
  {Sanfilippo}(2013)}]{Becirevic:2012pf}%
  \BibitemOpen
  \bibfield  {author} {\bibinfo {author} {\bibfnamefont {D.}~\bibnamefont
  {Becirevic}}\ and\ \bibinfo {author} {\bibfnamefont {F.}~\bibnamefont
  {Sanfilippo}},\ }\href {https://doi.org/10.1016/j.physletb.2013.03.004}
  {\bibfield  {journal} {\bibinfo  {journal} {Phys. Lett. B}\ }\textbf
  {\bibinfo {volume} {721}},\ \bibinfo {pages} {94} (\bibinfo {year} {2013})},\
  \Eprint {https://arxiv.org/abs/1210.5410} {arXiv:1210.5410 [hep-lat]}
  \BibitemShut {NoStop}%
\bibitem [{\citenamefont {Bernardoni}\ \emph {et~al.}(2015)\citenamefont
  {Bernardoni}, \citenamefont {Bulava}, \citenamefont {Donnellan},\ and\
  \citenamefont {Sommer}}]{Bernardoni:2014kla}%
  \BibitemOpen
  \bibfield  {author} {\bibinfo {author} {\bibfnamefont {F.}~\bibnamefont
  {Bernardoni}}, \bibinfo {author} {\bibfnamefont {J.}~\bibnamefont {Bulava}},
  \bibinfo {author} {\bibfnamefont {M.}~\bibnamefont {Donnellan}},\ and\
  \bibinfo {author} {\bibfnamefont {R.}~\bibnamefont {Sommer}} (\bibinfo
  {collaboration} {ALPHA}),\ }\href
  {https://doi.org/10.1016/j.physletb.2014.11.051} {\bibfield  {journal}
  {\bibinfo  {journal} {Phys. Lett. B}\ }\textbf {\bibinfo {volume} {740}},\
  \bibinfo {pages} {278} (\bibinfo {year} {2015})},\ \Eprint
  {https://arxiv.org/abs/1404.6951} {arXiv:1404.6951 [hep-lat]} \BibitemShut
  {NoStop}%
\bibitem [{\citenamefont {Flynn}\ \emph {et~al.}(2016)\citenamefont {Flynn},
  \citenamefont {Fritzsch}, \citenamefont {Kawanai}, \citenamefont {Lehner},
  \citenamefont {Samways}, \citenamefont {Sachrajda}, \citenamefont {Van~de
  Water},\ and\ \citenamefont {Witzel}}]{Flynn:2015xna}%
  \BibitemOpen
  \bibfield  {author} {\bibinfo {author} {\bibfnamefont {J.~M.}\ \bibnamefont
  {Flynn}}, \bibinfo {author} {\bibfnamefont {P.}~\bibnamefont {Fritzsch}},
  \bibinfo {author} {\bibfnamefont {T.}~\bibnamefont {Kawanai}}, \bibinfo
  {author} {\bibfnamefont {C.}~\bibnamefont {Lehner}}, \bibinfo {author}
  {\bibfnamefont {B.}~\bibnamefont {Samways}}, \bibinfo {author} {\bibfnamefont
  {C.~T.}\ \bibnamefont {Sachrajda}}, \bibinfo {author} {\bibfnamefont {R.~S.}\
  \bibnamefont {Van~de Water}},\ and\ \bibinfo {author} {\bibfnamefont
  {O.}~\bibnamefont {Witzel}} (\bibinfo {collaboration} {RBC, UKQCD}),\ }\href
  {https://doi.org/10.1103/PhysRevD.93.014510} {\bibfield  {journal} {\bibinfo
  {journal} {Phys. Rev. D}\ }\textbf {\bibinfo {volume} {93}},\ \bibinfo
  {pages} {014510} (\bibinfo {year} {2016})},\ \Eprint
  {https://arxiv.org/abs/1506.06413} {arXiv:1506.06413 [hep-lat]} \BibitemShut
  {NoStop}%
\bibitem [{\citenamefont {del Amo~Sanchez}\ \emph {et~al.}(2011)\citenamefont
  {del Amo~Sanchez} \emph {et~al.}}]{BaBar:2010efp}%
  \BibitemOpen
  \bibfield  {author} {\bibinfo {author} {\bibfnamefont {P.}~\bibnamefont {del
  Amo~Sanchez}} \emph {et~al.} (\bibinfo {collaboration} {BaBar}),\ }\href
  {https://doi.org/10.1103/PhysRevD.83.032007} {\bibfield  {journal} {\bibinfo
  {journal} {Phys. Rev. D}\ }\textbf {\bibinfo {volume} {83}},\ \bibinfo
  {pages} {032007} (\bibinfo {year} {2011})},\ \Eprint
  {https://arxiv.org/abs/1005.3288} {arXiv:1005.3288 [hep-ex]} \BibitemShut
  {NoStop}%
\bibitem [{\citenamefont {Boyd}\ and\ \citenamefont
  {Savage}(1997)}]{Boyd:1997qw}%
  \BibitemOpen
  \bibfield  {author} {\bibinfo {author} {\bibfnamefont {C.~G.}\ \bibnamefont
  {Boyd}}\ and\ \bibinfo {author} {\bibfnamefont {M.~J.}\ \bibnamefont
  {Savage}},\ }\href {https://doi.org/10.1103/PhysRevD.56.303} {\bibfield
  {journal} {\bibinfo  {journal} {Phys. Rev. D}\ }\textbf {\bibinfo {volume}
  {56}},\ \bibinfo {pages} {303} (\bibinfo {year} {1997})},\ \Eprint
  {https://arxiv.org/abs/hep-ph/9702300} {arXiv:hep-ph/9702300} \BibitemShut
  {NoStop}%
\bibitem [{\citenamefont {Colquhoun}\ \emph {et~al.}(2015)\citenamefont
  {Colquhoun}, \citenamefont {Davies}, \citenamefont {Dowdall}, \citenamefont
  {Kettle}, \citenamefont {Koponen}, \citenamefont {Lepage},\ and\
  \citenamefont {Lytle}}]{Colquhoun:2015oha}%
  \BibitemOpen
  \bibfield  {author} {\bibinfo {author} {\bibfnamefont {B.}~\bibnamefont
  {Colquhoun}}, \bibinfo {author} {\bibfnamefont {C.~T.~H.}\ \bibnamefont
  {Davies}}, \bibinfo {author} {\bibfnamefont {R.~J.}\ \bibnamefont {Dowdall}},
  \bibinfo {author} {\bibfnamefont {J.}~\bibnamefont {Kettle}}, \bibinfo
  {author} {\bibfnamefont {J.}~\bibnamefont {Koponen}}, \bibinfo {author}
  {\bibfnamefont {G.~P.}\ \bibnamefont {Lepage}},\ and\ \bibinfo {author}
  {\bibfnamefont {A.~T.}\ \bibnamefont {Lytle}} (\bibinfo {collaboration}
  {HPQCD}),\ }\href {https://doi.org/10.1103/PhysRevD.91.114509} {\bibfield
  {journal} {\bibinfo  {journal} {Phys. Rev. D}\ }\textbf {\bibinfo {volume}
  {91}},\ \bibinfo {pages} {114509} (\bibinfo {year} {2015})},\ \Eprint
  {https://arxiv.org/abs/1503.05762} {arXiv:1503.05762 [hep-lat]} \BibitemShut
  {NoStop}%
\bibitem [{\citenamefont {Ha}\ \emph {et~al.}(2011)\citenamefont {Ha} \emph
  {et~al.}}]{Belle:2010hep}%
  \BibitemOpen
  \bibfield  {author} {\bibinfo {author} {\bibfnamefont {H.}~\bibnamefont {Ha}}
  \emph {et~al.} (\bibinfo {collaboration} {Belle}),\ }\href
  {https://doi.org/10.1103/PhysRevD.83.071101} {\bibfield  {journal} {\bibinfo
  {journal} {Phys. Rev. D}\ }\textbf {\bibinfo {volume} {83}},\ \bibinfo
  {pages} {071101} (\bibinfo {year} {2011})},\ \Eprint
  {https://arxiv.org/abs/1012.0090} {arXiv:1012.0090 [hep-ex]} \BibitemShut
  {NoStop}%
\bibitem [{\citenamefont {Lees}\ \emph {et~al.}(2012)\citenamefont {Lees} \emph
  {et~al.}}]{BaBar:2012thb}%
  \BibitemOpen
  \bibfield  {author} {\bibinfo {author} {\bibfnamefont {J.~P.}\ \bibnamefont
  {Lees}} \emph {et~al.} (\bibinfo {collaboration} {BaBar}),\ }\href
  {https://doi.org/10.1103/PhysRevD.86.092004} {\bibfield  {journal} {\bibinfo
  {journal} {Phys. Rev. D}\ }\textbf {\bibinfo {volume} {86}},\ \bibinfo
  {pages} {092004} (\bibinfo {year} {2012})},\ \Eprint
  {https://arxiv.org/abs/1208.1253} {arXiv:1208.1253 [hep-ex]} \BibitemShut
  {NoStop}%
\bibitem [{\citenamefont {Sibidanov}\ \emph {et~al.}(2013)\citenamefont
  {Sibidanov} \emph {et~al.}}]{Belle:2013hlo}%
  \BibitemOpen
  \bibfield  {author} {\bibinfo {author} {\bibfnamefont {A.}~\bibnamefont
  {Sibidanov}} \emph {et~al.} (\bibinfo {collaboration} {Belle}),\ }\href
  {https://doi.org/10.1103/PhysRevD.88.032005} {\bibfield  {journal} {\bibinfo
  {journal} {Phys. Rev. D}\ }\textbf {\bibinfo {volume} {88}},\ \bibinfo
  {pages} {032005} (\bibinfo {year} {2013})},\ \Eprint
  {https://arxiv.org/abs/1306.2781} {arXiv:1306.2781 [hep-ex]} \BibitemShut
  {NoStop}%
\bibitem [{\citenamefont {Zyla}\ \emph {et~al.}(2020)\citenamefont {Zyla} \emph
  {et~al.}}]{ParticleDataGroup:2020ssz}%
  \BibitemOpen
  \bibfield  {author} {\bibinfo {author} {\bibfnamefont {P.~A.}\ \bibnamefont
  {Zyla}} \emph {et~al.} (\bibinfo {collaboration} {Particle Data Group}),\
  }\href {https://doi.org/10.1093/ptep/ptaa104} {\bibfield  {journal} {\bibinfo
   {journal} {PTEP}\ }\textbf {\bibinfo {volume} {2020}},\ \bibinfo {pages}
  {083C01} (\bibinfo {year} {2020})}\BibitemShut {NoStop}%
\bibitem [{\citenamefont {Biswas}\ \emph {et~al.}(2021)\citenamefont {Biswas},
  \citenamefont {Nandi}, \citenamefont {Patra},\ and\ \citenamefont
  {Ray}}]{Biswas:2021qyq}%
  \BibitemOpen
  \bibfield  {author} {\bibinfo {author} {\bibfnamefont {A.}~\bibnamefont
  {Biswas}}, \bibinfo {author} {\bibfnamefont {S.}~\bibnamefont {Nandi}},
  \bibinfo {author} {\bibfnamefont {S.~K.}\ \bibnamefont {Patra}},\ and\
  \bibinfo {author} {\bibfnamefont {I.}~\bibnamefont {Ray}},\ }\href
  {https://doi.org/10.1007/JHEP07(2021)082} {\bibfield  {journal} {\bibinfo
  {journal} {JHEP}\ }\textbf {\bibinfo {volume} {07}},\ \bibinfo {pages}
  {082}},\ \Eprint {https://arxiv.org/abs/2103.01809} {arXiv:2103.01809
  [hep-ph]} \BibitemShut {NoStop}%
\bibitem [{\citenamefont {Gonz\`alez-Sol\'\i{}s}\ \emph
  {et~al.}(2021)\citenamefont {Gonz\`alez-Sol\'\i{}s}, \citenamefont
  {Masjuan},\ and\ \citenamefont {Rojas}}]{Gonzalez-Solis:2021pyh}%
  \BibitemOpen
  \bibfield  {author} {\bibinfo {author} {\bibfnamefont {S.}~\bibnamefont
  {Gonz\`alez-Sol\'\i{}s}}, \bibinfo {author} {\bibfnamefont {P.}~\bibnamefont
  {Masjuan}},\ and\ \bibinfo {author} {\bibfnamefont {C.}~\bibnamefont
  {Rojas}},\ }\href {https://doi.org/10.1103/PhysRevD.104.114041} {\bibfield
  {journal} {\bibinfo  {journal} {Phys. Rev. D}\ }\textbf {\bibinfo {volume}
  {104}},\ \bibinfo {pages} {114041} (\bibinfo {year} {2021})},\ \Eprint
  {https://arxiv.org/abs/2110.06153} {arXiv:2110.06153 [hep-ph]} \BibitemShut
  {NoStop}%
\bibitem [{\citenamefont {Leljak}\ \emph {et~al.}(2021)\citenamefont {Leljak},
  \citenamefont {Meli\'c},\ and\ \citenamefont {van Dyk}}]{Leljak:2021vte}%
  \BibitemOpen
  \bibfield  {author} {\bibinfo {author} {\bibfnamefont {D.}~\bibnamefont
  {Leljak}}, \bibinfo {author} {\bibfnamefont {B.}~\bibnamefont {Meli\'c}},\
  and\ \bibinfo {author} {\bibfnamefont {D.}~\bibnamefont {van Dyk}},\ }\href
  {https://doi.org/10.1007/JHEP07(2021)036} {\bibfield  {journal} {\bibinfo
  {journal} {JHEP}\ }\textbf {\bibinfo {volume} {07}},\ \bibinfo {pages}
  {036}},\ \Eprint {https://arxiv.org/abs/2102.07233} {arXiv:2102.07233
  [hep-ph]} \BibitemShut {NoStop}%
\bibitem [{\citenamefont {Dingfelder}\ and\ \citenamefont
  {Mannel}(2016)}]{Dingfelder:2016twb}%
  \BibitemOpen
  \bibfield  {author} {\bibinfo {author} {\bibfnamefont {J.}~\bibnamefont
  {Dingfelder}}\ and\ \bibinfo {author} {\bibfnamefont {T.}~\bibnamefont
  {Mannel}},\ }\href {https://doi.org/10.1103/RevModPhys.88.035008} {\bibfield
  {journal} {\bibinfo  {journal} {Rev. Mod. Phys.}\ }\textbf {\bibinfo {volume}
  {88}},\ \bibinfo {pages} {035008} (\bibinfo {year} {2016})}\BibitemShut
  {NoStop}%
\bibitem [{\citenamefont {Detmold}\ \emph {et~al.}(2015)\citenamefont
  {Detmold}, \citenamefont {Lehner},\ and\ \citenamefont
  {Meinel}}]{Detmold:2015aaa}%
  \BibitemOpen
  \bibfield  {author} {\bibinfo {author} {\bibfnamefont {W.}~\bibnamefont
  {Detmold}}, \bibinfo {author} {\bibfnamefont {C.}~\bibnamefont {Lehner}},\
  and\ \bibinfo {author} {\bibfnamefont {S.}~\bibnamefont {Meinel}},\ }\href
  {https://doi.org/10.1103/PhysRevD.92.034503} {\bibfield  {journal} {\bibinfo
  {journal} {Phys. Rev. D}\ }\textbf {\bibinfo {volume} {92}},\ \bibinfo
  {pages} {034503} (\bibinfo {year} {2015})},\ \Eprint
  {https://arxiv.org/abs/1503.01421} {arXiv:1503.01421 [hep-lat]} \BibitemShut
  {NoStop}%
\bibitem [{\citenamefont {Aaij}\ \emph {et~al.}(2015)\citenamefont {Aaij} \emph
  {et~al.}}]{LHCb:2015eia}%
  \BibitemOpen
  \bibfield  {author} {\bibinfo {author} {\bibfnamefont {R.}~\bibnamefont
  {Aaij}} \emph {et~al.} (\bibinfo {collaboration} {LHCb}),\ }\href
  {https://doi.org/10.1038/nphys3415} {\bibfield  {journal} {\bibinfo
  {journal} {Nature Phys.}\ }\textbf {\bibinfo {volume} {11}},\ \bibinfo
  {pages} {743} (\bibinfo {year} {2015})},\ \Eprint
  {https://arxiv.org/abs/1504.01568} {arXiv:1504.01568 [hep-ex]} \BibitemShut
  {NoStop}%
\bibitem [{\citenamefont {Olive}\ \emph {et~al.}(2014)\citenamefont {Olive}
  \emph {et~al.}}]{ParticleDataGroup:2014cgo}%
  \BibitemOpen
  \bibfield  {author} {\bibinfo {author} {\bibfnamefont {K.~A.}\ \bibnamefont
  {Olive}} \emph {et~al.} (\bibinfo {collaboration} {Particle Data Group}),\
  }\href {https://doi.org/10.1088/1674-1137/38/9/090001} {\bibfield  {journal}
  {\bibinfo  {journal} {Chin. Phys. C}\ }\textbf {\bibinfo {volume} {38}},\
  \bibinfo {pages} {090001} (\bibinfo {year} {2014})}\BibitemShut {NoStop}%
\bibitem [{\citenamefont {Cossu}\ \emph {et~al.}(2013)\citenamefont {Cossu},
  \citenamefont {Noaki}, \citenamefont {Hashimoto}, \citenamefont {Kaneko},
  \citenamefont {Fukaya}, \citenamefont {Boyle},\ and\ \citenamefont
  {Doi}}]{Cossu:2013ola}%
  \BibitemOpen
  \bibfield  {author} {\bibinfo {author} {\bibfnamefont {G.}~\bibnamefont
  {Cossu}}, \bibinfo {author} {\bibfnamefont {J.}~\bibnamefont {Noaki}},
  \bibinfo {author} {\bibfnamefont {S.}~\bibnamefont {Hashimoto}}, \bibinfo
  {author} {\bibfnamefont {T.}~\bibnamefont {Kaneko}}, \bibinfo {author}
  {\bibfnamefont {H.}~\bibnamefont {Fukaya}}, \bibinfo {author} {\bibfnamefont
  {P.~A.}\ \bibnamefont {Boyle}},\ and\ \bibinfo {author} {\bibfnamefont
  {J.}~\bibnamefont {Doi}},\ }in\ \href@noop {} {\emph {\bibinfo {booktitle}
  {{31st International Symposium on Lattice Field Theory}}}}\ (\bibinfo {year}
  {2013})\ \Eprint {https://arxiv.org/abs/1311.0084} {arXiv:1311.0084
  [hep-lat]} \BibitemShut {NoStop}%
\bibitem [{\citenamefont {Gerosa}\ and\ \citenamefont
  {Vallisneri}(2017)}]{Gerosa:2017xrm}%
  \BibitemOpen
  \bibfield  {author} {\bibinfo {author} {\bibfnamefont {D.}~\bibnamefont
  {Gerosa}}\ and\ \bibinfo {author} {\bibfnamefont {M.}~\bibnamefont
  {Vallisneri}},\ }\href {https://doi.org/10.21105/joss.00222} {\bibfield
  {journal} {\bibinfo  {journal} {J. Open Source Softw.}\ }\textbf {\bibinfo
  {volume} {2}},\ \bibinfo {pages} {222} (\bibinfo {year} {2017})}\BibitemShut
  {NoStop}%
\end{thebibliography}%


\begin{thebibliography}{29}%
\makeatletter
\providecommand \@ifxundefined [1]{%
 \@ifx{#1\undefined}
}%
\providecommand \@ifnum [1]{%
 \ifnum #1\expandafter \@firstoftwo
 \else \expandafter \@secondoftwo
 \fi
}%
\providecommand \@ifx [1]{%
 \ifx #1\expandafter \@firstoftwo
 \else \expandafter \@secondoftwo
 \fi
}%
\providecommand \natexlab [1]{#1}%
\providecommand \enquote  [1]{``#1''}%
\providecommand \bibnamefont  [1]{#1}%
\providecommand \bibfnamefont [1]{#1}%
\providecommand \citenamefont [1]{#1}%
\providecommand \href@noop [0]{\@secondoftwo}%
\providecommand \href [0]{\begingroup \@sanitize@url \@href}%
\providecommand \@href[1]{\@@startlink{#1}\@@href}%
\providecommand \@@href[1]{\endgroup#1\@@endlink}%
\providecommand \@sanitize@url [0]{\catcode `\\12\catcode `\$12\catcode
  `\&12\catcode `\#12\catcode `\^12\catcode `\_12\catcode `\%12\relax}%
\providecommand \@@startlink[1]{}%
\providecommand \@@endlink[0]{}%
\providecommand \url  [0]{\begingroup\@sanitize@url \@url }%
\providecommand \@url [1]{\endgroup\@href {#1}{\urlprefix }}%
\providecommand \urlprefix  [0]{URL }%
\providecommand \Eprint [0]{\href }%
\providecommand \doibase [0]{https://doi.org/}%
\providecommand \selectlanguage [0]{\@gobble}%
\providecommand \bibinfo  [0]{\@secondoftwo}%
\providecommand \bibfield  [0]{\@secondoftwo}%
\providecommand \translation [1]{[#1]}%
\providecommand \BibitemOpen [0]{}%
\providecommand \bibitemStop [0]{}%
\providecommand \bibitemNoStop [0]{.\EOS\space}%
\providecommand \EOS [0]{\spacefactor3000\relax}%
\providecommand \BibitemShut  [1]{\csname bibitem#1\endcsname}%
\let\auto@bib@innerbib\@empty
\bibitem [{\citenamefont {Brower}\ \emph {et~al.}(2017)\citenamefont {Brower},
  \citenamefont {Neff},\ and\ \citenamefont {Orginos}}]{Brower:2012vk}%
  \BibitemOpen
  \bibfield  {author} {\bibinfo {author} {\bibfnamefont {R.~C.}\ \bibnamefont
  {Brower}}, \bibinfo {author} {\bibfnamefont {H.}~\bibnamefont {Neff}},\ and\
  \bibinfo {author} {\bibfnamefont {K.}~\bibnamefont {Orginos}},\ }\href
  {https://doi.org/10.1016/j.cpc.2017.01.024} {\bibfield  {journal} {\bibinfo
  {journal} {Comput. Phys. Commun.}\ }\textbf {\bibinfo {volume} {220}},\
  \bibinfo {pages} {1} (\bibinfo {year} {2017})},\ \Eprint
  {https://arxiv.org/abs/1206.5214} {arXiv:1206.5214 [hep-lat]} \BibitemShut
  {NoStop}%
\bibitem [{\citenamefont {Kaplan}(1992)}]{Kaplan:1992bt}%
  \BibitemOpen
  \bibfield  {author} {\bibinfo {author} {\bibfnamefont {D.~B.}\ \bibnamefont
  {Kaplan}},\ }\href {https://doi.org/10.1016/0370-2693(92)91112-M} {\bibfield
  {journal} {\bibinfo  {journal} {Phys. Lett. B}\ }\textbf {\bibinfo {volume}
  {288}},\ \bibinfo {pages} {342} (\bibinfo {year} {1992})},\ \Eprint
  {https://arxiv.org/abs/hep-lat/9206013} {arXiv:hep-lat/9206013} \BibitemShut
  {NoStop}%
\bibitem [{\citenamefont {Shamir}(1993)}]{Shamir:1993zy}%
  \BibitemOpen
  \bibfield  {author} {\bibinfo {author} {\bibfnamefont {Y.}~\bibnamefont
  {Shamir}},\ }\href {https://doi.org/10.1016/0550-3213(93)90162-I} {\bibfield
  {journal} {\bibinfo  {journal} {Nucl. Phys. B}\ }\textbf {\bibinfo {volume}
  {406}},\ \bibinfo {pages} {90} (\bibinfo {year} {1993})},\ \Eprint
  {https://arxiv.org/abs/hep-lat/9303005} {arXiv:hep-lat/9303005} \BibitemShut
  {NoStop}%
\bibitem [{\citenamefont {Furman}\ and\ \citenamefont
  {Shamir}(1995)}]{Furman:1994ky}%
  \BibitemOpen
  \bibfield  {author} {\bibinfo {author} {\bibfnamefont {V.}~\bibnamefont
  {Furman}}\ and\ \bibinfo {author} {\bibfnamefont {Y.}~\bibnamefont
  {Shamir}},\ }\href {https://doi.org/10.1016/0550-3213(95)00031-M} {\bibfield
  {journal} {\bibinfo  {journal} {Nucl. Phys. B}\ }\textbf {\bibinfo {volume}
  {439}},\ \bibinfo {pages} {54} (\bibinfo {year} {1995})},\ \Eprint
  {https://arxiv.org/abs/hep-lat/9405004} {arXiv:hep-lat/9405004} \BibitemShut
  {NoStop}%
\bibitem [{\citenamefont {Ginsparg}\ and\ \citenamefont
  {Wilson}(1982)}]{Ginsparg:1981bj}%
  \BibitemOpen
  \bibfield  {author} {\bibinfo {author} {\bibfnamefont {P.~H.}\ \bibnamefont
  {Ginsparg}}\ and\ \bibinfo {author} {\bibfnamefont {K.~G.}\ \bibnamefont
  {Wilson}},\ }\href {https://doi.org/10.1103/PhysRevD.25.2649} {\bibfield
  {journal} {\bibinfo  {journal} {Phys. Rev. D}\ }\textbf {\bibinfo {volume}
  {25}},\ \bibinfo {pages} {2649} (\bibinfo {year} {1982})}\BibitemShut
  {NoStop}%
\bibitem [{\citenamefont {Boyle}(2015)}]{Boyle:2015vda}%
  \BibitemOpen
  \bibfield  {author} {\bibinfo {author} {\bibfnamefont {P.~A.}\ \bibnamefont
  {Boyle}} (\bibinfo {collaboration} {UKQCD}),\ }\href
  {https://doi.org/10.22323/1.214.0087} {\bibfield  {journal} {\bibinfo
  {journal} {PoS}\ }\textbf {\bibinfo {volume} {LATTICE2014}},\ \bibinfo
  {pages} {087} (\bibinfo {year} {2015})}\BibitemShut {NoStop}%
\bibitem [{\citenamefont {Neuberger}(1998{\natexlab{a}})}]{Neuberger:1997fp}%
  \BibitemOpen
  \bibfield  {author} {\bibinfo {author} {\bibfnamefont {H.}~\bibnamefont
  {Neuberger}},\ }\href {https://doi.org/10.1016/S0370-2693(97)01368-3}
  {\bibfield  {journal} {\bibinfo  {journal} {Phys. Lett. B}\ }\textbf
  {\bibinfo {volume} {417}},\ \bibinfo {pages} {141} (\bibinfo {year}
  {1998}{\natexlab{a}})},\ \Eprint {https://arxiv.org/abs/hep-lat/9707022}
  {arXiv:hep-lat/9707022} \BibitemShut {NoStop}%
\bibitem [{\citenamefont {Neuberger}(1998{\natexlab{b}})}]{Neuberger:1998wv}%
  \BibitemOpen
  \bibfield  {author} {\bibinfo {author} {\bibfnamefont {H.}~\bibnamefont
  {Neuberger}},\ }\href {https://doi.org/10.1016/S0370-2693(98)00355-4}
  {\bibfield  {journal} {\bibinfo  {journal} {Phys. Lett. B}\ }\textbf
  {\bibinfo {volume} {427}},\ \bibinfo {pages} {353} (\bibinfo {year}
  {1998}{\natexlab{b}})},\ \Eprint {https://arxiv.org/abs/hep-lat/9801031}
  {arXiv:hep-lat/9801031} \BibitemShut {NoStop}%
\bibitem [{\citenamefont {Hashimoto}\ \emph {et~al.}(2014)\citenamefont
  {Hashimoto}, \citenamefont {Aoki}, \citenamefont {Cossu}, \citenamefont
  {Fukaya}, \citenamefont {Kaneko}, \citenamefont {Noaki},\ and\ \citenamefont
  {Boyle}}]{Hashimoto:2014gta}%
  \BibitemOpen
  \bibfield  {author} {\bibinfo {author} {\bibfnamefont {S.}~\bibnamefont
  {Hashimoto}}, \bibinfo {author} {\bibfnamefont {S.}~\bibnamefont {Aoki}},
  \bibinfo {author} {\bibfnamefont {G.}~\bibnamefont {Cossu}}, \bibinfo
  {author} {\bibfnamefont {H.}~\bibnamefont {Fukaya}}, \bibinfo {author}
  {\bibfnamefont {T.}~\bibnamefont {Kaneko}}, \bibinfo {author} {\bibfnamefont
  {J.}~\bibnamefont {Noaki}},\ and\ \bibinfo {author} {\bibfnamefont {P.~A.}\
  \bibnamefont {Boyle}},\ }\href {https://doi.org/10.22323/1.187.0431}
  {\bibfield  {journal} {\bibinfo  {journal} {PoS}\ }\textbf {\bibinfo {volume}
  {LATTICE2013}},\ \bibinfo {pages} {431} (\bibinfo {year} {2014})}\BibitemShut
  {NoStop}%
\bibitem [{\citenamefont {Tomii}\ \emph {et~al.}(2016)\citenamefont {Tomii},
  \citenamefont {Cossu}, \citenamefont {Fahy}, \citenamefont {Fukaya},
  \citenamefont {Hashimoto}, \citenamefont {Kaneko},\ and\ \citenamefont
  {Noaki}}]{Tomii:2016xiv}%
  \BibitemOpen
  \bibfield  {author} {\bibinfo {author} {\bibfnamefont {M.}~\bibnamefont
  {Tomii}}, \bibinfo {author} {\bibfnamefont {G.}~\bibnamefont {Cossu}},
  \bibinfo {author} {\bibfnamefont {B.}~\bibnamefont {Fahy}}, \bibinfo {author}
  {\bibfnamefont {H.}~\bibnamefont {Fukaya}}, \bibinfo {author} {\bibfnamefont
  {S.}~\bibnamefont {Hashimoto}}, \bibinfo {author} {\bibfnamefont
  {T.}~\bibnamefont {Kaneko}},\ and\ \bibinfo {author} {\bibfnamefont
  {J.}~\bibnamefont {Noaki}} (\bibinfo {collaboration} {JLQCD}),\ }\href
  {https://doi.org/10.1103/PhysRevD.94.054504} {\bibfield  {journal} {\bibinfo
  {journal} {Phys. Rev. D}\ }\textbf {\bibinfo {volume} {94}},\ \bibinfo
  {pages} {054504} (\bibinfo {year} {2016})},\ \Eprint
  {https://arxiv.org/abs/1604.08702} {arXiv:1604.08702 [hep-lat]} \BibitemShut
  {NoStop}%
\bibitem [{\citenamefont {Weisz}(1983)}]{Weisz:1982zw}%
  \BibitemOpen
  \bibfield  {author} {\bibinfo {author} {\bibfnamefont {P.}~\bibnamefont
  {Weisz}},\ }\href {https://doi.org/10.1016/0550-3213(83)90595-3} {\bibfield
  {journal} {\bibinfo  {journal} {Nucl. Phys. B}\ }\textbf {\bibinfo {volume}
  {212}},\ \bibinfo {pages} {1} (\bibinfo {year} {1983})}\BibitemShut {NoStop}%
\bibitem [{\citenamefont {Weisz}\ and\ \citenamefont
  {Wohlert}(1984)}]{Weisz:1983bn}%
  \BibitemOpen
  \bibfield  {author} {\bibinfo {author} {\bibfnamefont {P.}~\bibnamefont
  {Weisz}}\ and\ \bibinfo {author} {\bibfnamefont {R.}~\bibnamefont
  {Wohlert}},\ }\href {https://doi.org/10.1016/0550-3213(84)90563-7} {\bibfield
   {journal} {\bibinfo  {journal} {Nucl. Phys. B}\ }\textbf {\bibinfo {volume}
  {236}},\ \bibinfo {pages} {397} (\bibinfo {year} {1984})},\ \bibinfo {note}
  {[Erratum: Nucl.Phys.B 247, 544 (1984)]}\BibitemShut {NoStop}%
\bibitem [{\citenamefont {Luscher}\ and\ \citenamefont
  {Weisz}(1985)}]{Luscher:1984xn}%
  \BibitemOpen
  \bibfield  {author} {\bibinfo {author} {\bibfnamefont {M.}~\bibnamefont
  {Luscher}}\ and\ \bibinfo {author} {\bibfnamefont {P.}~\bibnamefont
  {Weisz}},\ }\href {https://doi.org/10.1007/BF01206178} {\bibfield  {journal}
  {\bibinfo  {journal} {Commun. Math. Phys.}\ }\textbf {\bibinfo {volume}
  {97}},\ \bibinfo {pages} {59} (\bibinfo {year} {1985})},\ \bibinfo {note}
  {[Erratum: Commun.Math.Phys. 98, 433 (1985)]}\BibitemShut {NoStop}%
\bibitem [{\citenamefont {Morningstar}\ and\ \citenamefont
  {Peardon}(2004)}]{Morningstar:2003gk}%
  \BibitemOpen
  \bibfield  {author} {\bibinfo {author} {\bibfnamefont {C.}~\bibnamefont
  {Morningstar}}\ and\ \bibinfo {author} {\bibfnamefont {M.~J.}\ \bibnamefont
  {Peardon}},\ }\href {https://doi.org/10.1103/PhysRevD.69.054501} {\bibfield
  {journal} {\bibinfo  {journal} {Phys. Rev. D}\ }\textbf {\bibinfo {volume}
  {69}},\ \bibinfo {pages} {054501} (\bibinfo {year} {2004})},\ \Eprint
  {https://arxiv.org/abs/hep-lat/0311018} {arXiv:hep-lat/0311018} \BibitemShut
  {NoStop}%
\bibitem [{\citenamefont {Hasenfratz}\ and\ \citenamefont
  {Knechtli}(2001)}]{Hasenfratz:2001hp}%
  \BibitemOpen
  \bibfield  {author} {\bibinfo {author} {\bibfnamefont {A.}~\bibnamefont
  {Hasenfratz}}\ and\ \bibinfo {author} {\bibfnamefont {F.}~\bibnamefont
  {Knechtli}},\ }\href {https://doi.org/10.1103/PhysRevD.64.034504} {\bibfield
  {journal} {\bibinfo  {journal} {Phys. Rev. D}\ }\textbf {\bibinfo {volume}
  {64}},\ \bibinfo {pages} {034504} (\bibinfo {year} {2001})},\ \Eprint
  {https://arxiv.org/abs/hep-lat/0103029} {arXiv:hep-lat/0103029} \BibitemShut
  {NoStop}%
\bibitem [{\citenamefont {Geles}\ and\ \citenamefont
  {Lang}(2011)}]{Geles:2011hh}%
  \BibitemOpen
  \bibfield  {author} {\bibinfo {author} {\bibfnamefont {F.}~\bibnamefont
  {Geles}}\ and\ \bibinfo {author} {\bibfnamefont {C.~B.}\ \bibnamefont
  {Lang}},\ }\href@noop {} {\  (\bibinfo {year} {2011})},\ \Eprint
  {https://arxiv.org/abs/1103.5368} {arXiv:1103.5368 [hep-lat]} \BibitemShut
  {NoStop}%
\bibitem [{\citenamefont {Cossu}\ \emph {et~al.}(2016)\citenamefont {Cossu},
  \citenamefont {Fukaya}, \citenamefont {Hashimoto},\ and\ \citenamefont
  {Tomiya}}]{Cossu:2015kfa}%
  \BibitemOpen
  \bibfield  {author} {\bibinfo {author} {\bibfnamefont {G.}~\bibnamefont
  {Cossu}}, \bibinfo {author} {\bibfnamefont {H.}~\bibnamefont {Fukaya}},
  \bibinfo {author} {\bibfnamefont {S.}~\bibnamefont {Hashimoto}},\ and\
  \bibinfo {author} {\bibfnamefont {A.}~\bibnamefont {Tomiya}} (\bibinfo
  {collaboration} {JLQCD}),\ }\href
  {https://doi.org/10.1103/PhysRevD.93.034507} {\bibfield  {journal} {\bibinfo
  {journal} {Phys. Rev. D}\ }\textbf {\bibinfo {volume} {93}},\ \bibinfo
  {pages} {034507} (\bibinfo {year} {2016})},\ \Eprint
  {https://arxiv.org/abs/1510.07395} {arXiv:1510.07395 [hep-lat]} \BibitemShut
  {NoStop}%
\bibitem [{\citenamefont {L\"uscher}(2010)}]{Luscher:2010iy}%
  \BibitemOpen
  \bibfield  {author} {\bibinfo {author} {\bibfnamefont {M.}~\bibnamefont
  {L\"uscher}},\ }\href {https://doi.org/10.1007/JHEP08(2010)071} {\bibfield
  {journal} {\bibinfo  {journal} {JHEP}\ }\textbf {\bibinfo {volume} {08}},\
  \bibinfo {pages} {071}},\ \bibinfo {note} {[Erratum: JHEP 03, 092 (2014)]},\
  \Eprint {https://arxiv.org/abs/1006.4518} {arXiv:1006.4518 [hep-lat]}
  \BibitemShut {NoStop}%
\bibitem [{\citenamefont {Borsanyi}\ \emph {et~al.}(2012)\citenamefont
  {Borsanyi} \emph {et~al.}}]{Borsanyi:2012zs}%
  \BibitemOpen
  \bibfield  {author} {\bibinfo {author} {\bibfnamefont {S.}~\bibnamefont
  {Borsanyi}} \emph {et~al.},\ }\href {https://doi.org/10.1007/JHEP09(2012)010}
  {\bibfield  {journal} {\bibinfo  {journal} {JHEP}\ }\textbf {\bibinfo
  {volume} {09}},\ \bibinfo {pages} {010}},\ \Eprint
  {https://arxiv.org/abs/1203.4469} {arXiv:1203.4469 [hep-lat]} \BibitemShut
  {NoStop}%
\bibitem [{\citenamefont {Bar}\ and\ \citenamefont
  {Golterman}(2014)}]{Bar:2013ora}%
  \BibitemOpen
  \bibfield  {author} {\bibinfo {author} {\bibfnamefont {O.}~\bibnamefont
  {Bar}}\ and\ \bibinfo {author} {\bibfnamefont {M.}~\bibnamefont
  {Golterman}},\ }\href {https://doi.org/10.1103/PhysRevD.89.034505} {\bibfield
   {journal} {\bibinfo  {journal} {Phys. Rev. D}\ }\textbf {\bibinfo {volume}
  {89}},\ \bibinfo {pages} {034505} (\bibinfo {year} {2014})},\ \bibinfo {note}
  {[Erratum: Phys.Rev.D 89, 099905 (2014)]},\ \Eprint
  {https://arxiv.org/abs/1312.4999} {arXiv:1312.4999 [hep-lat]} \BibitemShut
  {NoStop}%
\bibitem [{\citenamefont {Aoki}\ \emph {et~al.}(2021)\citenamefont {Aoki} \emph
  {et~al.}}]{Aoki:2021kgd}%
  \BibitemOpen
  \bibfield  {author} {\bibinfo {author} {\bibfnamefont {Y.}~\bibnamefont
  {Aoki}} \emph {et~al.},\ }\href@noop {} {\  (\bibinfo {year} {2021})},\
  \Eprint {https://arxiv.org/abs/2111.09849} {arXiv:2111.09849 [hep-lat]}
  \BibitemShut {NoStop}%
\bibitem [{\citenamefont {Aoki}\ \emph {et~al.}(2020)\citenamefont {Aoki} \emph
  {et~al.}}]{FlavourLatticeAveragingGroup:2019iem}%
  \BibitemOpen
  \bibfield  {author} {\bibinfo {author} {\bibfnamefont {S.}~\bibnamefont
  {Aoki}} \emph {et~al.} (\bibinfo {collaboration} {Flavour Lattice Averaging
  Group}),\ }\href {https://doi.org/10.1140/epjc/s10052-019-7354-7} {\bibfield
  {journal} {\bibinfo  {journal} {Eur. Phys. J. C}\ }\textbf {\bibinfo {volume}
  {80}},\ \bibinfo {pages} {113} (\bibinfo {year} {2020})},\ \Eprint
  {https://arxiv.org/abs/1902.08191} {arXiv:1902.08191 [hep-lat]} \BibitemShut
  {NoStop}%
\bibitem [{\citenamefont {Bruno}\ \emph {et~al.}(2014)\citenamefont {Bruno},
  \citenamefont {Schaefer},\ and\ \citenamefont {Sommer}}]{Bruno:2014ova}%
  \BibitemOpen
  \bibfield  {author} {\bibinfo {author} {\bibfnamefont {M.}~\bibnamefont
  {Bruno}}, \bibinfo {author} {\bibfnamefont {S.}~\bibnamefont {Schaefer}},\
  and\ \bibinfo {author} {\bibfnamefont {R.}~\bibnamefont {Sommer}} (\bibinfo
  {collaboration} {ALPHA}),\ }\href {https://doi.org/10.1007/JHEP08(2014)150}
  {\bibfield  {journal} {\bibinfo  {journal} {JHEP}\ }\textbf {\bibinfo
  {volume} {08}},\ \bibinfo {pages} {150}},\ \Eprint
  {https://arxiv.org/abs/1406.5363} {arXiv:1406.5363 [hep-lat]} \BibitemShut
  {NoStop}%
\bibitem [{\citenamefont {Aoki}\ \emph {et~al.}(2018)\citenamefont {Aoki},
  \citenamefont {Cossu}, \citenamefont {Fukaya}, \citenamefont {Hashimoto},\
  and\ \citenamefont {Kaneko}}]{Aoki:2017paw}%
  \BibitemOpen
  \bibfield  {author} {\bibinfo {author} {\bibfnamefont {S.}~\bibnamefont
  {Aoki}}, \bibinfo {author} {\bibfnamefont {G.}~\bibnamefont {Cossu}},
  \bibinfo {author} {\bibfnamefont {H.}~\bibnamefont {Fukaya}}, \bibinfo
  {author} {\bibfnamefont {S.}~\bibnamefont {Hashimoto}},\ and\ \bibinfo
  {author} {\bibfnamefont {T.}~\bibnamefont {Kaneko}} (\bibinfo {collaboration}
  {JLQCD}),\ }\href {https://doi.org/10.1093/ptep/pty041} {\bibfield  {journal}
  {\bibinfo  {journal} {PTEP}\ }\textbf {\bibinfo {volume} {2018}},\ \bibinfo
  {pages} {043B07} (\bibinfo {year} {2018})},\ \Eprint
  {https://arxiv.org/abs/1705.10906} {arXiv:1705.10906 [hep-lat]} \BibitemShut
  {NoStop}%
\bibitem [{\citenamefont {Brower}\ \emph {et~al.}(2003)\citenamefont {Brower},
  \citenamefont {Chandrasekharan}, \citenamefont {Negele},\ and\ \citenamefont
  {Wiese}}]{Brower:2003yx}%
  \BibitemOpen
  \bibfield  {author} {\bibinfo {author} {\bibfnamefont {R.}~\bibnamefont
  {Brower}}, \bibinfo {author} {\bibfnamefont {S.}~\bibnamefont
  {Chandrasekharan}}, \bibinfo {author} {\bibfnamefont {J.~W.}\ \bibnamefont
  {Negele}},\ and\ \bibinfo {author} {\bibfnamefont {U.~J.}\ \bibnamefont
  {Wiese}},\ }\href {https://doi.org/10.1016/S0370-2693(03)00369-1} {\bibfield
  {journal} {\bibinfo  {journal} {Phys. Lett. B}\ }\textbf {\bibinfo {volume}
  {560}},\ \bibinfo {pages} {64} (\bibinfo {year} {2003})},\ \Eprint
  {https://arxiv.org/abs/hep-lat/0302005} {arXiv:hep-lat/0302005} \BibitemShut
  {NoStop}%
\bibitem [{\citenamefont {Aoki}\ \emph {et~al.}(2007)\citenamefont {Aoki},
  \citenamefont {Fukaya}, \citenamefont {Hashimoto},\ and\ \citenamefont
  {Onogi}}]{Aoki:2007ka}%
  \BibitemOpen
  \bibfield  {author} {\bibinfo {author} {\bibfnamefont {S.}~\bibnamefont
  {Aoki}}, \bibinfo {author} {\bibfnamefont {H.}~\bibnamefont {Fukaya}},
  \bibinfo {author} {\bibfnamefont {S.}~\bibnamefont {Hashimoto}},\ and\
  \bibinfo {author} {\bibfnamefont {T.}~\bibnamefont {Onogi}},\ }\href
  {https://doi.org/10.1103/PhysRevD.76.054508} {\bibfield  {journal} {\bibinfo
  {journal} {Phys. Rev. D}\ }\textbf {\bibinfo {volume} {76}},\ \bibinfo
  {pages} {054508} (\bibinfo {year} {2007})},\ \Eprint
  {https://arxiv.org/abs/0707.0396} {arXiv:0707.0396 [hep-lat]} \BibitemShut
  {NoStop}%
\bibitem [{\citenamefont {Leutwyler}\ and\ \citenamefont
  {Smilga}(1992)}]{Leutwyler:1992yt}%
  \BibitemOpen
  \bibfield  {author} {\bibinfo {author} {\bibfnamefont {H.}~\bibnamefont
  {Leutwyler}}\ and\ \bibinfo {author} {\bibfnamefont {A.~V.}\ \bibnamefont
  {Smilga}},\ }\href {https://doi.org/10.1103/PhysRevD.46.5607} {\bibfield
  {journal} {\bibinfo  {journal} {Phys. Rev. D}\ }\textbf {\bibinfo {volume}
  {46}},\ \bibinfo {pages} {5607} (\bibinfo {year} {1992})}\BibitemShut
  {NoStop}%
\bibitem [{\citenamefont {Colangelo}\ \emph {et~al.}(2001)\citenamefont
  {Colangelo}, \citenamefont {Gasser},\ and\ \citenamefont
  {Leutwyler}}]{Colangelo:2001df}%
  \BibitemOpen
  \bibfield  {author} {\bibinfo {author} {\bibfnamefont {G.}~\bibnamefont
  {Colangelo}}, \bibinfo {author} {\bibfnamefont {J.}~\bibnamefont {Gasser}},\
  and\ \bibinfo {author} {\bibfnamefont {H.}~\bibnamefont {Leutwyler}},\ }\href
  {https://doi.org/10.1016/S0550-3213(01)00147-X} {\bibfield  {journal}
  {\bibinfo  {journal} {Nucl. Phys. B}\ }\textbf {\bibinfo {volume} {603}},\
  \bibinfo {pages} {125} (\bibinfo {year} {2001})},\ \Eprint
  {https://arxiv.org/abs/hep-ph/0103088} {arXiv:hep-ph/0103088} \BibitemShut
  {NoStop}%
\bibitem [{\citenamefont {Noaki}\ \emph {et~al.}(2008)\citenamefont {Noaki}
  \emph {et~al.}}]{JLQCD:2008zxm}%
  \BibitemOpen
  \bibfield  {author} {\bibinfo {author} {\bibfnamefont {J.}~\bibnamefont
  {Noaki}} \emph {et~al.} (\bibinfo {collaboration} {JLQCD, TWQCD}),\ }\href
  {https://doi.org/10.1103/PhysRevLett.101.202004} {\bibfield  {journal}
  {\bibinfo  {journal} {Phys. Rev. Lett.}\ }\textbf {\bibinfo {volume} {101}},\
  \bibinfo {pages} {202004} (\bibinfo {year} {2008})},\ \Eprint
  {https://arxiv.org/abs/0806.0894} {arXiv:0806.0894 [hep-lat]} \BibitemShut
  {NoStop}%
\end{thebibliography}%
\end{document}


\title{Supplementary material}
\maketitle
\nopagebreak 
\section{Lattice formulations and ensembles}
\label{sec:Lattices}

\subsection{M\"obius domain-wall fermions}
In this study, we use M\"obius domain-wall fermions for both sea
and valence quarks.
This is a generalized version~\cite{Brower:2012vk} of the domain-wall
fermion formulation~\cite{Kaplan:1992bt,Shamir:1993zy,Furman:1994ky}, 
which is a five-dimensional (5D) implementation of the Ginsparg-Wilson 
fermion~\cite{Ginsparg:1981bj}.
In this paper, we follow the notation of Ref.~\cite{Boyle:2015vda}.

The generalized domain-wall fermion action is of the form
\begin{equation}
  S_{\GDW}=\sum_x \bar{\psi} D_{\GDW}^{(5)}(m) \psi,
\end{equation}
where a suffix ``$x$'' for the 5D fields $\psi$ and $\bar{\psi}$ is
implicit. 
It is a 5D vector of length $L_s$, the depth in the fifth dimension,
on which the 5D Dirac operator $D_{\GDW}^{(5)}(m)$ for a fermion of mass
$m$ is applied.
In matrix form for the coordinate $s$ of the fifth direction, it can be
written as 
\begin{equation}
  D_{\GDW}^{(5)}(m) = \left(
    \begin{array}[c]{cccccc}
      \tilde{D} & -P_- & 0 & \cdots & 0 & mP_+\\
      -P_+ & \tilde{D} & -P_- & 0 & \cdots & 0\\
      0 & -P_+ & \tilde{D} & \ddots & 0 & \vdots\\
      \vdots & 0 & \ddots & \ddots & \ddots & \vdots\\
      0 & \cdots & \ddots & -P_+ & \tilde{D} & -P_-\\
      mP_- & 0 & \cdots & 0 & -P_+ & \tilde{D}
    \end{array}
  \right),
\end{equation}
where $\tilde{D}=(D_-)^{-1}D_+$,
$D_+=1+bD_W(-M)$ and $D_-=1-cD_W(-M)$.
The conventional 4D Wilson-Dirac operator $D_W(-M)$ enters in this
definition with a large negative mass term, $-M$.
$P_\pm$ denotes the chirality projector: $P_\pm=(1\pm\gamma_5)/2$.
The parameters $b$ and $c$ control the kernel and approximation of the
Ginsparg-Wilson fermions, where the choice $(b,c)=(1,0)$ corresponds to
the standard domain-wall fermion action.
One can also make these parameters $s$-dependent to further improve
the chiral symmetry, but we do not consider such a possibility in this
work. 
For practical reasons we multiply each row by $D_-$ and use the 5D
operator
\begin{equation}
  D_-D_{\GDW}^{(5)}(m) = \left(
    \begin{array}[c]{cccccc}
      D_+ & -D_-P_- & 0 & \cdots & 0 & mD_-P_+\\
      -D_-P_+ & D_+ & -D_-P_- & 0 & \cdots & 0\\
      0 & -D_-P_+ & D_+ & \ddots & 0 & \vdots\\
      \vdots & 0 & \ddots & \ddots & \ddots & \vdots\\
      0 & \cdots & \ddots & -D_-P_+ & D_+ & -D_-P_-\\
      mD_-P_- & 0 & \cdots & 0 & -D_-P_+ & D_+
    \end{array}
  \right)
\end{equation}
rather than $D_{\GDW}^{(5)}$ to avoid the need for the inverse term $(D_-)^{-1}$ in our simulations.

We identify the physical four-dimensional quark fields 
$q$ and $\bar{q}$ as the surface modes of the 5D fields 
$\psi$ and $\bar{\psi}$:
$q_R=P_+\psi_{L_s}$, $q_L=P_-\psi_1$; 
$\bar{q}_R=\bar{\psi}_{L_s}P_-$, $\bar{q}_L=\bar{\psi}_1P_+$,
where the subscripts $L$ and $R$ denote the chirality components of
the fermion fields.
Then, one can show that the quark propagator $\langle q\bar{q}\rangle$
can be written in terms of the inverse of the 5D Dirac operator as
\begin{equation}
  \tilde{D}_{\ov}^{-1}(m) = \left[ {\cal P}^{-1} 
    D_{\GDW}^{(5)}(m)^{-1} R_5 {\cal P}
    \right]_{11},
\end{equation}
where $R_5$ denotes an operator to reverse the ordering in the fifth
direction and the permutation operators ${\cal P}$ and ${\cal P}^{-1}$
are defined as follows:
\begin{eqnarray}
  {\cal P} & = &
  \left(
    \begin{array}[c]{ccccc}
      P_- & P_+ & 0 & \cdots & 0\\
      0 & P_- & P_+ & 0 & \vdots\\
      \vdots & 0 & \ddots & \ddots & \vdots\\
      0 & \cdots & 0 & P_- & P_+\\
      P_+ & 0 & \cdots & 0 & P_-
    \end{array}
  \right),
  \\
  {\cal P}^{-1}={\cal P}^\dagger & = &
  \left(
    \begin{array}[c]{ccccc}
      P_- & 0 & \cdots & 0 & P_+\\
      P_+ & P_- & 0 & \cdots & 0\\
      \vdots & \ddots & \ddots & \ddots & \vdots\\
      0 & \cdots & P_+ & P_- & 0\\
      0 & \cdots & 0 & P_+ & P_-
    \end{array}
  \right).
\end{eqnarray}
The 4D propagator corresponds to an inverse of the 4D effective
operator 
\begin{eqnarray}
  D^{(4)}(m) & = & 
            \left[ {\cal P}^{-1}
            (D_{\GDW}^5(m=1))^{-1} D_{\GDW}^5(m)
            {\cal P} \right]_{11}
            \nonumber\\
      & = & \frac{1+m}{2}+\frac{1-m}{2}
            \gamma_5 \epsilon(H_M)
            \label{eq:D_eff4D}
\end{eqnarray}
after subtracting a contribution from the contact term: 
\begin{equation}
  \tilde{D}_{\ov}^{-1}(m) = \frac{1}{1-m}
  \left[ (D^{(4)}(m))^{-1} - 1 \right].
\end{equation}
In Eq.~\eqref{eq:D_eff4D}, another 5D operator $D_{\GDW}^{(5)}(m=1)$ is
introduced to cancel the irrelevant eigenmodes of $D_{\GDW}^{(5)}(m)$
living in the bulk of 5D by multiplying by its inverse
$(D_{\GDW}^{(5)}(m=1))^{-1}$.
The 4D effective operator $D^{(4)}(m)$ thus constructed has a form
similar to the overlap-Dirac operator~\cite{Neuberger:1997fp,Neuberger:1998wv} 
with a polar approximation of the matrix sign function
\begin{equation}
  \epsilon(H_M) =
  \frac{(1+H_M)^{L_s}-(1-H_M)^{-L_s}}{(1+H_M)^{L_s}+(1-H_M)^{-L_s}},
\end{equation}
where the kernel operator $H_M$ is given by
\begin{equation}
  H_M = \gamma_5\frac{(b+c)D_W(-M)}{2+(b-c)D_W(-M)}.
  \label{eq:H_M}
\end{equation}
In the limit of large $L_s$, the approximation of the sign function
becomes exact and the 4D effective operator reduces to the
overlap-Dirac operator.
The kernel in Eq.~\eqref{eq:H_M} is different from that in Refs.~\cite{Neuberger:1997fp,Neuberger:1998wv}
by the presence of its denominator.
The standard domain-wall fermion action corresponds to the special case where
$b+c=b-c=1$.
In this work we choose a different parameter set, $b+c=2$ and $b-c=1$,
motivated by a detailed study of the residual chiral symmetry
violation~\cite{Hashimoto:2014gta}, which is also partly described in 
Sec.~\ref{sec:residual}. 

We construct the fermion bilinear operator with the 4D
quark fields $q$ and $\bar{q}$ as $\bar{q}\Gamma q$ with an appropriate
$\gamma$-matrix $\Gamma$.
The renormalization factors for (axial-)vector and (pseudo-)scalar
operators are separately determined through short-distance current
correlators~\cite{Tomii:2016xiv}.

\subsection{Action and generation parameters}
For the gauge part of the lattice action, we adopt the tree-level
improved Symanzik gauge action~\cite{Weisz:1982zw,Weisz:1983bn,Luscher:1984xn}
\begin{equation}
  S_{\mathrm{Sym}} = \frac{\beta}{3}
  \left\{ \frac{5}{3}\sum_{x,\mu<\nu} P_{x,\mu,\nu} 
    - \frac{1}{12} \sum_{x,\mu\not=\nu} R_{x,\mu,\nu} \right\},
\end{equation}
with plaquette $P_{x,\mu,\nu}$ and a $1\times 2$ rectangular
Wilson loop $R_{x,\mu,\nu}$
summed over the lattice sites $x$ and directions $\mu$ and $\nu$.
The $\beta$ values in our simulations are $\beta = 4.17$, $4.35$
and $4.47$, corresponding to three different lattice spacings.
The fermion action is the M\"obius domain-wall fermion action with the
M\"obius parameters $b+c=2$ and $b-c=1$, as discussed in the previous
subsection.
The kernel operator $H_M$ is defined with the Wilson-Dirac operator
with a mass parameter $M=1$.
Three iterations of the stout smearing procedure~\cite{Morningstar:2003gk} are applied to the gauge links appearing in the Wilson-Dirac operator,
with which short-distance fluctuations of gauge fields are suppressed and
the chiral properties of the M\"obius domain-wall fermions are improved.
The parameter $\rho$, used to control the strength of the smearing, is taken
to be 0.1.
The effect of our choice of link smearing roughly corresponds to
that of a single application of the HYP smearing~\cite{Hasenfratz:2001hp}, which is another popular choice of smearing
procedure~\cite{Geles:2011hh}.
The depth in the fifth dimension is $L_s=12$ on the coarsest lattice
at $\beta=4.17$ and $L_s=8$ for finer latices.

\begin{table}[!htbp]
  \centering
  \begin{tabular}{ccccccccc}
    \hline\hline
    $\beta$ & $a$ & $L^3\times T$ & $L$ & $am_{ud}$ & $am_s$ & $M_\pi$ & $M_\pi L$ & ID\\ 
    & [fm] & & [fm] & & & [MeV] & & \\
    \hline
    4.17 & 0.080 & $32^3\times64$ & 2.6 & 0.0035 & 0.040 & 230(1) & 3.0 & C-$ud2$-$s$a\\
    & & & & 0.007 & 0.030 & 310(1) & 4.0 & C-$ud3$-$s$b\\
    & & & & 0.007 & 0.040 & 309(1) & 4.0 & C-$ud3$-$s$a\\
    & & & & 0.012 & 0.030 & 397(1) & 5.2 & C-$ud4$-$s$b\\
    & & & & 0.012 & 0.040 & 399(1) & 5.2 & C-$ud4$-$s$a\\
    & & & & 0.019 & 0.030 & 498(1) & 6.5 & C-$ud5$-$s$b\\
    & & & & 0.019 & 0.040 & 499(1) & 6.5 & C-$ud5$-$s$a\\
    & & $48^3\times 96$ & 3.9 & 0.0035 & 0.040 & 230 & 4.4 & C-$ud2$-$s$a-L\\
    \hline
    4.35 & 0.055 & $48^3\times 96$ & 2.6 & 0.0042 & 0.018 & 296(1) & 3.9 & M-$ud3$-$s$b\\
    & & & & 0.0042 & 0.025 & 300(1) & 3.9 & M-$ud3$-$s$a\\
    & & & & 0.0080 & 0.018 & 407(1) & 5.4 & M-$ud4$-$s$b\\
    & & & & 0.0080 & 0.025 & 408(1) & 5.4 & M-$ud4$-$s$a\\
    & & & & 0.0120 & 0.018 & 499(1) & 6.6 & M-$ud5$-$s$b\\
    & & & & 0.0120 & 0.025 & 501(1) & 6.6 & M-$ud5$-$s$a\\
    \hline
    4.47 & 0.044 & $64^3\times 128$ & 2.7 & 0.0030 & 0.015 & 284(1) & 4.0 & F-$ud3$-$s$a\\
    \hline\hline
  \end{tabular}
  \caption{
    Lattice ensembles generated for this study.
    The ensemble with  $M_\pi L\approx 3.0$ is excluded in the final
    analysis to avoid possible finite volume effects. We use a subset of these
    ensembles to determine the $B\to\pi\ell\nu$ form factors and $|V_{ub}|$.
  }
  \label{tab:lattices}
\end{table}

We generated 15 lattice ensembles with three different lattice spacings
and various combinations of sea quark masses as listed in Table~\ref{tab:lattices}.
We assign an ID name for each ensemble, which distinguishes coarse (C), medium (M) and fine (F) lattices, as well
as the masses of up and down ($ud$) and strange ($s$) quarks.
In the C and M ensembles, we use two values of the strange quark mass
that sandwich the physical value from above (a) or from below (b).
The number given after ``$ud$'' represents the corresponding pion
mass, {\it e.g.} ``$ud3$'' for a $300\;\mathrm{MeV}$ pion, and so on.
The lattice size $L$ is taken such that the physical lattice extent is
kept approximately constant at $\sim 2.6\;\mathrm{fm}$.
There is a single coarse ensemble with a larger volume:
$48^3\times 96$ compared to the regular size of $32^3\times 64$.
This is indicated by ``-L'' in the ID.

Each ensemble of the ``C'' lattices has 10,000 Hybrid Monte Carlo (HMC)
updates, where each update is of 0.5 molecular dynamics time in length.
The ``M'' and ``F'' lattices use 1.0 and 2.0 molecular dynamics
time units with 5,000 and 2,500 updates, respectively.

\subsection{Residual masses}
\label{sec:residual}
Although M\"obius domain-wall fermions offer precise chiral
symmetry, this is slightly violated due to the finite extent of the fifth
direction, $L_s$.
This violation may be represented by an operator $\Delta_L$ defined
through 
\begin{equation}
  2\gamma_5\Delta_L =
  \gamma_5 D^{(4)}(0) + D^{(4)}(0)\gamma_5 - 
  2 D^{(4)}(0)\gamma_5D^{(4)}(0),
  \label{eq:Delta_L}
\end{equation}
which can also be written as
\begin{equation}
  \Delta_L = \frac{1}{4} \left[ 1-\epsilon^2(H_M) \right].
\end{equation}
By measuring some matrix elements of this operator, which probes the
violation of the Ginsparg-Wilson relation, one can characterize the
size of such effects.

A global measure of the chiral symmetry violation can be constructed~\cite{Brower:2012vk} as 
\begin{equation}
  \bar{m}_{\mathrm{res}} = \frac{
    \left\langle
      \mathrm{Tr} 
      \left[ (\tilde{D}_{\ov}^{-1})^\dagger \Delta_L
        \tilde{D}_{\ov}^{-1} \right]
    \right\rangle
  }{
    \left\langle
      \mathrm{Tr} 
      \left[ (\tilde{D}_{\ov}^{-1})^\dagger
        \tilde{D}_{\ov}^{-1} \right]
    \right\rangle
  },
  \label{eq:resmass}
\end{equation}
where the trace runs over all indices, including $x$ and the color/spinor
indices of the 4D quark propagator.
This is called the ``residual mass'' since the operator $\Delta_L$ plays
the same role as the mass term, as can be seen from the definition of
Eq.~\eqref{eq:Delta_L}.
As discussed in~\cite{Cossu:2015kfa}, we may consider a decomposition
of $m_{\mathrm{res}}$ in terms of the eigenmodes of $D^{(4)}(0)$.
Since the eigenvalue $\lambda$ of $D^{(4)}(0)$ is distributed more densely
for high modes (as $\sim\lambda^3$), $\bar{m}_{\mathrm{res}}$ as defined in
Eq.~\eqref{eq:resmass} is most sensitive to the high end of the eigenvalue
spectrum, which is of the order of the lattice cutoff.
It therefore probes the violation of chiral symmetry at short distances.

In addition to Eq.~\eqref{eq:resmass} we consider a measurement on a given
time slice,
\begin{equation}
  m_{\mathrm{res}}(t) = \frac{
    \sum_{\mathbf{x},\mathbf{y}} 
    \langle \bar{q}_{\mathbf{x},t}\gamma_5\Delta_L q_{\mathbf{x},t} 
    \bar{q}_{\mathbf{y},0}\gamma_5 q_{\mathbf{y},0} \rangle
  }{
    \sum_{\mathbf{x},\mathbf{y}} 
    \langle \bar{q}_{\mathbf{x},t}\gamma_5 q_{\mathbf{x},t} 
    \bar{q}_{\mathbf{y},0}\gamma_5 q_{\mathbf{y},0} \rangle
  },
  \label{eq:mres_sliced}
\end{equation}
which can be evaluated by calculating the 4D propagator
$\tilde{D}_{\ov}^{-1}$ from randomly generated $Z_2$ noise spread over
points $\mathbf{y}$ on time slice ``0'' and contracting at another
time slice ``$t$''.

\begin{figure}
  \centering
  \includegraphics[width=12cm]{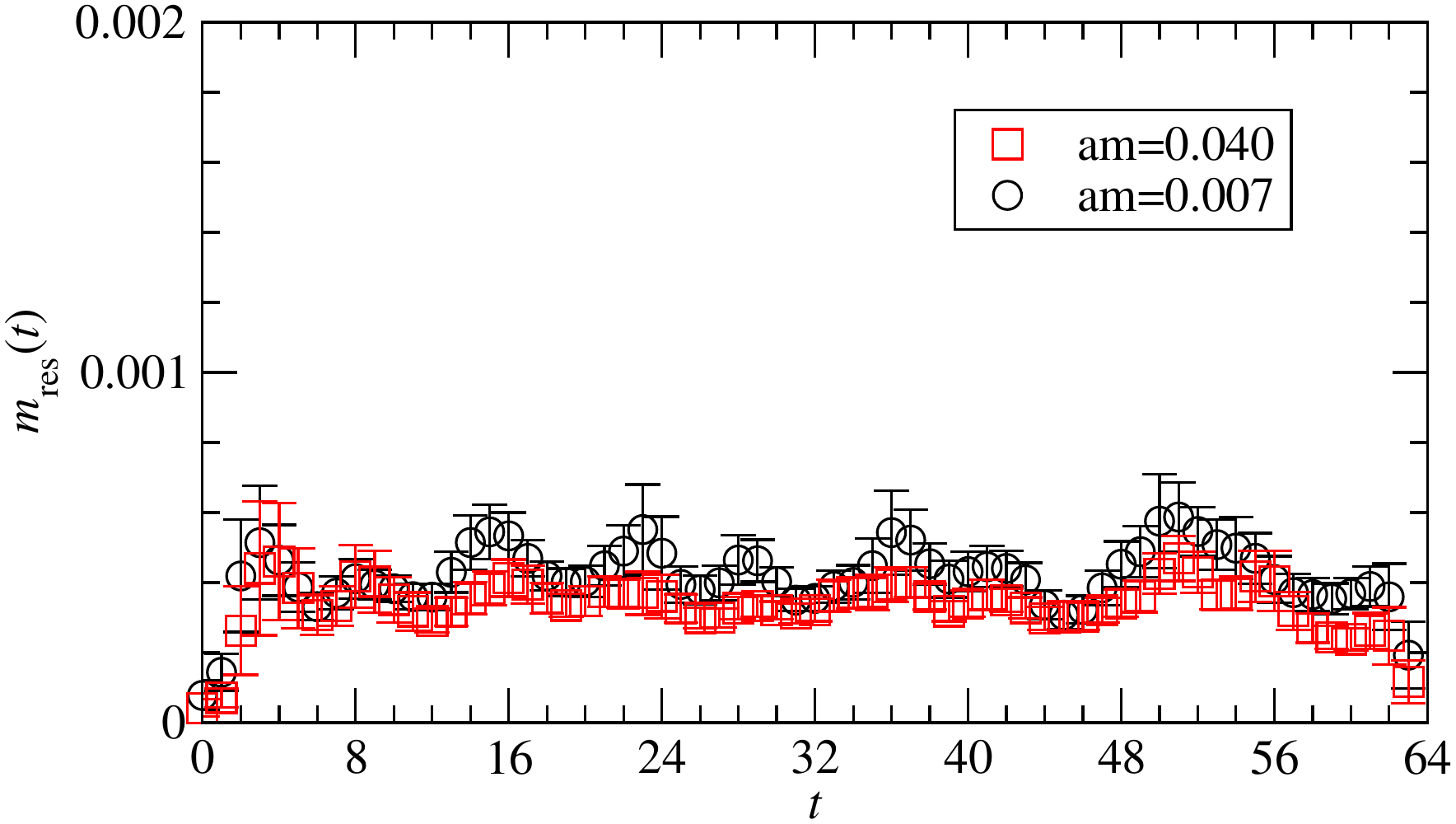}
  \includegraphics[width=12cm]{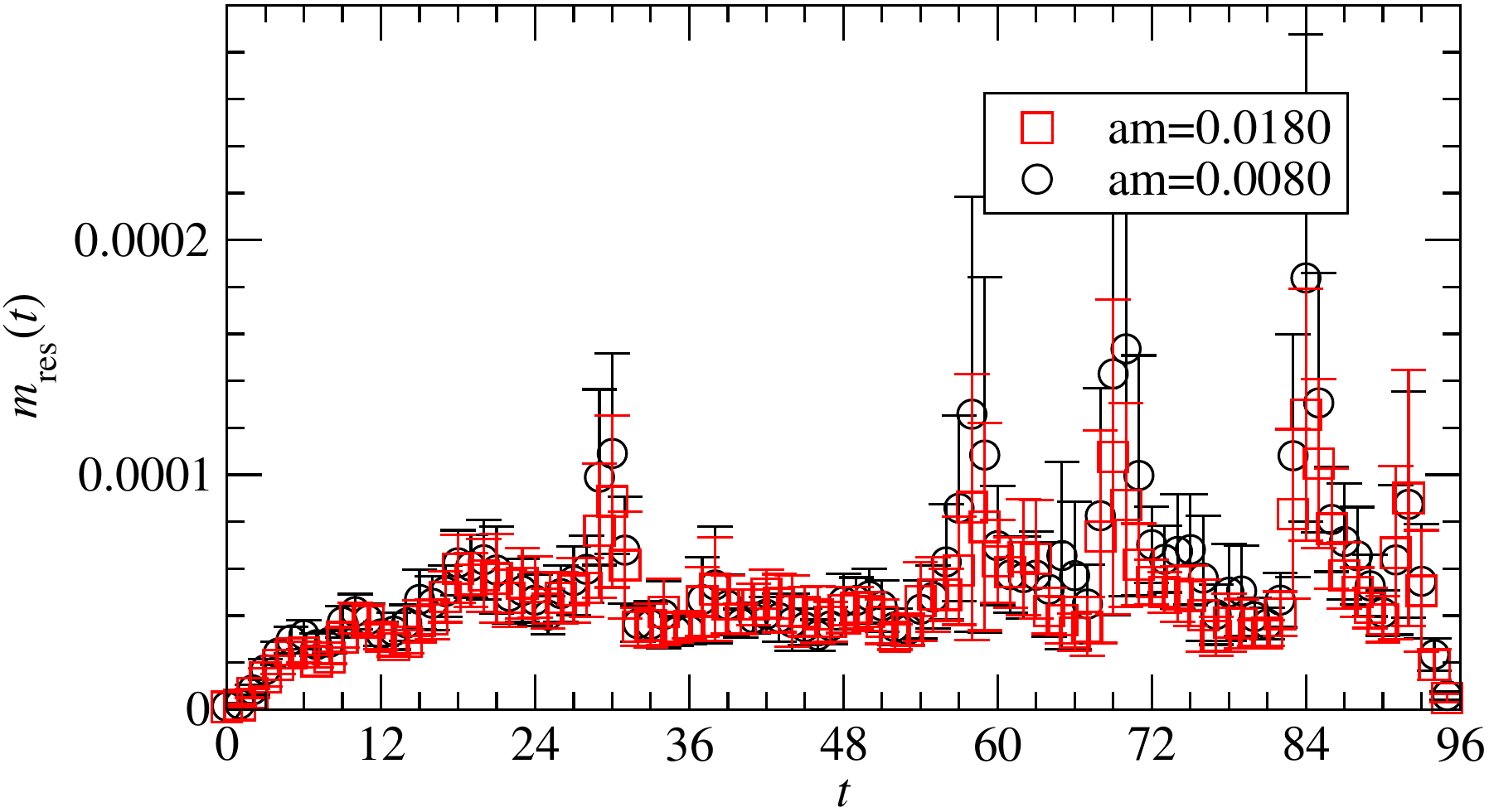}
  \caption{
    Residual mass calculated with finite separation $t$ as defined in
    Eq.~\eqref{eq:mres_sliced}. 
    Lattice data is from ensembles C-$ud3$-$s$a (top panel) and
    M-$ud4$-$s$b (bottom panel).
    The valence quark mass is set to the same as that of the sea quarks:
    $ud$ (circles) or $s$ (squares).
    Results are given in lattice units.
  }
  \label{fig:mres_sliced}
\end{figure}

Some examples are shown in Fig.~\ref{fig:mres_sliced}. 
We find that the residual mass $m_{\res}(t)$ as defined in
Eq~\eqref{eq:mres_sliced} is nearly independent of $t$.
On the coarse lattice at $\beta = 4.17$ (upper panel), it shows a
plateau at about $0.4\times10^{-3}$, which corresponds to $1\;\mathrm{MeV}$ in physical
units.
On the medium lattice at $\beta = 4.35$ (lower panel), it is reduced by
an order of magnitude to $\sim 0.5\times 10^{-4}$, which is about
$0.2\;\mathrm{MeV}$ and therefore almost negligible in the analysis
of physical quantities.
We expect $m_{\res}(t)$ to be even smaller on the finest lattice.
Numerical results are summarized in Tables~\ref{tab:mres_b4.17} and~\ref{tab:mres_b4.35}.
Measurements with $ud$ and $s$ valence quarks are separately
averaged between $t=T/4$ and $3T/4$ where a clear plateau is found on the plots.
The results for the global measurement $\bar{m}_{\res}$ are also listed. 

When measured with different valence quark masses, the residual mass values
are consistent within their statistical uncertainties.
Going to even heavier valence quark masses, such as those of the charm and bottom 
quark masses, we find that $m_{\res}(t)$ is slightly smaller but has
a larger statistical error.
In any case, the residual mass for heavy quarks does not have any impact on their
analysis because $m_{\res}(t)$ is $10^{-3}$ of their
bare quark mass. We also find that the results are insensitive to the sea quark masses.

Another point to notice from Fig.~\ref{fig:mres_sliced} is that $m_{\res}(t)$ turns out to be smaller
at short distances, {\it i.e.} up to $t = 2$--$3$ in lattice units.
This is consistent with the observation of~\cite{Cossu:2015kfa} in
which the matrix elements of $\Delta_L$ were calculated on individual
eigenmodes of $D^{(4)}(0)$ and enhancement of the violation was found for
low-lying eigenmodes.
Although there is no unique definition of the ``residual mass'', we
can take the plateau value since this is used when we analyze low-energy
physical observables for which the violation at long distances is most
relevant.
Alternatively, we can take $M_\pi^2$ in place of the quark mass when we perform a chiral extrapolation of the lattice data.

We also reiterate that the extent of the fifth dimension $L_s$ in our work is not
large: $L_s=12$ on the coarse lattices and $L_s=8$
on the fine lattices.
This suggests that precise chiral symmetry may be achieved by rather
modest computational overhead compared to the conventional Wilson
fermion formulation. 

\begin{table}[!htbp]
  \centering
  \begin{tabular}{cccccccc}
    \hline
    ID & $m_{ud}$ & $m_s$ & $N_{\meas}$ / $N_{\meas}^{(s)}$ & 
    $m_{\res}(t)\times 10^3$ & $\bar{m}_{\res}\times 10^3$ &
    $m_{\res}^{(s)}(t)\times 10^3$ & $\bar{m}^{(s)}_{\res}\times 10^3$\\
    \hline
    C-$ud2$-$s$a & 0.0035 & 0.040 & 100/200 & 0.39(08) & 0.22(2) & 0.29(6) & 0.06(1)\\
    C-$ud3$-$s$b & 0.007 & 0.030 & 200/200 & 0.45(07) & 0.17(1) & 0.39(10) & 0.07(1)\\
    C-$ud3$-$s$a & 0.007 & 0.040 & 200/200 & 0.43(07) & 0.23(4) & 0.35(8) & 0.11(3)\\
    C-$ud4$-$s$b & 0.012 & 0.030 & 100/200 & 0.40(08) & 0.15(2) & 0.36(07) & 0.08(1)\\
    C-$ud4$-$s$a & 0.012 & 0.040 & 100/200 & 0.44(09) & 0.12(8) & 0.35(7) & 0.06(1)\\
    C-$ud5$-$s$b & 0.019 & 0.030 & 200/200 & 0.45(08) & 0.15(3) & 0.41(08) & 0.11(3)\\
    C-$ud5$-$s$a & 0.019 & 0.040 & 100/200 & 0.41(10) & 0.15(3) & 0.38(8) & 0.09(1)\\
    \hline
  \end{tabular}
  \caption{
    Residual mass on the coarse ($\beta = 4.17$) lattices.
    The ensemble IDs are those defined in Table~\ref{tab:lattices},
    and the number of measurements $N_{\meas}$ are listed separately
    for each ensemble.
    Time-dependent measurements of $m_{\res}(t)$ are averaged over time
    slices between $T/4$ and $3T/4$.
    Results for the global measurement $\bar{m}_{\res}$ are also listed.
    The measurements with strange valence quarks are denoted by a
    superscript $(s)$, {\it e.g.}, $\bar{m}_{\res}^{(s)}$.
    All results are in lattice units, and $m_{\res}$ is multiplied
    by $10^3$.
  }
  \label{tab:mres_b4.17}
\end{table}

\begin{table}[!htbp]
  \centering
  \begin{tabular}{cccccccc}
    \hline
    ID & $m_{ud}$ & $m_s$ & $N_{\meas}$ / $N_{\meas}^{(s)}$ & 
    $m_{\res}(t)\times 10^3$ & $\bar{m}_{\res}\times 10^3$ &
    $m_{\res}^{(s)}(t)\times 10^3$ & $\bar{m}^{(s)}_{\res}\times 10^3$ \\
    \hline
    M-$ud3$-$s$b & 0.0042 & 0.0180 & 50/50 & 0.061(40) & 0.013(1) & 0.049(26) & 0.006(1)\\
    M-$ud3$-$s$a & 0.0042 & 0.0250 & 50/50 & 0.067(51) & 0.019(4) & 0.058(62) & 0.008(2)\\
    M-$ud4$-$s$b & 0.0080 & 0.0180 & 50/50 & 0.067(56) & 0.011(2) & 0.057(39) & 0.007(1)\\
    M-$ud4$-$s$a & 0.0080 & 0.0250 & 50/50 & 0.051(30) & 0.012(3) & 0.040(27) & 0.006(2)\\
    M-$ud5$-$s$b & 0.0120 & 0.0180 & 90/50 & 0.053(22) & 0.011(2) & 0.047(20) & 0.012(3)\\
    M-$ud5$-$s$a & 0.0120 & 0.0250 & 50/50 & 0.056(23) & 0.010(1) & 0.056(30) & 0.005(1)\\
    \hline
  \end{tabular}
  \caption{
    Same as Table~\ref{tab:mres_b4.17}, but for
    medium ($\beta$ = 4.35) lattices.
  }
  \label{tab:mres_b4.35}
\end{table}

\subsection{Scale setting}
In this work the lattice scale is set by the Yang-Mills gradient-flow
time $t_0$, defined by 
$t^2\langle E(t)\rangle_{t=t_0}=0.3$~\cite{Luscher:2010iy},
with the energy density operator $E(t)$ of the gluon field evaluated
at flow time $t$.
(The symbol ``$t$'' is used here for the flow time and should not be confused with
the time coordinate on the lattice. This usage is restricted to this subsection.)
We adopt the gradient flow defined by the conventional Wilson gauge
action and discretize the flow by a step size $\Delta t/a^2 = 0.01$.
For each ensemble, we perform 700 (500) measurements of the
gradient flow on the coarse (medium/fine)
lattices. 

\begin{table}
  \centering
  \begin{tabular}{ccccc}
    \hline\hline
    ID & $m_{ud}$ & $m_s$ & $t_0^{1/2}/a$ & $w_0/a$\\
    \hline
    C-$ud2$-$s$a & 0.0035 & 0.0400 & 1.8156(19) & 2.1746(48)\\
    C-$ud3$-$s$b & 0.007 & 0.030 & 1.8081(16) & 2.1544(44)\\
    C-$ud3$-$s$a & 0.007 & 0.040 & 1.8035(18) & 2.1455(45)\\
    C-$ud4$-$s$b & 0.012 & 0.030 & 1.7953(17) & 2.1216(42)\\
    C-$ud4$-$s$a & 0.012 & 0.040 & 1.7911(18) & 2.1083(40)\\
    C-$ud5$-$s$b & 0.019 & 0.030 & 1.7693(16) & 2.0532(36)\\
    C-$ud5$-$s$a & 0.019 & 0.040 & 1.7721(17) & 2.0597(34)\\
    \hline
    M-$ud3$-$s$b & 0.0042 & 0.0180 & 2.6639(32) & 3.1849(92)\\
    M-$ud3$-$s$a & 0.0042 & 0.0250 & 2.6516(42) & 3.155(10)\\
    M-$ud4$-$s$b & 0.0080 & 0.0180 & 2.6411(38) & 3.1126(77)\\
    M-$ud4$-$s$a & 0.0080 & 0.0250 & 2.6320(31) & 3.0979(79)\\
    M-$ud5$-$s$b & 0.0120 & 0.0180 & 2.6130(27) & 3.0458(55)\\
    M-$ud5$-$s$a & 0.0120 & 0.0250 & 2.6007(32) & 3.0153(78)\\
    \hline
    F-$ud3$-$s$a & 0.0030 & 0.0150 & 3.3167(55) & 3.938(11)\\
    \hline
  \end{tabular}
  \caption{
    Gradient flow time $t_0^{1/2}/a$ and $w_0/a$ measured on each
    ensemble. 
  }
  \label{tab:wflow}
\end{table}

We list the numerical results for $t_0^{1/2}/a$ in Table~\ref{tab:wflow}.
The lattice data is averaged taking into account autocorrelations
using the jackknife method with a bin size of 140 or 250 HMC
trajectories for the coarse or medium/fine ensembles, respectively.

The flow time $w_0/a$, which is defined by a slope,
$t(d/dt)\{t^2\langle E(t)\rangle\}|_{t=w_0^2}=0.3$~\cite{Borsanyi:2012zs},
is also calculated and the results are summarized in the same table.
The statistical signal is slightly worse for $w_0/a$ and so we only use
$t_0^{1/2}/a$ in the following analysis.

We extrapolate the lattice data to the physical point assuming a
linear dependence on the quark masses, or equivalently on the square of the
pseudoscalar meson mass.
We take the form
\begin{equation}
  \label{eq:wflow_chext}
  \frac{t_0^{1/2}}{a} =
  \left(\frac{t_0^{1/2}}{a}\right)^{\mathrm{phys}}
  \left[
    1 + 
    c_\pi \left( t_0 M_\pi^2 - (t_0 M_\pi^2)^{\mathrm{phys}} \right)
    + c_s
    \left( t_0(2M_K^2-M_\pi^2)-t_0(2M_K^2-M_\pi^2)^{\mathrm{phys}} \right)
  \right],
\end{equation}
where the superscripts ``phys'' denote values at the physical point.
The form of Eq.~\eqref{eq:wflow_chext}, which does not contain chiral
logarithms is justified in~\cite{Bar:2013ora}.
For inputs from experimental data we use 
the pion mass $M_\pi^{\mathrm{phys}} = 134.8\;\mathrm{MeV}$ and the kaon mass
$M_K^{\mathrm{phys}} = 494.2\;\mathrm{MeV}$, which are the recommended values in
the isospin limit~\cite{Aoki:2021kgd,*FlavourLatticeAveragingGroup:2019iem}.
For the reference value of $t_0$, we use the value $t_0^{1/2} =0.1465(21)(13)\;\mathrm{fm}$ from~\cite{Borsanyi:2012zs}
as an input. 
We assume that the slope parameters $c_\pi$ and $c_s$ are common among
different lattice spacings (or $\beta$ values), 
and fit the results of all ensembles in Table~\ref{tab:wflow}
together. 
In this way we can take account of the sea quark mass dependence for
the finest lattice ``F-$ud3$-$s$a'', for which only one value of each sea quark mass is available.

\begin{figure}[!htbp]
  \centering
  \includegraphics[width=12cm]{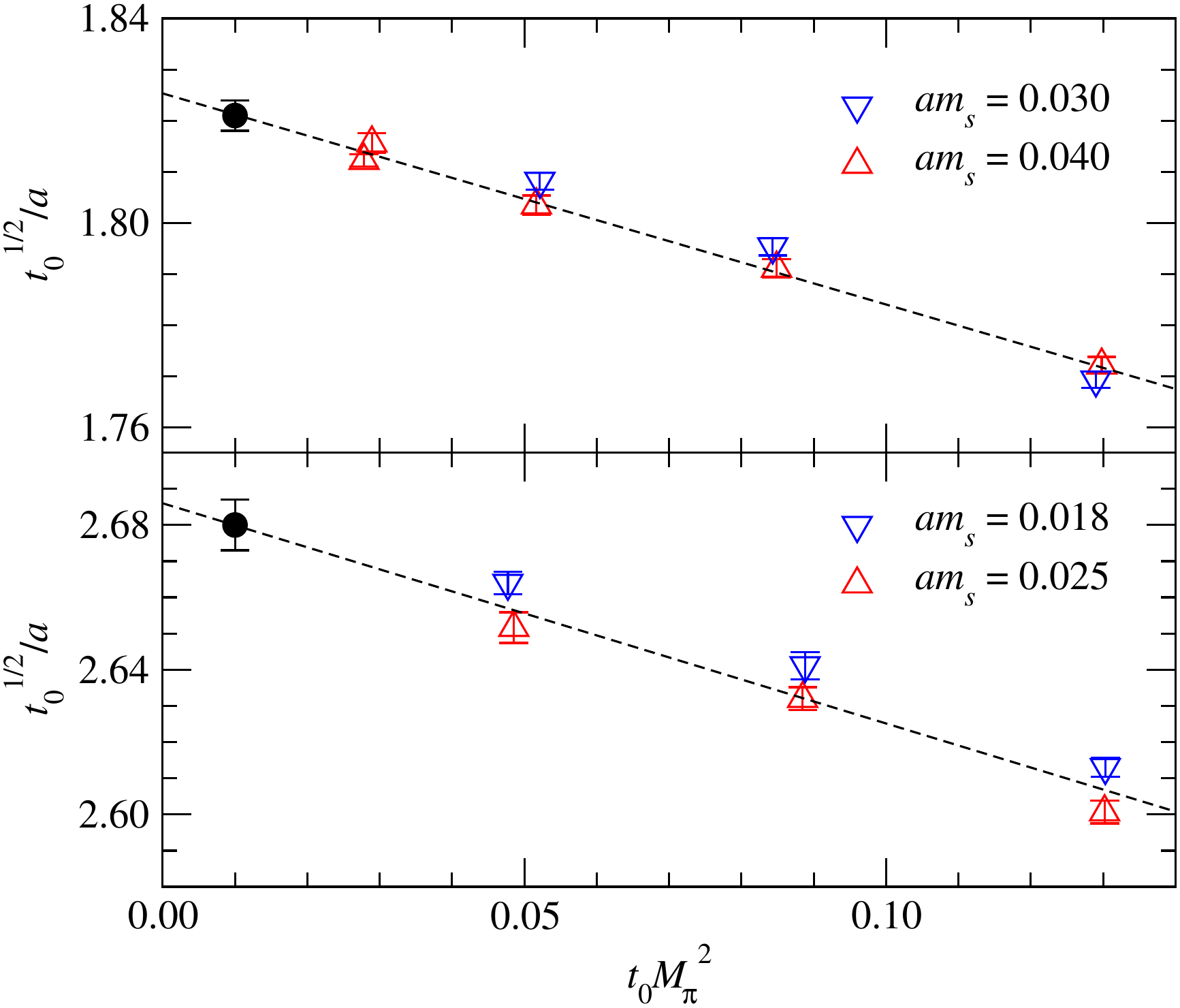}
  \caption{
    Extrapolation of the gradient flow time $t_0^{1/2}/a$ towards the
    physical pion mass (black circles).
    The lattice data are those of two different strange quark masses that sandwich the physical strange mass:
    above physical strange (upward triangles) and below physical strange (downward
    triangles).
    Results on the coarse lattices ($\beta = 4.17$, top panel) and
    medium lattices ($\beta = 4.35$, bottom panel) are shown.
  }
  \label{fig:t0}
\end{figure}

The extrapolation is shown in Fig.~\ref{fig:t0} for the coarse
($\beta$ = 4.17) and the medium ($\beta$ = 4.35) lattices.
The lattice data for two strange quark masses are plotted; 
the extrapolation is represented by a straight line which corresponds to the 
physical strange quark mass.
The numerical results are
$(t_0^{1/2}/a)^{\mathrm{phys}} = 1.821(3)$, $2.680(7)$ and $3.338(7)$ for
the $\beta$ values $4.17$, $4.35$ and $4.47$, respectively.
The corresponding inverse lattice spacings are $2.453(4)$, $3.610(9)$ and
$4.496(9)\;\mathrm{GeV}$, respectively.
The slope parameters are determined as
$c_\pi=-0.23(2)$ and $c_s=-0.03(3)$.

\subsection{Topological charge distribution}
Autocorrelations between gauge configurations generated by
Monte Carlo methods for lattice QCD is a potentially serious
problem. 
In particular, the expectation value of the
$F_{\mu\nu}\tilde{F}_{\mu\nu}$ operator summed over space-time may
have a long autocorrelation time, as it is related to a 
topological quantity of the SU(3) gauge field.
In continuum QCD, the definition of the topological charge $Q$ is such that it becomes an
integer and may not change its value under continuous deformations of
the gauge field.
Its lattice counterpart tends to have the same property as we
approach the continuum limit.
Since the Hybrid Monte Carlo algorithm relies on continuous
evolution of the gauge fields, the topological charge would have
long autocorrelation times when the continuum limit is approached. 
This is a problem because the vacuum of QCD should have a distribution
of the topological charge characterized by the topological
susceptibility $\chi_t=\langle Q^2\rangle/V$, where $V$ is the 4D volume.

\begin{figure}[!htbp]
  \centering
  \includegraphics[width=9cm]{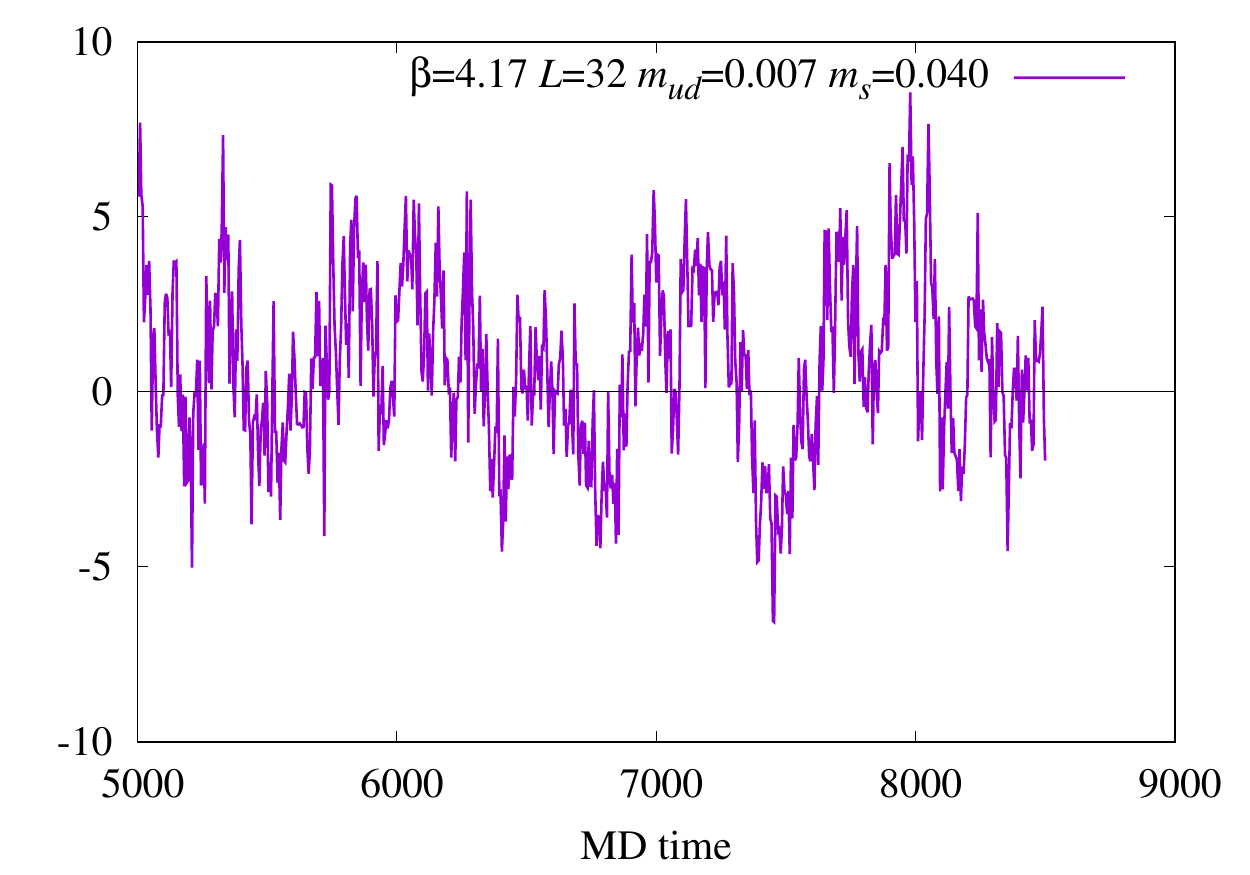}
  \includegraphics[width=9cm]{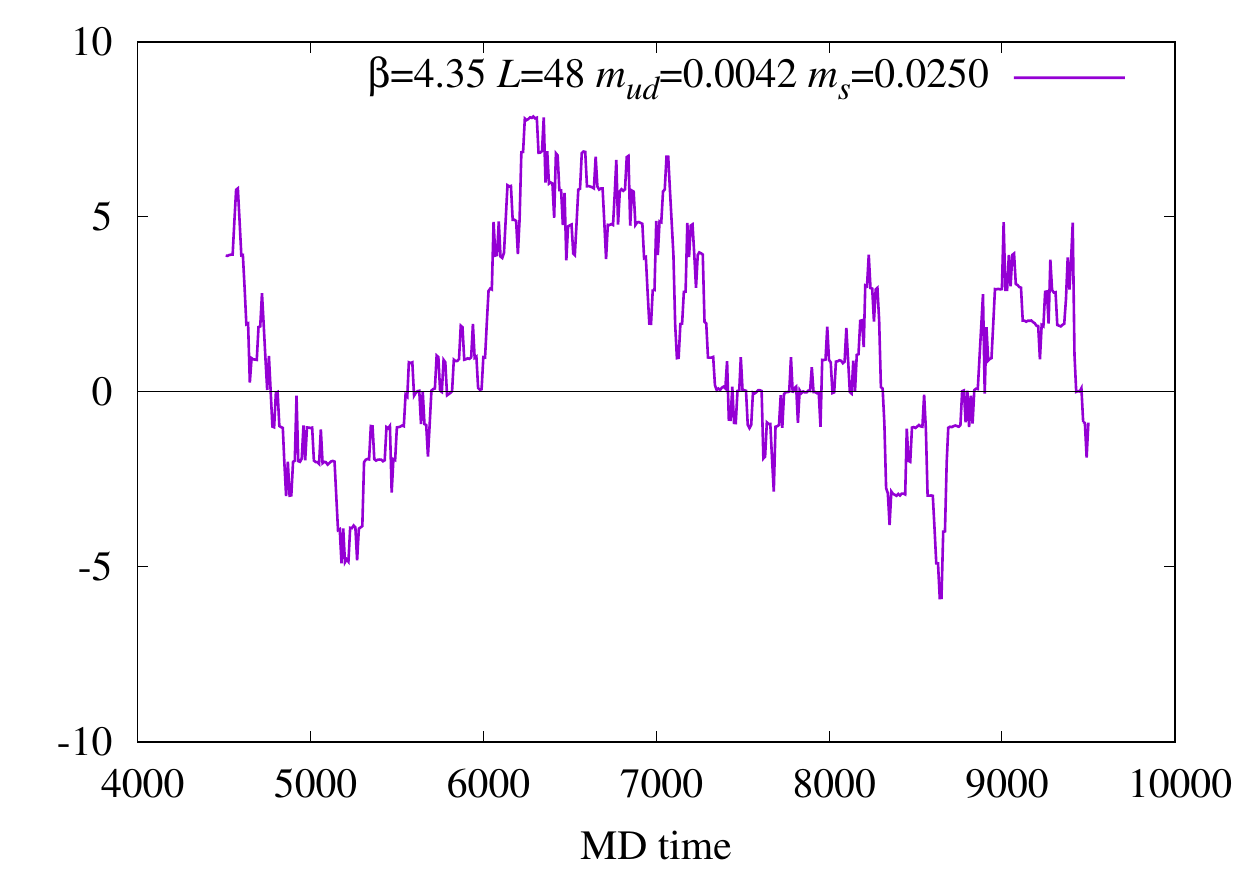}
  \includegraphics[width=9cm]{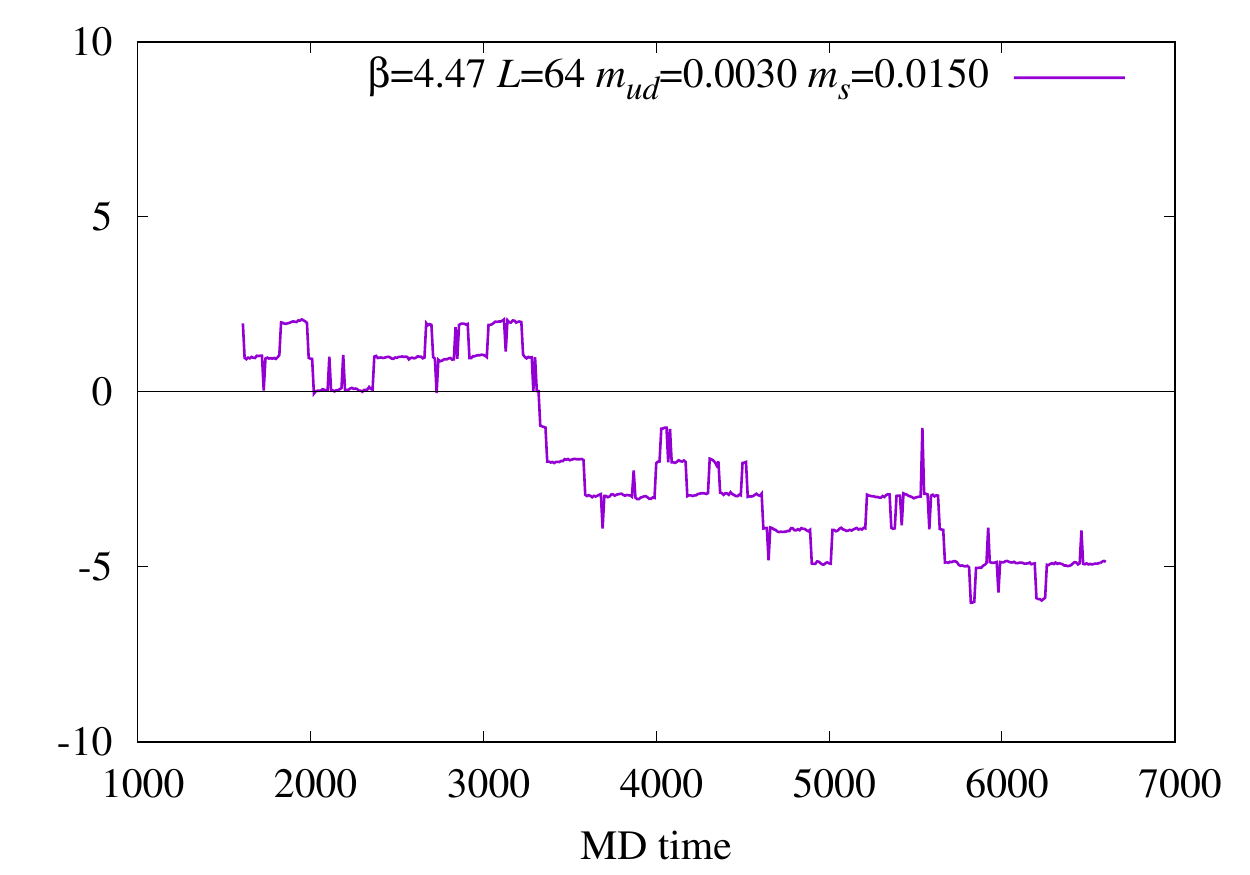}
  \caption{
    Monte Carlo history of the global topological charge $Q$.
    It is monitored on the ensembles C-$ud3$-$s$a (top panel),
    M-$ud3$-$s$a (middle), F-$ud3$-$s$a (bottom).
    The topological charge is calculated on the gauge configurations
    after applying the gradient flow.
  }
  \label{fig:Qhistory}
\end{figure}

\begin{figure}[!htbp]
  \centering
  \includegraphics[width=9cm]{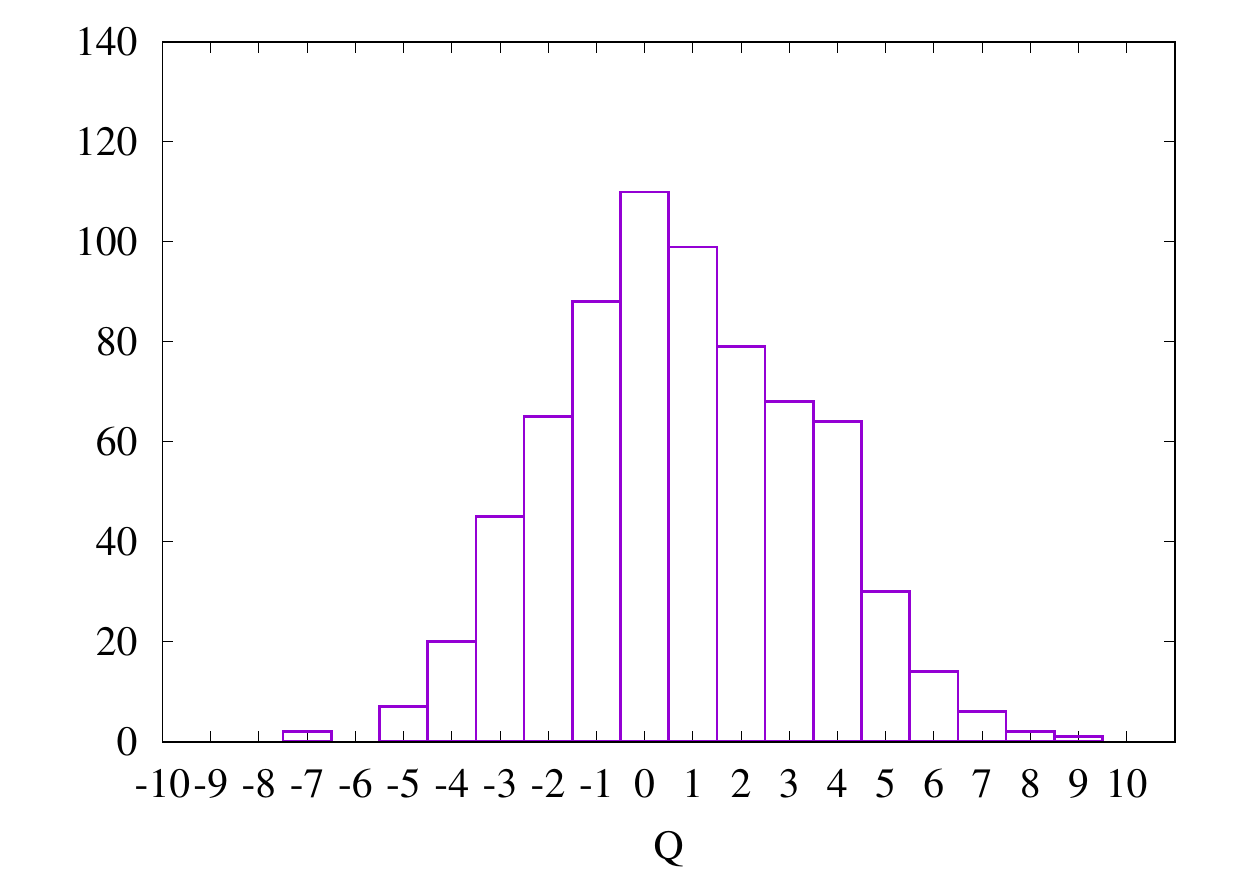}
  \includegraphics[width=9cm]{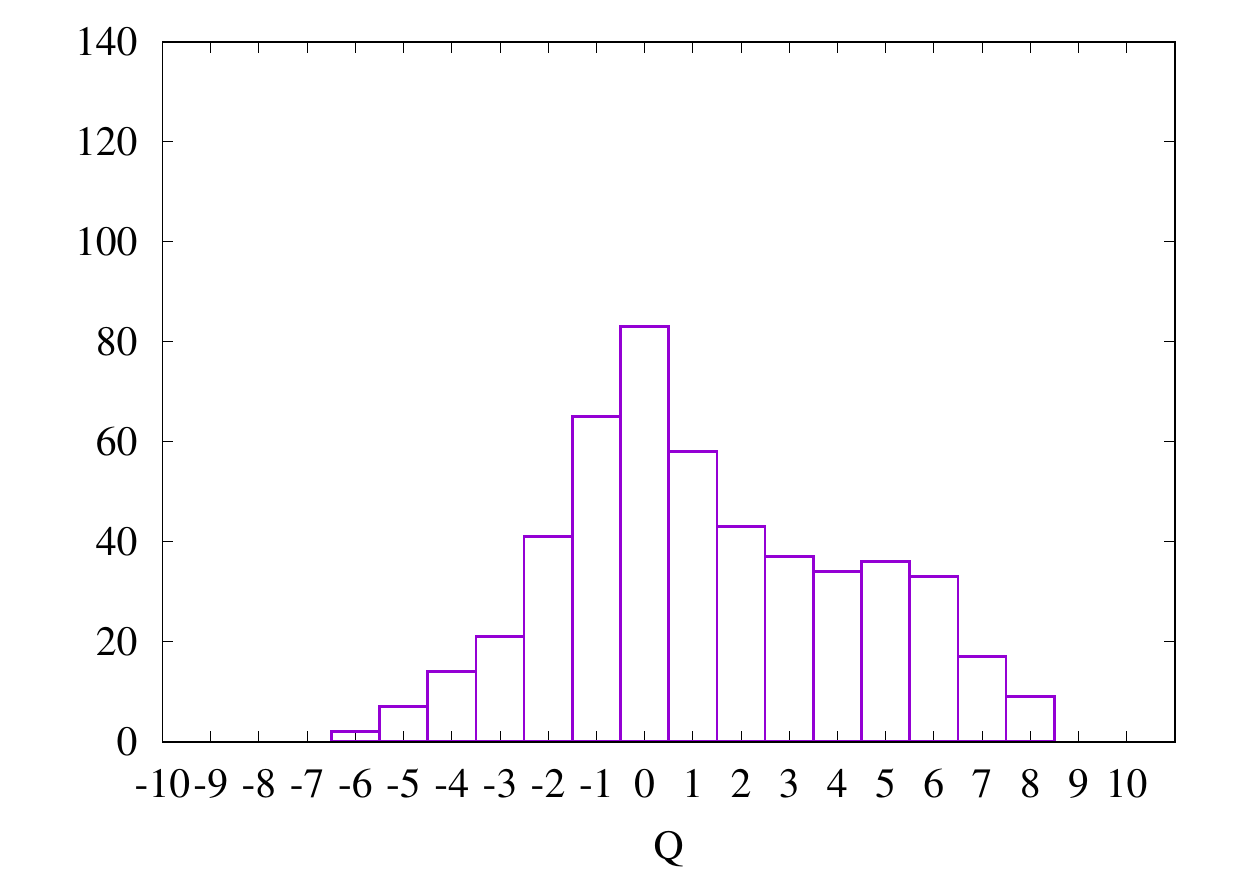}
  \includegraphics[width=9cm]{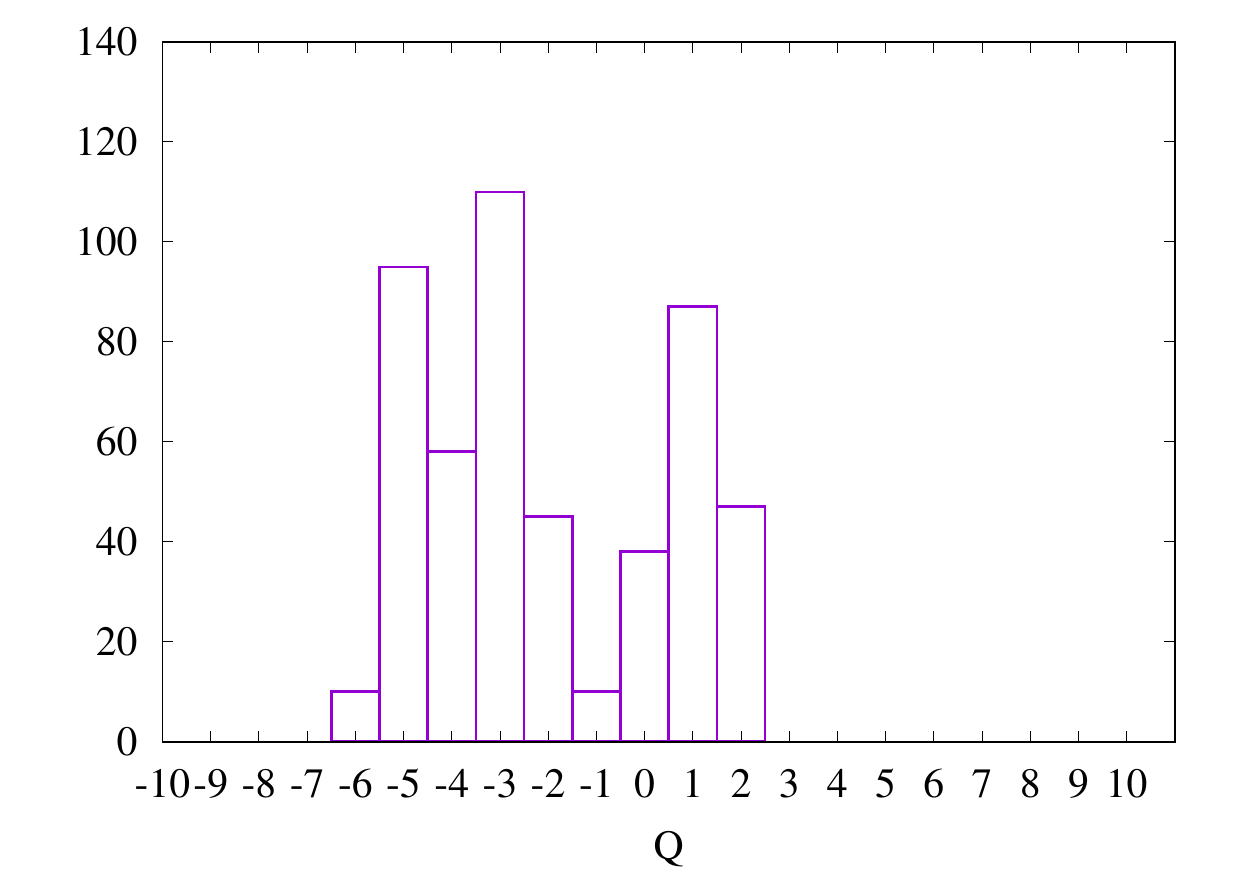}
  \caption{
    Histograms of the global topological charge $Q$.
    It is monitored on the ensembles C-$ud3$-$s$a (top panel),
    M-$ud3$-$s$a (middle), F-$ud3$-$s$a (bottom).
    The topological charge is calculated on the gauge configurations
    after applying the gradient flow.
  }
  \label{fig:Qhistgram}
\end{figure}

Some examples of the Monte Carlo history of the global topological
charge $Q$ are shown in Fig.~\ref{fig:Qhistory}.
We adopt the conventional definition of the topological charge density~\cite{Bruno:2014ova}
constructed from the gauge link variables after
performing the gradient flow as described previously.
After some gradient flow time, the global topological charge
tends to become unchanging, and we take this frozen value for
this study.
These examples are on the ensembles with pion mass around $300\;\mathrm{MeV}$ at
three values of the lattice spacing.
They are therefore expected to show a similar variation of $Q$.
In fact, the range of fluctuations is very similar among the different
lattice spacings, while it is evident that
$Q$ changes much less frequently on finer lattices.
We estimate the autocorrelation time as $14(3)$ and $243(153)$ on the
coarser two lattices at $\beta = 4.17$ and $4.35$, respectively, while
it is $\mathcal{O}(2000)$ or larger on the finest lattice~\cite{Aoki:2017paw}.
Histograms of $Q$ on these ensembles are shown in Fig.~\ref{fig:Qhistgram}. 
The distribution of $Q$ on the
coarsest lattice adheres to a Gaussian-like form, while it is highly distorted on fine lattices.
It is therefore important to address the question of how such a
distorted topological distribution affects the measurements of other
physical quantities.

First, let us consider the worst-case scenario in which the global 
topological charge is frozen at a certain integer value $Q$ on 
consecutive gauge configurations.
Estimates of physical quantities calculated on these configurations
may be biased since the topological charge is not sampled according to
the correct distribution, which should satisfy 
$\langle Q^2\rangle=\chi_tV$.

The bias may be understood in a systematic way as being a series in $1/V$. 
The formula for a CP-even observable $G^{\mathrm{even}}$ is available
from~\cite{Brower:2003yx,Aoki:2007ka} as
\begin{equation}
  \label{eq:G_Q}
  G_Q^{\mathrm{even}} = G(0) + G^{(2)}(0)
  \frac{1}{2\chi_tV} \left[ 1 - \frac{Q^2}{\chi_tV} \right]
  + O(1/V^2),
\end{equation}
where the left-hand side is the measurement in a fixed topological
sector $Q$.
The right-hand side concerns quantities defined in the
$\theta=0$ vacuum:
$G(0)$ is the observable at $\theta=0$, and $G^{(2)}(0)$ is its
second derivative with respect to $\theta$ evaluated at $\theta=0$. 
If we sum over all possible values of $Q$, the second term on the
right-hand side of Eq.~\eqref{eq:G_Q} vanishes and the correct value
$G(0)$ is recovered.

The formula in Eq.~\eqref{eq:G_Q} can be used to estimate the size of the
distortion due to the biased distribution of $Q$, provided that the
derivative $G^{(2)}(0)$ is known.
There is no general rule available for the $\theta$-dependence of
physical quantities, but for pions and kaons one can use an
estimate based on chiral effective theory.
The result for the pion mass $M_\pi$ is 
$G^{(2)}(0)/G(0)=-1/2N_f^2$~\cite{Brower:2003yx}.
We use the estimate for $\chi_t$ from chiral effective theory,
$\chi_t=m_q\Sigma/N_f$~\cite{Leutwyler:1992yt}, which can also be
written as $\chi_t=M_\pi^2 F_\pi^2/2N_f^2$.
We thus obtain
$-(1-Q^2/\chi_t V)\times 1/(2 M_\pi^2 F_\pi^2 V)$.
For our choice of the parameters
($M_\pi\simeq 300\;\mathrm{MeV}$, $F_\pi\simeq 100\;\mathrm{MeV}$, $L\simeq 2.6\;\mathrm{fm}$),
this gives an estimate 
$-(1-Q^2/\chi_tV)\times 0.45\%$ for the worst case.

\begin{figure}[!htbp]
  \centering
  \includegraphics[width=9cm]{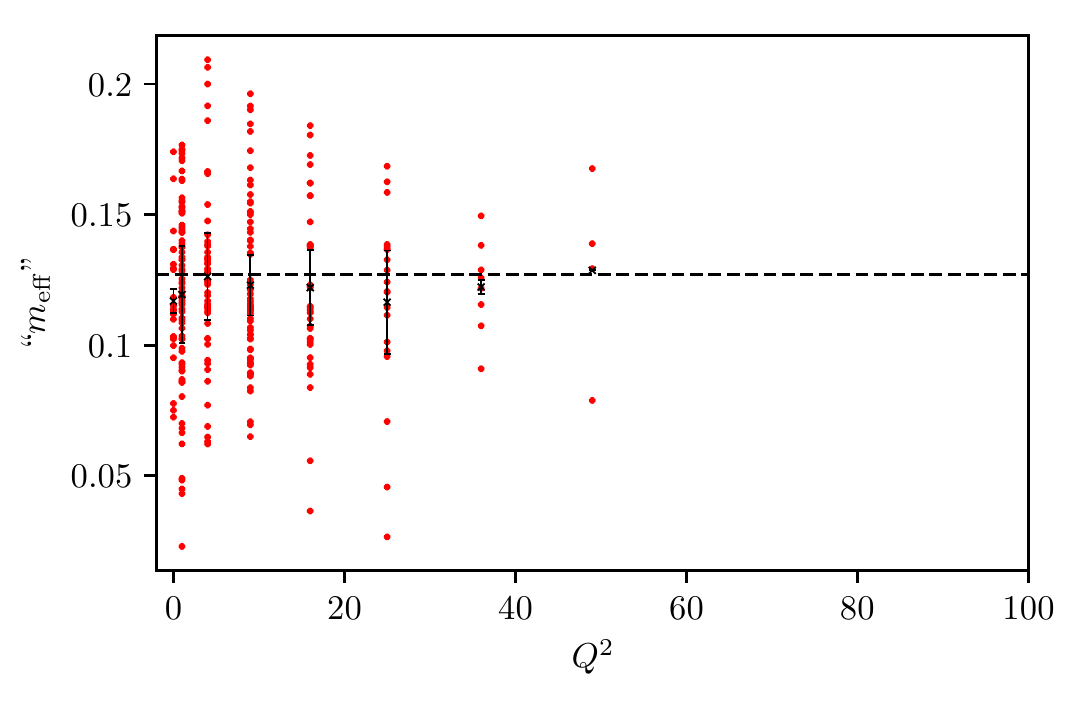}
  \includegraphics[width=9cm]{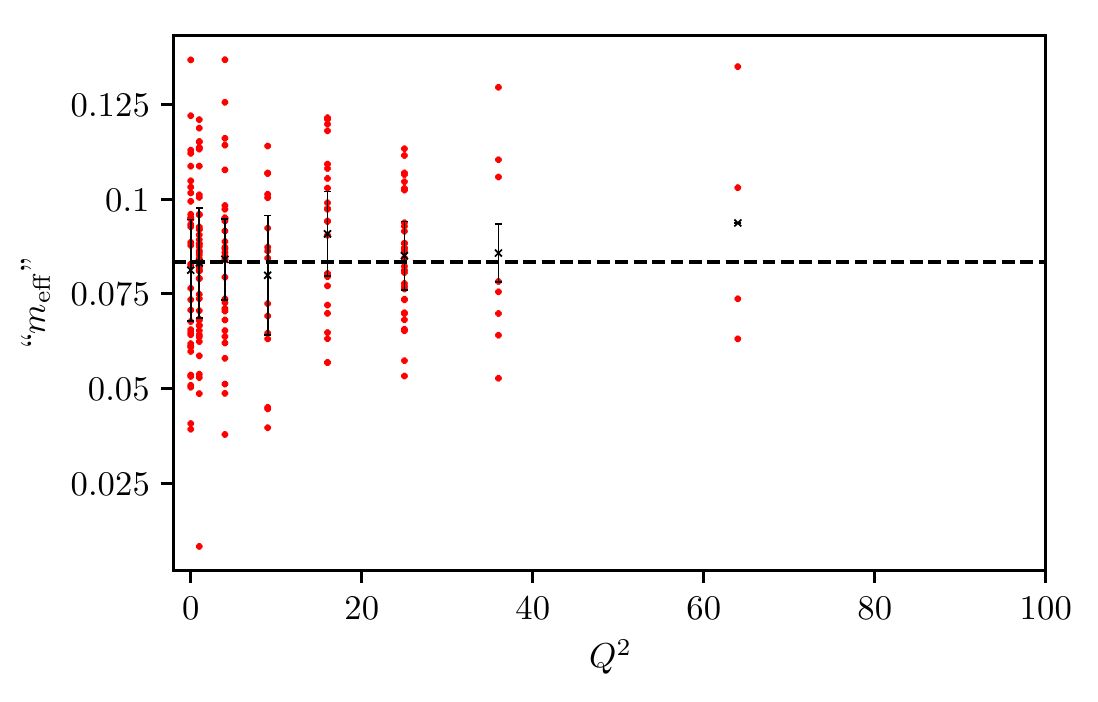}
  \includegraphics[width=9cm]{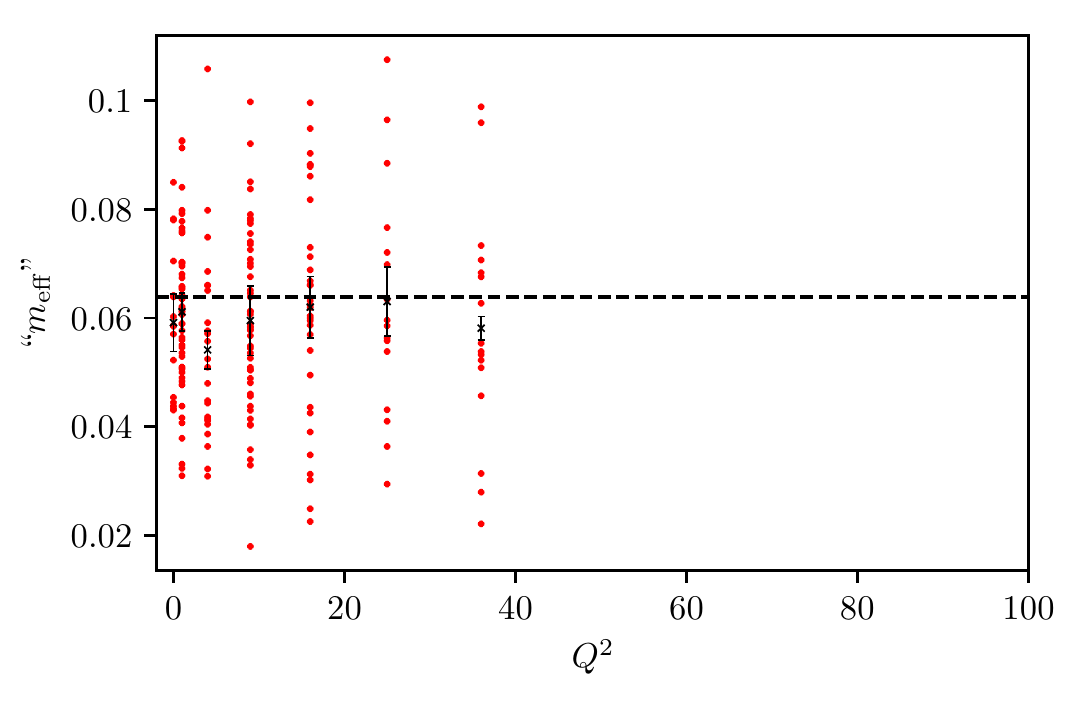}
  \caption{
    Pion ``effective mass'' estimated on each configuration plotted
    as a function of $Q^2$.
    Results on the ensembles C-$ud3$-$s$a (top panel),
    M-$ud3$-$s$a (middle), and F-$ud3$-$s$a (bottom) are shown.
    Red dots represent the calculation of
    $M_{\mathrm{eff}}$ at $t/a = 15$, $20$, and $30$, respectively,
    calculated on each configuration.
    Their averages for individual topological sectors are shown by
    black points with error bars.
    The dashed horizontal line is the actual pion mass calculated on these
    ensembles. 
  }
  \label{fig:meff_vs_Q}
\end{figure}

Using our lattice data, we attempt to confirm the prediction discussed
above. 
For each configuration with a known $Q$, we calculate the pion
``effective mass'' $M_{\mathrm{eff}}$ from a ratio of the
zero-momentum pion correlators at neighboring time separations,
{\it i.e.}  $M_{\mathrm{eff}}=\ln[C(t)/C(t+1)]$.
This does not give the correct pion mass, because the correlators are
not averaged over configurations, but would reflect the effect of the
background gauge field.
In particular, if the pion mass is affected by the topological charge
of the background gauge field, that should be visible in this
quantity.
Results for the ensembles with $M_\pi\simeq$ 300~MeV are shown in
Fig.~\ref{fig:meff_vs_Q} as a function of $Q^2$.
Red dots measured on each configuration do not show any significant
dependence on $Q^2$,
and their averages within individual $Q^2$ values are also independent of
$Q^2$. 
This is consistent with our expectation that the $Q^2$-dependence is
small ($\lesssim 1\%$).

When we average over different $Q$s, the prefactor $(1-Q^2/\chi_tV)$
in the above estimate gives another suppression factor even when the
distribution of $Q$ is slightly biased.
On the finest lattice, F-$ud3$-$s$a, it turns out that this factor is
about $1/4$, so that the actual bias is reduced to $\sim 0.1\%$, which
is below the statistical error ($\sim 0.2\%$) on this lattice.
We may therefore ignore the potential finite volume effect from this
source.
Going to smaller pion masses, this effect would become more
significant due to the suppression factor $1/\chi_tV$ containing 
a term that scales as $1/M_\pi^2$.

The small effect, as estimated above, is not a surprise because local
topological fluctuations are active even when the global charge gets
stuck. 
Some methods to probe such local fluctuations exist, and it turns out
that the topological susceptibility $\chi_t$ measured locally is
consistent with chiral perturbation theory~\cite{Aoki:2017paw}.
Most physical quantities, such as the hadron correlation functions, are
measured in a small subvolume of the whole lattice.
For the pions, the size of the subvolume is $\sim 1/M_\pi^4$, which is 
much smaller than the total volume $2L^4$ by about a factor of 
$1/2(M_\pi L)^4\sim 0.002$.
The correlation functions for heavier hadrons are even more local, and
the estimate above gives a conservative upper limit for the bias.

\section{Light hadron spectra and decay constants}
\label{sec:Light hadron}
Here we describe the measurements of basic physical quantities, {\it i.e.} the masses and decay constants of pions and kaons. Note that this study is separate from the $B\to\pi\ell\nu$ analysis in which we calculate form factors and $|V_{ub}|$. In that case we once again fit all relevant two-point and three-point correlators simultaneously, but get fit results that are consistent with the independent analysis described here. Some of the simulation details, such as the number of time sources used are different between the two analyses.

In this work, the correlators are computed using $Z_2$ noise sources, which are then smeared with a gauge-invariant Gaussian smearing.
The Gaussian smearing is defined by the operator
$(1-(\alpha/N) \Delta)^N$ where
$\Delta$ is the 3D Laplacian and the parameters are
$\alpha= 20.0$ and $N=200$.
On the coarsest lattice, $\beta = 4.17$, the correlators are measured on $100$ configurations, repeated $8$ times with sources distributed on different time slices for each lattice.
On the $\beta = 4.35$ and $\beta = 4.47$ lattices $50$ configurations are
used with sources on $12$ and $8$ time slices respectively. 

Two-point correlation functions $\braket{P^L(x){P^G}^\dagger(0)}$ are fit to
a single exponential function
simultaneously with correlators $\braket{P^G(x){P^G}^\dagger(0)}$,
where $L$ indicates an unsmeared local operator
while $G$ denotes Gaussian smeared operators.
The fit range is determined by setting $t_{\text{max}}$ to half the
lattice temporal extent minus one and $t_{\text{min}}$ is set to the first point in which three consecutive values of
the effective mass are constant within $5\%$ of their standard
deviations. The $5\%$ is chosen after visually inspecting the results
and ensuring that $t_{\text{min}}$ appears to be in the plateau
region.
All fits are then performed again with
$t_{\text{min}}+2$ to check that values remain constant.
These differences in fitted mass are of the order of $0.1\%$ and the
differences in amplitudes $0.5\%$.

Thanks to the good chiral symmetry of our M\"obius domain-wall action,
we compute the pseudoscalar decay constants directly from the pseudoscalar current using
$Z_A\partial_\mu A_\mu=(m_{q_1}+m_{q_2})P$,
where $A_\mu$ is the lattice axial current, $P$ is the pseudoscalar
current and $m_q$ values are the unrenormalized quark masses.
The unrenormalized quark masses include the residual masses:
$m= m_\text{bare} + m_\text{res}$.
We thus obtain the decay constant from
$F_P = (m_{q_1} + m_{q_2})\sqrt{2A_{PP}/M_\pi^3}$,
where $A_{PP}$ denotes the amplitude of the $\langle P^LP^{L\dagger}\rangle$ correlator.

\begin{table}
  \centering
  \begin{tabular}{ccccccc}
    \hline
    ID & $m_{ud}$ & $m_s$ & $M_\pi$ & $F_\pi$ & $M_K$ & $F_K$ \\
    \hline
    C-$ud2$-s$a$ & 0.0035 & 0.040 & 0.09206(13) & 0.0541(11) & 0.21383(15) & 0.06424(36)\\
    C-$ud3$-s$b$ & 0.007  & 0.030 & 0.12627(42) & 0.05792(69) & 0.19798(39) & 0.06419(49)\\
    C-$ud3$-s$a$ & 0.007  & 0.040 & 0.12602(40) & 0.05802(77) & 0.22280(40) & 0.06621(50)\\
    C-$ud4$-s$b$ & 0.012  & 0.030 & 0.16175(28) & 0.06188(60) & 0.21122(29) & 0.06626(49)\\
    C-$ud4$-s$a$ & 0.012  & 0.040 & 0.16265(31) & 0.06271(62) & 0.23538(32) & 0.06907(47)\\
    C-$ud5$-s$b$ & 0.019  & 0.030 & 0.20301(29) & 0.06742(50) & 0.22948(29) & 0.06981(48)\\
    C-$ud5$-s$a$ & 0.019  & 0.040 & 0.20330(29) & 0.06789(55) & 0.25193(29) & 0.07235(48)\\
    \hline
    M-$ud3$-s$b$ & 0.0042 & 0.018 & 0.08200(33) & 0.03916(60) & 0.13123(33) & 0.04321(45)\\
    M-$ud3$-s$a$ & 0.0042 & 0.025 & 0.08308(36) & 0.03944(55) & 0.15142(34) & 0.04512(45)\\
    M-$ud4$-s$b$ & 0.0080 & 0.018 & 0.11283(32) & 0.04280(40) & 0.14288(33) & 0.04539(39)\\
    M-$ud4$-s$a$ & 0.0080 & 0.025 & 0.11301(28) & 0.04294(39) & 0.16101(27) & 0.04716(25)\\
    M-$ud5$-s$b$ & 0.0120 & 0.018 & 0.13813(25) & 0.04612(32) & 0.15414(24) & 0.04754(31)\\
    M-$ud5$-s$a$ & 0.0120 & 0.025 & 0.13875(28) & 0.04631(40) & 0.17186(28) & 0.04930(40)\\
    \hline
    F-$ud3$-s$a$ & 0.0030 & 0.015 & 0.06323(15) & 0.03140(23) & 0.10805(18) & 0.03508(23)\\
    \hline
  \end{tabular}
  \caption{Numerical results for the masses and decay constants of pions and kaons.
    The results are in lattice units.
  }
  \label{tab:mass_and_decayconst}
\end{table}

Numerical results for the masses and decay constants obtained on each ensemble are summarized in Table~\ref{tab:mass_and_decayconst}.

\subsection{Chiral fits}
We perform a global fit of the data for the pion mass $M_\pi$ and decay constant $F_\pi$ using the $SU(2)$ chiral perturbation theory ($\chi$PT) formulae
(see, {\it e.g.}~\cite{Aoki:2021kgd,*FlavourLatticeAveragingGroup:2019iem})
\begin{align}
  \frac{M_\pi^2}{\bar{m}_q } &= 2B\s{1-\frac{1}{2}x \ln\frac{\Lambda_3^2}{M^2} + \frac{17}{8} x^2\p{\ln\frac{\Lambda_M^2}{M^2}}^2 + k_M x^2 +\mathcal{O}\p{x^3} }, \label{mx-expand}\\
  F_\pi &= F\s{1 + x \ln\frac{\Lambda_4^2}{M^2}-\frac{5}{4} x^2\p{\ln\frac{\Lambda_F^2}{M^2} }^2 + k_F x^2 +\mathcal{O}\p{x^3}}.
          \label{fx-expand}
\end{align}
These are written as an expansion in the parameter
$x = M^2/(4\pi F)^2$ where
$M^2 = 2B \bar{m}_q = 2\bar{m}_q \Sigma/F^2$.
The light quark mass $\bar{m}_q$ is the averaged up and down quark mass,
which is $m_{ud}$.
The parameters 
$\Lambda_3$ and $\Lambda_4$ are related to the effective coupling
constants of $\chi$PT through
$\bar{\ell}_n = \ln
{\Lambda_n^2/\p{M^{\text{phys}}_\pi}^2}$. $\Lambda_M$ and $\Lambda_F$
are linear combinations of different
$\Lambda_n$'s:
\begin{align}
  \ln \frac{\Lambda_M^2}{M^2} &= \frac{1}{51}\p{60\ln\frac{\lambda_{12}^2}{M^2} - 9 \ln \frac{\Lambda_3^2}{M^2}  + 49},\\
  \ln \frac{\Lambda_F^2}{M^2} &= \frac{1}{30}\p{30\ln\frac{\lambda_{12}^2}{M^2} +6 \ln \frac{\Lambda_3^2}{M^2} -6 \ln \frac{\Lambda_4^2}{M^2} + 23},
\end{align}
with $\Lambda_{12}$ being the combination
$\Lambda_{12}^2 = \p{7\ln\Lambda_1^2 + 8 \ln \Lambda_2^2}/15$.

The chiral expansions above are fit to the data for $F_\pi$ and
$M_\pi^2/\bar{m}_q$ simultaneously at both next-to-leading order (NLO)
and next-to-next-to-leading order (NNLO).
At NLO only terms up to $\mathcal{O}{\p{x^2}}$ in Eqs.~\eqref{mx-expand} and~\eqref{fx-expand} are included,
leaving the free parameters $F$, $B$, $\Lambda_3$, and $\Lambda_4$.
At NNLO there are additional free parameters $k_M$, and $k_F$, while the values of $\Lambda_{1}$ and $\Lambda_{2}$ are fixed
to the phenomenological values from Ref.~\cite{Colangelo:2001df}.

Our lattices are produced at two bare strange quark masses which are on either side of the physical strange quark mass.
Thus, the extrapolation to the physical point includes a term to interpolate to the strange quark mass,
$\gamma_{s}^{\p{n}}(M_{s\bar{s}} -  M_{s\bar{s}}^{\text{phys}})$,
where $M_{s\bar{s}}= 2M_K^2 - M_\pi^2$.

An alternative method is to express these $\chi$PT
expansions using the parameter
$\xi =M_\pi^2/(4\pi  F_\pi)^2$ as
\begin{align}
            M^2  &= M_{\pi}^2 \s{1+\frac{1}{2} \xi \ln\frac{\Lambda_3^2}{M_\pi^2} - \frac{5}{8} \xi^2\p{\ln\frac{\Omega_M^2}{M_\pi^2}}^2 + c_M \xi^2 +\mathcal{O}\p{\xi^3} },\\
F &= F_\pi \s{1 -  \xi \ln\frac{\Lambda_4^2}{M_\pi^2}-\frac{1}{4} \xi^2\p{\ln\frac{\Omega_F^2}{M_\pi^2} }^2 + c_F \xi^2 +\mathcal{O}\p{\xi^3}},
          \label{eq:xi-expand}
\end{align}
where similarly the values $\Omega_M$ and $\Omega_F$ are the combinations
of other LEC's:
\begin{align}
  \ln \frac{\Omega_M^2}{M_\pi^2} &= \frac{1}{15}\p{60\ln\frac{\lambda_{12}^2}{M_\pi^2} - 33 \ln \frac{\Lambda_3^2}{M_\pi^2} - 12 \ln \frac{\Lambda_4^2}{M_\pi^2} + 52},\\
  \ln \frac{\Omega_F^2}{M_\pi^2} &= \frac{1}{3}\p{-15\ln\frac{\lambda_{12}^2}{M_\pi^2} +18 \ln \frac{\Lambda_4^2}{M_\pi^2} - \frac{29}{2}}.
\end{align}
Reorganizing the expansion in $\xi$, one expects a better convergence of the chiral expansion~\cite{JLQCD:2008zxm}.

To perform these fits we simultaneously fit the expressions for
$F_\pi$ and $M_\pi^2/\bar{m_q}$.
In order to get reasonable $\chi^2$ values for NLO fits
we include only those ensembles with pion masses below $450$ MeV, while for NNLO all 14 ensembles are included.
To account for the strange-quark mass dependence the fit function is interpolated to the physical value using $M_{s\bar{s}}^2 = 2M_K^2 - M_\pi^2$.
We therefore multiply each global fit ansatz by a prefactor
$\Delta^{\p{n}} = (1+ \gamma_{a}^{\p{n} }a^2 + \gamma_{s}^{\p{n}}(M^2_{s\bar{s}} -
  {M^2_{s\bar{s}}}^{\text{phys}}) )$,
with $\gamma_a^{(1)}$ and $\gamma_a^{(2)}$ fit parameters for the $a^2$
dependence of $M_\pi$ and $F_\pi$, respectively, and similarly for
$\gamma_s^{(1)}$ and $\gamma_s^{(2)}$ for the strange quark dependence.

\begin{figure}[!htbp]
  \centering
  \includegraphics[width=0.55\textwidth]{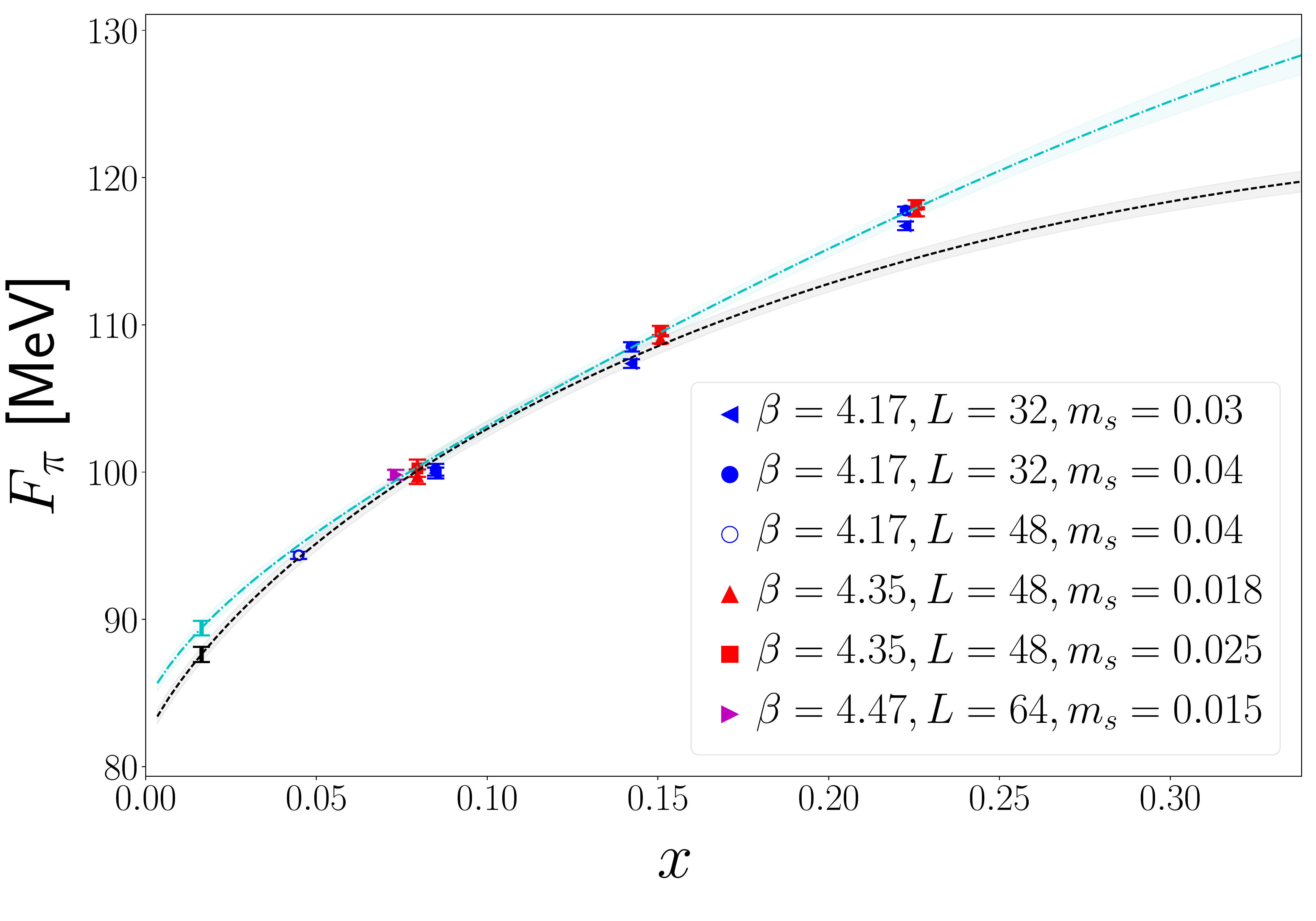}\\
  \includegraphics[width=0.55\textwidth]{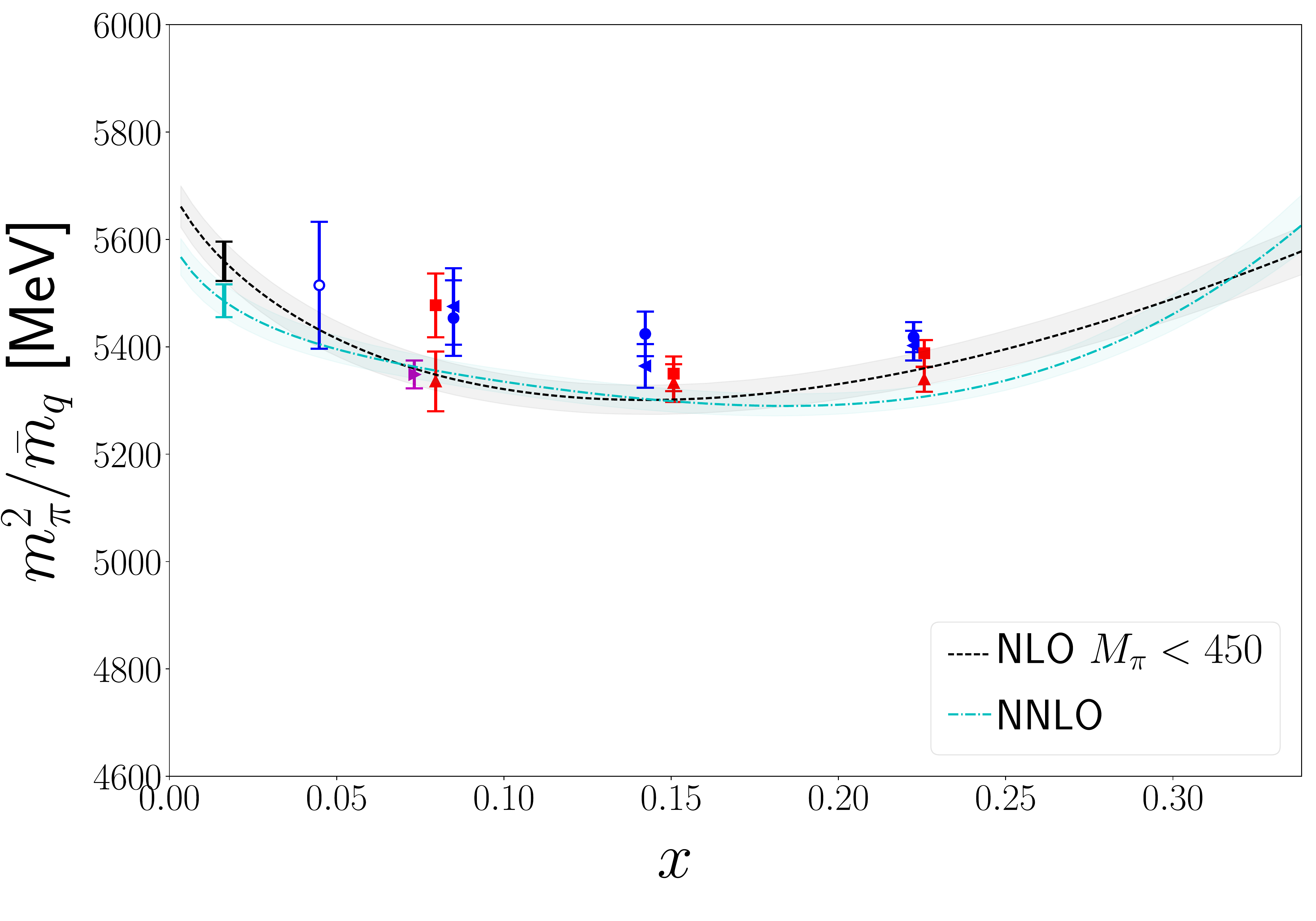}
  \caption{Plots of $M_\pi^2/\bar{m}_q $ (left panel) and $F_\pi$
    (right panel), using the expansion in $x=2\bar{m}_q B/(4\pi
    F)^2$. Fit lines show the best NLO (blue) and NNLO (green) fits in
    the continuum limit and interpolated to physical strange quark mass. The NLO fits
    only include the ensembles for $M_\pi < 450\;\mathrm{MeV}$.}
  \label{fig:Fpi_x}
\end{figure}

\begin{table}[ptb]
  \centering
  \begin{tabular}{|r|r|r|r|r|r|}\hline
                              & NLO       & NLO       & NLO              & NNLO      & NNLO      \\ \hline \hline
    $M_\pi$ Max         & 350  MeV     & 450 MeV      & -                & 450 MeV      & -         \\ \hline
    $\chi^2$/dof              & 1.22      & 1.81      & 5.4 & 1.24      & 0.96      \\ \hline    \hline
    $B$  [MeV]                     & 2789(38)  & 2856(20)  & 2884(15)         & 2793(39)  & 2808(17)  \\ \hline
    $F$  [MeV]                     & 84.12(92) & 81.70(50) & 79.70(39)        & 84.1(1.1) & 84.04(56) \\ \hline
    $\Lambda_3$ [MeV]              & 508(69)   & 637(18)   & 688(10)          & 663(159)  & 711(42)   \\ \hline
    $\Lambda_4$ [MeV]              & 1040(53)  & 1190(15)  & 1283(7.9)        & 1028(115) & 1001(37)  \\ \hline
    $\gamma^{(1)}$                & 3.8(1.5)  & 2.79(88)  & 3.01(65)         & 3.04(88)  & 2.83(65)  \\ \hline
    $\gamma^{(2)}$                & -2.67(89) & -2.14(58) & -1.79(44)        & -2.11(58) & -1.65(45) \\ \hline
    $\gamma_s^{(1)}\times 10^{8}$ & 6.8(5.2)  & 5.3(2.4)  & 3.4(1.7)         & 7.0(2.6)  & 5.2(1.7)  \\ \hline
    $\gamma_s^{(2)}\times 10^{8}$ & 2.1(3.2)  & 5.7(1.7)  & 7.4(1.3)         & 4.5(1.8)  & 4.4(1.3)  \\ \hline
    $k_F$                     & -         & -         & -                & 1.6(1.2)  & 1.81(30)  \\ \hline
    $k_M$                     & -         & -         & -                & -0.2(1.3) & 0.38(20)  \\ \hline
  \end{tabular}
  \caption{Fits to the $x$ expansion at NLO and NNLO. The error on each
    fit parameter comes from the statistical uncertainty.
    \label{tab:fitx}}
\end{table}

\begin{table}[ptb]
  \centering
  \begin{tabular}{|r|r|r|r|r|r|r|}\hline
                              & $M^2$ NLO & $F$ NLO   & $M^2$ NLO & $F$ NLO   & NNLO       & NNLO       \\ \hline \hline
    $M_\pi $ Max         & 350 MeV   & 350 MeV   & 450 MeV   & 450 MeV   & 450 MeV    & -          \\ \hline
    $\chi^2$/dof              & 3.75      & 2.33      & 1.49      & 6.2       & 2.28       & 2.17       \\ \hline    \hline
    $B$   [MeV]                    & 2776(40)  &           & 2836(21)  &           & 2764(54)   & 2817(23)   \\ \hline
    $F$   [MeV]                    &           & 84.2(1.1) &           & 81.35(62) & 86.5(1.7)  & 85.85(95)  \\ \hline
    $\Lambda_3$ [MeV]              & 631(173)  &           & 902(71)   &           & 499(240)   & 698(89)    \\ \hline
    $\Lambda_4$ [MeV]              &           & 1380(134) &           & 1853(55)  & 727(173)   & 765(70)    \\ \hline
    $\gamma^{(1)}$                & 4.2(1.5)  &           & 3.17(94)  &           & 3.40(96)   & 3.01(69)   \\ \hline
    $\gamma^{(2)}$                &           & -4.3(1.4) &           & -3.9(1.0) & -4.0(1.0)  & -3.73(92)  \\ \hline
    $\gamma_s^{(1)}\times 10^{8}$ & 6.9(5.3)  &           & 5.7(2.6)  &           & 6.8(2.9)   & 4.5(1.8)   \\ \hline
    $\gamma_s^{(2)}\times 10^{8}$ &           & 2.2(5.0)  &           & 8.2(3.1)  & 6.5(3.5)   & 7.7(2.8)   \\ \hline
    $c_F$                     &           &           &           &           & -12.8(7.3) & -12.1(2.7) \\ \hline
    $c_M$                     &           &           &           &           & 9.3(7.7)   & 2.27(46)   \\ \hline
  \end{tabular}
  \caption{Fits to the $\xi$ expansion at NLO and NNLO. At NLO the
    parameters for $F_\pi$ and $M^2$ decouple and can be fit
    independently. Even when fitting to only ensembles with pion
    masses below $450\;\mathrm{MeV}$ the fit for $F$ at NLO is still poor. So for
    the $\xi$ expansion especially, only the NNLO fit should be valid
    for our data. The uncertainty listed is again 
    statistical.\label{tab:fitxi}}
\end{table}

The results of the NLO and NNLO fits alongside our lattice data are shown in
Fig.~\ref{fig:Fpi_x}. Details of the fit parameters and their
statistical uncertainties are given in Table~\ref{tab:fitx}. The blue
line shows the NLO fit and the black line shows the NNLO fit. Bands of the same color represent a statistical standard deviation from the global fit.
Each
line has a point with an error bar drawn at the physical light quark mass
representing the value and statistical error of the global fit.
The resulting values for $F_{\pi}$ are 
\begin{align}
  F_{\pi}^{x-\text{NLO}} &= 87.6 \pm 0.5^\text{stat} \pm 1.1^\text{scale fit} \pm  1.477^\text{scale} \text{ MeV},\\
  F_{\pi}^{x-\text{NNLO}} &= 89.4 \pm 0.5^\text{stat} \pm 0.9^\text{scale fit} \pm 1.5^\text{scale} \text{ MeV},
\end{align}
where the first error is the statistical error from the fit.
The second error is the uncertainty estimated from how the fit changes by
varying the scale by $1\sigma$, and the third error is the systematic
error directly from the scale setting. 
Similarly, estimates for $F_\pi$ from the $\xi$ parametrization are
\begin{align}
  F_{\pi}^{\xi-\text{NLO}} &= 87.6 \pm 0.6^\text{stat} \pm 1.3^\text{scale fit} \pm  1.5^\text{scale} \text{ MeV},\\
  F_{\pi}^{\xi-\text{NNLO}} &= 90.4 \pm 0.8^\text{stat} \pm 1.1^\text{scale fit} \pm 1.5^\text{scale} \text{ MeV}.
\end{align}
In both cases the NLO fits required that we limited the data to
ensembles with lower pion masses and they still have poor $\chi^2$ per degree of
freedom compared to the NNLO fits.

It is also useful to look at the ratio of the physical pion decay
constant $F_{\pi}$ and the value of the $SU(2)$ constant $F$ in the
chiral limit. This ratio should mitigate systematic effects such as the scale
setting uncertainty. We get
$F_\pi/F$ = 1.0635(13)(10) from the $x$-fit at NNLO and
1.0532(28)(16) from the $\xi$ fit at NNLO.

The values of the $SU(2)$ low energy constants are also listed in Tables~\ref{tab:fitx} and~\ref{tab:fitxi}.
The values of $\Lambda_3$ and $\Lambda_4$ can be expressed as the low energy
constants to $\ell_3$ and $\ell_4$. The values of the NLO low-energy
constants from both expansions and fits are shown in
Table~\ref{tab:lecs}.
Estimates of the uncertainties in the fit due to varying the scale setting
and from the uncertainty of the values directly from the scale in the case
of dimensionful constants are given.
Due to the reasons above, we prefer the values of the
NNLO fits.

\begin{table}[!htbp]
  \centering
  \begin{tabular}{|l|c|c|c|c|}
    \hline
    & mean    & stat err & scale fit & scale err \\ \hline
    \hline
    $\ell_3$ $x$ NLO          & 3.105 & 0.058  & 0.082    & -         \\ \hline
    $\ell_3$ $x$ NNLO         & 3.324 & 0.117  & 0.058    & -         \\ \hline
    $\ell_3$ $\xi$ NLO        & 3.795 & 0.157  & 0.156    & -         \\ \hline
    $\ell_3$ $\xi$ NNLO       & 3.273 & 0.254  & 0.096    & -         \\ \hline
    \hline
    $\ell_4$ $x$ NLO          & 4.356 & 0.025  & 0.024    & -         \\ \hline
    $\ell_4$ $x$ NNLO         & 4.009 & 0.074  & 0.042    & -         \\ \hline
    $\ell_4$ $\xi$ NLO        & 5.241 & 0.060  & 0.076    & -         \\ \hline
    $\ell_4$ $\xi$ NNLO       & 3.465 & 0.182  & 0.101    & -         \\ \hline
    \hline
    $\Sigma^{1/3}$ $x$ NLO    & 267.1 & 0.8  & 3.5    & 3.8     \\ \hline
    $\Sigma^{1/3}$ $x$ NNLO   & 270.7 & 1.1  & 2.8    & 3.9     \\ \hline
    $\Sigma^{1/3}$ $\xi$ NLO  & 265.7 & 1.1  & 4.0    & 3.8     \\ \hline
    $\Sigma^{1/3}$ $\xi$ NNLO & 274.8 & 1.7  & 3.3    & 3.9     \\ \hline
    \hline
    F $x$ NLO                 & 81.7 & 0.5  & 1.1    & 1.2     \\ \hline
    F $x$ NNLO                & 84.0 & 0.6  & 0.9    & 1.2     \\ \hline
    F $\xi$ NLO               & 81.3 & 0.6  & 1.3    & 1.2     \\ \hline
    F $\xi$ NNLO              & 85.9 & 0.9  & 1.2    & 1.2     \\ \hline
  \end{tabular}
  \caption{LECs computed from the fits.}
  \label{tab:lecs}
\end{table}

\subsection{Kaon Decay Constant}

\begin{figure}[!htbp]
  \centering
  \includegraphics[width=0.55\textwidth]{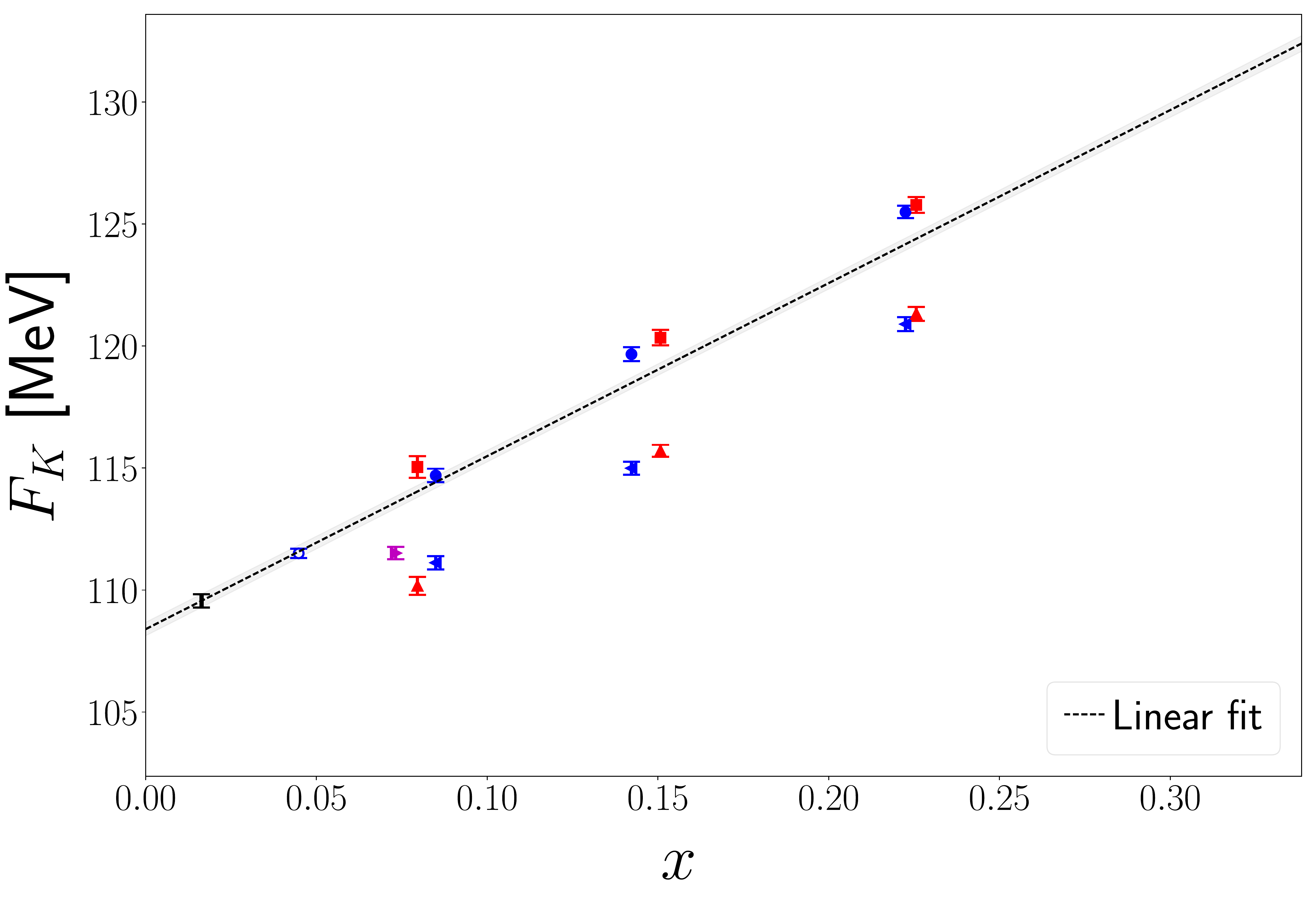}
  \includegraphics[width=0.55\textwidth]{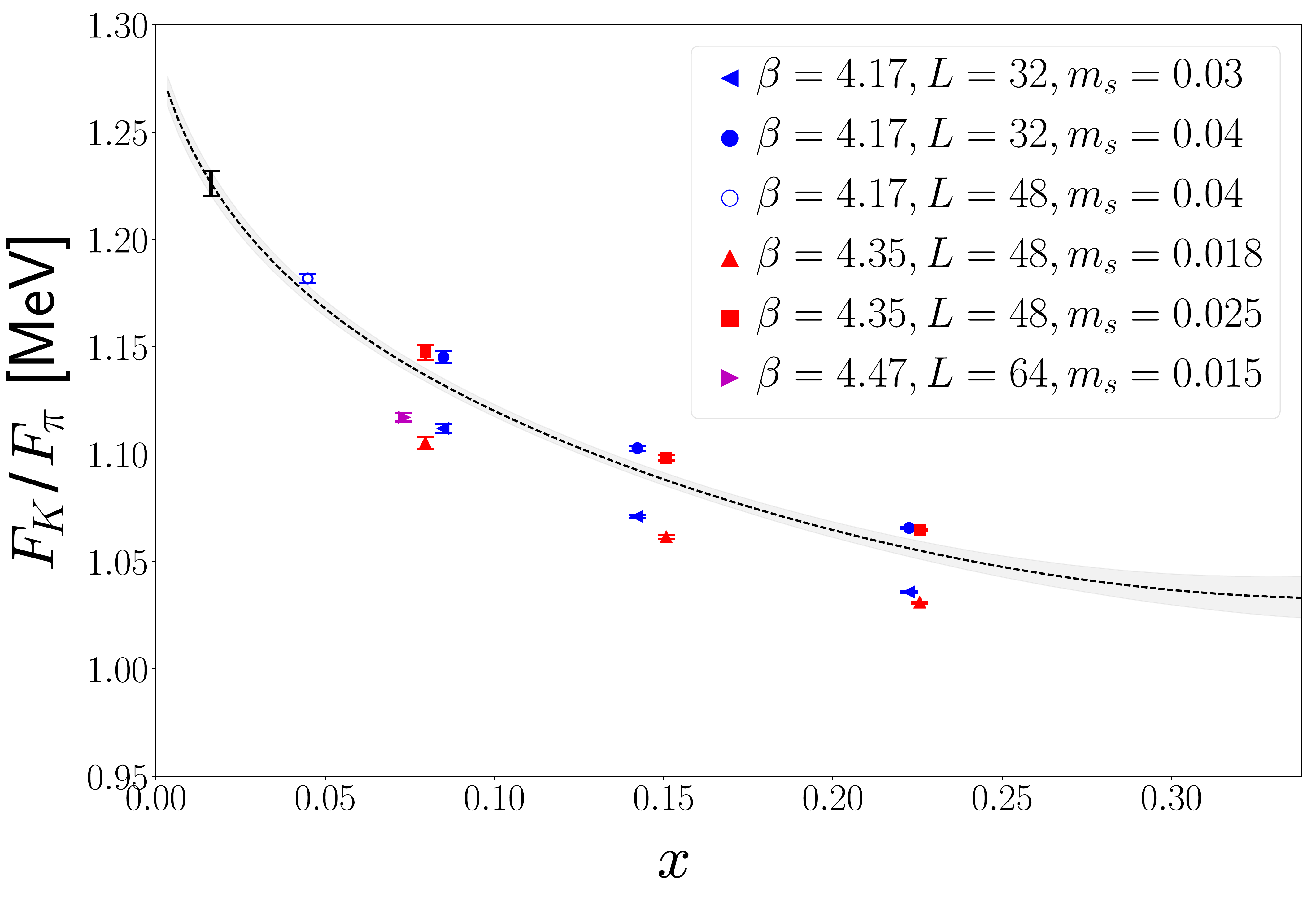}
  \caption{$F_K$ (top) and $F_K/F_\pi$ (bottom) with linear
    extrapolations to the physical pion mass point. The fit to $F_k$
    is a simple linear extrapolation in the chiral $x$ expansion
    parameter to the chiral limit. The extrapolation in $F_K/F_\pi$ uses
    a linear extrapolation for $F_K$ over the NNLO fit for
    the $x$ expansion for $F_\pi$.}
  \label{fig:FK}
\end{figure}

We also compute the kaon decay constant $F_K$.
Here the results appear incredibly linear in $M_\pi$
in the region covered by our lattice data.
We tested a fit ansatz that includes the $SU(3)$ chiral logarithms, but we have insufficient data at small pion masses to obtain a reliable fit. So for the kaon we
simply perform a linear fit in $m_{ud}$ using $x = M^2/(4\pi F)^2$,
fixing $M$ and $F$ to the value obtained in the NNLO fit (see Fig.~\ref{fig:FK}). Evaluating this extrapolation at the physical point for
$m_{ud}$ gives
\begin{align}
  \label{eq:fk}
  F_K &= 154.96 \pm 0.37^\text{stat} \pm 1.13^\text{scale fit} \pm 2.61^\text{scale}.
\end{align}
To extract $F_K/F_\pi$ we use a linear extrapolation for $F_K$ and the NNLO $x$ expansion for $F_\pi$. Evaluating this at the physical point gives
\begin{align}
  \label{eq:fkbyfpi}
  F_K/F_\pi = 1.2260 \pm 0.0057^\text{stat}  \pm 0.0022^\text{scale fit}.
\end{align}
This ratio may be a better quantity as it does not suffer from a
strong uncertainty due to the scaling, although it is a biased comparison since we
use only a linear extrapolation for $F_K$ and a full NNLO fit for
$F_\pi$.
\bibliographystyle{apsrev4-2}
\bibliography{supplementary}